\documentclass[10pt,letterpaper]{article}
\usepackage[top=0.85in,left=2.75in,footskip=0.75in]{geometry}

\usepackage{amsmath,amssymb}

\usepackage{changepage}
\usepackage{algorithm}
\usepackage{algorithmic}
\usepackage[utf8x]{inputenc}
\usepackage[spaces,hyphens]{url}
\usepackage{textcomp,marvosym}

\usepackage{cite}
\usepackage{amsmath,amssymb,amsfonts,amsthm,bbm,mathrsfs,verbatim,bm} % Typical maths resource packages
 \usepackage{graphicx}
\usepackage{nameref,hyperref}
\usepackage[right]{lineno}

\usepackage{microtype}
\DisableLigatures[f]{encoding = *, family = * }

\usepackage[table]{xcolor}

\usepackage{array}

\newcolumntype{+}{!{\vrule width 2pt}}

\newlength\savedwidth

\newcommand\thickhline{\noalign{\global\savedwidth\arrayrulewidth\global\arrayrulewidth 2pt}%
\hline
\noalign{\global\arrayrulewidth\savedwidth}}

\raggedright
\usepackage[aboveskip=1pt,labelfont=bf,labelsep=period,justification=raggedright,singlelinecheck=off]{caption}
\usepackage{subcaption}
 
\makeatletter
\renewcommand{\@biblabel}[1]{\quad#1.}
\makeatother

\usepackage{lastpage,fancyhdr,graphicx}
\usepackage{epstopdf}
\usepackage[numbers, square, comma, sort&compress]{natbib}
\fancyheadoffset[L]{2.25in}
\fancyfootoffset[L]{2.25in}
\begin{document}
\vspace*{0.2in}

\begin{flushleft}
{\Large
\textbf\newline{ Age-stratified epidemic model using a latent marked Hawkes process} % Please use "sentence case" for title and headings (capitalize only the first word in a title (or heading), the first word in a subtitle (or subheading), and any proper nouns).
}
\newline
\\
Stamatina Lamprinakou\textsuperscript{1*},
Axel Gandy\textsuperscript{1*}
\\
\bigskip
\textbf{1} Department of Mathematics, Imperial College London, London,  United Kingdom
\\
\bigskip

% Insert additional author notes using the symbols described below. Insert symbol callouts after author names as necessary.
% 
% Remove or comment out the author notes below if they aren't used.
%
% Primary Equal Contribution Note
%\Yinyang These authors contributed equally to this work.

% Additional Equal Contribution Note
% Also use this double-dagger symbol for special authorship notes, such as senior authorship.
%\ddag These authors also contributed equally to this work.

%\textcurrency Current Address: Dept/Program/Center, Institution Name, City, State, Country % change symbol to "\textcurrency a" if more than one current address note
% \textcurrency b Insert second current address 
% \textcurrency c Insert third current address


% Group/Consortium Author Note
%\textpilcrow Membership list can be found in the Acknowledgments section.

% Use the asterisk to denote corresponding authorship and provide email address in note below.
* s.lamprinakou18@imperial.ac.uk, a.gandy@imperial.ac.uk

\end{flushleft}
\section*{Abstract}
We extend the unstructured homogeneously mixing epidemic model introduced by Lamprinakou et al.~\cite{https://doi.org/10.48550/arxiv.2208.07340} considering a finite population stratified by age bands. We model the actual unobserved infections using a latent marked Hawkes process and the reported aggregated infections as random quantities driven by the underlying Hawkes process. We apply a Kernel Density Particle Filter (KDPF) to infer the marked counting process, the instantaneous reproduction number for each age group and forecast the epidemic's future trajectory in the near future; considering the age bands and the population size does not increase the computational effort. We demonstrate the performance of the proposed inference algorithm on synthetic data sets and COVID-19 reported cases in various local authorities in the UK. We illustrate that taking into account the individual heterogeneity in age decreases the uncertainty of estimates and provides a real-time measurement of interventions and behavioural changes.

% Please keep the Author Summary between 150 and 200 words
% Use first person. PLOS ONE authors please skip this step. 

%\linenumbers

\section{Introduction}
Modelling the spread of an infectious disease must take into account the mechanism of its transmission, the individual heterogeneities, and the nature and duration of interactions among the population~\cite{isham1996models,wallinga1999perspective,farrington2001estimation}. 

Diseases are often spread via social contacts. This induces that the rate at which a disease is spread is dependent on the number of contacts between infectious and susceptibles. Empirical studies (e.g.~\cite{mossong2008social, leung2017social, de2018characteristics, beraud2015french})  quantified via matrices, known as contact matrices, the contact patterns relevant for infections transmitted by the respiratory or close-contact route in several countries.

Individuals vary in their tendency to interact with others; personal hygiene is a key factor in the propagation of diseases; individuals' community structure and location might be significant in spreading epidemics. The simplest assumption of individual heterogeneity is to consider that contact rates vary with only one characteristic of an individual, such as age. Age as behavioural and physiological factor is highly correlated to the risk of infection in many diseases like influenza-like diseases~\cite{eames2012measured, apolloni2013age, luca2018impact, worby2015relative}, pertussis~\cite{rohani2010contact}, tuberculosis~\cite{arregui2018data, guzzetta2011modeling}, varicella~\cite{marangi2017natural} and COVID-19~\citep{davies2020age}. Understanding the impact of age on the transmission of disease is critical for determining and implementing social-distancing and interventions, especially closing schools~\cite{ cauchemez2008estimating, hens2009estimating, davies2020age}.

Wallinga et al.~\cite{wallinga2006using} showed that school-aged children and young adults are more likely to get infected and contribute most to the spread of infection due to their high number of contacts. The 65 study participants of Edmunds et al.~\cite{edmunds1997mixes} showed that the mean age of contacts increased with the age of participants, and older participants ($\geq 40$ years) had more contacts with older adults and a larger variability in the age of their contacts than younger participants ($<40$ years).

Balabdaoui and Mohr~\cite{balabdaoui2020age} have proposed a compartmental model to capture the dynamic of highly age-sensitive epidemics and evaluate the effect of social contact patterns on a load of hospitals and their intensive care units. Stocks et al.~\cite{stocks2020model} have introduced a model selection process using transmission models subdividing the population into age classes. Pellis et al.~\cite{pellis2020systematic} have also suggested a mathematical approach to select between age and household structure in designing a model for an initial, rapid assessment of potential epidemic severity.

Lamprinakou et al.~\cite{https://doi.org/10.48550/arxiv.2208.07340} have introduced a novel epidemic model using a latent Hawkes process with temporal covariates for the infections and a probability distribution with a mean driven by the underlying Hawkes process for the reported infection cases. A Kernel Density Particle Filter (KDPF) \citep{sheinson2014comparison, liu2001combined} is proposed for inference of both latent cases and instantaneous reproduction number and for predicting the new infections over short time horizons. Modelling the infections via a Hawkes process allows us to estimate by whom an infected individual was infected~\cite{bertozzi2020challenges}.

The epidemic model proposed by Lamprinakou et al.~\cite{https://doi.org/10.48550/arxiv.2208.07340} can be viewed as a new approach in deriving epidemic models that consider individual heterogeneities and provide insight into underlying dynamics. Here, we extend that model by considering a finite population and the individual heterogeneity in age groups by using a multidimensional Hawkes process for modelling the infections. The aggregated reported cases per age group, in turn, have a probability distribution with a mean driven by the underlying Hawkes process. We apply a KDPF for inferring the latent infections and the instantaneous reproduction number for each age group and for forecasting the epidemic's future trajectory in the near future. We demonstrate the performance of the proposed model on COVID-19 data in several London boroughs published by the government in the UK~\cite{link_GOVUK} using the empirical contact matrix derived from Jarvis et al.~\cite{jarvis2020quantifying} within the framework of the latent multidimensional Hawkes process.

\section*{Methods}
\subsection* {Model} \label{AModel}
We introduce a novel age-stratified epidemic model by extending the epidemic model of Lamprinakou et al.~\cite{https://doi.org/10.48550/arxiv.2208.07340} considering a finite population stratified by age bands $\mathcal{A}$.

We restrict our attention to an epidemic process over a horizon $[T_0, T)$, $T_0<T$, in which we assume immunity to re-infection that is a reasonable assumption over the time scales we consider. We break the horizon $[T_0,T)$ into $k$ subintervals $\mathcal{T}_j=[T_{j-1},T_j)$ for $j=1,..,k$ with $T_k=T$. We assume that the epidemic is triggered by a set of infectious individuals at the beginning of the process, the times of their infections denoted by a finite set $\mathcal{H}_0$.

The epidemic process is seen as a marked counting process $N(t)$ with a set of jump times $\mathcal{T}^N=\{t_0<t_1<t_2<...\}$
 and a set of associated age groups, $A^{N}=\{a_i\}$ where $a_i \in \mathcal{A}$ is the age group of infection at time $t_i$. The intensity of latent cases in age group $a\in \mathcal{A}$ at time $t$ is given by 

\begin{equation*}
\lambda^N(t,a)=\frac{S_{t,a}}{N_a}\gamma(t, a)\sum_{t_i\in h_t^0}h(t-t_i)m_{aa_i}
\end{equation*} for $t>T_0$ with $h_t^0=\{t_i|t_i<t\}\cup \mathcal{H}_0$ being the set of all infection events prior to time $t$, $S_{t,a}$ being the number of susceptibles in age group $a$ at time $t$, $\gamma(t,a)$ being a process dependent on age group $a$, $N_a$ being the size of the population of age group $a$ and $m_{aa_i}$ being the average number of contacts per unit time of a person in age group $a$ with people in age group $a_i$. The kernel $h(t-t_i)$ represents the relative infectiousness at time $t$ of an infection at time $t_i$. The transition kernel $h$ is a probability density function with non-negative real-valued support: $ h:[0,\infty)\rightarrow [0,\infty)$ and  $\int\limits_{0}^{\infty}h(s)ds=1$

In the manner of Lamprinakou et al.~\cite{https://doi.org/10.48550/arxiv.2208.07340}, we model the observed cases of age $a$ falling in $\mathcal{T}_n$, denoted by $Y_{na}$, via  a probability distribution $G$ having mean $\mu_{na}$ equal to the expected observed cases of age $a$ in $\mathcal{T}_n$ given by 
\begin{equation*}
\mu_{na}=\beta\sum\limits_{t_w\in [T_0,T_n), a_w\equiv a\ }\int\limits_{\max(t_w,T_{n-1})}^{T_n}{g(s-t_w)}ds.
\end{equation*}The kernel $g(s -t_i)$ represents the relative delay between the infection at time $t_i$ and the time at $s$ the infection is detected. Similar to the transition kernel of latent cases $h$, we specify the transition kernel of observed cases $g$ to be a probability density function with non-negative real-valued support.

We model $\gamma(t, a)$ as a stepwise function having as many weights as the number of subintervals, that is
\begin{equation*}
\gamma(t, a)=\prod_{n=1}^k \gamma_{na}^{\mathbbm{1}(t\in \mathcal{T}_n)},
\end{equation*} where $\{\gamma_{na}\}_{n=1}^k$ is assumed to be a Markov process. The average number of secondary cases of age $a'\in \mathcal{A}$ each infected individual of age $a$ would infect is given by
\begin{equation*}
R_{aa'}(t)=\frac{S_{t,a'}}{N_{a'}}\gamma(t,a')m_{a'a} 
\end{equation*} if the conditions remained as they were at $t$. The instantaneous reproduction number of age group $a$ is 
\begin{equation*}
R_{a}(t)=\sum_{a'}R_{aa'}(t),
\end{equation*}
which is the number of newly infected people that each infected individual aged $a$ would infect if the conditions remained as they were at $t$. We derive the instantaneous reproduction number $R(t)$ as the infected population-weighted average of $R_{a}(t)$ at time $t$.

The age-stratified model is described by the equations: 

\begin{align}
\label{MdsAM1}
   \lambda^{N}(t,a)&=\frac{S_{t,a}}{N_a}\gamma(t,a)\sum\limits_{t_i\in h_t^0}h(t-t_i)m_{aa_i},\ t\in[T_0,T),\  \forall \  \ a \in \mathcal{A} \ ; \\
 Y_{na} &\sim \mbox{G} \text{ with mean } E(Y_{na})=\mu_{na},\ n=1,..,k,\  \forall \  \ a \in \mathcal{A}\ ; \\
 \gamma(t,a)&=\prod\limits_{n=1}^{k}\gamma_{na}^{\mathbbm{1}\{t\in\mathcal{T}_n\}},\ t\in[T_0,T),\  \forall \  \ a \in \mathcal{A}\ ; \\
  \{\gamma_{na}\}_{n=1}^k& \text{is a Markov process},\  \forall \  \ a \in \mathcal{A}\ ; \\ 
  \label{MdsAM2}
\mu_{na}&=\beta\sum\limits_{t_w\in [T_0,T_n), a_w\equiv a\ }\int\limits_{\max(t_w,T_{n-1})}^{T_n}{g(s-t_w)}ds,\ n=1,..,k \ \text{and}\  \forall \  \ a \in \mathcal{A} \ .
\end{align}

\subsection*{Inference algorithm} \label{AInference}
Given a set of observed infections with their associated age groups $\{\{Y_{na}\}_{a\in \mathcal{A}}\}_{n=1}^k$, we seek to infer the marked counting process $N(t)$ and the processes $\{ \gamma(t, a)\}_{a\in \mathcal{A}}$.

Following the inference approach of Lamprinakou et al.~\cite{https://doi.org/10.48550/arxiv.2208.07340}, the proposed epidemic model described by the equations (\ref{MdsAM1})-(\ref{MdsAM2}) is seen as a state-space model with a latent state process $\{X_n:1\ \leq n\leq k\}$ and an observed process $\{Y_n=(Y_{na})_{a\in \mathcal{A}}:\  1\leq n \leq k\}$. Each hidden state $X_n$  consists of the weights $\{\gamma_{na}\}_{a\in \mathcal{A}}$ associated to $\mathcal{T}_n$ and the set of latent cases $S_n^N$ falling into $\mathcal{T}_n$ along with their associated age groups $A_n^N$. The time-constant parameters are the parameters associated with the distribution $G$ and the prior imposed on the weights $\{\{\gamma_{na}\}_{n=1}^k\}_{a\in \mathcal{A}}$. We apply a KDPF (Algorithm \ref{APAlg}) for inferring the marked counting process $N(t)$, the weights $\{\{\gamma_{na}\}_{n=1}^k\}_{a\in \mathcal{A}}$, and the time-constant parameters.

We focus on illustrating the performance of the model on COVID-19. We model the observed cases $Y_{na}$ via a negative binomial distribution (NB) with mean $\mu_{na}$ and dispersion $v>0$. Before we proceed with the simulation analysis, we define the transition kernels of the observed and latent cases, the prior on the weights $\{\{\gamma_{na}\}_{n=1}^k\}_{a\mathcal{A}}$, an algorithm for sampling the hidden latent cases, the complexity of Algorithm \ref{APAlg} and a simple method to initialize $\mathcal{H}_0$.

\begin{figure}[!h] 
  \begin{subfigure}{6cm}
    \centering\includegraphics[width=6cm]{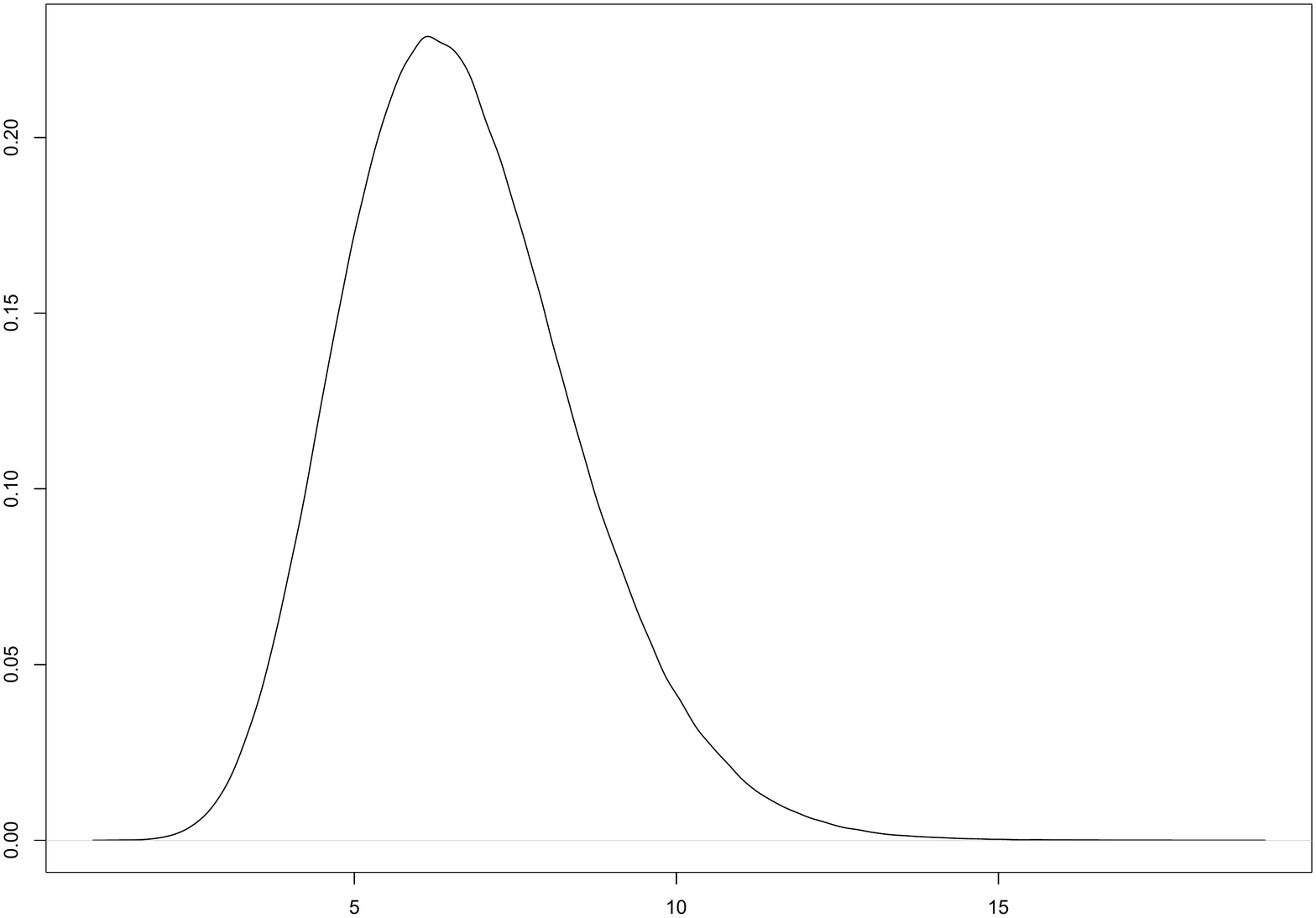}
  \end{subfigure}
  \begin{subfigure}{6cm}
    \centering\includegraphics[width=6cm]{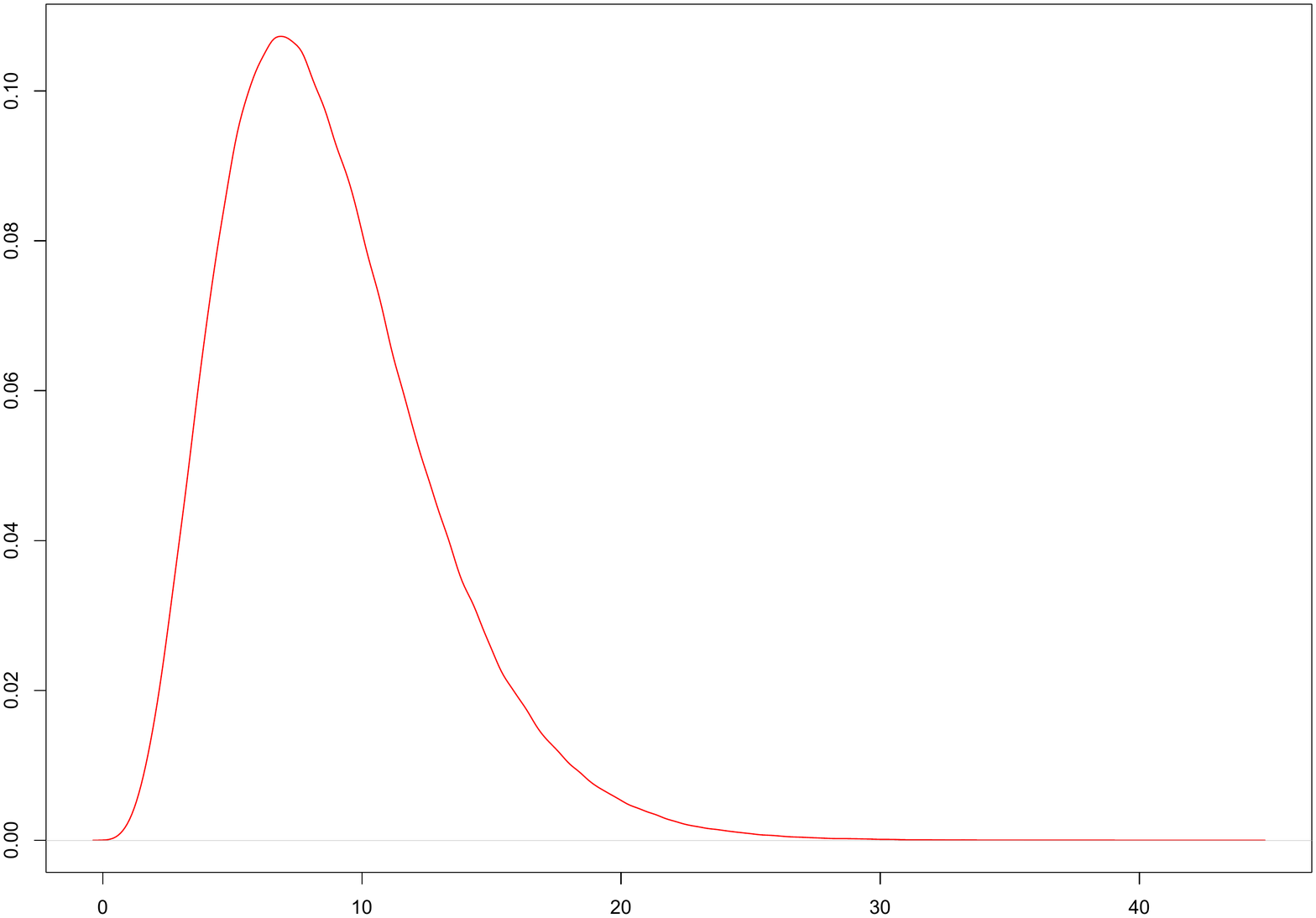}
  \end{subfigure}
  
  \caption{\bf The generation interval (GI) (black curve) and the period between observed and actual infection times (red curve).}
  \label{TK_10ma}
\end{figure}

\paragraph*{Transition Kernels} 
The dynamics of the latent and observed cases are determined by the generation interval (GI) and incubation period (IP) \citep{fine2003interval}. The generation interval is the time interval between the time of infection of the infector (the primary case) and that of the infectee (the secondary case generated by the primary case). The incubation period is the time interval between the infection and the onset of symptoms in a specific case. Zhao et al.~\cite{zhao2021estimating} assume that the GI and IP follow a gamma distribution. They infer that the mean and SD of GI are equal at 6.7 days and 1.8 days and those of IP at 6.8 and 4.1 days by using a maximum likelihood estimation approach and contact tracing data of COVID-19 cases. We follow the same assumption for the GI (namely, the transition kernel of latent cases is a gamma density with a mean at 6.7 days and  SD of 1.8 days). We model the time interval between the observed time and actual time of infection as a gamma density with a mean at 8.8 days and SD of 4.1 days  (that is, the transition kernel of observed cases is a gamma density having mean equal at 8.8 days and  SD of 4.1 days). For the transition kernel of the observed events, we adopt the values inferred by Zhao et al.~\cite{zhao2021estimating}  for IP with a slightly increased mean to consider the necessary time for conducting a test against COVID-19. Figure \ref{TK_10ma} illustrates the transition kernels. 

\paragraph*{Set of infectious at the beginning of the process, $\mathcal{H}_0$} We adopt a heuristic approach to initialize $\mathcal{H}_0$.  The transition kernel of latent cases illustrated in Figure \ref{TK_10ma} shows that a latent case at $t_w$ can influence the latent intensity at $t$ if $t_w$ has occurred at most 21 days before $t$. Otherwise, the influence of $t_w$ is negligible. Therefore, as the history of the process, we consider the latent cases of 21 days/3 weeks before the beginning of the process. The transition kernel of observed cases shown in Figure \ref{TK_10ma} demonstrates that an event is most likely to be observed seven days after the actual infection time. Considering the observed cases are daily, we initialize the history of latent cases in age group $a$, $\mathcal{H}_{0a}$ by uniformly spreading on the day $-i$ the number of cases of age $a$ occurred on the day $(-i+7)$ times $1/\beta$. The times of their infections at the beginning of the process, $\mathcal{H}_0$ is given by the union of the sets $\mathcal{H}_{0a}$, that is, $\mathcal{H}_{0}=\cup_a\mathcal{H}_{0a}$. We denote by $A_0^N$ the age groups associated with the time infections in $\mathcal{H}_0$. In simulation analysis, we propose initialization of $\mathcal{H}_0$ when we deal with weekly reported cases.

\paragraph*{Imposed prior on weights $\mathbf{\{\{\gamma_{na}\}_{n=1}^k}\}_a$}  A geometric random walk (RW) is imposed as prior on the weights $\{\{\gamma_{na}\}_{n=1}^k\}_a$ :
\begin{align*}
\log \gamma_{na}&=\log \gamma_{n-1,a}+\log\epsilon_n, \  \epsilon_n \sim\mbox{Gamma}(d,d),\ n=2,..,k,\ \forall \  \ a \in \mathcal{A}\ ; \\
\gamma_{1a}&\sim \mbox{Uniform}(\alpha,b),\ \forall \  \ a \in \mathcal{A} .
\end{align*}We impose a gamma prior on the noise of RW $\epsilon_n$ with equal shape and rate at $d$. This induces that the time-varying number $\gamma_{na}$ is gamma distributed with a mean equal to $\gamma_{n-1,a}$ and standard deviation $\gamma_{n-1,a}/ \sqrt{d}$. The stronger fluctuations in the observed data, the more flexible modelling we need.  Smaller values of $d$  have higher standard deviation and lead to a wider range of possible values of $\gamma_{n-1,a}$ increasing the flexibility of the model.

\paragraph*{Sampling the hidden latent cases and associated age groups}
We sample the latent cases $S_n^N$ along with their associated age groups $A_n^N$ falling into the subinterval $\mathcal{T}_n$ by applying Algorithm \ref{MdsAlg1}, which is a simulation procedure based on the branching structure of the Hawkes process \citep{laub2015hawkes}. The proposed algorithm is a superposition of Poisson processes in the interval $\mathcal{T}_n$; the descendants of each latent event at $t_i$ form an inhomogeneous Poisson process with intensity \begin{equation*}\lambda_i(t)=h(t-t_i)\sum\limits_a\gamma_{na}\frac{S_{t,a}}{N_a}m_{aa_i}\end{equation*}
for $t>t_i$ and $t\in[T_{n-1},T_n)$. This induces that: 
\begin{itemize}
\item{The number of events $n_i$ triggered by an event at $t_i$ in the interval $\mathcal{T}_n$ is Poisson distributed with parameter  
\begin{equation*}\lambda=\left(\sum\limits_a\frac{\gamma_{na}S_{max\left(T_{n-1},t_i\right),a}}{N_a}m_{aa_i}\right)\int\limits_{\max(t_i,T_{n-1})}^{T_n}h(s-t_i)ds.\end{equation*}}
\item{ The arrival times of the $n_i$ descendants are  $t_i+E_i$ with $E_i$ being iid random variables with pdf the truncated distribution $h(t)$ in $[\max(t_i,T_{n-1}),T_n)$.}
\item{ Sample the associated age groups w.p. $P(a)=\frac{\frac{S_{max\left(T_{n-1},t_i\right),a}}{N_a}m_{aa_i}}{\sum\limits_a\frac{S_{max\left(T_{n-1},t_i\right),a}}{N_a}m_{aa_i}}$, $\forall a \in \mathcal{A}$.} 
\end{itemize}

 \begin{algorithm}[!h] % enter the algorithm environment
\caption{Sample $S_n^N,\  A_n^N|S_{1:(n-1)}^N,\ \mathcal{H}_0,\ A_0^N,\ \{\gamma_{na}\}_a$} 
\label{MdsAlg1}
\begin{algorithmic}[1]
\STATE{Input: $S_{1:(n-1)}^N$,\ $A_{1:(n-1)}^N$,\ $\mathcal{H}_0$,\ $A_0^N$ ,\ $\{\gamma_{na}\}_a$ }\\
\STATE{Initialize two empty queues: $Q_t$ and $Q_a$.}\\
\STATE{$Q_t=\mathcal{H}_0 \cup \{S_{v}^N\}_{v=n-\eta}^{n-1}$ and $Q_a= A_0^N \cup \{A_v^N\}_{v=n-\eta}^{n-1}$ with $n-\eta\geq 1$ and $\eta$ being the number of former subintervals we consider (the value of $\eta$ is determined by the transition kernel of latent cases)}.
\WHILE{$Q$ is not empty }

\STATE{Remove the first element $t_i$ from $Q_t$ and the first element $a_i$ from $Q_a$.}\\
\STATE{Draw the number of events $n_{i}$ triggered by an event $i$ from a Poisson distribution with parameter  $\lambda=\left(\sum\limits_a\frac{\gamma_{na}S_{max\left(T_{n-1},t_i\right),a}}{N_a}m_{aa_i}\right)\int\limits_{\max(t_i,T_{n-1})}^{T_n}h(s-t_i)ds$ that is the average number of offsprings generated by an event at $t_i$ in $\mathcal{T}_n$.}
\STATE{Generate $n_{i}$ events from the truncated distribution $h(t)$ in $[\max(t_i,T_{n-1}),T_n)$, and add the new elements to the back of queue $Q_t$. }\\
\STATE{Sample the associated age groups w.p.   $P(a)=\frac{\frac{S_{max\left(T_{n-1},t_i\right),a}}{N_a}m_{aa_i}}{\sum\limits_a\frac{S_{max\left(T_{n-1},t_i\right),a}}{N_a}m_{aa_i}}$, and add the asociated age groups to the back of queue $Q_a$. }
\ENDWHILE
\STATE{Return $Q_t$ and $Q_a$.}
\end{algorithmic}
\end{algorithm}

\paragraph*{Who infected whom} The proposed model can capture the process's branching pattern by saving the parent of each latent infection at step 7 of Algorithm \ref{MdsAlg1}. Alternatively, the parent of each infection $j$ aged $a$ is assumed to be sampled from a multinomial distribution parameterized by $\pi_{ja}$, where $\pi_{ja}=\{\pi_{jia}\}_{i\in h_j}$ with 
\begin{equation*}
    \pi_{jia}=\frac{h(t_j-t_i)m_{aa_i}}{\sum_{t_w\in h_{t_j}^0}h(t_j-t_w)m_{aa_w}}
\end{equation*} being the probability of secondary infection $j$ aged $a$ having been caused by primary infection $i$ and
$h_j=\{i: t_i\in h_{t_j}^0, \hspace{0.1 cm} t_i \in \cup_{v=j-\eta}^j\mathcal{T}_v,\hspace{0.1 cm} t_j\in \mathcal{T}_j\}$.

\paragraph*{Complexity} Compared to the algorithm of Lamprinakou et al.~\cite{https://doi.org/10.48550/arxiv.2208.07340}, the computational cost of the propagation step (step 12 of KDPF) increases, taking into account the age band of an individual due to the cost of finding the probabilities $\{P(a)\}_a$, that is, $O(|\mathcal{A}|)$ with $|\mathcal{A}|$ being the cardinality of the set of age bands $\mathcal{A}$. However, the increase is negligible even if we consider 200 age groups. For this reason, the computational costs of propagation (step 12 of KDPF) and finding weights (step 14 of KDPF) at state (interval) $j$ remain equal at $O\left(N\sum\limits_{v=j-\eta}^j|S_v^N|\right)$. The computational cost of finding auxiliary weights (step 9 of KDPF) at state (interval) $j$ is the combined costs of propagation and finding weights. Hence, the computational cost of the algorithm over all states (intervals) is $O\left(N\left(\eta+1\right)|\mathcal{T}^N|\right)$. $N$ is the number of particles, $S_j^N$ the set of latent cases falling into subinterval $\mathcal{T}_j$, $\mathcal{T}^N=\cup_{j}S_j^N$ and $\eta$ the number of former subintervals that influence the latent cases falling into $\mathcal{T}_j$ determined by the transition kernel of latent cases. The $O$-notation denotes the asymptotic upper bound \citep{cormen2022introduction}. The algorithm is easily parallelized over $N$. We note that accounting for each person's age band does not increase the algorithm's complexity.

\begin{algorithm}[!h] % enter the algorithm environment
\algsetup{linenosize=\tiny}
\tiny %\small, \footnotesize, \scriptsize, or \tiny
\caption{\bf Kernel density particle filter} 
\label{APAlg}
\begin{algorithmic}[1]
\STATE{Initialize the parameters $\{\theta_{j1}\}_{j=1}^N$, $\theta_{j1}=(d_{j1},v_{j1})$ with $d_{min}\leq d \leq d_{max}$ and $v_{min}\leq v \leq v_{max}$:\\
\hspace*{0.5em} \textbf{for} $j$ in $1:N$ \textbf{do} \\
\hspace*{1.5em}$\log d_{j1}=\mathcal{N}(\mu_d, \sigma_d^2)$ with $\mu_d=\frac{\log d_{max} +\log d_{min}}{2}$ and $\sigma_d=\frac{\log d_{max} -\log d_{min}}{8}$  \\
\hspace*{1.5em}$\log v_{j1}=\mathcal{N}(\mu_v, \sigma_v^2)$ with $\mu_v=\frac{\log v_{max} +\log v_{min}}{2}$ and $\sigma_v=\frac{\log v_{max} -\log v_{min}}{8}$\\
\hspace*{0.5em}\textbf{end for} \\
}
\STATE{Sample $N$ particles $\{X_{j1}\}_{j=1}^N$, $X_{j1}=\left(\{\gamma_{j1a}\}_a,\ S_{j1}^N,\ A_{j1}^N\right)$: \\
\hspace*{0.5em} \textbf{for} $j$ in $1:N$ \textbf{do} \\
\hspace*{1.5em}$\gamma_{j1a}\sim \mbox{Uniform}(\alpha,\beta)$ $\forall$ age group $a$ \\
\hspace*{1.5em}$\left(S_{j1}^N,\ A_{j1}^N\right) \sim P\left(S_1^N, A_1^N|\{\gamma_{j1a}\}_a,\mathcal{H}_0, A_0^N\right)$\\
\hspace*{0.5em}\textbf{end for}
}

\STATE{Find the weights, $\tilde{w}_1=\{\tilde{w}_{j1}\}_{j=1}^N$:\\
\hspace*{0.5em} \textbf{for} $j$ in $1:N$ \textbf{do} \\
\hspace*{1.5em}$\tilde{w}_{j1}=P\left(Y_1|S_{j1}^N,A_{j1}^N,\beta,\mathcal{H}_0,A_0^N,v_{j1}\right)=\prod_a P\left(Y_{1a}|S_{j1a}^N,\beta, \mathcal{H}_0,A_0^N,v_{j1}\right)$ \\
\hspace*{0.5em}\textbf{end for}\\
where $S_{j1a}^N$ is the set of latent cases of age $a$ in $\mathcal{T}_1^N$ associated to particle $j$.
}

\STATE{Normalize the weights, $w_1=\{w_{j1}\}_{j=1}^N$:\\
\hspace*{0.5em} \textbf{for} $j$ in $1:N$ \textbf{do} \\
\hspace*{1.5em}$w_{j1}=\frac{\tilde{w}_{j1}}{\sum\limits_{j=1}^N\tilde{w}_{j1}}$ \\
\hspace*{0.5em}\textbf{end for}
}

\FOR{$n=1,..,k$}
\STATE{
\hspace*{0.5em} \textbf{for} $j$ in $1:N$ \textbf{do} \\
\hspace*{1.5em}$m_{j,dn}^{(L)}=a\log d_{jn} + (1-a)\bar{d}_{Ln}, \hspace{0.5cm} \bar{d}_{Ln}=\sum\limits_{j=1}^Nw_{jn}\log d_{jn}$ \\
\hspace*{1.5em}$m_{j,vn}^{(L)}=a\log v_{jn} + (1-a)\bar{v}_{Ln}, \hspace{0.5cm} \bar{v}_{Ln}=\sum\limits_{j=1}^Nw_{jn}\log v_{jn}$ \\
\hspace*{1.5em}$m_{j,dn}=a d_{jn} + (1-a)\bar{d}_{n}, \hspace{0.5cm} \bar{d}_{n}=\sum\limits_{j=1}^Nw_{jn} d_{jn}$ \\
\hspace*{1.5em}$m_{j,vn}=a  v_{jn} + (1-a)\bar{v}_{n}, \hspace{0.5cm} \bar{v}_{n}=\sum\limits_{j=1}^N w_{jn} v_{jn}$ \\
\hspace*{0.5em}\textbf{end for}

}
\STATE{ For each particle $j$, we calculate an estimate of $X_{j,n+1}$ called $\tilde{X}_{j,n+1}$ by drawing a sample from $P(X_{n+1}|X_n,\mathcal{H}_0)$: \\
\hspace*{0.5em} \textbf{for} $j$ in $1:N$ \textbf{do} \\
\hspace*{1.5em}$\tilde{\gamma}_{j,n+1,a}\sim P\left(\gamma_{n+1,a}|\gamma_{jna},m_{j,dn}\right)$ $\forall$ age group $a$ \\
\hspace*{1.5em}$\left(\tilde{S}_{j,n+1}^N, \tilde{A}_{j,n+1}^N\right)\sim P\left(S_{n+1}^N, A_{n+1}^N|S_{j,1:n}^N,A_{j,1:n}^N,\tilde{\gamma}_{j,n+1,a},\mathcal{H}_0, A_0^N\right)$\\
\hspace*{0.5em}\textbf{end for}
}
\STATE{Find the auxiliary weights, $\tilde{g}_{n+1}=\{\tilde{g}_{j,n+1}\}_{j=1}^N$:\\
\hspace*{0.5em} \textbf{for} $j$ in $1:N$ \textbf{do} \\
\hspace*{1.5em}$\tilde{g}_{j,n+1}=g_{jn}w_{jn}P\left(Y_{n+1}|S_{j,1:n}^N,A_{j,1:n}^N,\tilde{S}_{j,n+1}^N,\tilde{A}_{j,n+1}^N,\beta,\mathcal{H}_0,A_0^N, m_{j,vn}\right)$\\ \hspace*{4.5em}$=g_{jn}w_{jn}\prod_a P\left(Y_{n+1,a}|\{S_{jka}\}_{k=1}^n,\beta,\mathcal{H}_0,A_0^N,m_{j,vn}\right)$\\
\hspace*{0.5em}\textbf{end for}\\
where $S_{jka}^N$ is the set of latent cases of age $a$ in $\mathcal{T}_k$ associated to particle $j$.
}
\STATE{Normalize the auxiliary weights, $g_{n+1}=\{g_{j,n+1}\}_{j=1}^N$:\\
\hspace*{0.5em} \textbf{for} $j$ in $1:N$ \textbf{do} \\
\hspace*{1.5em}$g_{j,n+1}=\frac{\tilde{g}_{j,n+1}}{\sum\limits_{j=1}^N\tilde{g}_{j,n+1}}$ \\
\hspace*{0.5em}\textbf{end for}}

\STATE{
\textbf{if}\Big($ESS(g_{n+1})=1/\sum\limits_{j=1}^Ng_{j,n+1}^2<0.8N$\Big) \textbf{then}
resample and form $N$ equally weighted particles, $\Bar{X}_{n}=\{\Bar{X}_{n}^i\}_{i=1}^N$:\\
\hspace{0.5em}\textbf{for} $j$ in $1:N$ \textbf{do} \\
\hspace*{1em}(i) sample index $i_j$ from a multinomial distribution with probabilities $g_{n+1}$\\
\hspace*{1em}(ii) $\bar{X}_{n}^j=X_{i_j,n}$\\
\hspace*{1em}(iii) $g_{j,n+1}=1$\\
\hspace{0.5em}\textbf{end for}\\
\textbf{end if}
}

\STATE{Regenerate the fixed parameters: \\\hspace*{0.5em} \textbf{for} $j$ in $1:N$ \textbf{do} \\
\hspace*{1.5em}$ \log v_{j,n+1}\sim \mathcal{N}(m_{i_j,vn}^{(L)},h^2V_{nv}^{(L)})$ \\
\hspace*{1.5em}$\log d_{j,n+1}\sim \mathcal{N}(m_{i_j,dn}^{(L)},h^2V_{nd}^{(L)})$\\
\hspace*{0.5em}\textbf{end for}\\
where $V^{(L)}_{nv}$ is the weighted variance of $\{\log v_{jn}\}_{n=1}^N$ and $V^{(L)}_{nd}$ the weighted variance of $\{\log d_{jn}\}_{n=1}^N$.
}
\STATE{Using $\bar{X}_{n}$ propagate: \\
\hspace*{0.5em} \textbf{for} $j$ in $1:N$ \textbf{do} \\
\hspace*{1.5em}$\gamma_{j,n+1,a}\sim P(\gamma_{n+1,a}|\gamma_{jna},d_{j,n+1})$ \\
\hspace*{1.5em}$\left(S_{j,n+1}^N, A_{j,n+1}^N\right) \sim P\left(S_{n+1}^N,A_{n+1}^N|S_{j,1:n}^N,A_{j,1:n}^N,\{\gamma_{jna}\}_a,\mathcal{H}_0,A_0^N\right)$\\
\hspace{1.5em}Set $X_{j,n+1}=(\bar{X}_{n}^j,\left(\{\gamma_{j,n+1,a}\}_a,S_{j,n+1}^N)\right)$\\
\hspace*{0.5em}\textbf{end for}
}

\STATE{Find the weights, $\tilde{w}_{n+1}=\{\tilde{w}_{j,n+1}\}_{j=1}^N$:\\
\hspace*{0.5em} \textbf{for} $j$ in $1:N$ \textbf{do} \\
\hspace*{1.5em}$\tilde{w}_{j,n+1}=\frac{P\left(Y_{n+1}|S_{j,1:n+1}^N,A_{j,1:n+1}^N,\beta,\mathcal{H}_0,A_0^N,v_{j,n+1}\right)}{P\left(Y_{n+1}|S_{j,1:n}^N,A_{j,1:n}^N,\tilde{S}_{i_j,n+1}^N,\tilde{A}_{i_j,n+1}^N\beta,\mathcal{H}_0,A_0^N,m_{i_j,vn}\right)}$ \\
\hspace*{0.5em}\textbf{end for}
}

\STATE{Normalize the weights, $w_{n+1}=\{w_{j,n+1}\}_{j=1}^N$:\\
\hspace*{0.5em} \textbf{for} $j$ in $1:N$ \textbf{do} \\
\hspace*{1.5em}$w_{j,n+1}=\frac{\tilde{w}_{j,n+1}}{\sum\limits_{j=1}^N\tilde{w}_{j,n+1}}$ \\
\hspace*{0.5em}\textbf{end for}
}

\STATE{ To draw a sample from $P\left(X_{1:n+1},d_{n+1},v_{n+1}|Y_{1:n+1}\right)$. We do resampling with weights $\{w_{j,n+1}\}_{j=1}^N\}$ if resampling was performed at step 10. Otherwise, we do resampling with weights $k_{j,n+a}\propto \tilde{w}_{j,n+1}g_{j,n+1}$.
}

\ENDFOR

\end{algorithmic}
\end{algorithm}

\section*{Results}
\subsection*{Simulation Analysis} \label{ASSimulation}
We carried out a simulation study to illustrate the performance of the KDPF (Algorithm \ref{APAlg}) for inferring the latent cases and the weights $\{\gamma_{na}\}_{n=1}^k$ per age group $a$ over various group numbers. 

In the simulation concepts, we adopt the demographic features in Leicester published by the Office for National Statistics (ONS)~\cite{ONS::Pop}. We deal with 16 hidden states $\{X_n\}_{n=1}^{16}$ and 16 subintervals $\{\mathcal{T}_n\}_{n=1}^{16}$; each subinterval corresponds to the duration of one week. We infer the latent intensity $\lambda^N(t, a)$, the weights $\{\gamma_{na}\}_{n=1}^{16}$ and the weekly latent cases per age group $a$ via the particle sample derived by drawing samples from the smoothing density with lag equal to 4. 

To confirm the convergence of posterior estimates of weights and weekly hidden cases per age group concerning the number of particles, we find the associated Monte Carlo Standard Errors (MCSEs). The MCSEs of posterior means of weights $\gamma_a=\{\gamma_{ia}\}_{i=1}^k$ and weekly latent cases of age $a$, $Y_a=\{Y_{ia}\}_{i=1}^k$ are given by
\begin{equation*}
    MCSE(\gamma_a)=\frac{1}{k}\sum_{i=1}^k\left(\frac{\mathrm{var}(\gamma_{ia})}{N} \right)^{1/2}
\end{equation*} and
\begin{equation*}
    MCSE(Y_a)=\frac{1}{k}\sum_{i=1}^k\left(\frac{\mathrm{var}(Y_{ia})}{N} \right)^{1/2}
\end{equation*}
where $\mathrm{var}(z)$ is the variance of $z$ and $Y_{ia}$ the aggregated latent cases of age $a$ in $i_{th}$ week.

We illustrate a simulation study for 2 age groups. Appendix \ref{Appendix4G} also includes a simulation study for 4 age groups. The simulation analysis showed that the KDPF (Algorithm \ref{APAlg}) approaches well the ground truth. The reported infections carry information about the progress of the epidemic with a maximum delay of 21 days between the reported and actual infection time for the COVID-19 pandemic. For this reason, the uncertainty of estimates increases in the last days. The MCSEs verified the convergence of posterior estimates concerning the number of particles.

\subsubsection*{2 age groups}
ONS shows that $83.1\%$ of the population is aged under 60 years (hereafter 0-59) and $16.9\%$ aged 60 years and over (60+). We coarse the age groups of the contact matrix for reopening schools \citep{jarvis2020quantifying} and get the matrix: \begin{equation*}
m=\begin{bmatrix}
6.81 & 0.66\\
2.14 & 1.27 
\end{bmatrix}. \end{equation*}
The process is triggered by 4963 infectious. The times of their infections, $\mathcal{H}_0$ , are uniformly allocated in 21 days ($[0,21)$) with a day being the time unit. We generate weekly latent and observed cases according to the model equations (\ref{MdsAM1})-(\ref{MdsAM2}) for weeks -$2-17$ ($[21,161)$) given $\mathcal{H}_0$, $v=0.004$, $d=15.22$, $\beta=0.5$, $\gamma_{-2,0-59}=0.2$ and $\gamma_{-2,60+}=0.17$. We consider that about 70\% of the population is susceptible at the beginning of week -$2$ ; 249073 susceptibles $(0-59:\ 206969,\ 60+:\  42104)$. 

We are interested in inferring the latent cases in weeks $1-16$. That induces that $\mathcal{H}_0$ is the set of times of latent infections in weeks -$2-0$. Using the generated observed cases in weeks -$1-1$ as described above, we estimate the latent infections with their associated age groups in weeks -$2-0$ as follows: The latent cases of age group $a_v$ on the week $i$ is equal to the number of events in age $a_v$ occurred on the week $(i + 1)$ times $1/\beta$, and are spread uniformly in $[(i+2)\times7 + 21, (i+3)\times7 +21)$ for $-2\leq i \leq 0$. We assume $\alpha=0$, $b=0.5$, $d_{min}=10$, $d_{max}=20$, $v_{min}=0.0001$ and $v_{max}=0.5$.    

\begin{figure}[!h] 
  \begin{subfigure}{7cm}
    \centering\includegraphics[width=6cm]{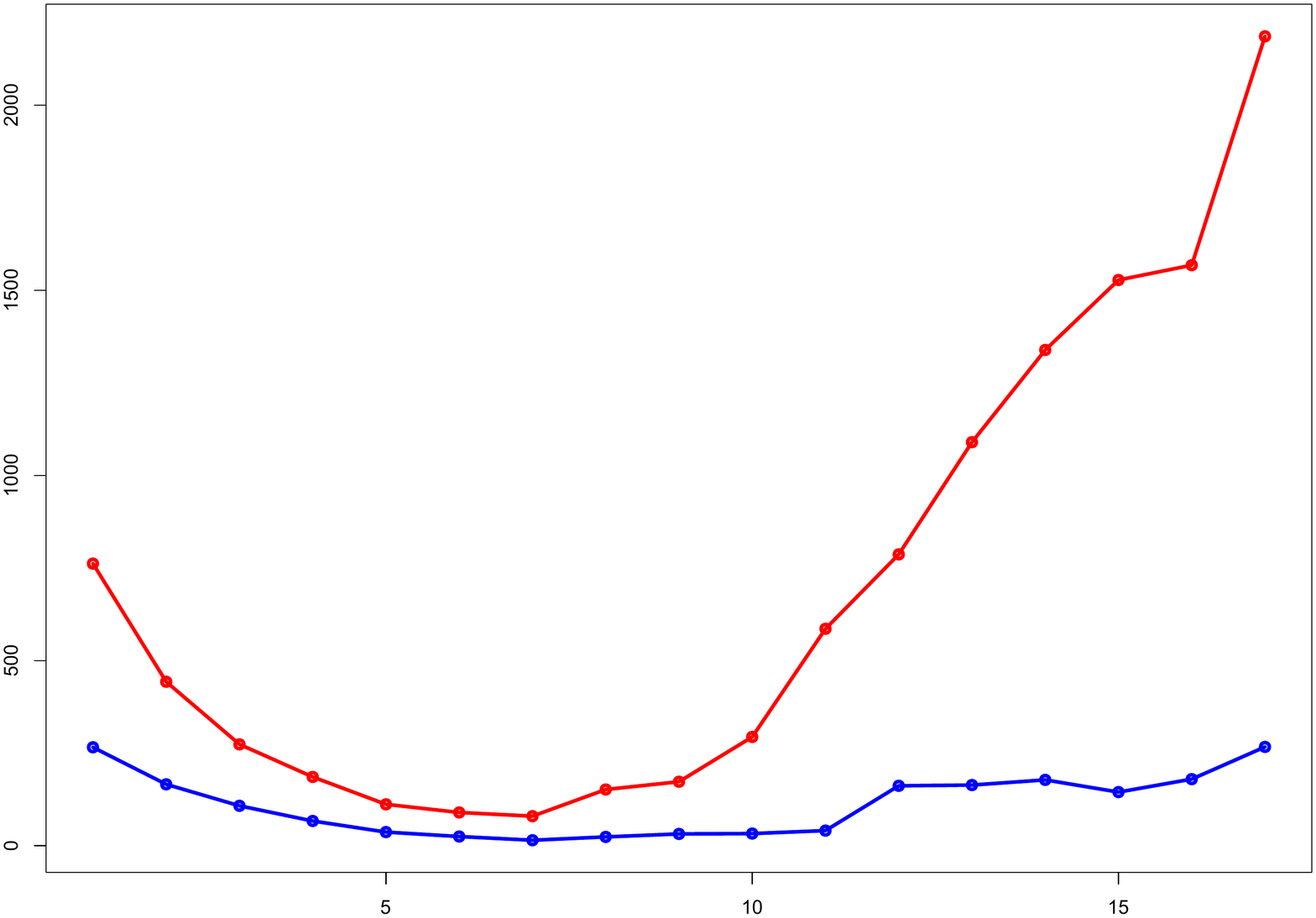}
    \caption{The weekly observed cases aged 0-59 (red\\ line) and 60+ (blue line).} 
  \end{subfigure}
  \begin{subfigure}{7cm}
    \centering\includegraphics[width=6cm]{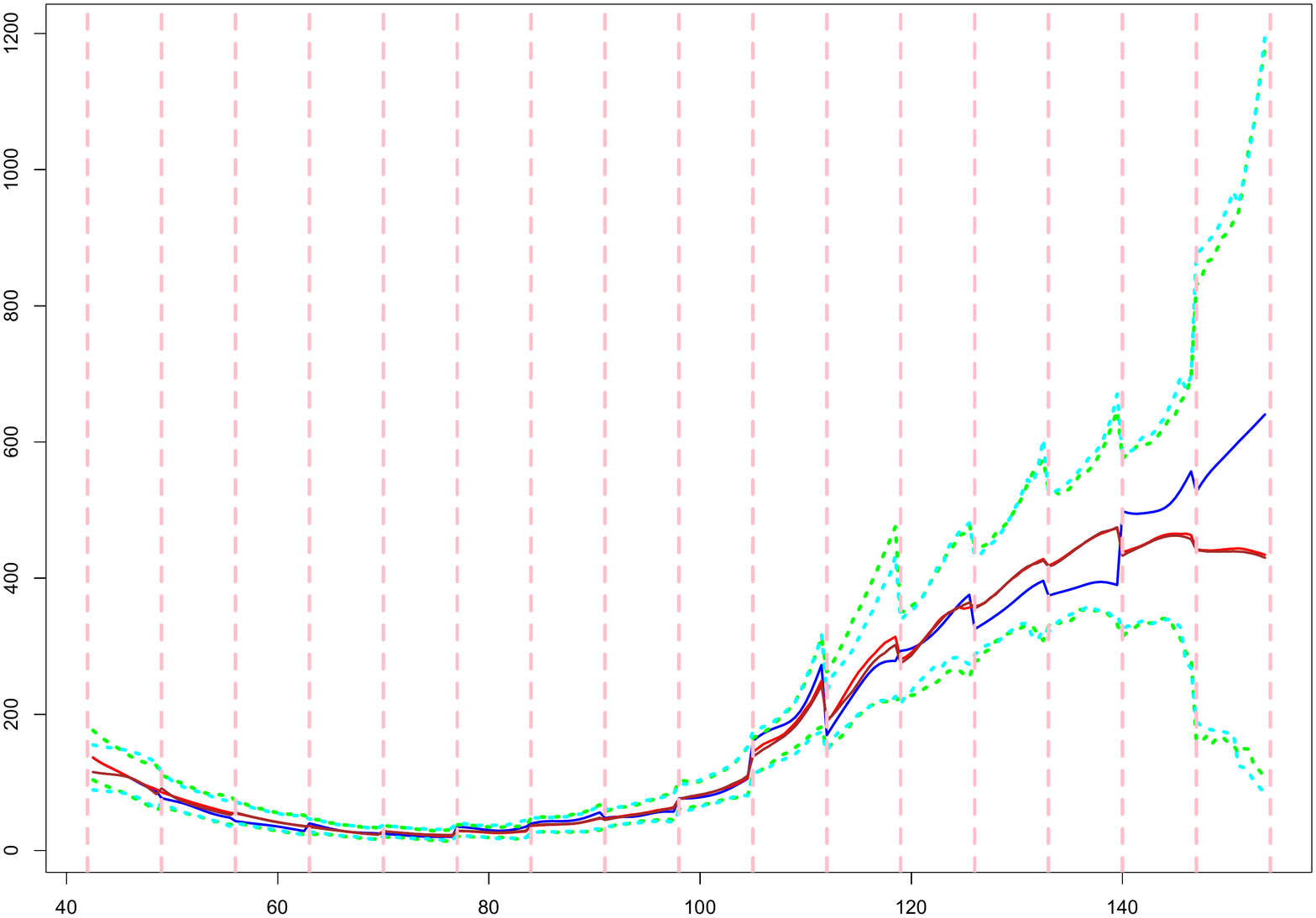}
   \caption{The estimated and true intensity of latent cases aged 0-59.}
  \end{subfigure}
  \begin{subfigure}{7cm}
    \centering\includegraphics[width=6cm]{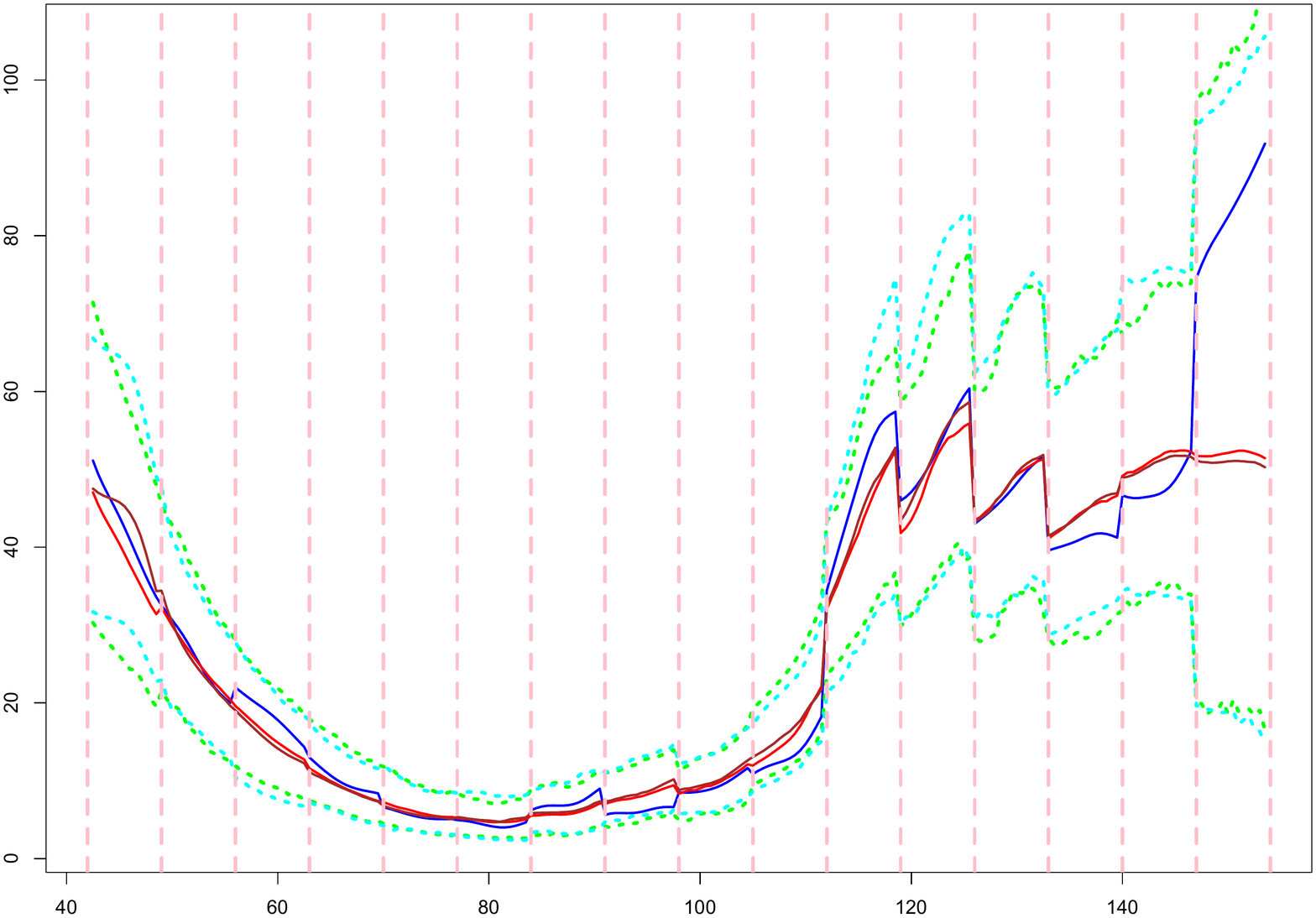}
   \caption{The estimated and true intensity of latent cases aged 60+.}
  \end{subfigure}
  \begin{subfigure}{7.2cm}
    \centering\includegraphics[width=6cm]{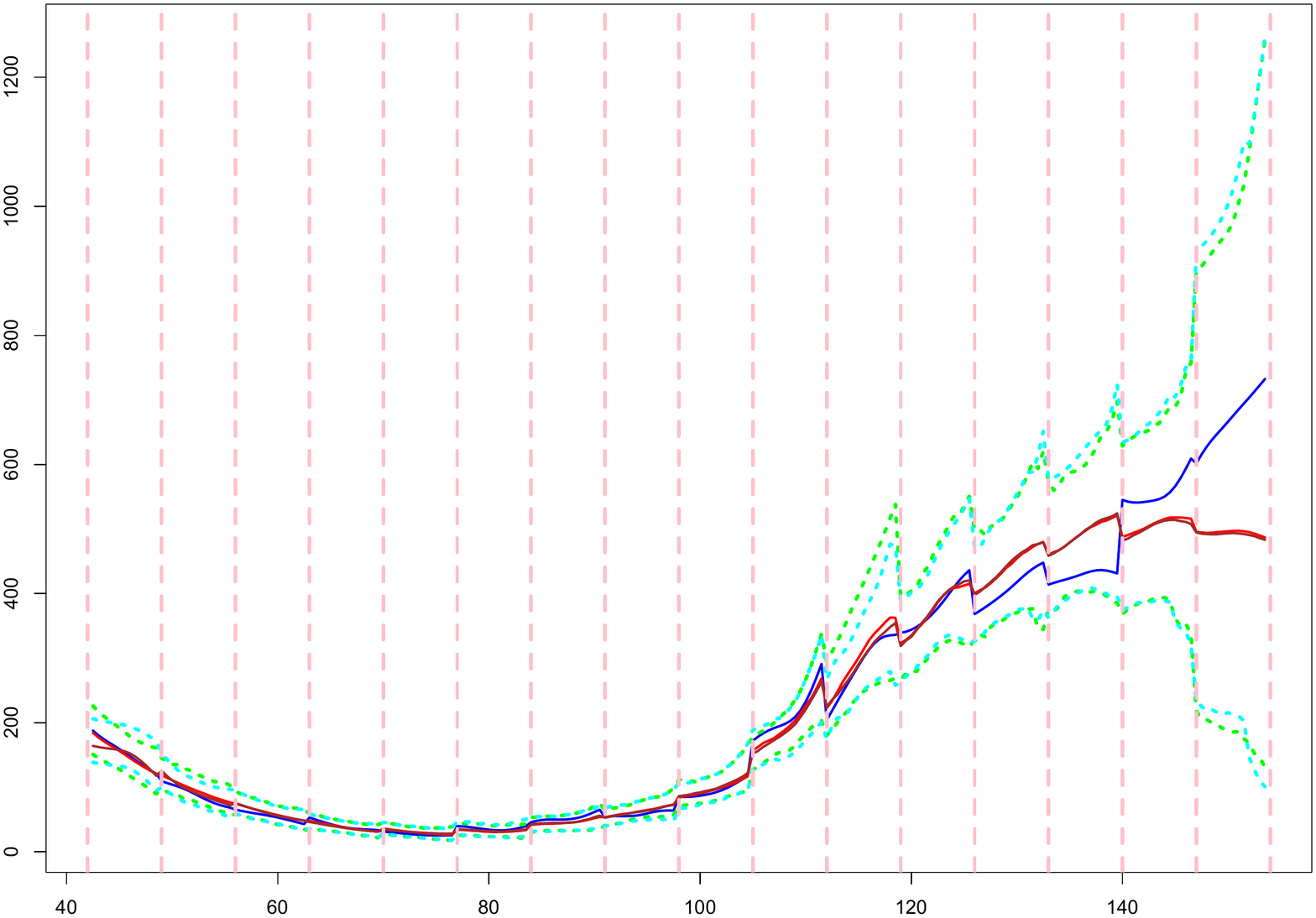}
   \caption{The aggregated estimated and true intensity of latent cases.}
  \end{subfigure}
  \caption{\bf{The weekly observed cases, the true latent intensities (blue line), and the estimated latent intensities considering 2 age groups (with estimated seeds (posterior median (brown line) ; 99\% CI (cyan line)), and true seeds (posterior median (red line) ; 99\% CI (green line))).}}
  \label{EstInt_AG2}
\end{figure}

The ground truth is characterized by $\mathcal{H}_0$ consisting of 6644 infections $(0-59:\ 4974,\ 60+:\ 1670)$, and 242429 susceptibles $(0-59:\ 201995,\ 60+:\ 40434)$ at the beginning of week 1. The estimated seeds and susceptibles are 6624 $(0-59:\ 5018,\ 60+:\ 1606)$, and 242449 $(0-59: 201951,\ 60+:\ 40498)$, respectively. The observed cases in weeks $1-17$ are 13560 $(0-59:\ 11650,\ 60+:\ 1910)$ (Figure \ref{EstInt_AG2}).  

\begin{figure}[!h] 
  \begin{subfigure}{7cm}
    \centering\includegraphics[width=6cm]{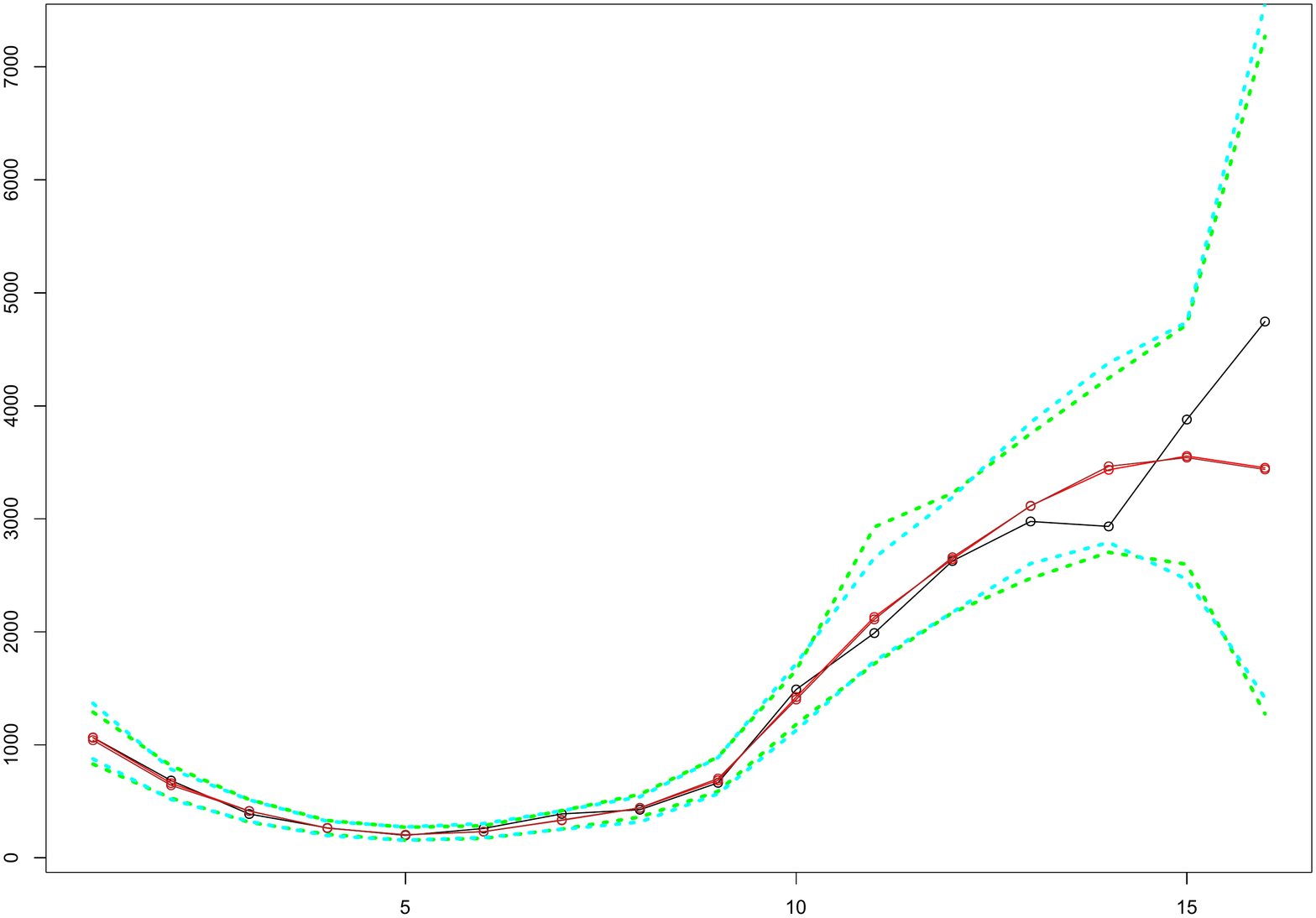}
   \caption{The aggregated estimated weekly hidden cases.}
  \end{subfigure}
  \begin{subfigure}{7cm}
    \centering\includegraphics[width=6cm]{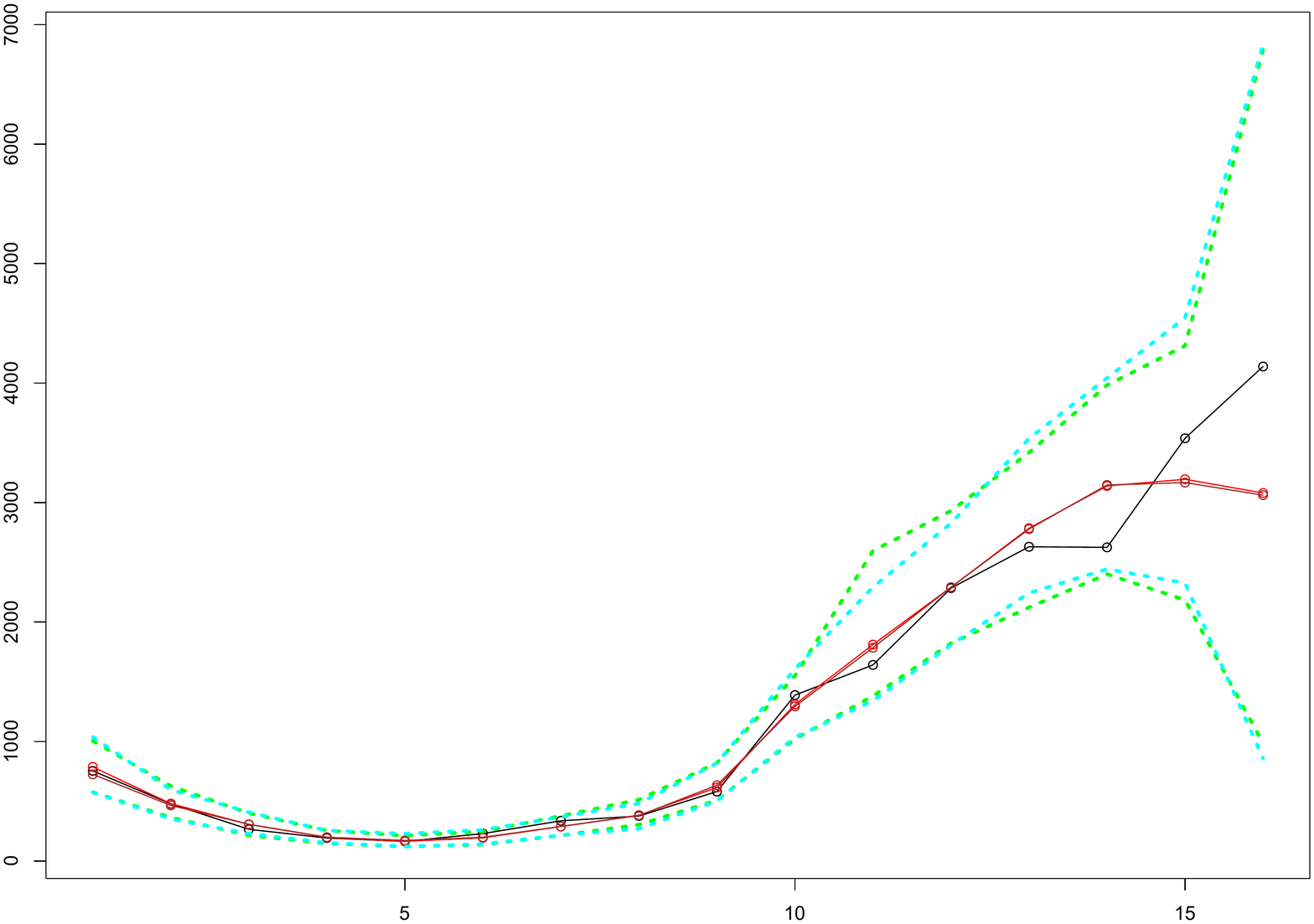}
    \caption{The estimated weekly infected cases aged 0-59.}
  \end{subfigure}
  \begin{subfigure}{7cm}
    \centering\includegraphics[width=6cm]{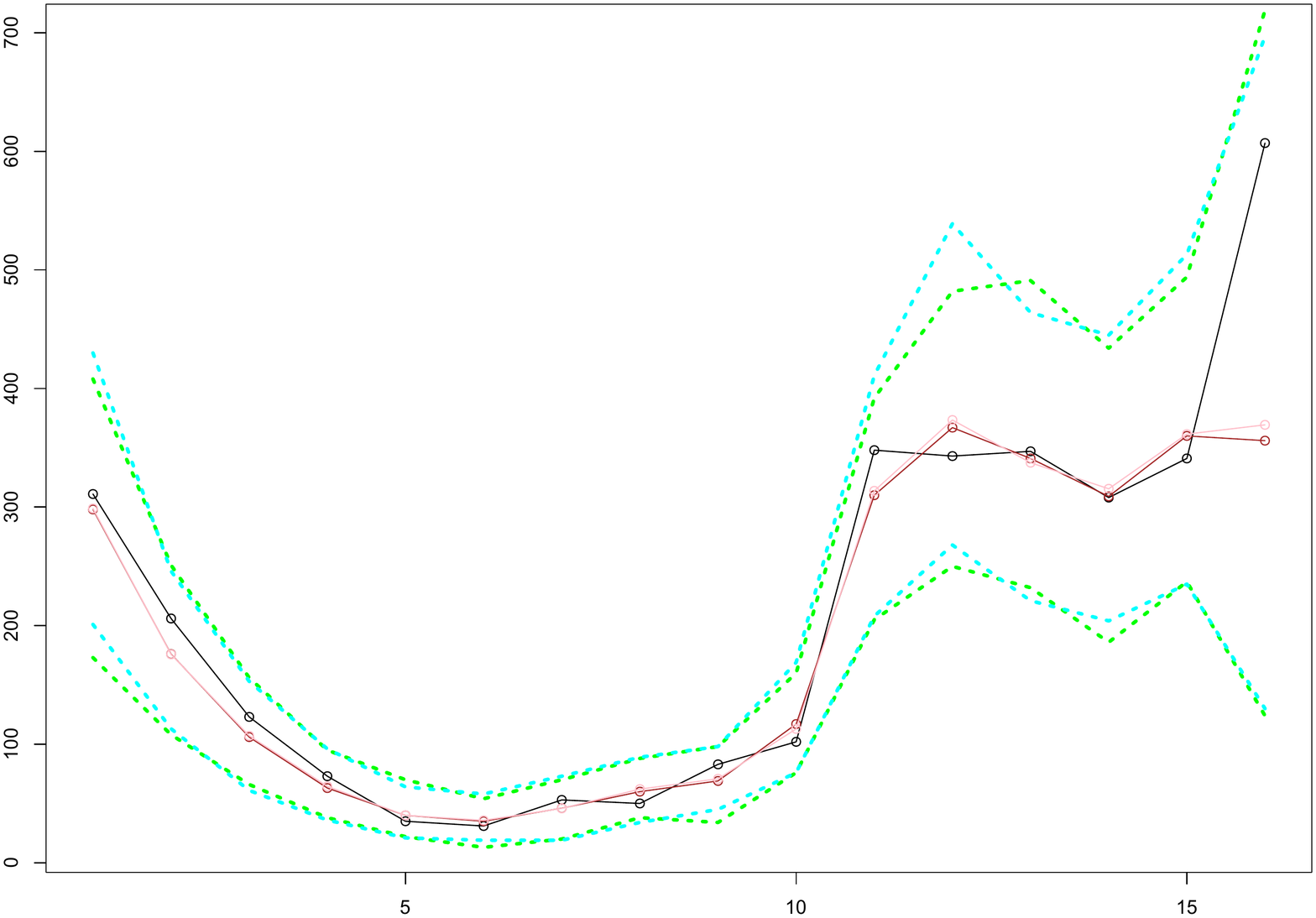}
    \caption{The estimated weekly infected cases aged 60+.}
  \end{subfigure}
   \caption{\bf{The estimated weekly latent cases (with estimated seeds (posterior median (brown line); 99\% CI (cyan line)) and  true seeds (posterior median (red line); 99\% CI (green line)))  and the true weekly hidden cases (black line) considering 2 age groups.}}
   \label{EHC_AG2}
\end{figure}

\begin{figure}[!h] 
  \begin{subfigure}{7cm}
    \centering\includegraphics[width=6cm]{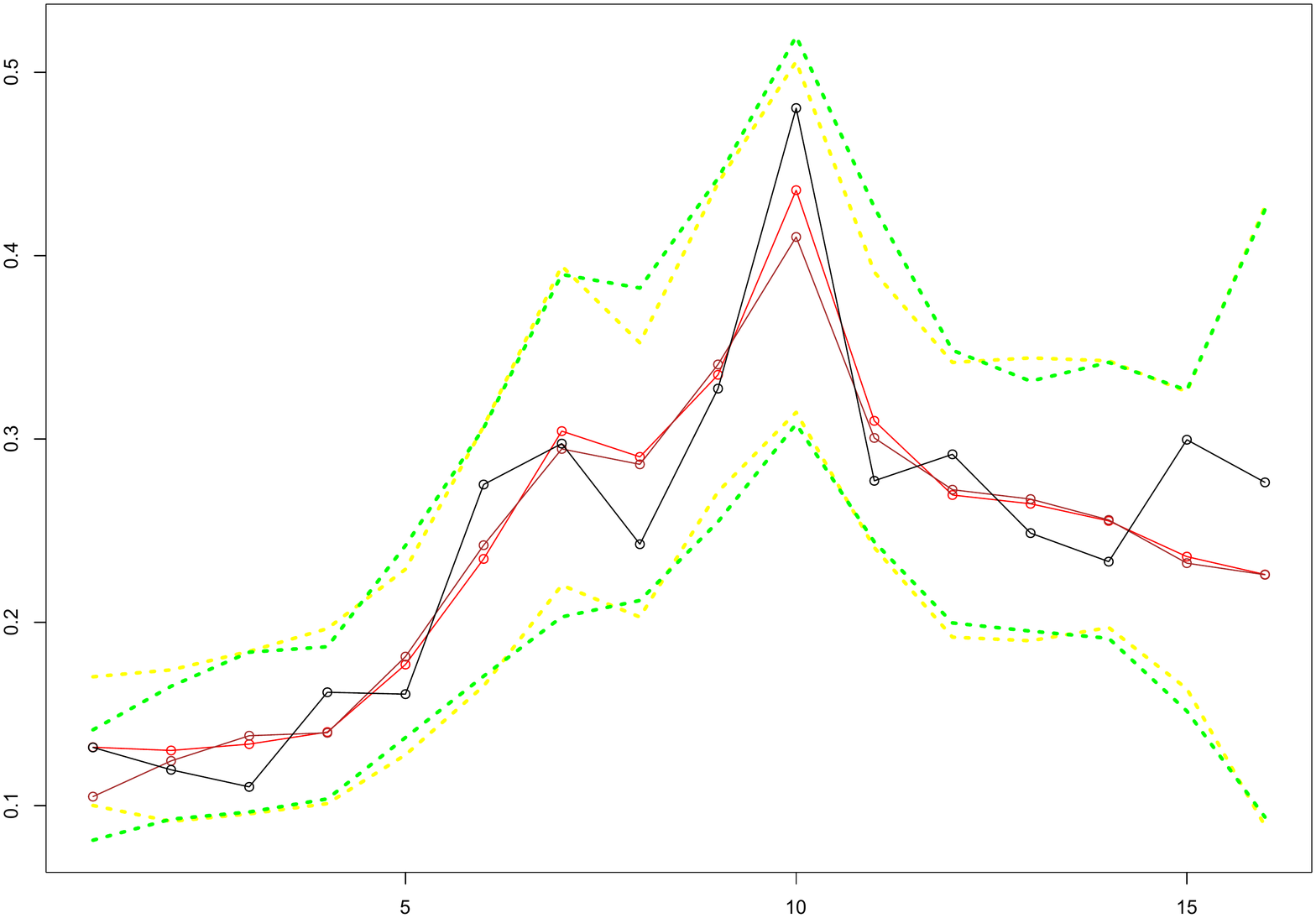}
   \caption{The estimated weights $\{\gamma_{i,0-59}\}_{i=1}^{16}$.}
  \end{subfigure}
  \begin{subfigure}{7cm}
    \centering\includegraphics[width=6cm]{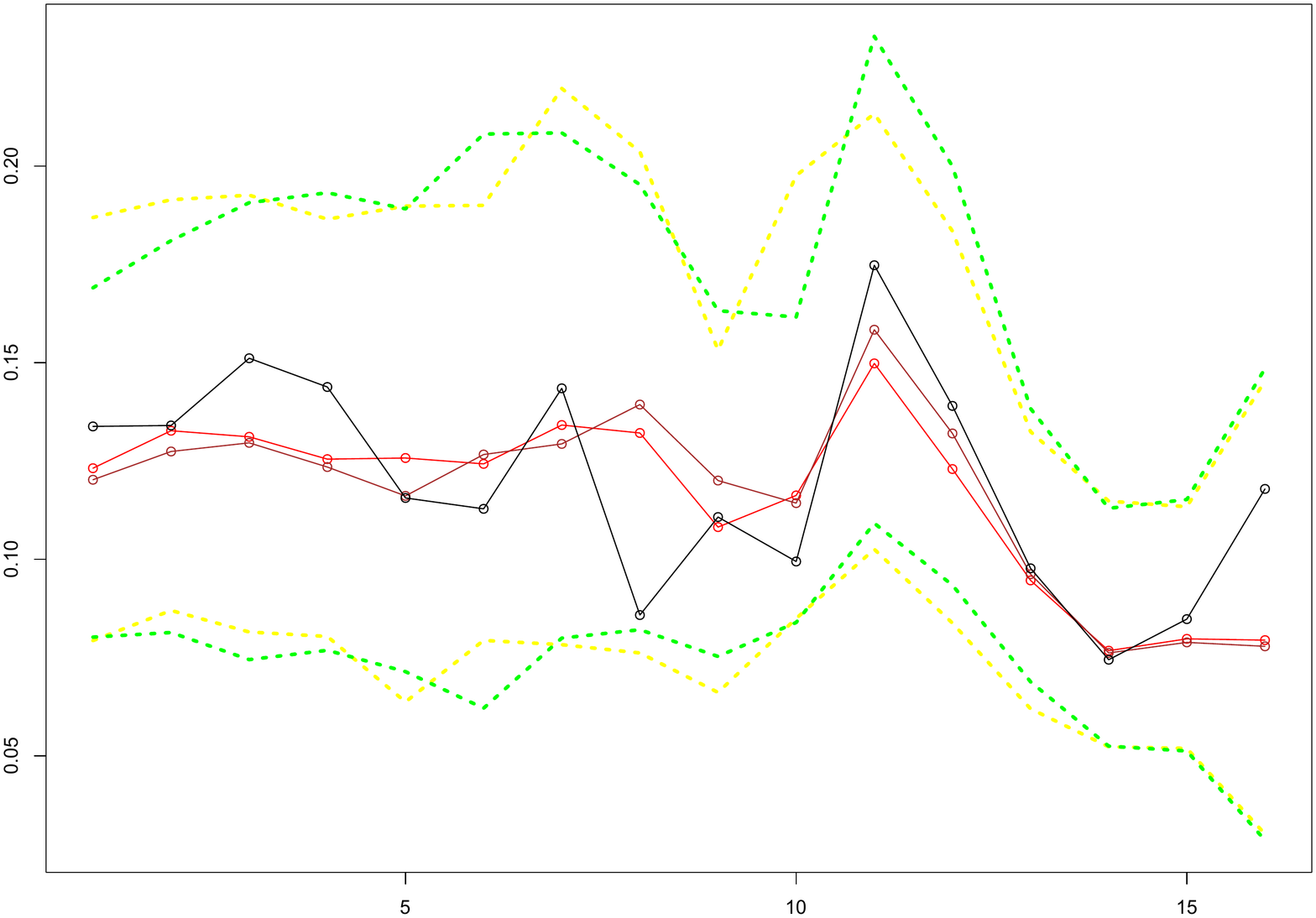}
     \caption{The estimated weights $\{\gamma_{i,60+}\}_{i=1}^{16}$.}
  \end{subfigure}

   \caption{\bf{The estimated weights $\{\gamma_{na}\}_a$ (with estimated seeds (posterior median (brown line); 99\% CI (green line)) and true seeds (red line; 99\% CI (yellow line)))  and the true values (black line) considering 2 groups.}}
   \label{ER_AG2}
\end{figure}

 The figures \ref{EstInt_AG2}-\ref{ER_AG2} show the estimated intensities, the estimated weekly hidden cases and the estimated weights $\{\{\gamma_{na}\}_{n=1}^{16}\}_a$ for 30000 particles. The Credible Intervals (CIs) widen in the last weeks as we infer the associated random quantities without being aware of the reported infections three weeks ahead.  Table \ref{tableM2} verifies the convergence of posterior estimates of weights and weekly hidden cases per age group concerning the number of particles. We note that the 99\% CIs of the time-constant parameters include the actual values of the parameters.
 
\begin{table}[!ht]
\centering
\caption{\bf{MCSEs of posterior means of weights $\{\gamma_a\}_a$ and  weekly hidden cases considering 2 groups.}}
\begin{tabular}{ |l|l|l|l| }
 \hline
 \multicolumn{4}{|l|}{\bf{Convergence of the posterior estimates} } \\
 \thickhline
  $MCSE$ & $N=10000$ & $N=20000$ & $N=30000$\\
 \hline
 $\gamma_1$& 0.000173 & 0.000189 & 0.000182\\ \hline
 $\gamma_2$ & 0.000123 & 0.000119 & 0.000118   \\ \hline
 $Y_1$ & 1.05305 & 1.114323 & 1.077618  \\ \hline
 $Y_2$ & 0.193806 & 0.191016 & 0.193658  \\
 \hline
\end{tabular}
\label{tableM2}
\end{table}
  
\subsection*{Real Data}
\begin{figure}[!h] 
\begin{subfigure}{7cm}
    \centering\includegraphics[width=6cm]{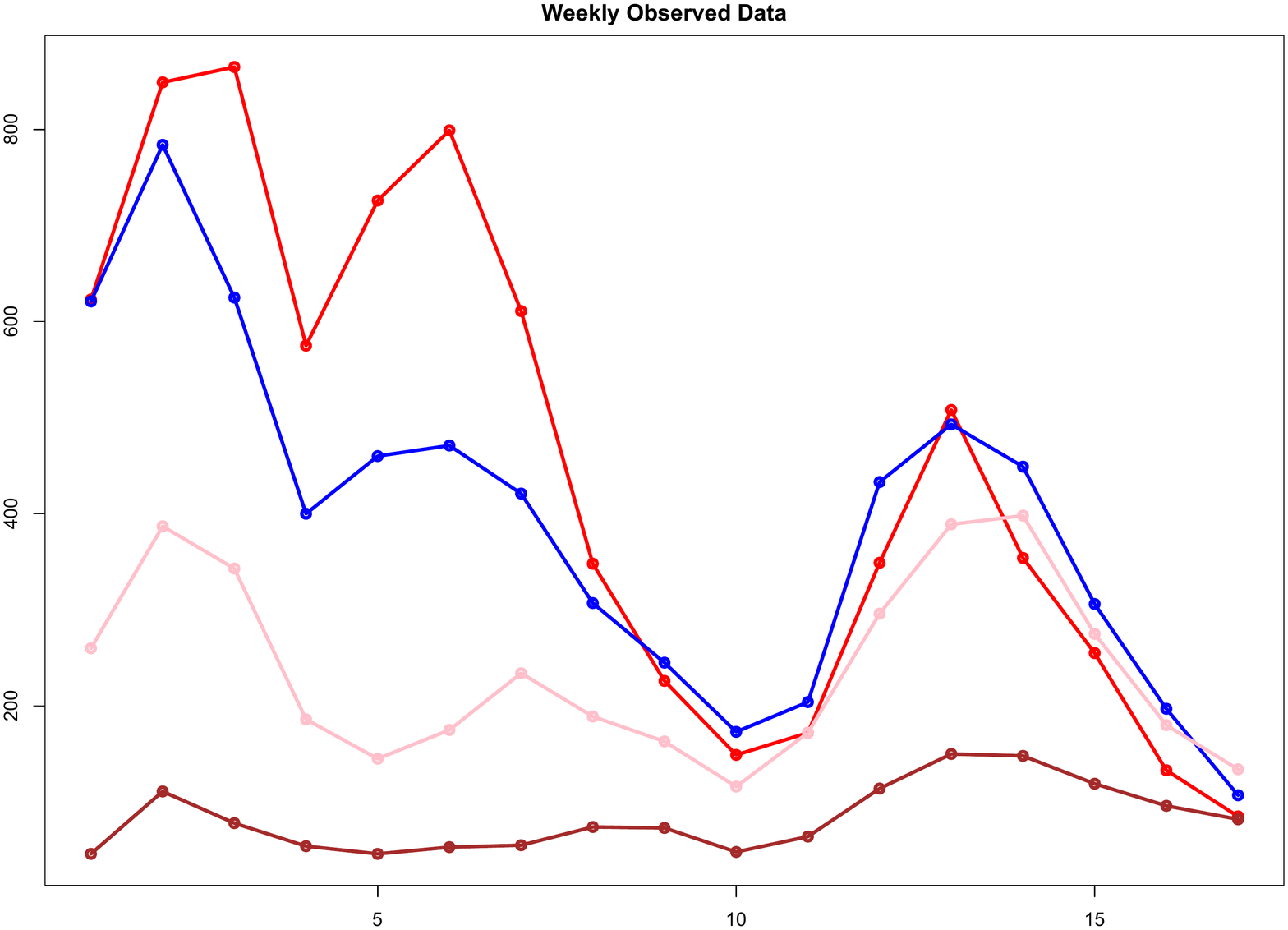}
    \caption{The weekly observed cases in Ashford } 
  \end{subfigure}
  \begin{subfigure}{7cm}
    \centering\includegraphics[width=6cm]{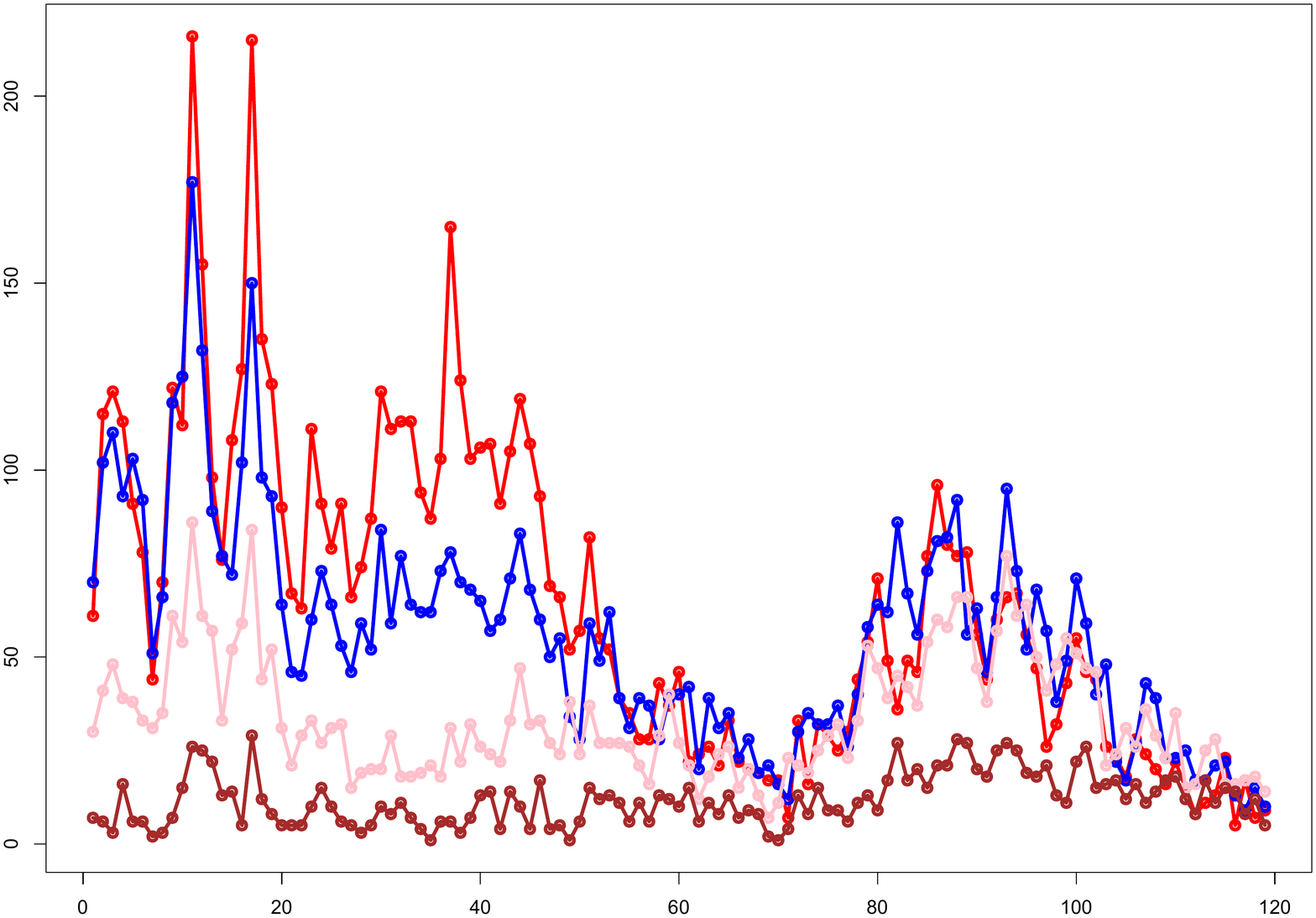}
   \caption{The daily observed cases in Ashford}
  \end{subfigure}
  \begin{subfigure}{7cm}
    \centering\includegraphics[width=6cm]{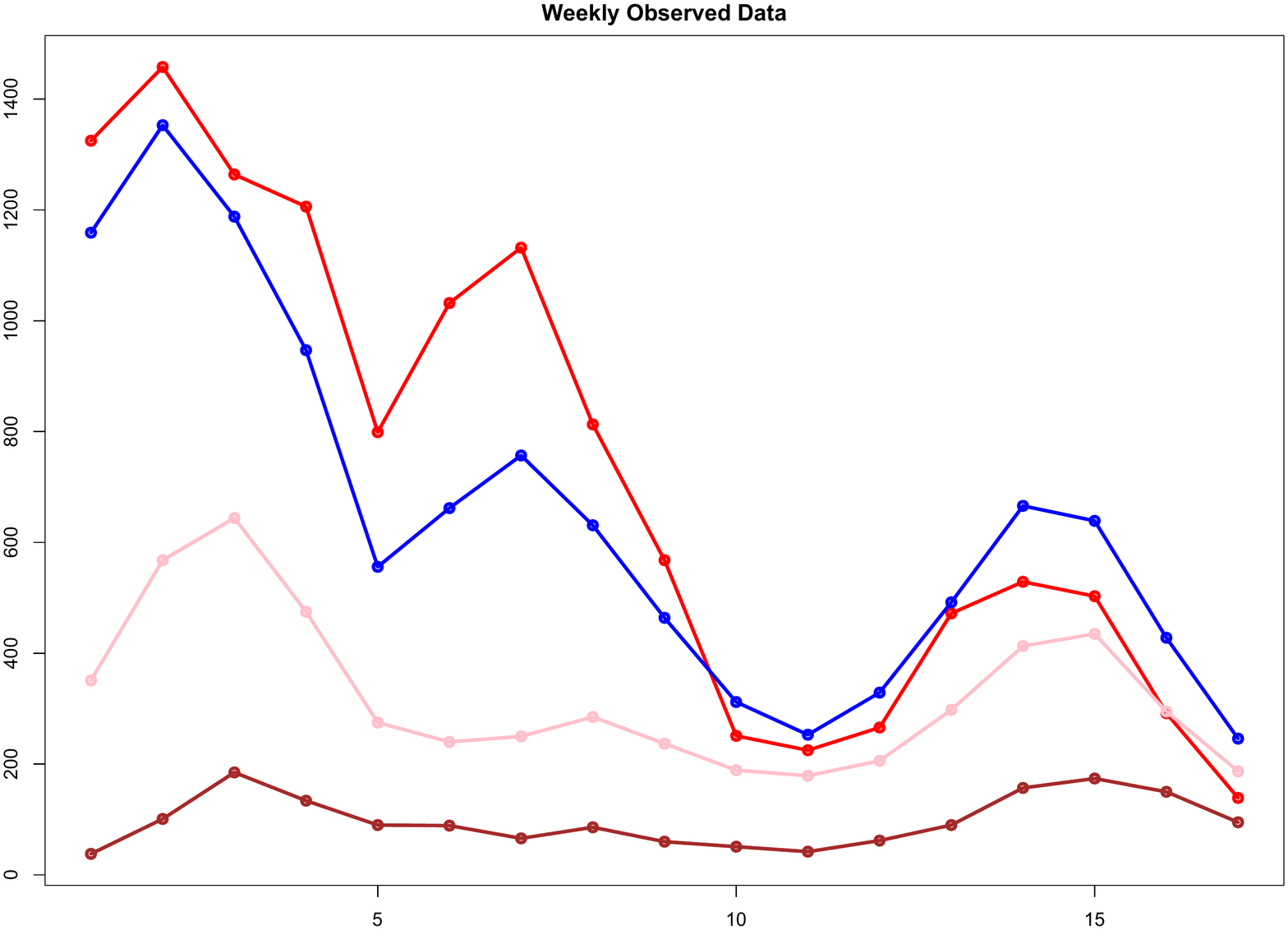}
    \caption{The weekly observed cases in Kingston } 
  \end{subfigure}
  \begin{subfigure}{7cm}
    \centering\includegraphics[width=6cm]{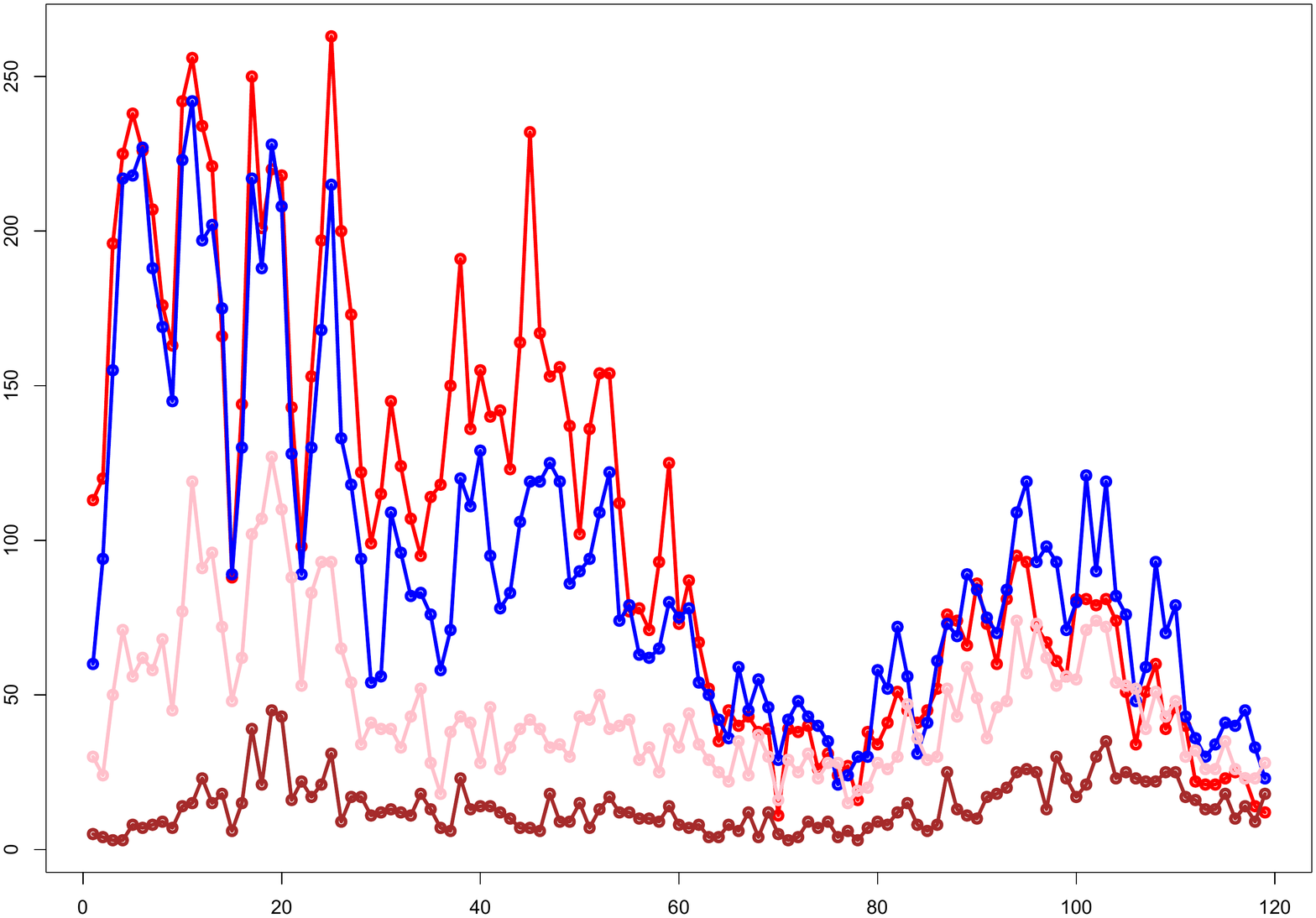}
   \caption{The daily observed cases in Kingston}
  \end{subfigure}
  \begin{subfigure}{7cm}
    \centering\includegraphics[width=6cm]{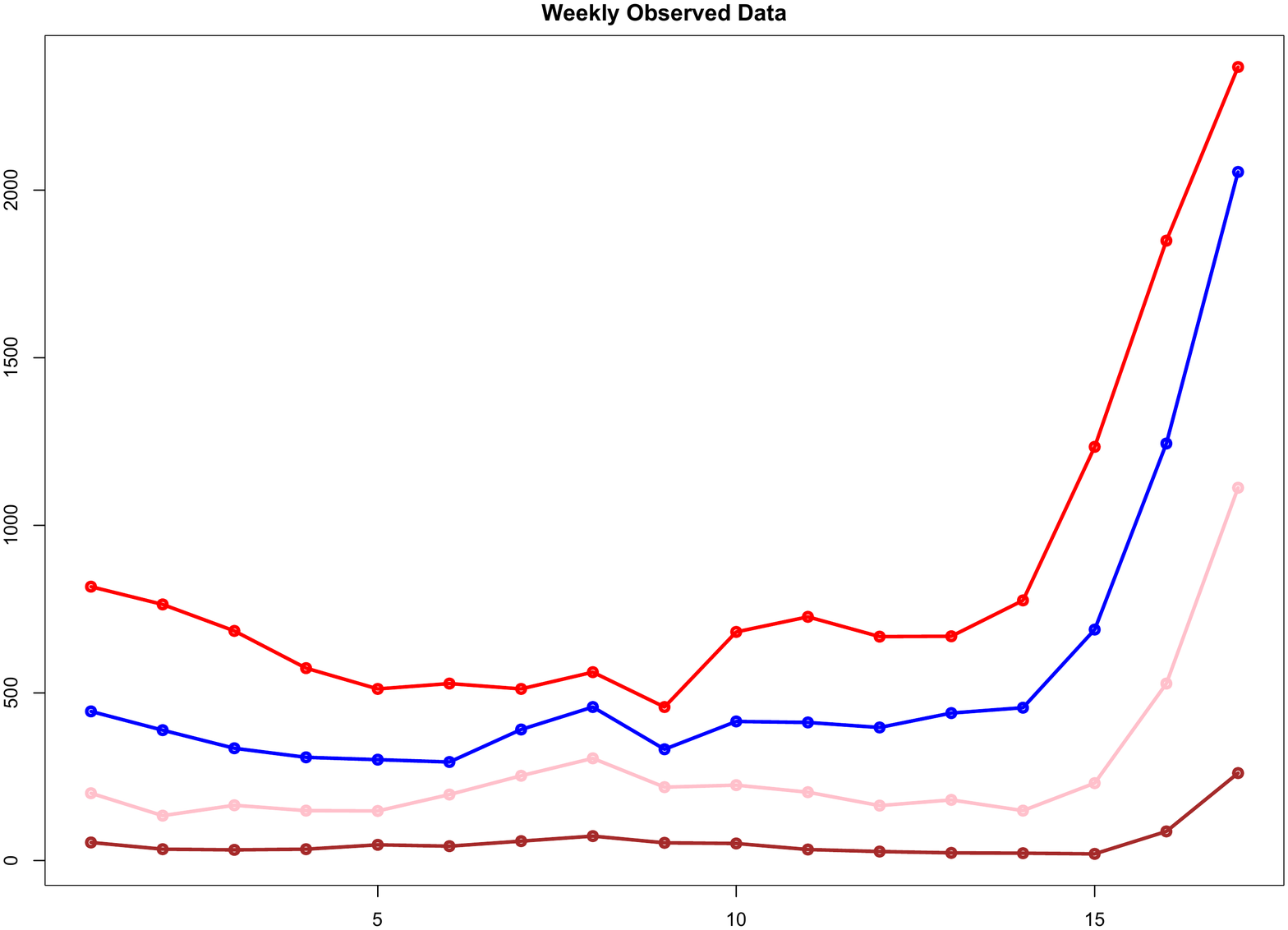}
    \caption{The weekly observed cases in Leicester } 
  \end{subfigure}
  \begin{subfigure}{7cm}
    \centering\includegraphics[width=6cm]{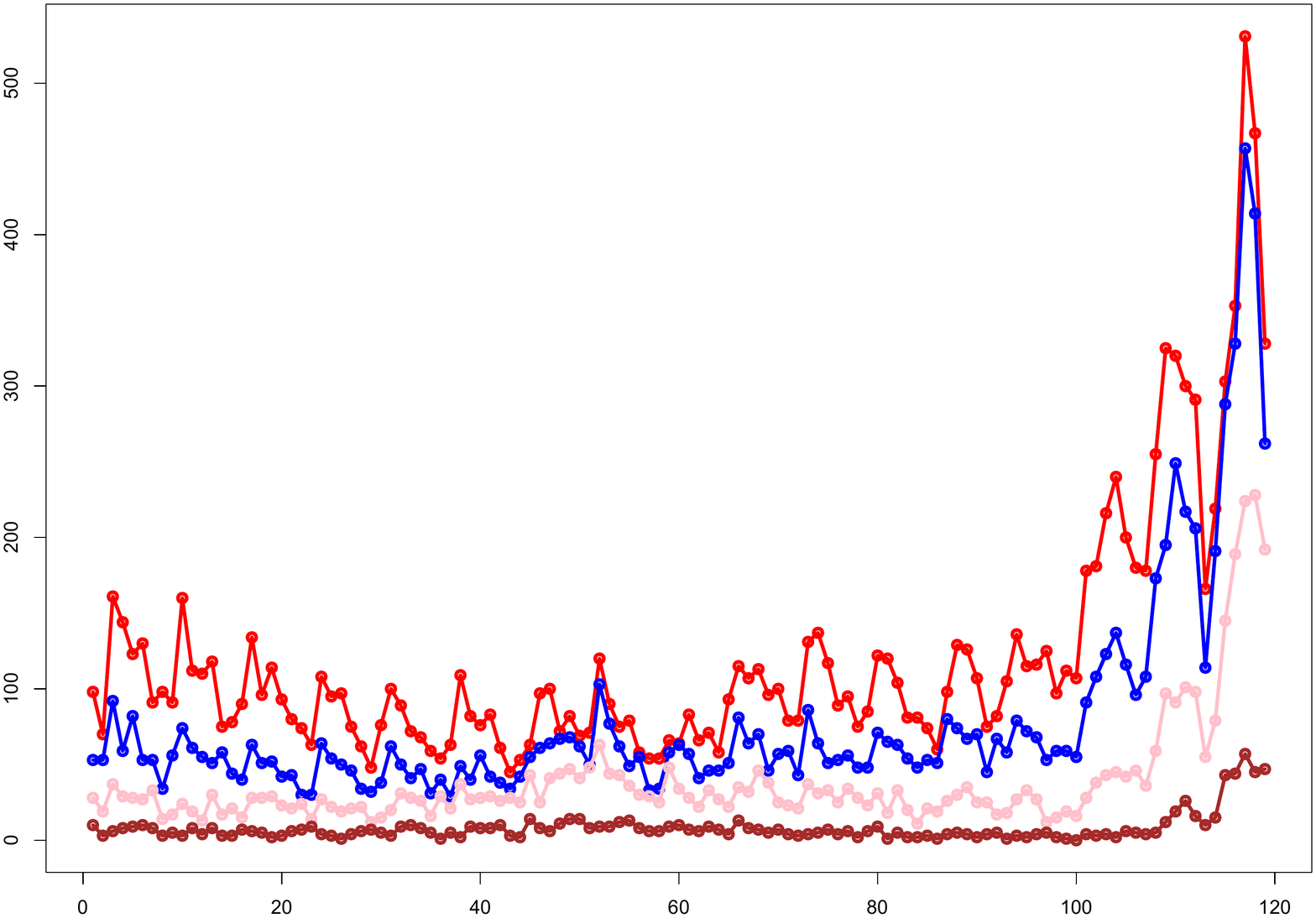}
   \caption{The daily observed cases in Leicester}
  \end{subfigure}
   \caption{\bf{The weekly and observed cases in age groups 0-29 (red line), 30-49 (blue line), 50-69 (pink line) and  70+ (brown line) in the local authorities.}}
  \label{ObsData_RDAM}
\end{figure}

We apply the KDPF (Algorithm \ref{APAlg}) to real cases among individuals aged 0 to 29 ($0-29$), 30 to 49 ($30-49$), 50 to 69 ($50-69$) and $70+$ years in the local authorities: Leicester (4/9/2021 -24/12/2021), Kinston upon the Thames (11/12/2021 -1/4/2022) and Ashford (19/12/2021 - 9/4/2022) (see Figure \ref{ObsData_RDAM})~\cite{link_GOVUK}. Estimates of the local authorities' population are available from ONS~\cite{ONS::Pop}. We deal with 16 hidden states $\{X_n\}_{n=1}^{16}$ and 16 subintervals $\{ \mathcal{T}_n \}_{n=1}^{16}$; each subinterval corresponds to the duration of one week. We infer the latent intensity $\lambda^N(t,a)$ and the instantaneous reproduction number, $R_a(t)$ per age group $a$ and the weights, $\{\gamma_{na}\}_a$ as well as the weekly and daily latent cases via the  particle sample derived by drawing samples from the smoothing density with lag equal to 4. We also infer the instantaneous reproduction number, $R(t)$ as the infected population-weighted average of $R_a(t)$. Appendix \ref{LA} includes the simulation study for the local authorities: Leicester and Ashford.
 
 We assume that the average number of secondary cases an individual would infect at the beginning of the first week is uniformly distributed over $[0.5,2]$ that includes the 90\% Confidence Interval published from the government in the UK: 0.9-1.1 on 4/9/2021 and 11/12/2021, 1-1.2 on 19/12/2021~\cite{link_GOVR}. Under this assumption, the weights $\{\gamma_{1a}\}_a$ are uniformly distributed over the interval from $2\left(\sum_{a_i}\sum_a\frac{S_{t,a}}{N_a}m_{aa_i}\right)^{-1}$ to $8\left(\sum_{a_i}\sum_a\frac{S_{t,a}}{N_a}m_{aa_i}\right)^{-1}$.  We also assume that 50\% of the infections are reported ($\beta=0.5$), $d_{min}=1$, $d_{max}=10$, $v_{min}=0.0001$ and $v_{max}=0.5$. We coarse the age groups of the contact matrix for reopening schools \citep{jarvis2020quantifying} to find the matrix $m$.

%We use the percentage of the population aged $a$ with levels of antibodies against SARS-CoV-2 at or above a threshold of 179 nanograms per millilitre (ng/ml), denoted by $p_a$, available from the ONS \citep{ONS::Antibodies}, to initialize the number of susceptibles aged $a$ at the beginning of the first week. The percentage of people 0-15 having antibodies at or above the threshold is not available. We assume the percentage is the same as that of 16-24. Following this methodology, $(1-p_a)N_a$ gives the susceptible population aged $a$ at the start of the first week. However, the estimated susceptible population is less than the total number of reported infections in Ashford and Kingston for some age bands. The antibodies might not fully protect against infection in December 2021; this could be due to declining immunity or immune escape (new variants being different from old variants and thus the previous infection being less protective against a new one). For this reason, we assume that a smaller percentage of the population aged $a$, given by $k_ap_a$ with $0<k_a<1$, has enough antibodies against a new infection in December 2021. We run Algorithm 10 for various values of $k_a$ and choose that value leading to the narrower 95\% CIs of the predicted reported cases in week 17. Summarizing the number of susceptibles aged $a$ at the beginning of the first week in Leicester (4/9/2021), Kingston upon Thames (11/12/2021) and Ashford (19/12/2021) are given by $(1-p_a)N_a$, $(1-0.4p_a)N_a$ and $(1-0.5p_a)N_a$, respectively.

We use the percentage of the population aged $a$ with levels of antibodies against SARS-CoV-2 at or above a threshold of 179 nanograms per millilitre (ng/ml), denoted by $p_a$, estimated by the lower 95\% credible interval available from~\cite{ONS::Antibodies}, to initialize the number of susceptibles aged $a$ at the beginning of the first week. The percentage of people 0-15 having antibodies at or above the threshold is not available. We assume the percentage is the same as that of 16-24. Following this methodology, $(1-p_a)N_a$ gives the susceptible population aged $a$ at the start of the first week. However, the estimated susceptible population is less than the total number of reported infections in Ashford and Kingston for some age bands. The antibodies might not fully protect against infection in December 2021; this could be due to declining immunity or immune escape (new variants being different from old variants and thus the previous infection being less protective against a new one). For this reason, we assume that a smaller percentage of the population aged $a$, given by $p_a-z$ with $0<z<\min_a{p_a}$, has enough antibodies against a new infection in December 2021. We choose the minimum value of $z$ so that the susceptible population is at least twice as large as the reported cases for each age group to maintain consistency with our assumption that we see 50\% of the infections. Summarizing the number of susceptibles aged $a$ at the beginning of the first week in Leicester (4/9/2021), Kingston upon Thames (11/12/2021) and Ashford (19/12/2021) are given by $(1-p_a)N_a$, $(1-p_a+0.2)N_a$ and $(1-p_a+0.2)N_a$, respectively. 

\begin{figure}[!h] 
  \begin{subfigure}{7cm}
    \centering\includegraphics[width=6cm]{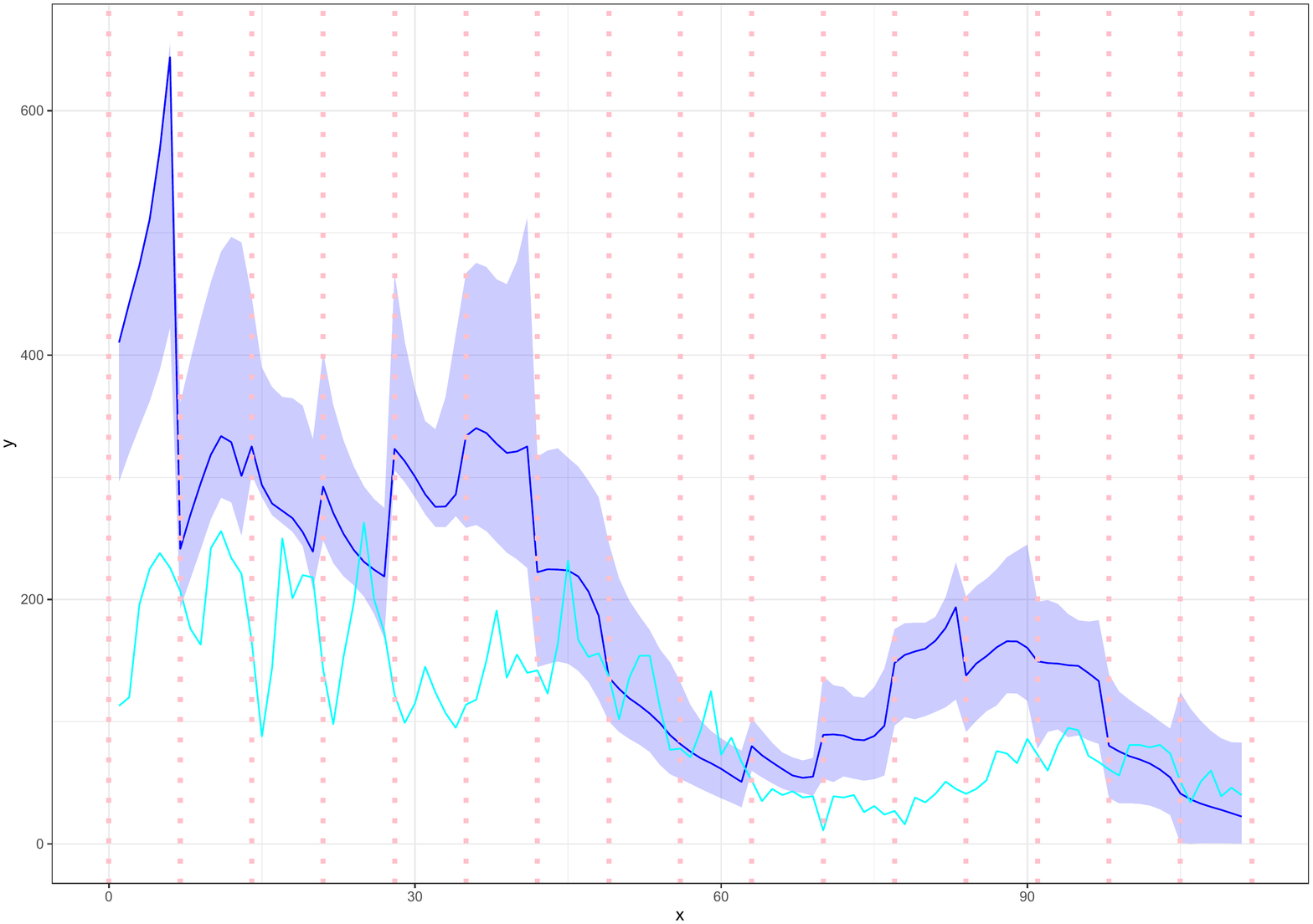}
   \caption{The estimated intensity of latent cases aged 0-29.}
  \end{subfigure}
  \begin{subfigure}{7cm}
    \centering\includegraphics[width=6cm]{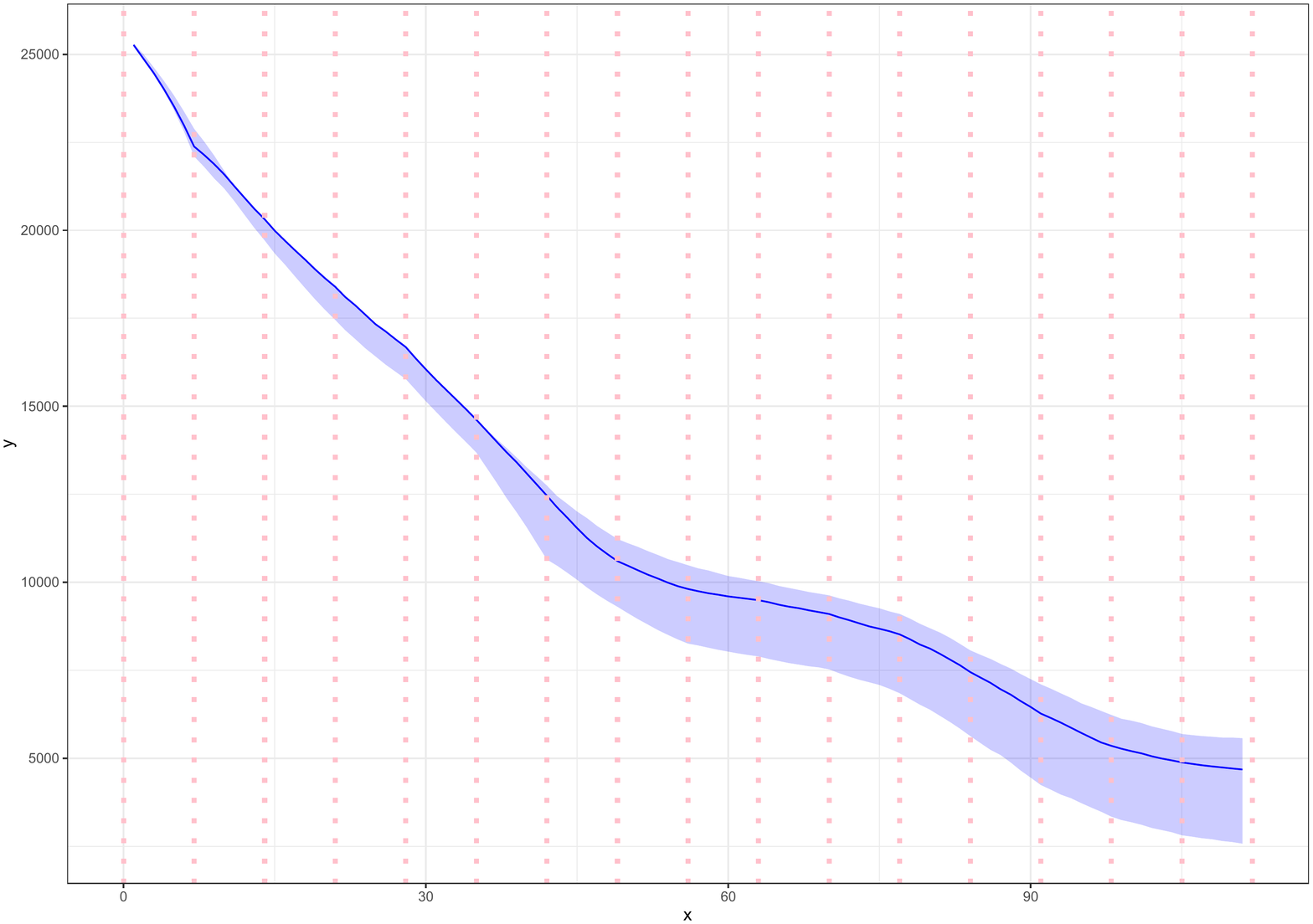}
   \caption{The estimated susceptibles aged 0-29.}
  \end{subfigure}
  \begin{subfigure}{7cm}
    \centering\includegraphics[width=6cm]{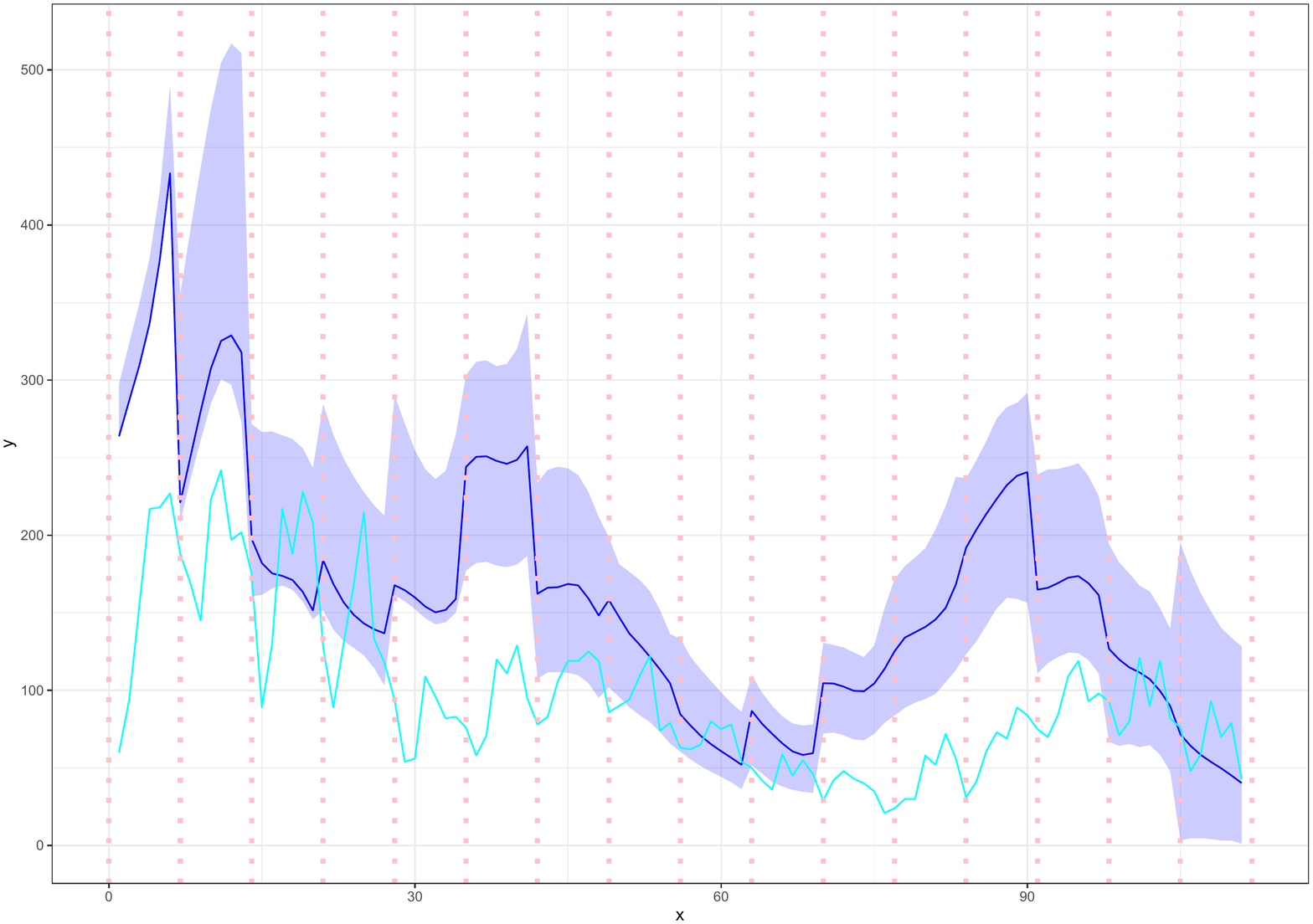}
   \caption{The estimated intensity of latent cases aged 30-49.}
  \end{subfigure}
  \begin{subfigure}{7cm}
    \centering\includegraphics[width=6cm]{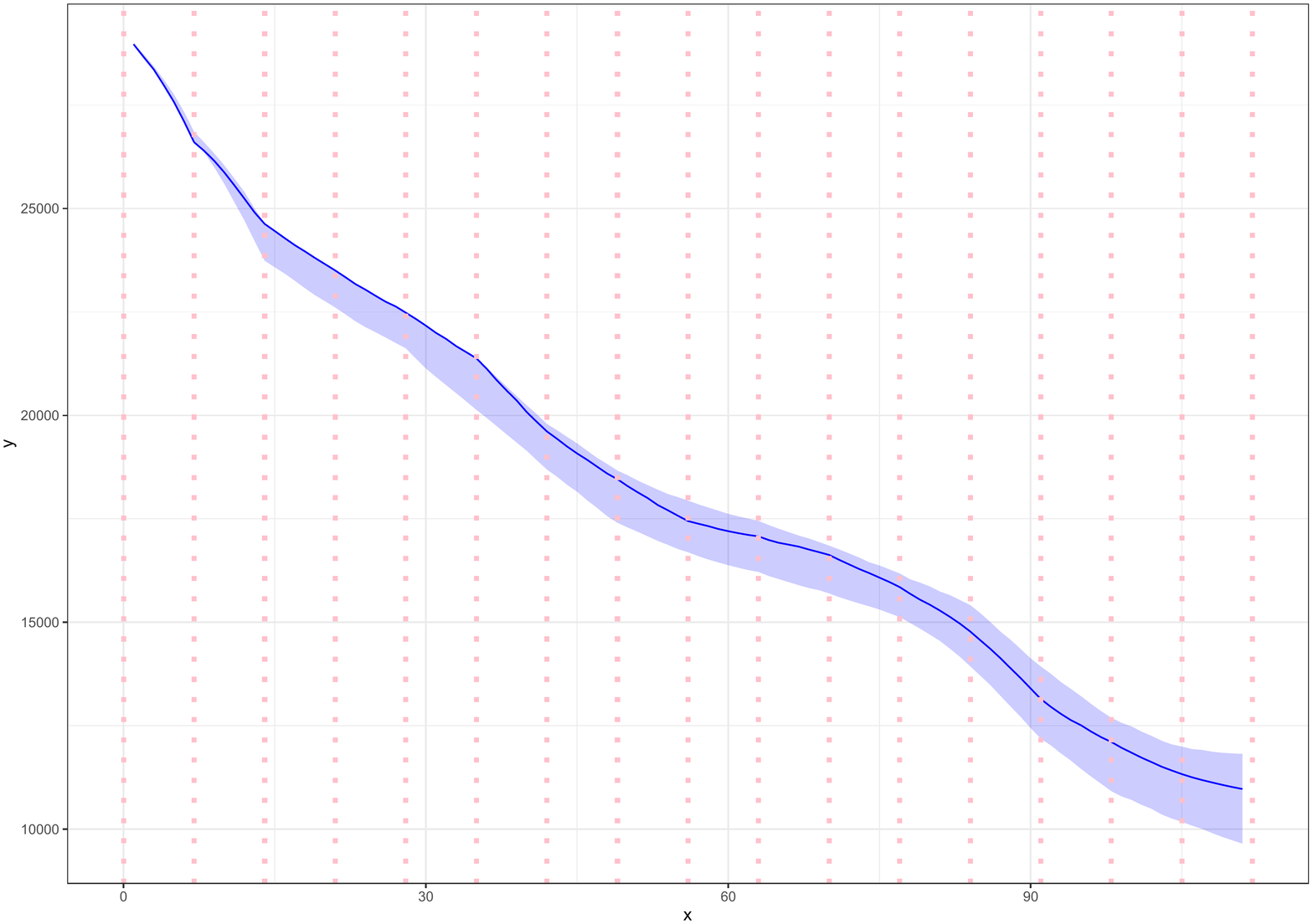}
   \caption{The estimated susceptibles aged 30-49.}
  \end{subfigure}
   \begin{subfigure}{7cm}
    \centering\includegraphics[width=6cm]{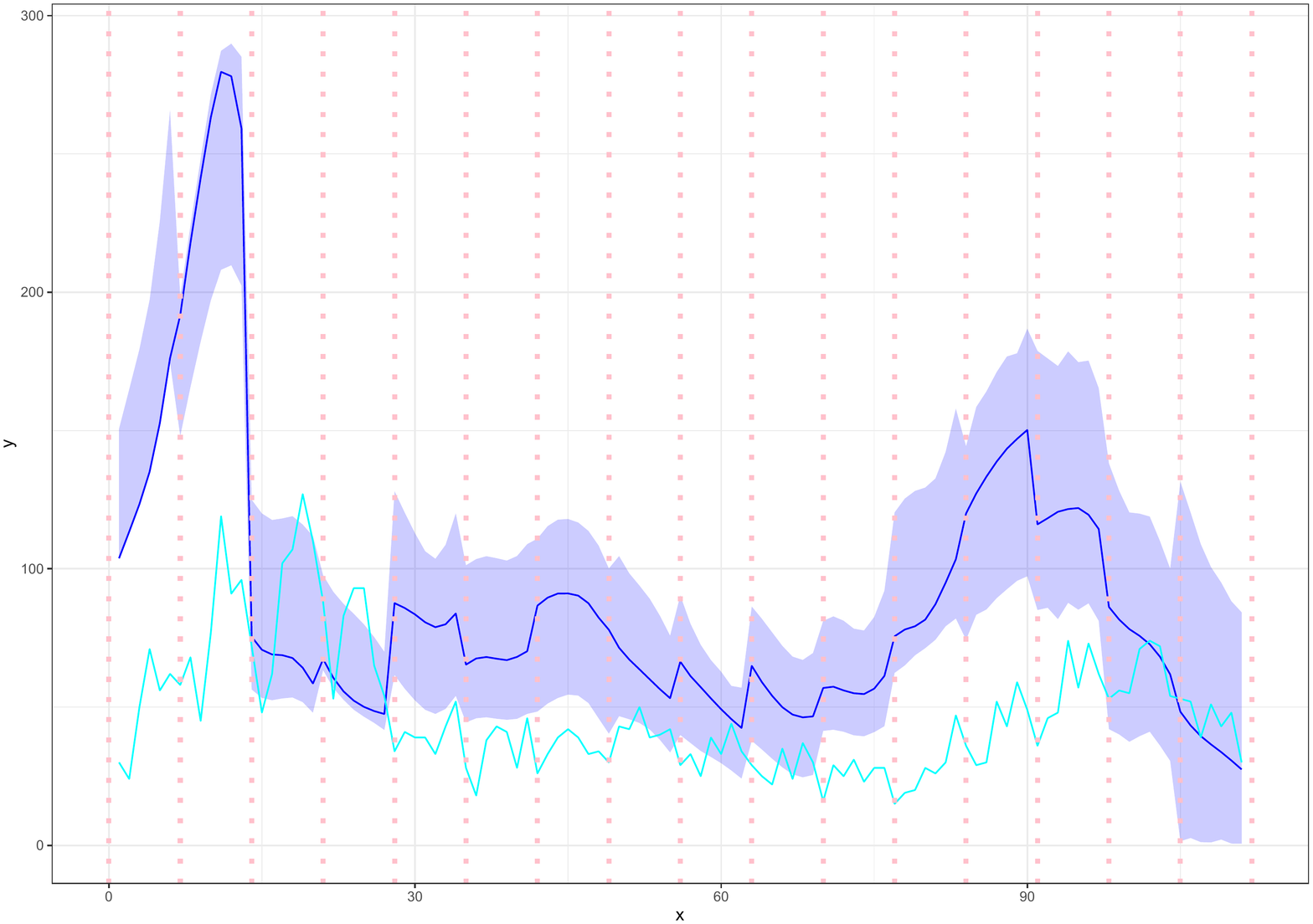}
   \caption{The estimated intensity of latent cases aged 50-69.}
  \end{subfigure}
  \begin{subfigure}{7cm}
    \centering\includegraphics[width=6cm]{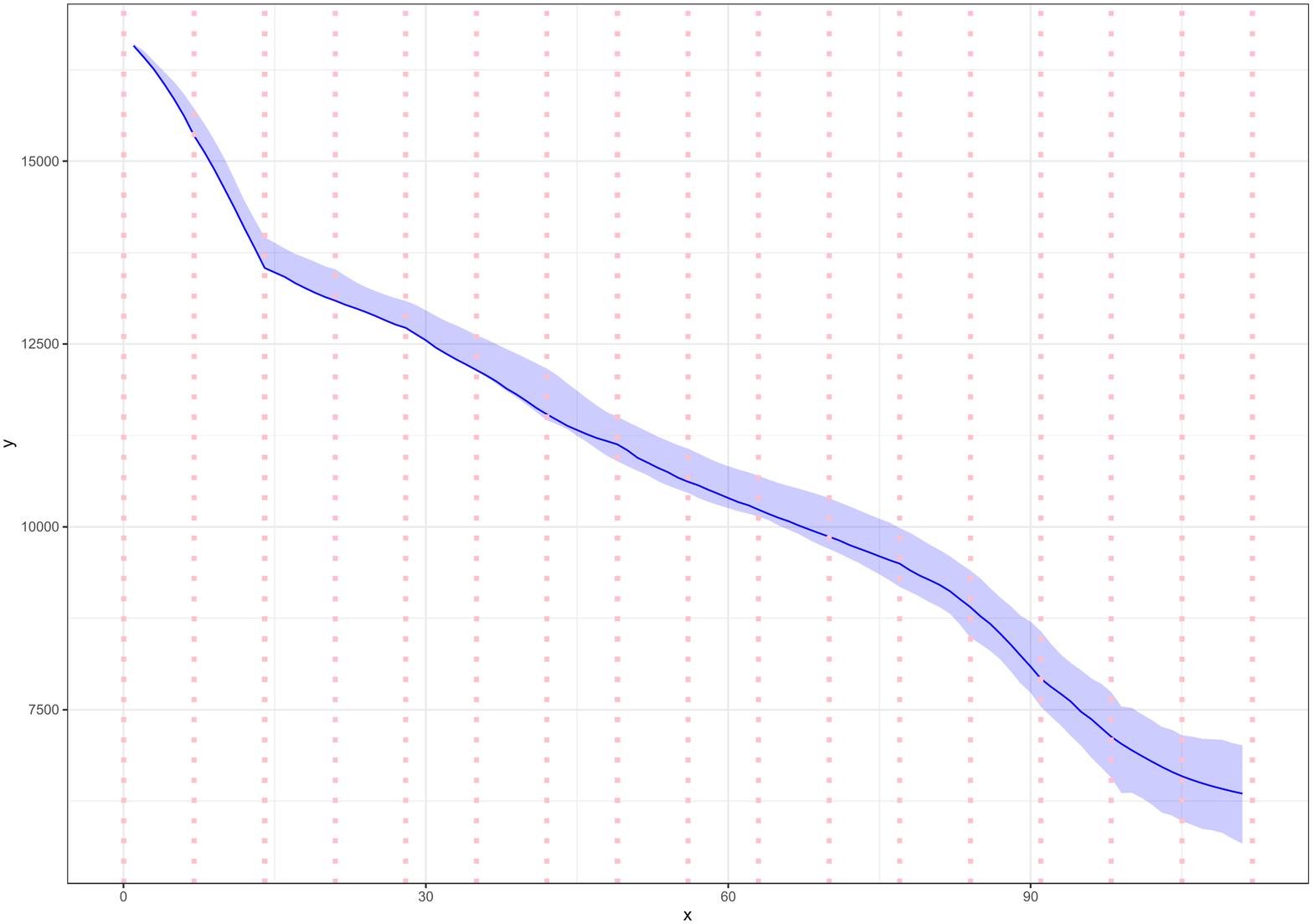}
   \caption{The estimated susceptibles aged 50-69.}
  \end{subfigure}
   \begin{subfigure}{7cm}
    \centering\includegraphics[width=6cm]{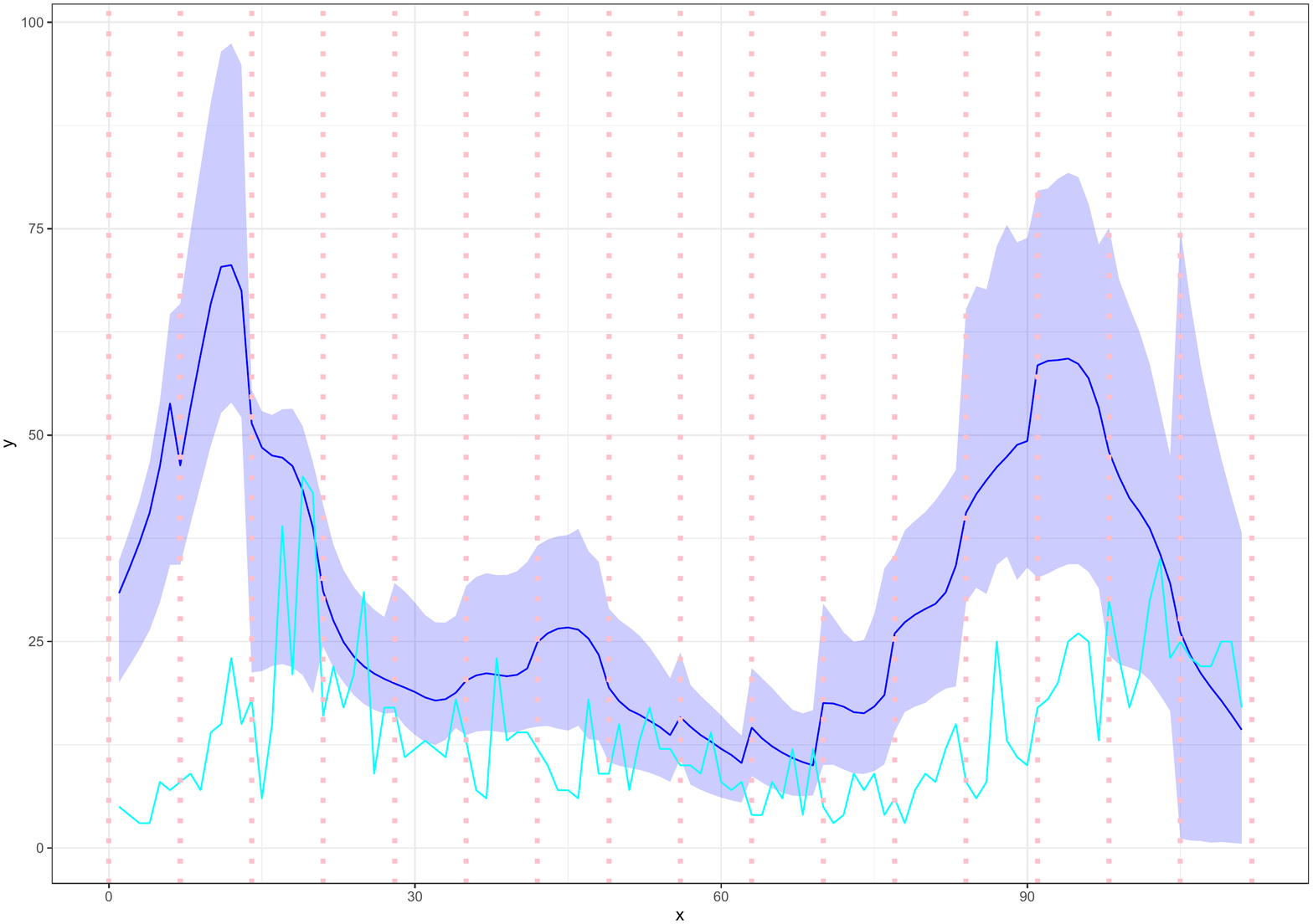}
   \caption{The estimated intensity of latent cases aged 70+.}
  \end{subfigure}
  \begin{subfigure}{7cm}
    \centering\includegraphics[width=6cm]{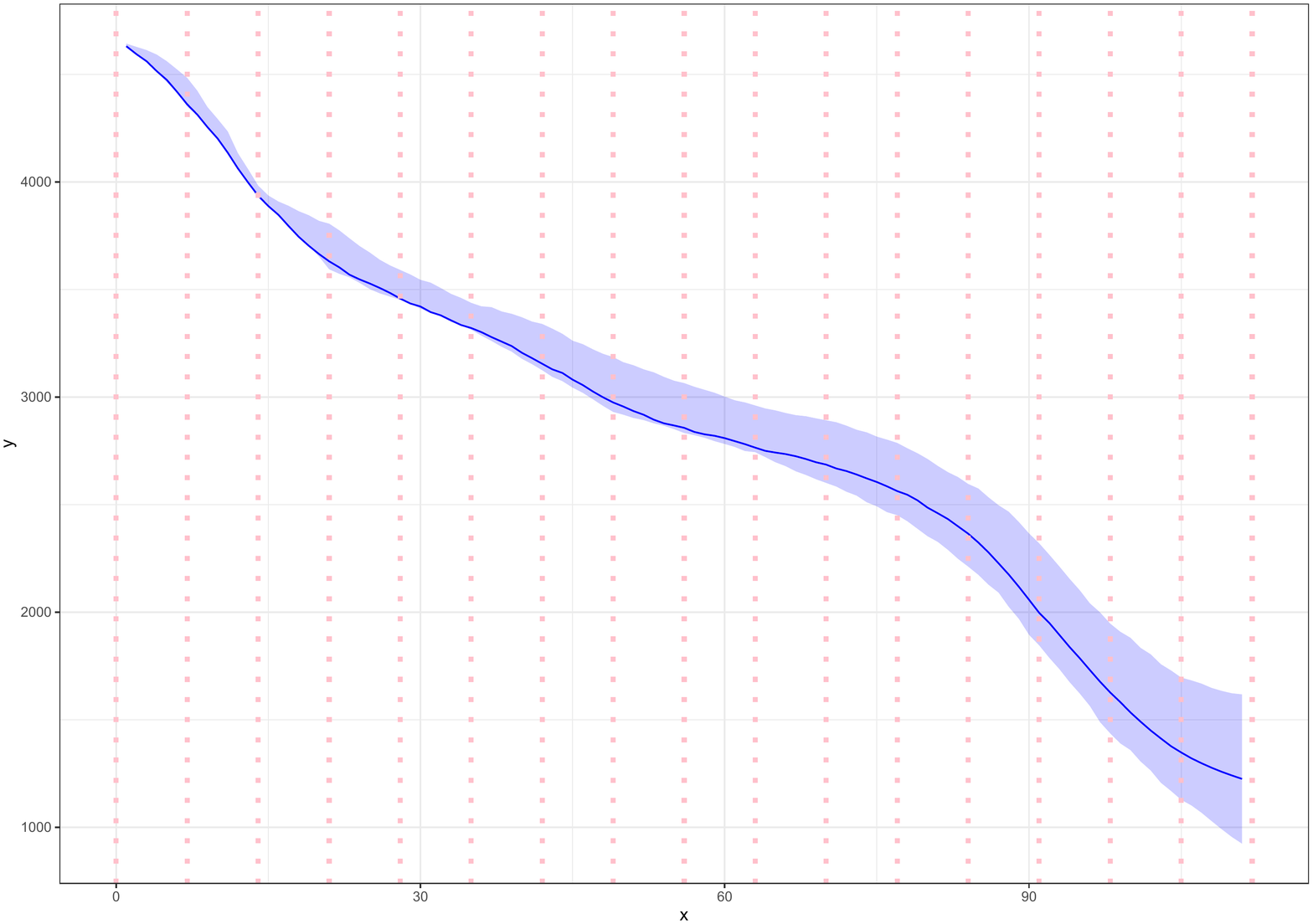}
   \caption{The estimated susceptibles aged 70+.}
  \end{subfigure}
  
  \caption{{\bf The estimated latent intensity and susceptibles (posterior median (blue line) ; 99\% CI (ribbon)) and the daily observed cases (cyan line) in Kingston.} The vertical dotted lines show the beginning of each week in the period we examine. }
  \label{EstInt_Kings4G}
\end{figure}

\begin{figure}[!h] 
  \begin{subfigure}{7cm}
    \centering\includegraphics[width=6cm]{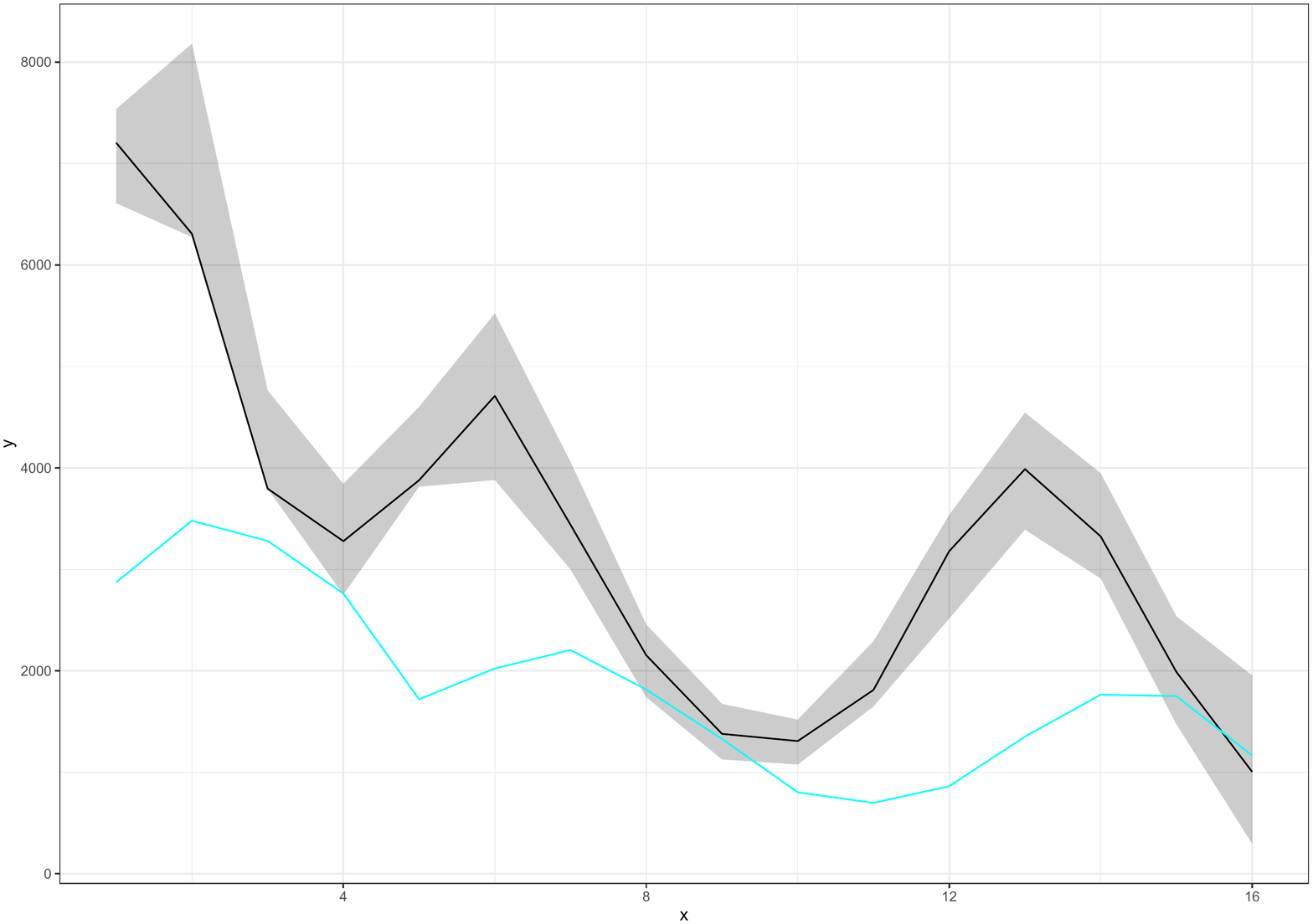}
   \caption{The estimated aggregated weekly hidden \\cases.}
  \end{subfigure}
  \begin{subfigure}{7cm}
    \centering\includegraphics[width=6cm]{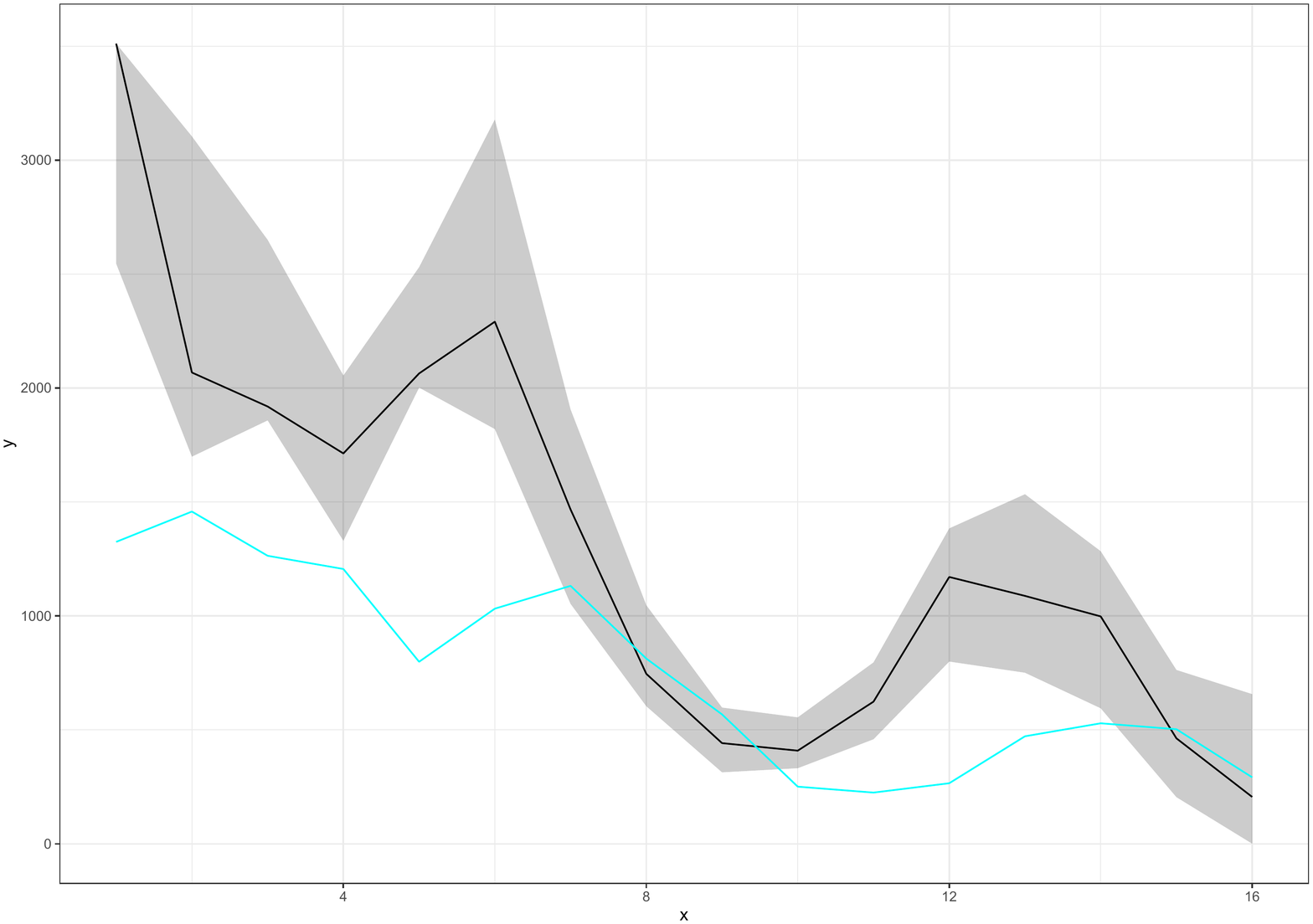}
    \caption{The estimated weekly hidden cases aged 0-29.}
  \end{subfigure}
  \begin{subfigure}{7cm}
    \centering\includegraphics[width=6cm]{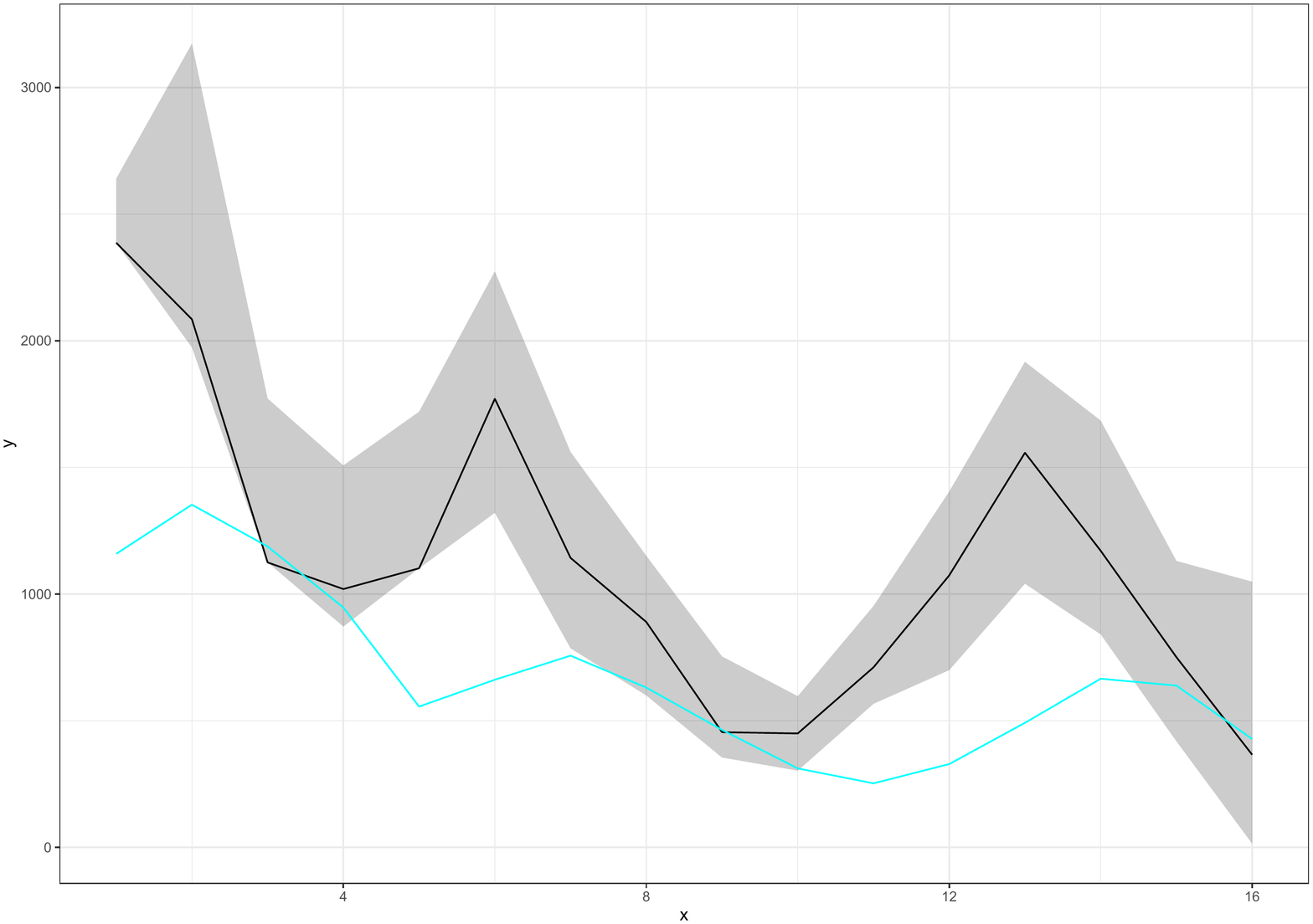}
    \caption{The estimated weekly hidden cases aged 30-49.}
  \end{subfigure}
  \begin{subfigure}{7cm}
    \centering\includegraphics[width=6cm]{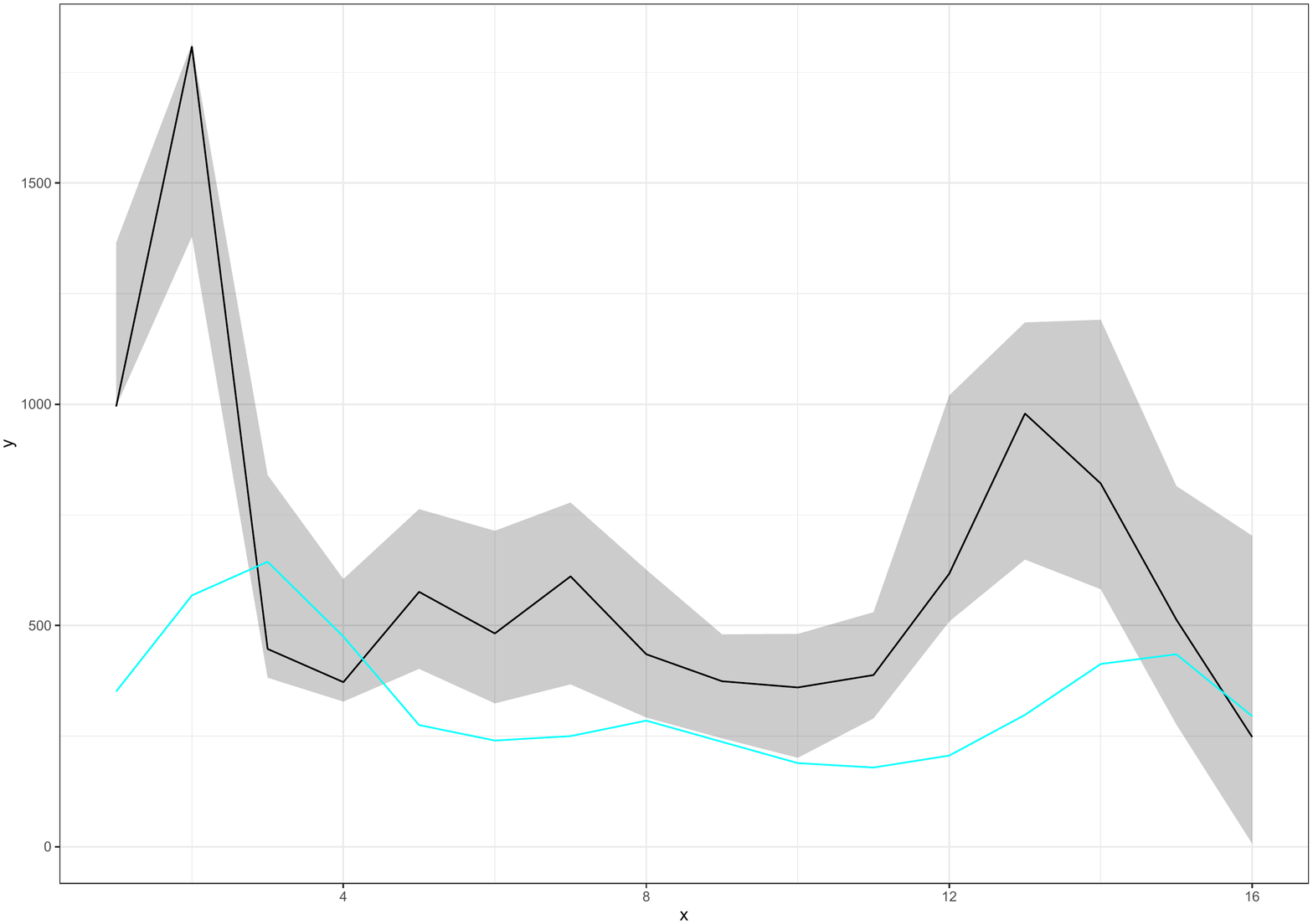}
    \caption{The estimated weekly hidden cases aged 50-69.}
  \end{subfigure}
  \begin{subfigure}{7cm}
    \centering\includegraphics[width=6cm]{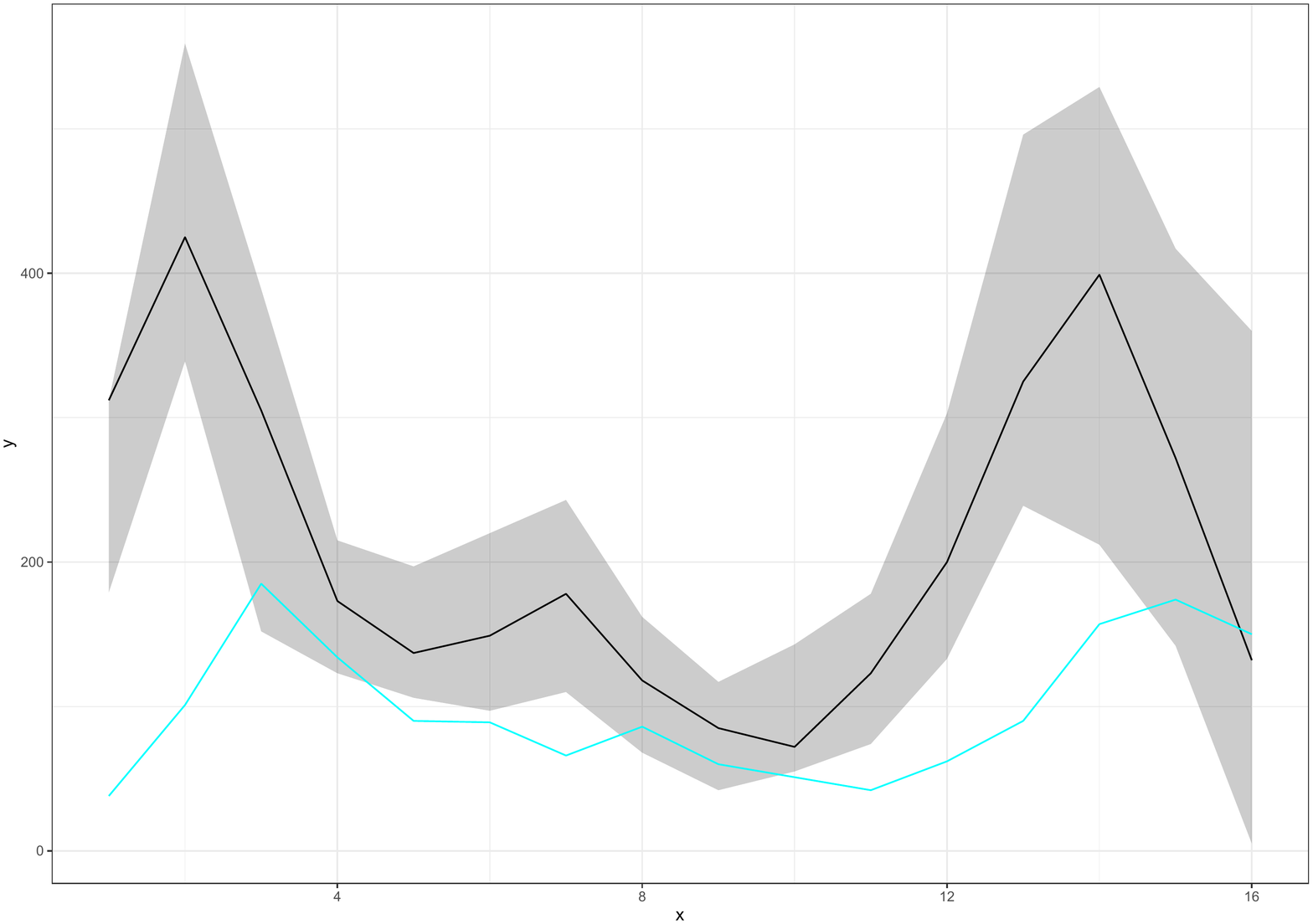}
    \caption{The estimated weekly hidden cases aged 70+.}
  \end{subfigure}
   \caption{\bf {The estimated weekly latent cases (black line; 99\% CI (ribbon)) and the weekly observed cases (cyan line) in Kingston.}}
   \label{EHC_Kings4G}
\end{figure}

\begin{figure}[!h] 
   \begin{subfigure}{7cm}
    \centering\includegraphics[width=6cm]{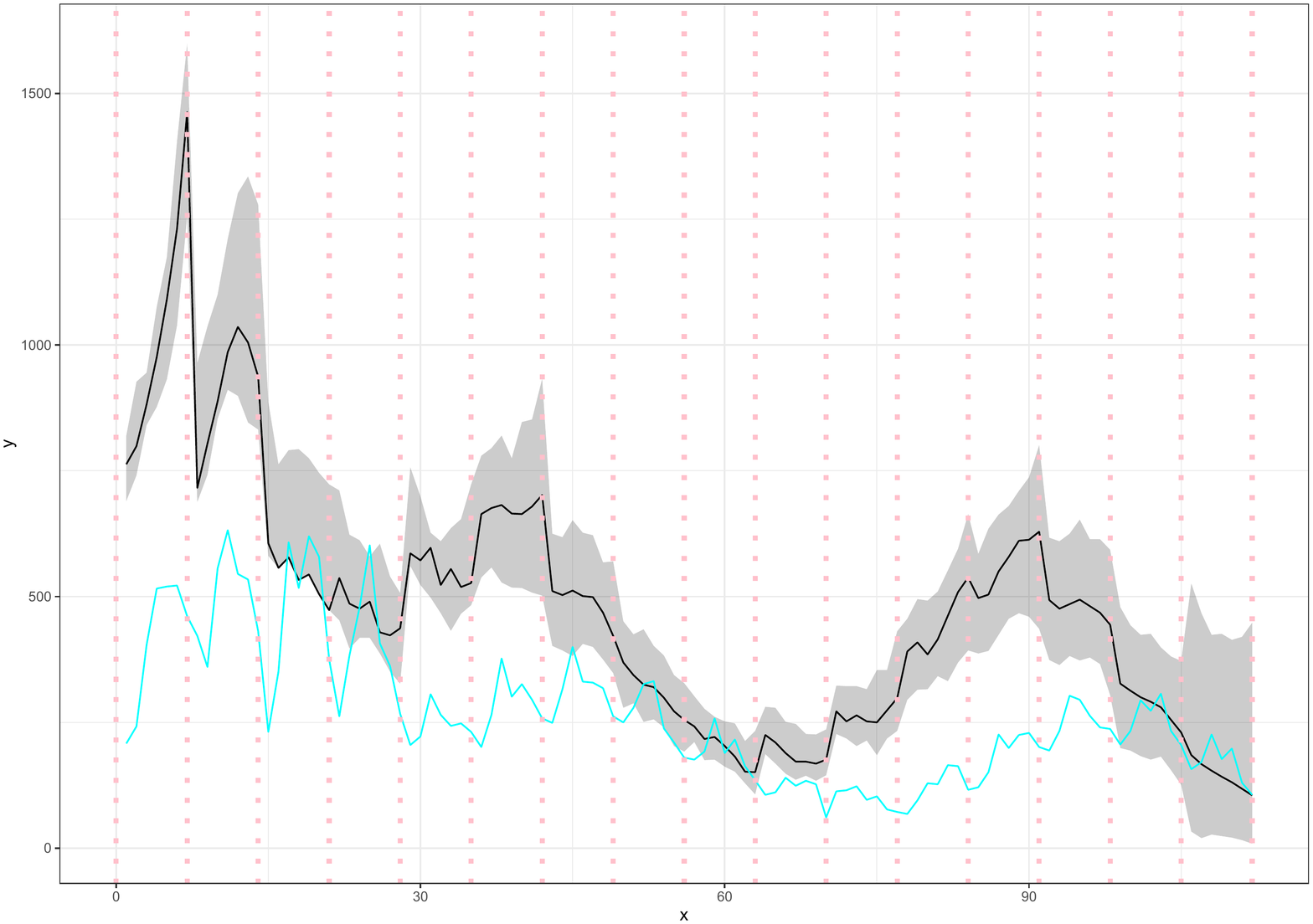}
    \caption{The estimated aggregated daily hidden cases.}
  \end{subfigure}
   \begin{subfigure}{7cm}
    \centering\includegraphics[width=6cm]{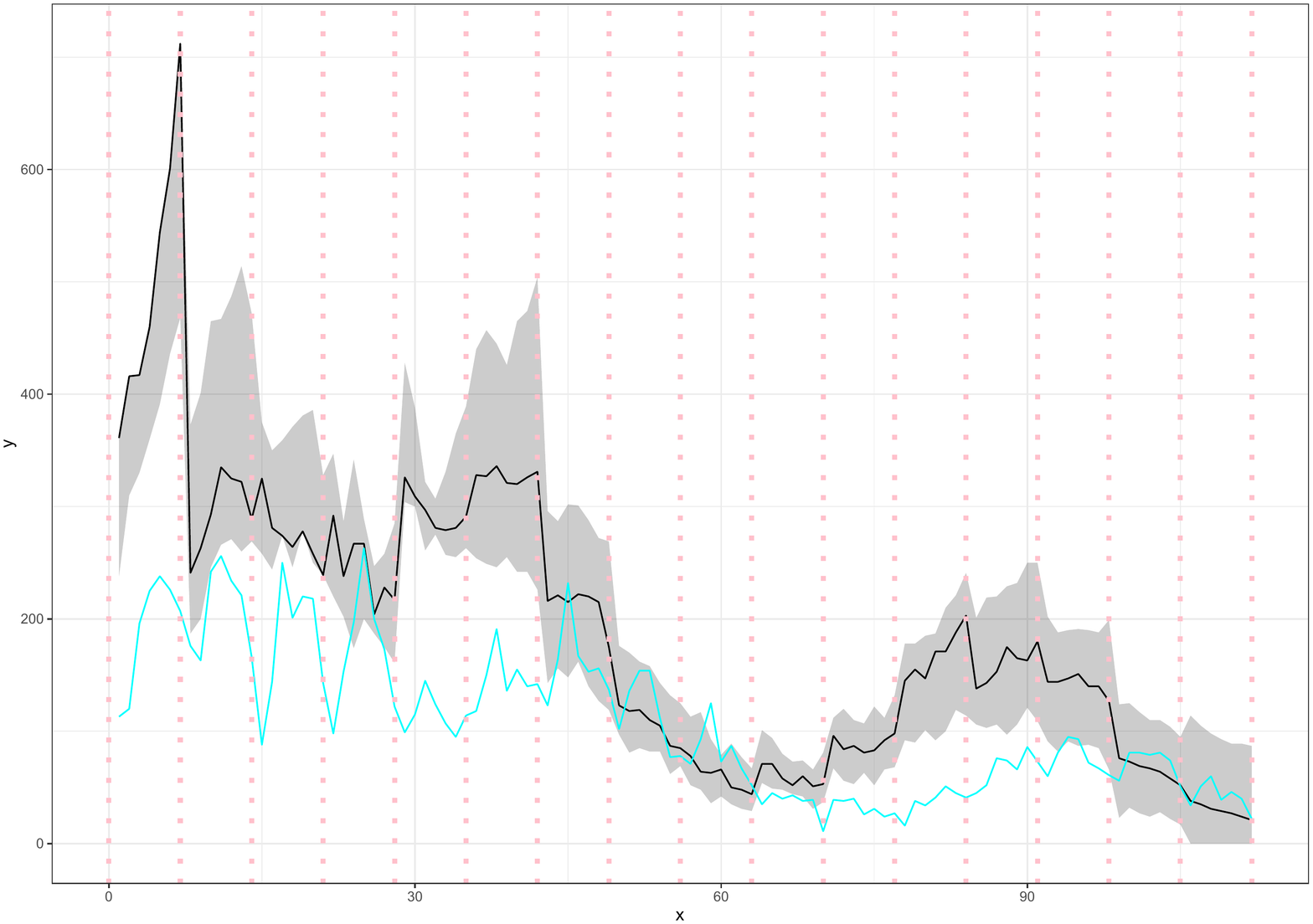}
    \caption{The estimated daily hidden cases aged 0-29.}
  \end{subfigure}
  \begin{subfigure}{7cm}
    \centering\includegraphics[width=6cm]{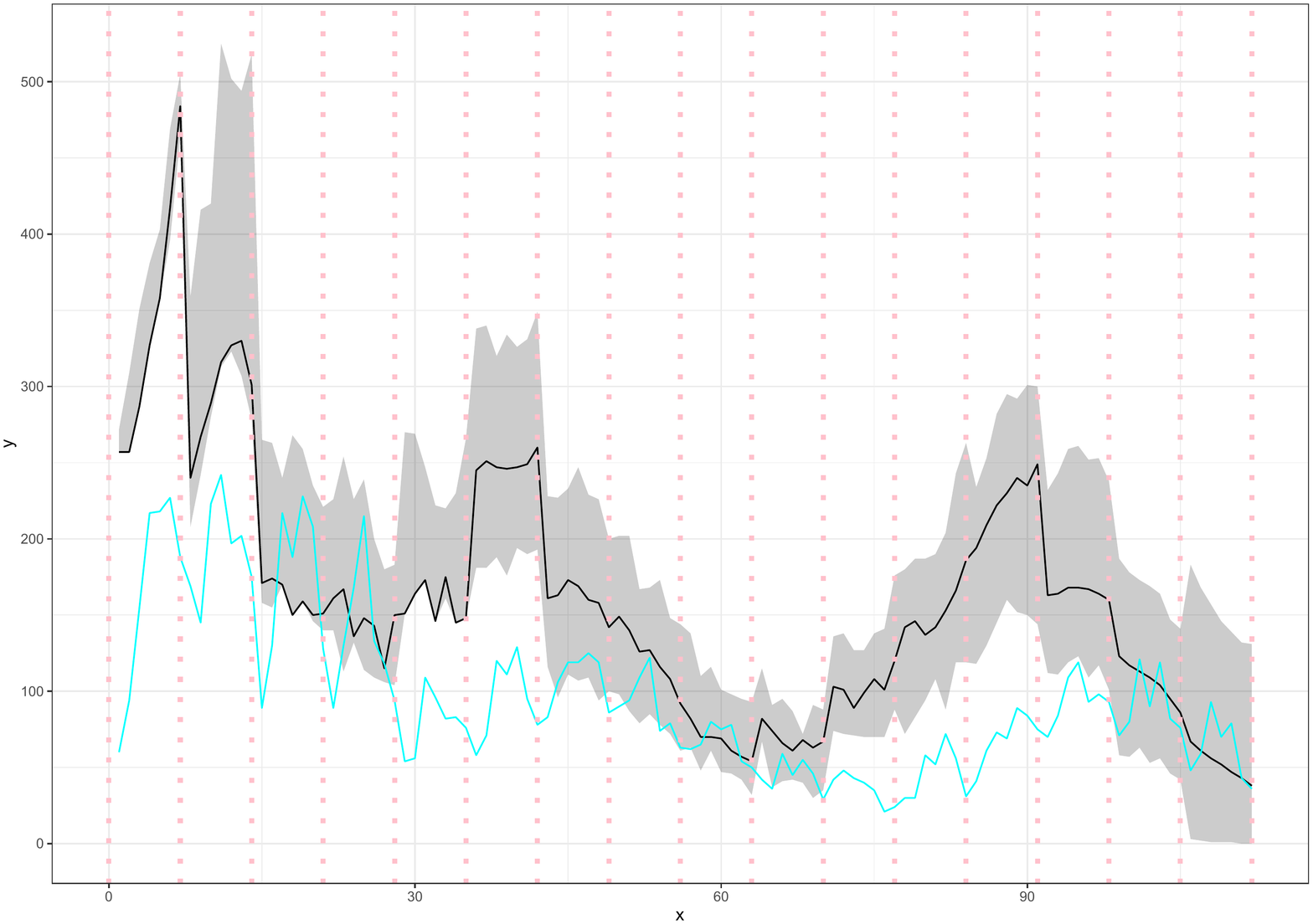}
    \caption{The estimated daily hidden cases aged 30-49.}
  \end{subfigure}
  \begin{subfigure}{7cm}
    \centering\includegraphics[width=6cm]{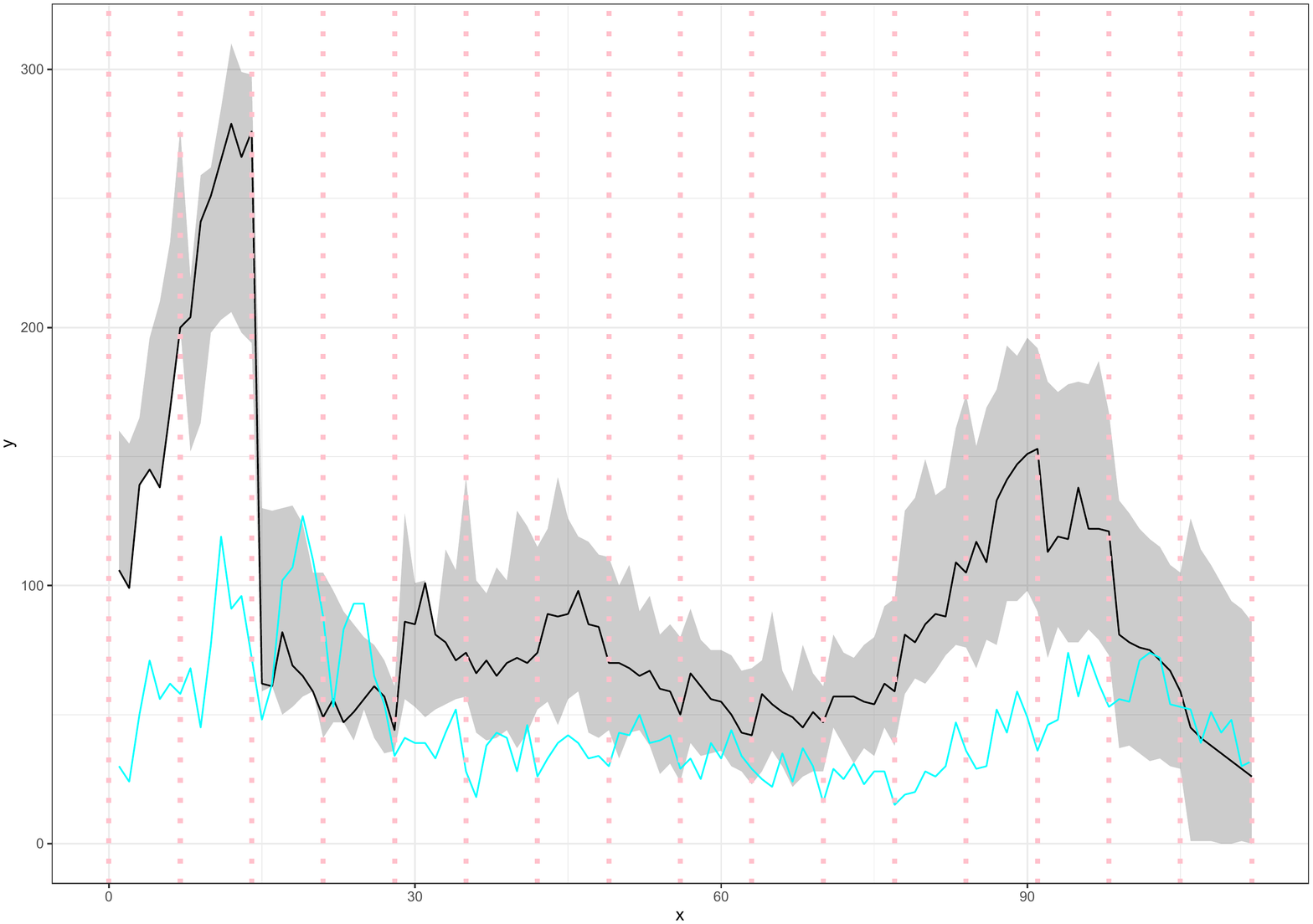}
    \caption{The estimated daily hidden cases aged 50-69.}
  \end{subfigure}
  \begin{subfigure}{7cm}
    \centering\includegraphics[width=6cm]{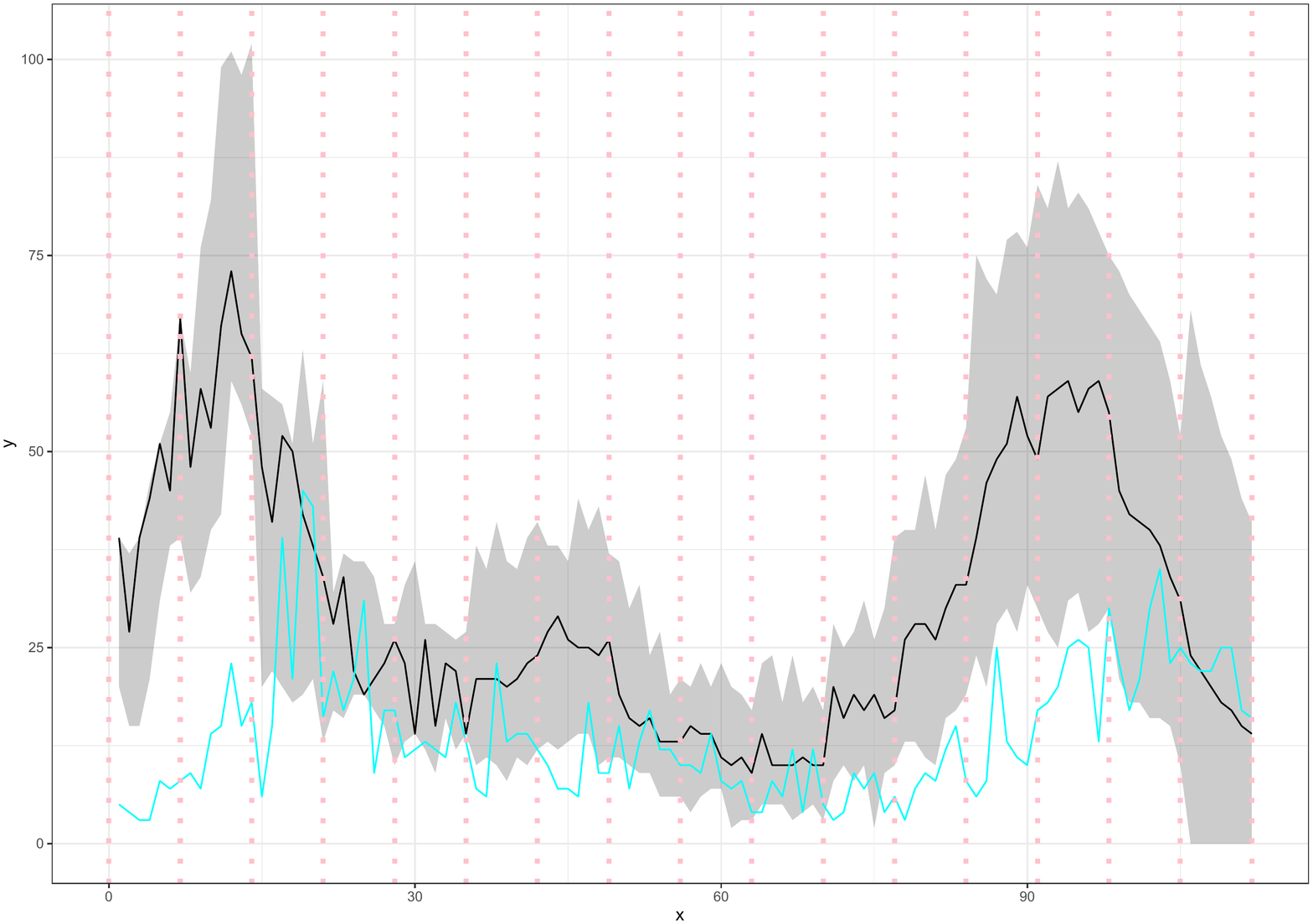}
    \caption{The estimated daily hidden cases aged 70+.}
  \end{subfigure}
   \caption{{\bf The estimated daily latent cases (posterior median (black line); 99\% CI (ribbon)) and the daily observed cases (cyan line) in Kingston.} The vertical dotted lines show the beginning of each week in the period we examine.}
   \label{EHDC_Kings4G}
\end{figure}

\begin{figure}[!h] 
 \begin{subfigure}{7cm}
    \centering\includegraphics[width=6cm]{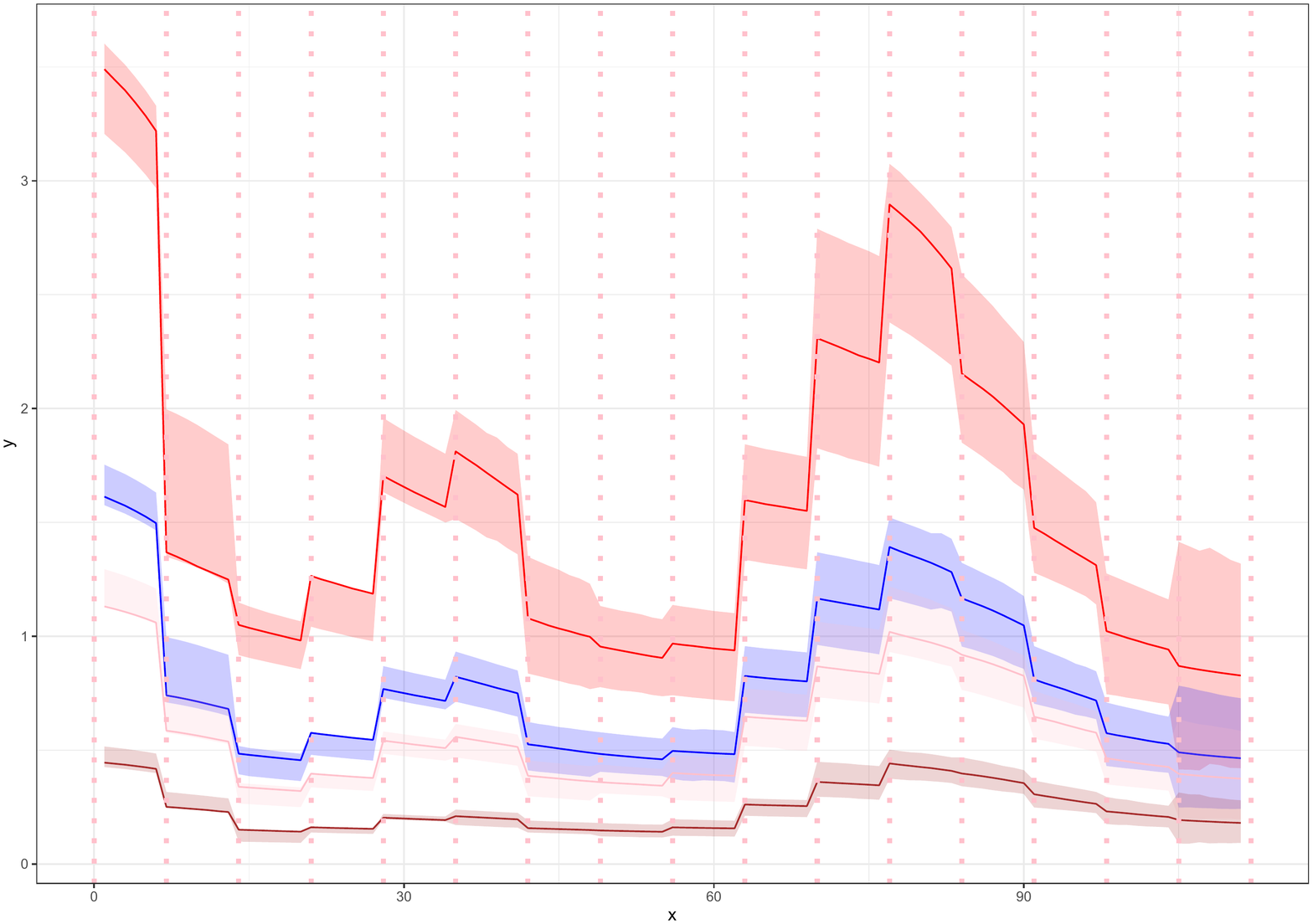}
     \caption{The per age group instantaneous reproduction numbers.}
  \end{subfigure}
  \begin{subfigure}{7cm}
    \centering\includegraphics[width=6cm]{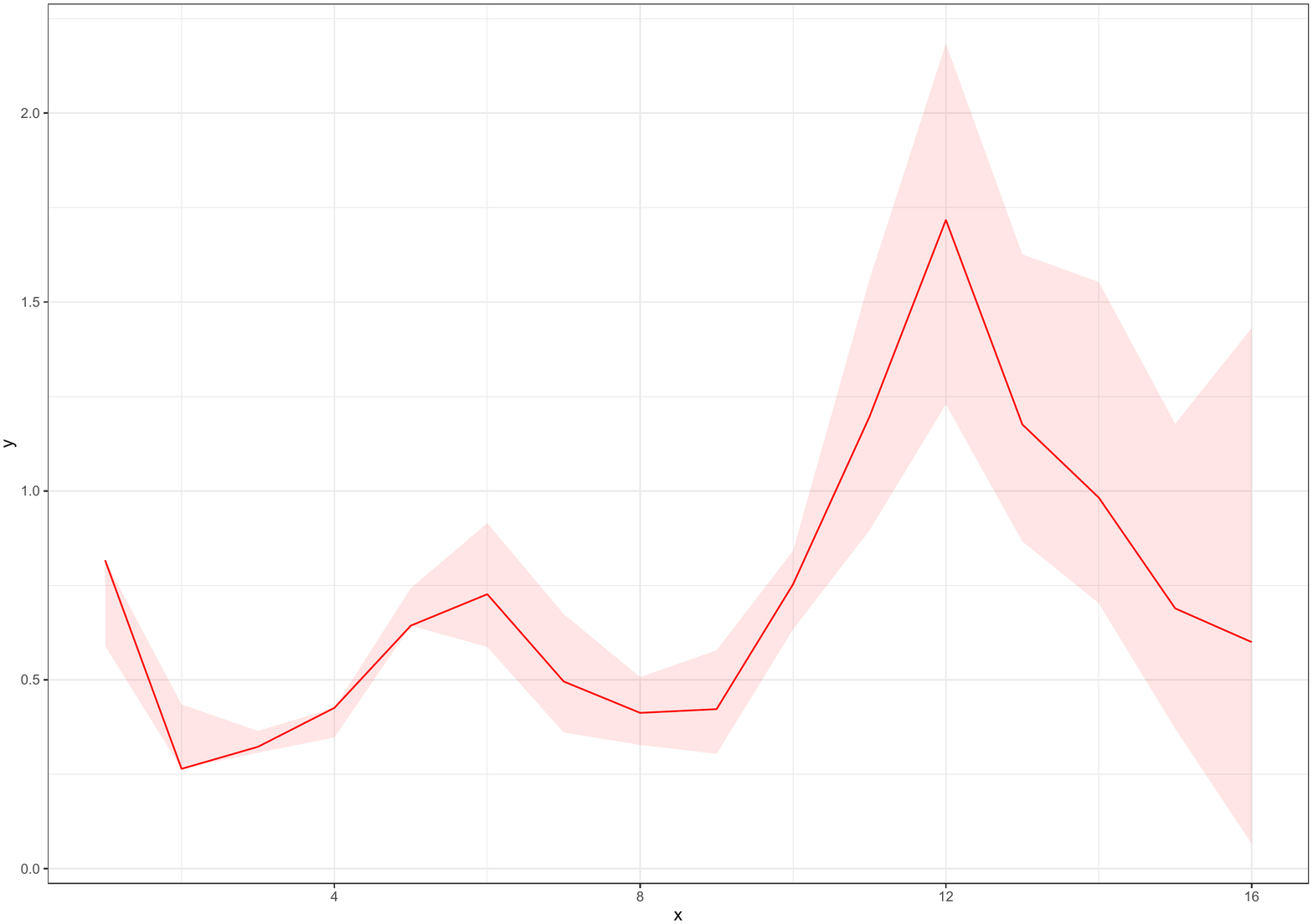}
   \caption{The estimated weights $\{\gamma_{i,0-29}\}_{i=1}^{16}$.}
  \end{subfigure}
  \begin{subfigure}{7cm}
    \centering\includegraphics[width=6cm]{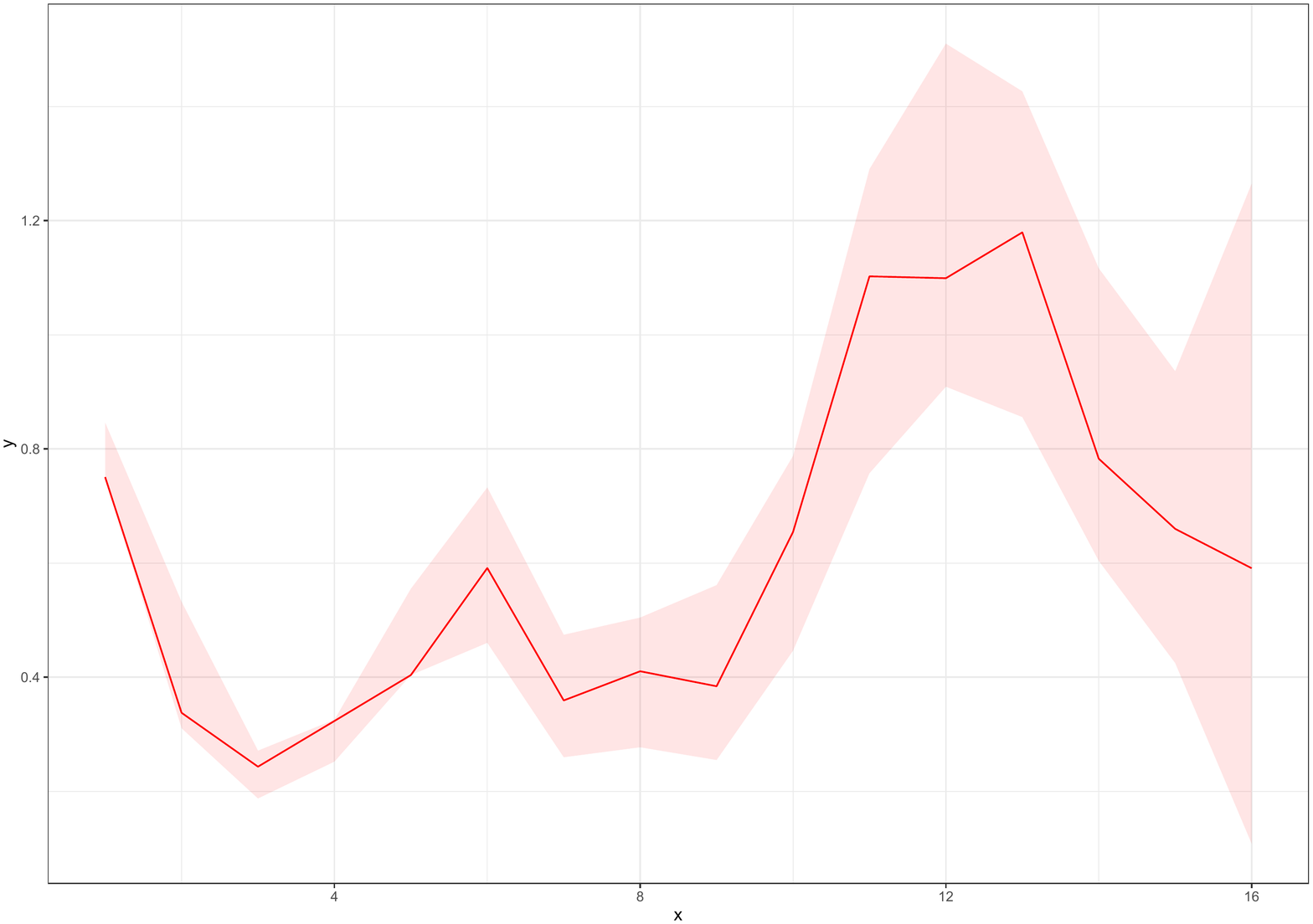}
     \caption{The estimated weights $\{\gamma_{i,30-49}\}_{i=1}^{16}$.}
  \end{subfigure}
  \begin{subfigure}{7cm}
    \centering\includegraphics[width=6cm]{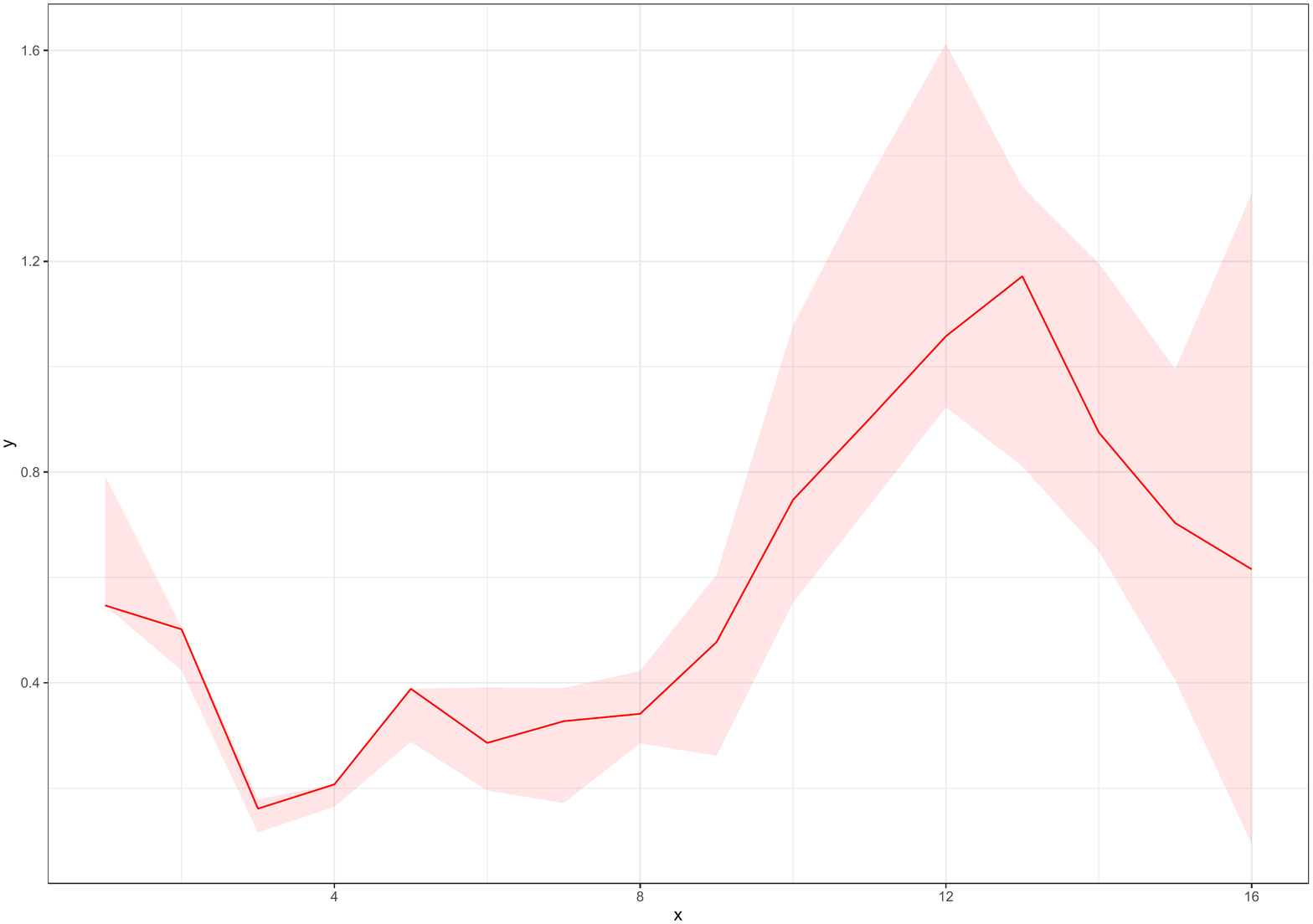}
    \caption{The estimated weights $\{\gamma_{i,50-69}\}_{i=1}^{16}$.}
  \end{subfigure}
   \begin{subfigure}{7cm}
    \centering\includegraphics[width=6cm]{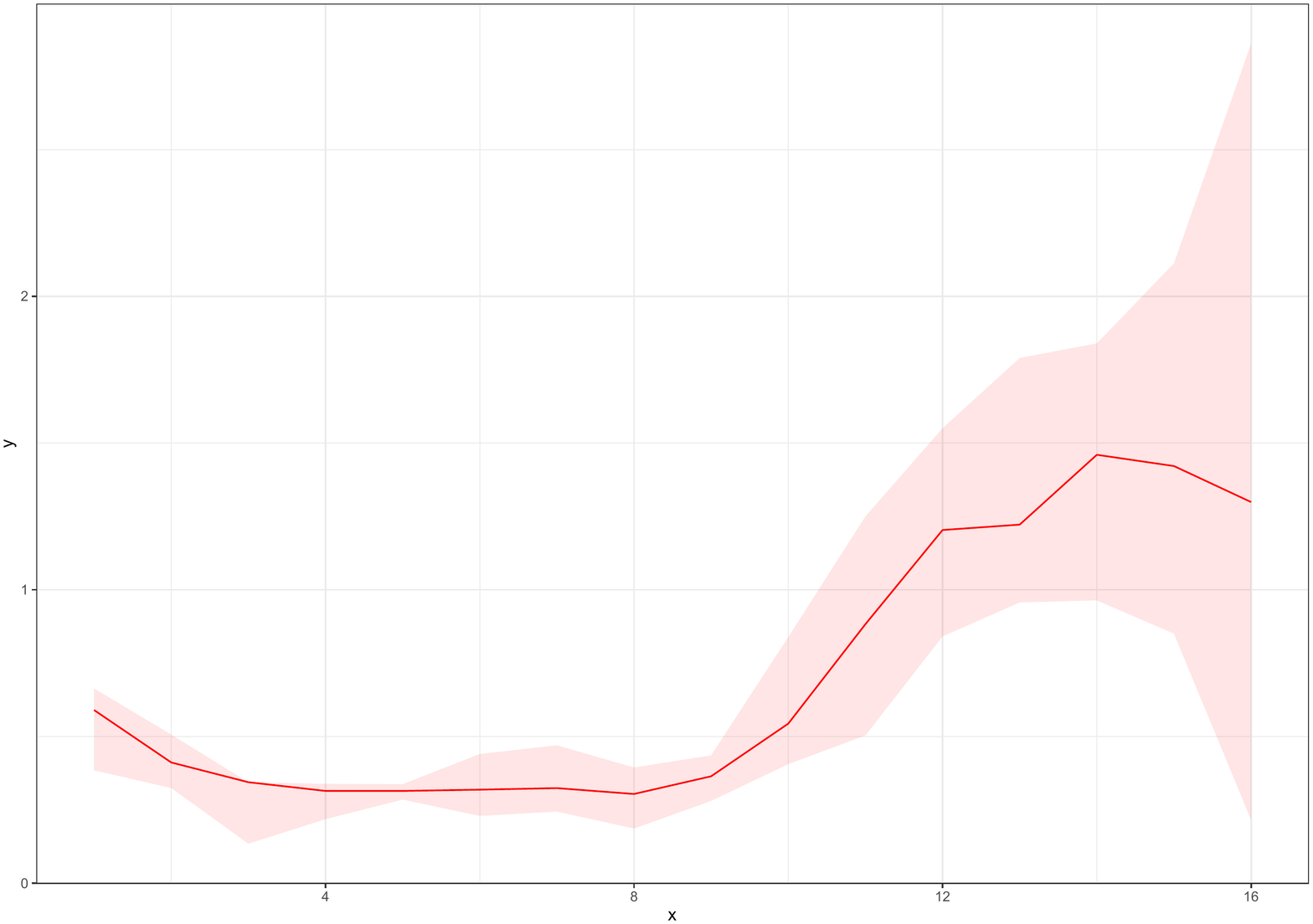}
     \caption{The estimated weights $\{\gamma_{i,70+}\}_{i=1}^{16}$.}
  \end{subfigure}

   \caption{\bf The posterior median estimate of instantaneous reproduction number per age group (0-29 (red line), 30-49 (blue line), 50-69 (pink line) and 70+ (brown line)), the posterior median estimate of weights $\mathbf{\{\{\gamma_{na}\}_{n=1}^{16}\}_a}$ (red line) and the 99\% CIs (ribbon) for Kingston.}
   \label{ER_Kings4G}
\end{figure}

\begin{figure}[!h] 
  \begin{subfigure}{7cm}
    \centering\includegraphics[width=6cm]{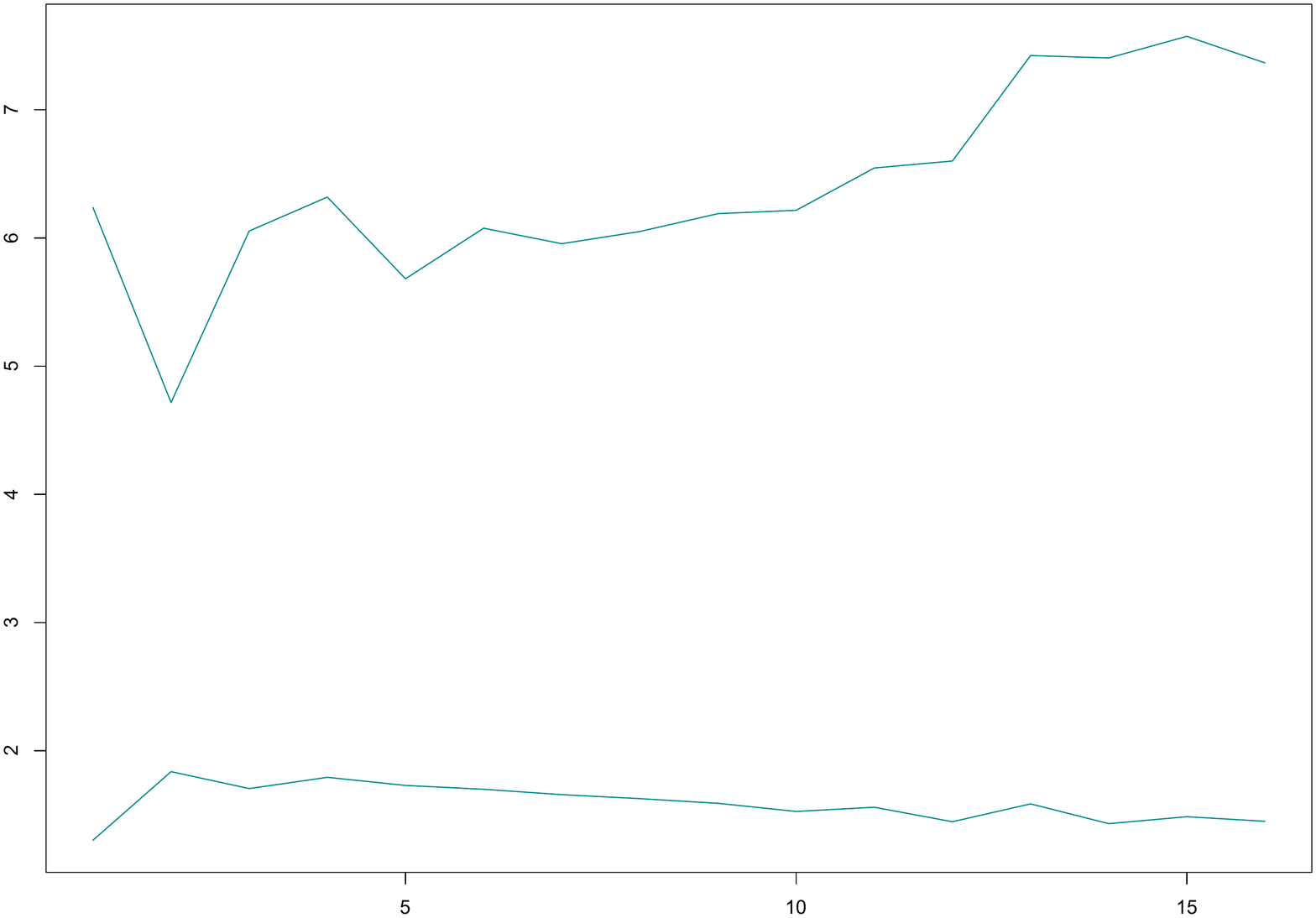}
    \caption{$d$}
  \end{subfigure}
  \begin{subfigure}{7cm}
    \centering\includegraphics[width=6cm]{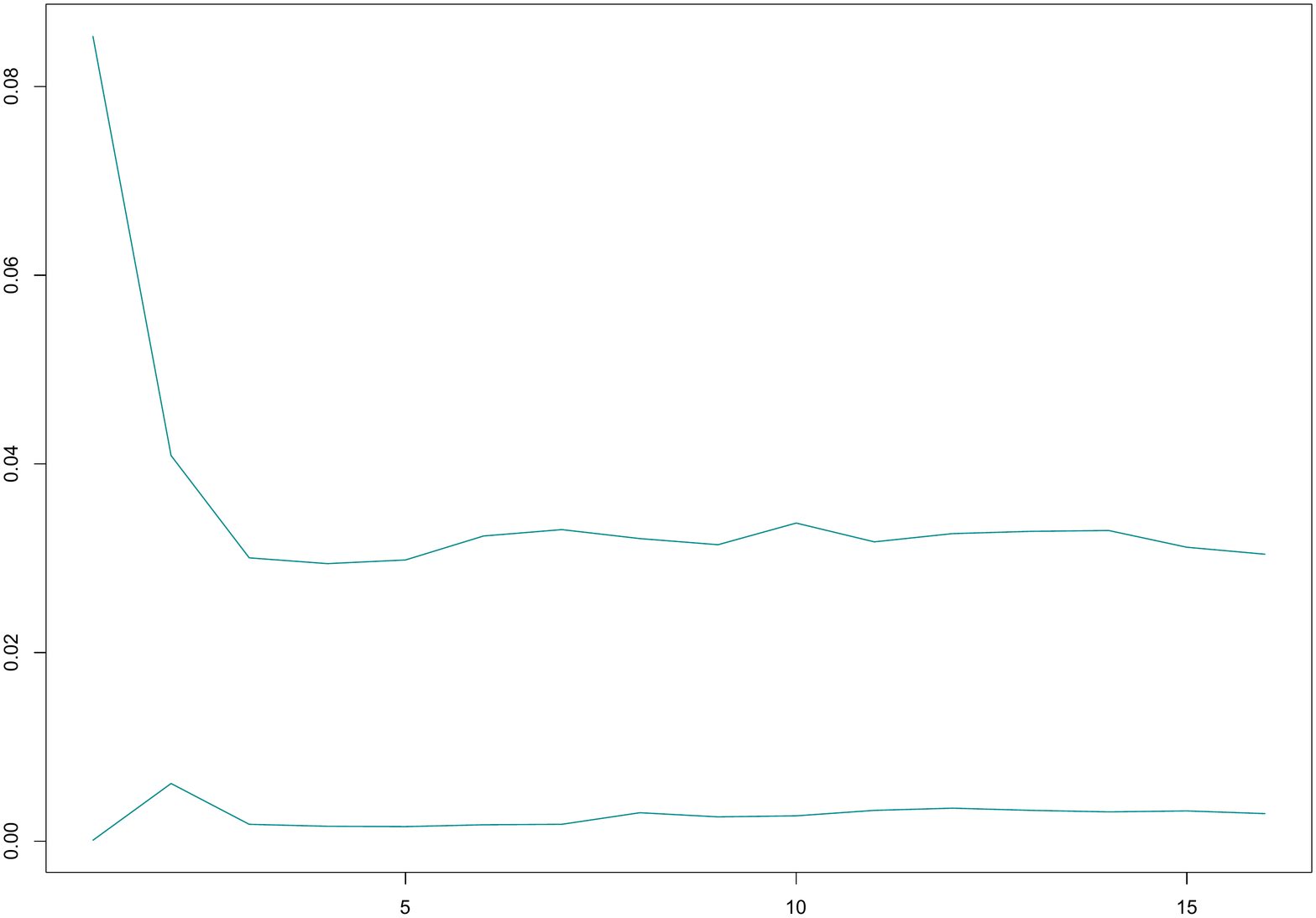}
    \caption{v}
  \end{subfigure}
   \caption{\bf The $99\%$ CIs of time-constant parameters for Kingston upon the Thames.}
   \label{EHC2_Kings4G}
\end{figure}

\begin{figure}[!h] 
  \begin{subfigure}{7cm}
    \centering\includegraphics[width=6cm]{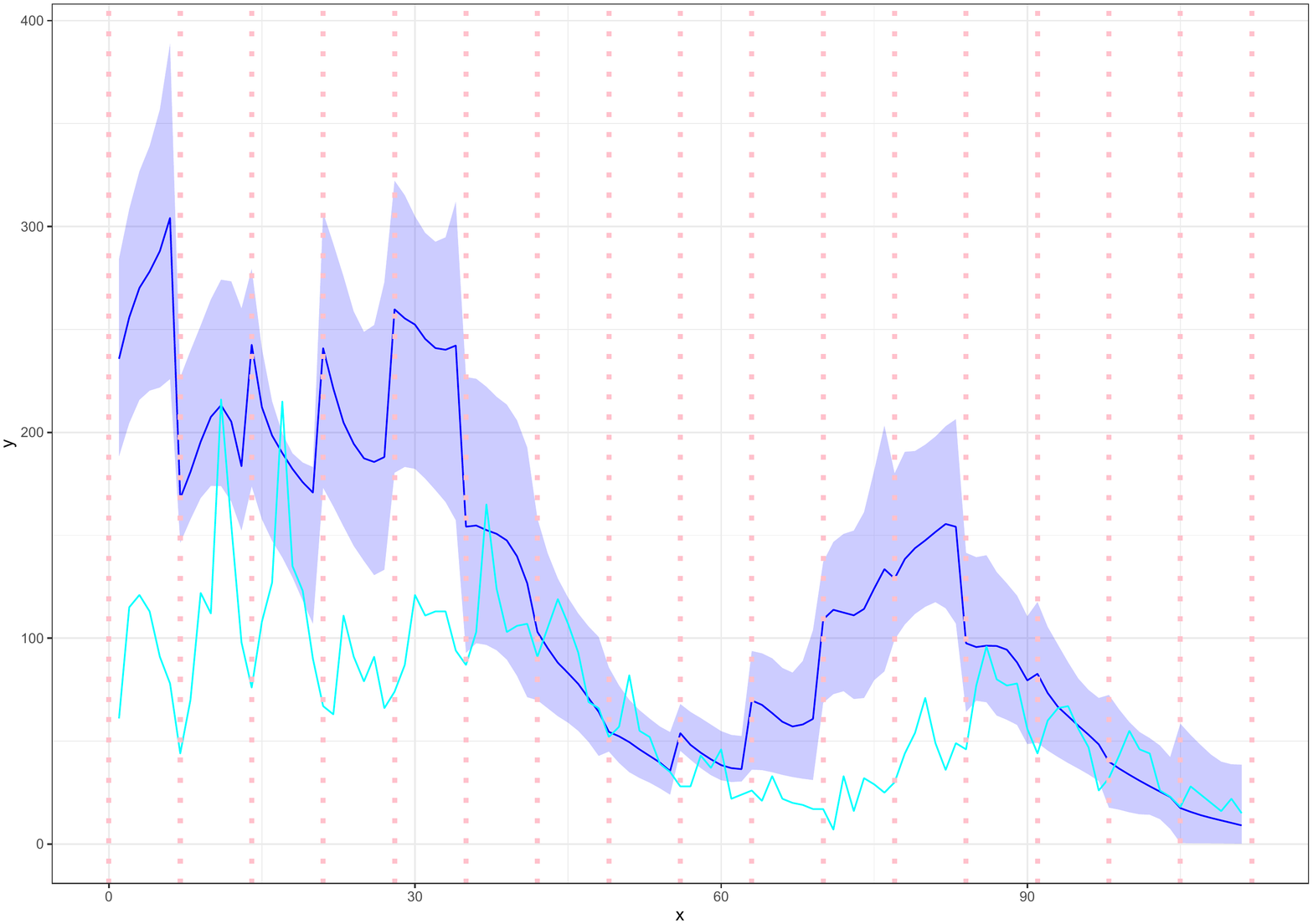}
   \caption{The estimated intensity of latent cases aged 0-29.}
  \end{subfigure}
  \begin{subfigure}{7cm}
    \centering\includegraphics[width=6cm]{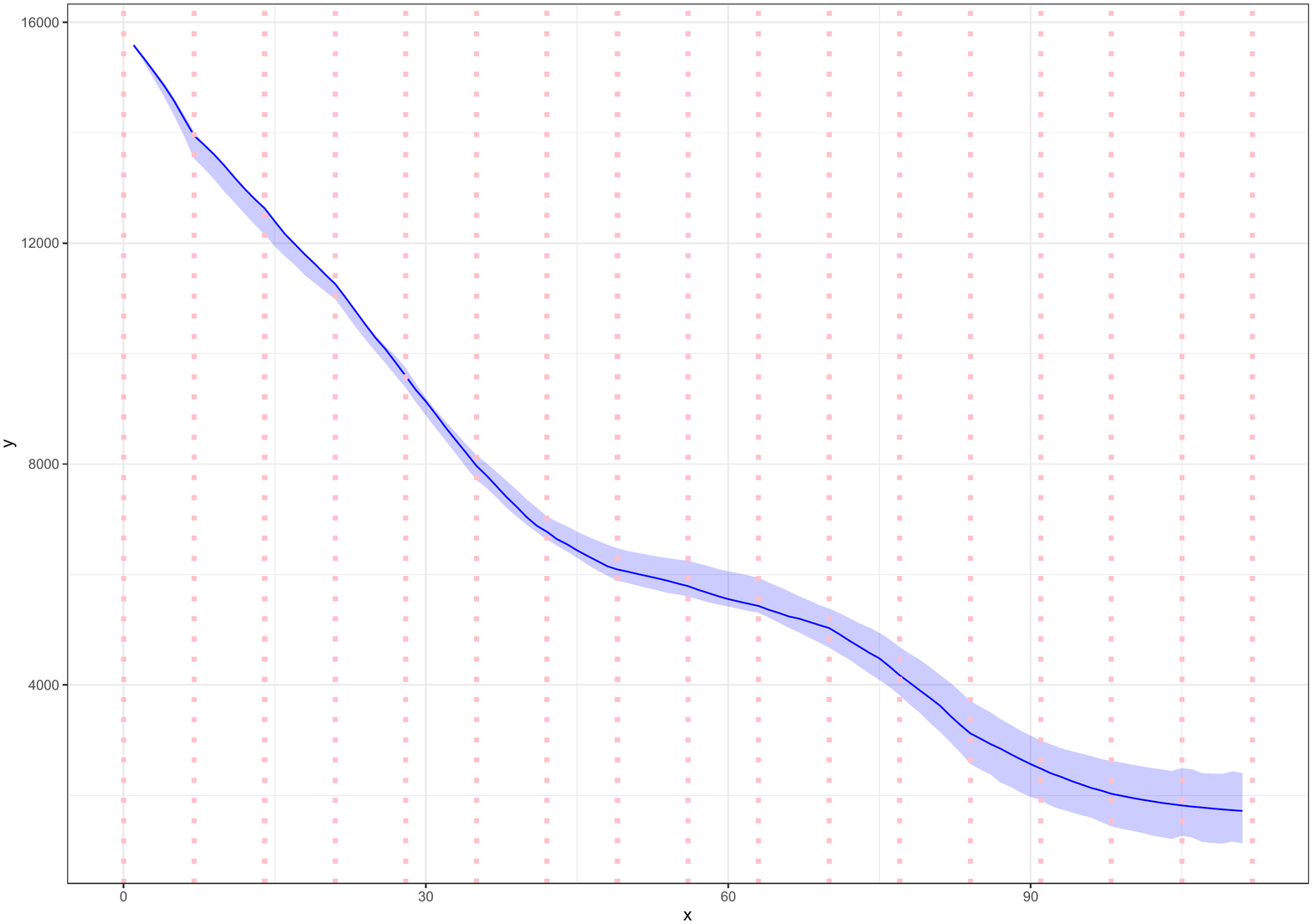}
   \caption{The estimated susceptibles aged 0-29.}
  \end{subfigure}
  \begin{subfigure}{7cm}
    \centering\includegraphics[width=6cm]{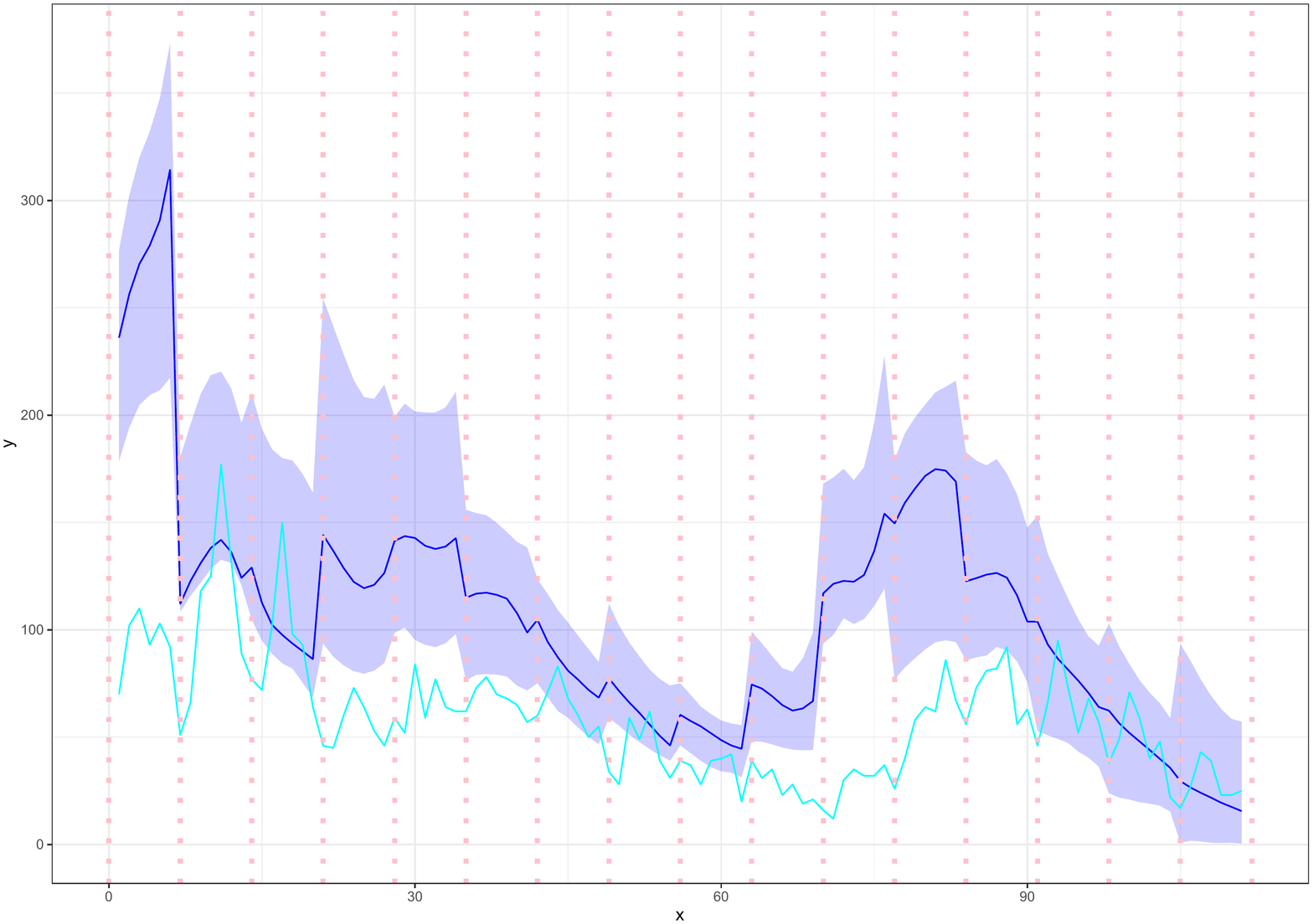}
   \caption{The estimated intensity of latent cases aged 30-49.}
  \end{subfigure}
  \begin{subfigure}{7cm}
    \centering\includegraphics[width=6cm]{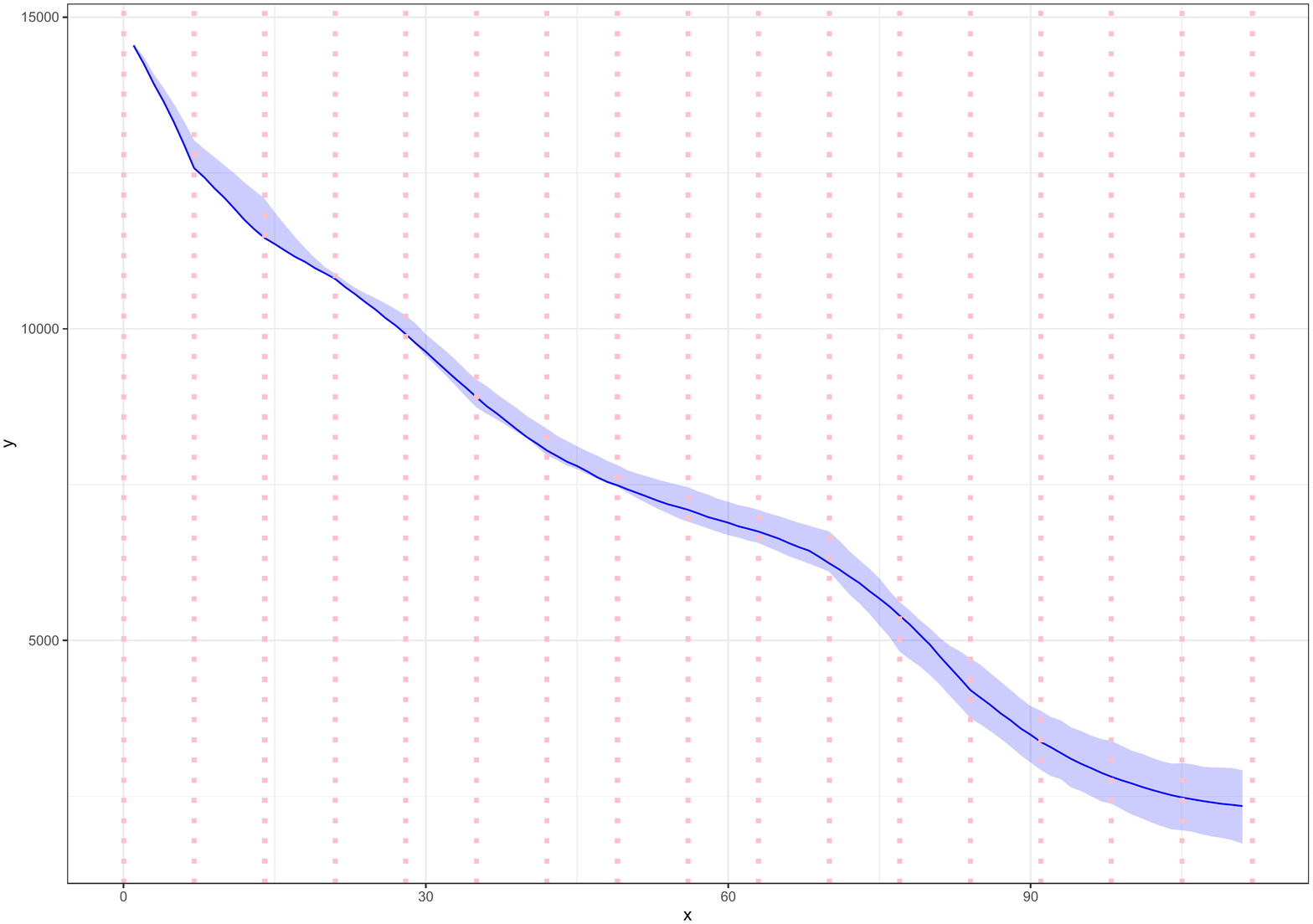}
   \caption{The estimated susceptibles aged 30-49.}
  \end{subfigure}
   \begin{subfigure}{7cm}
    \centering\includegraphics[width=6cm]{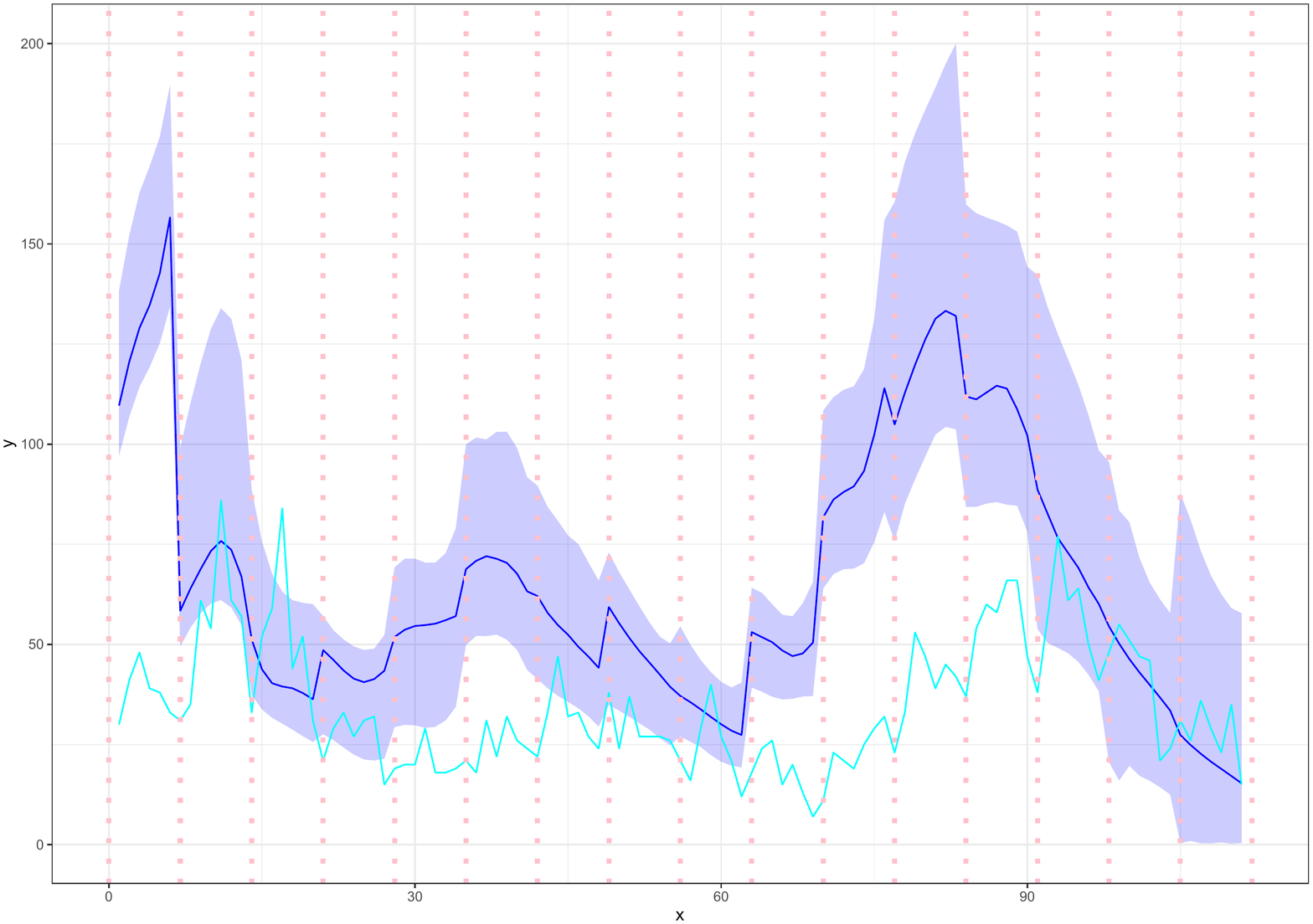}
   \caption{The estimated intensity of latent cases aged 50-69.}
  \end{subfigure}
  \begin{subfigure}{7cm}
    \centering\includegraphics[width=6cm]{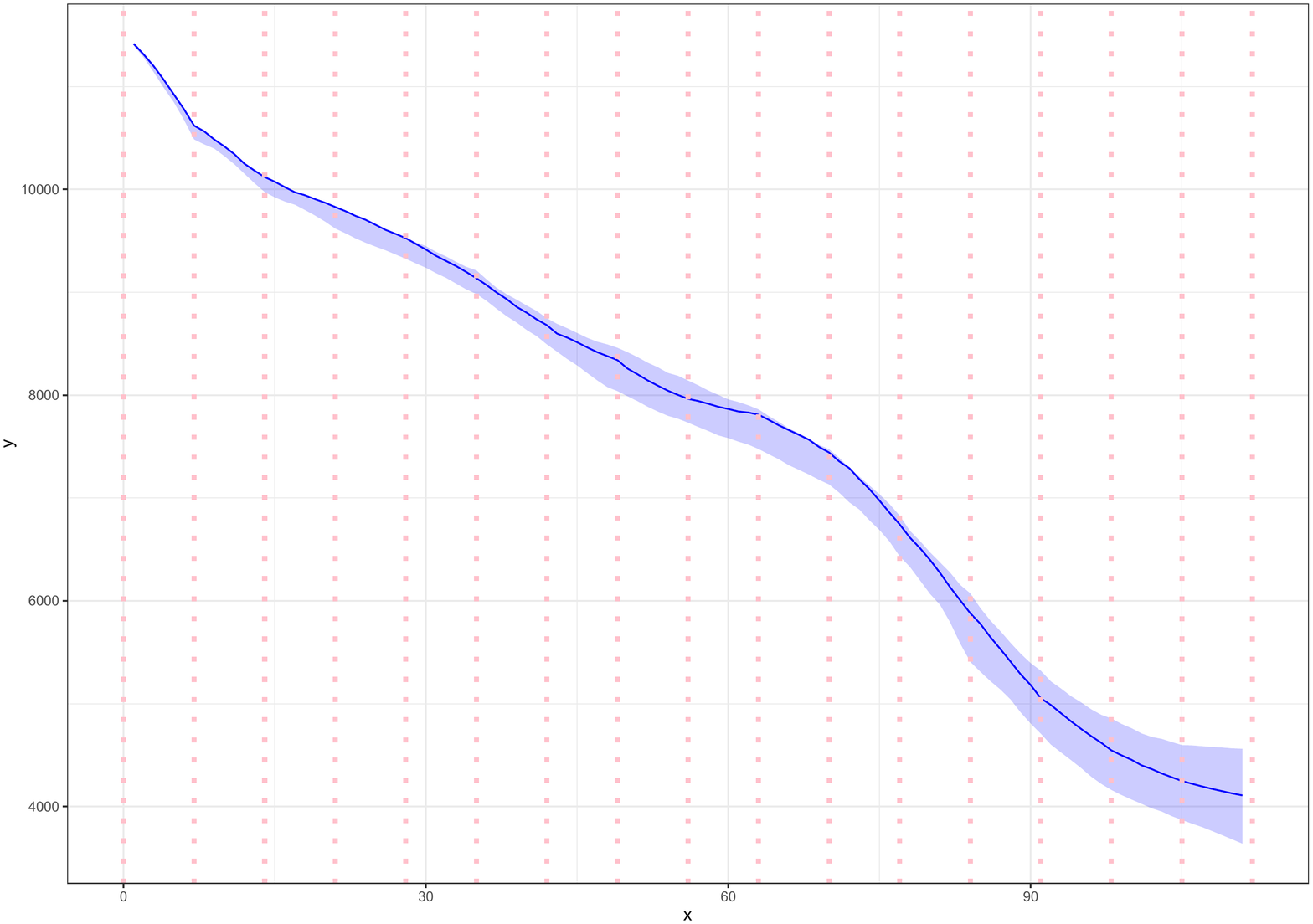}
   \caption{The estimated susceptibles aged 50-69.}
  \end{subfigure}
   \begin{subfigure}{7cm}
    \centering\includegraphics[width=6cm]{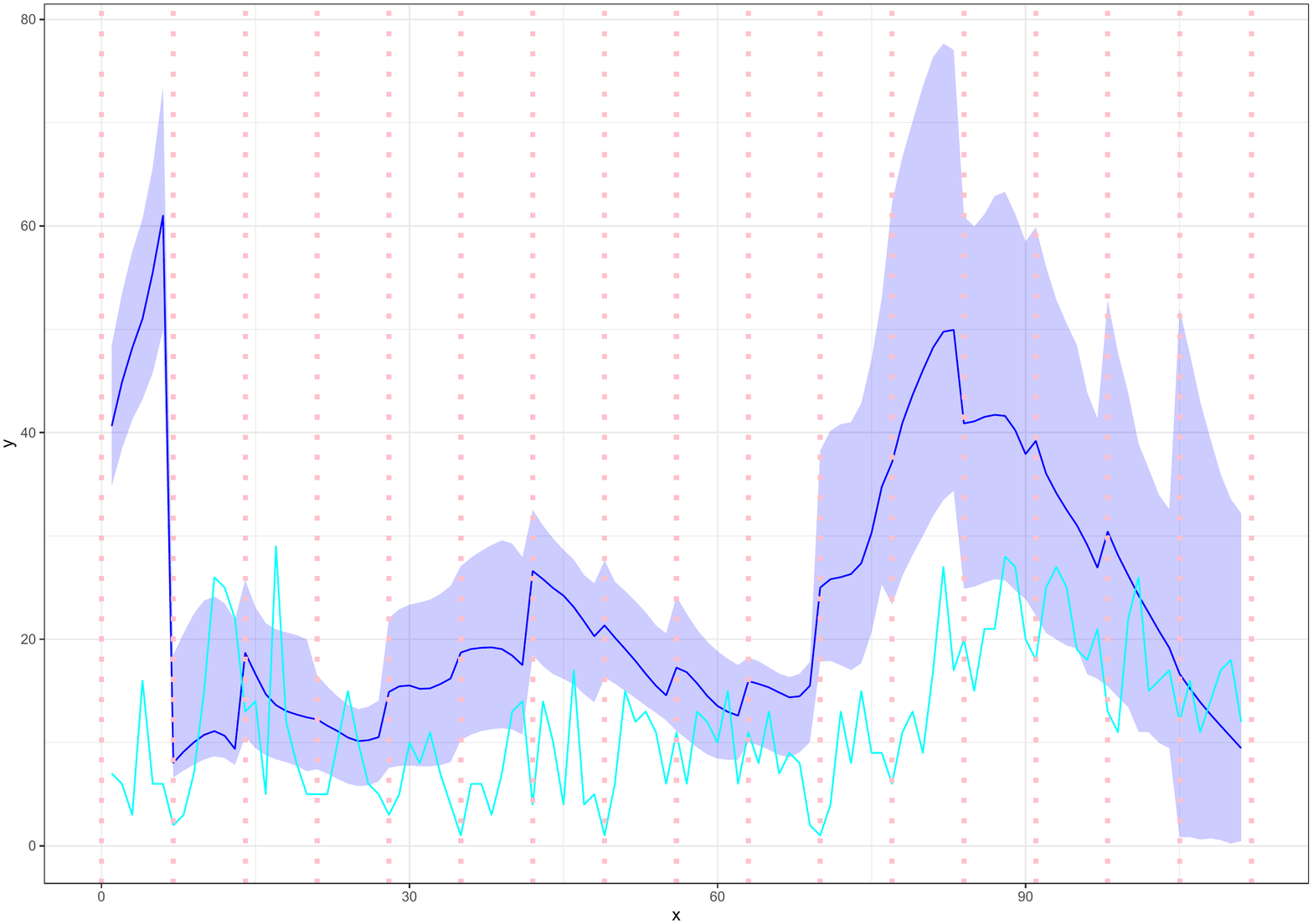}
   \caption{The estimated intensity of latent cases aged 70+.}
  \end{subfigure}
  \begin{subfigure}{7cm}
    \centering\includegraphics[width=6cm]{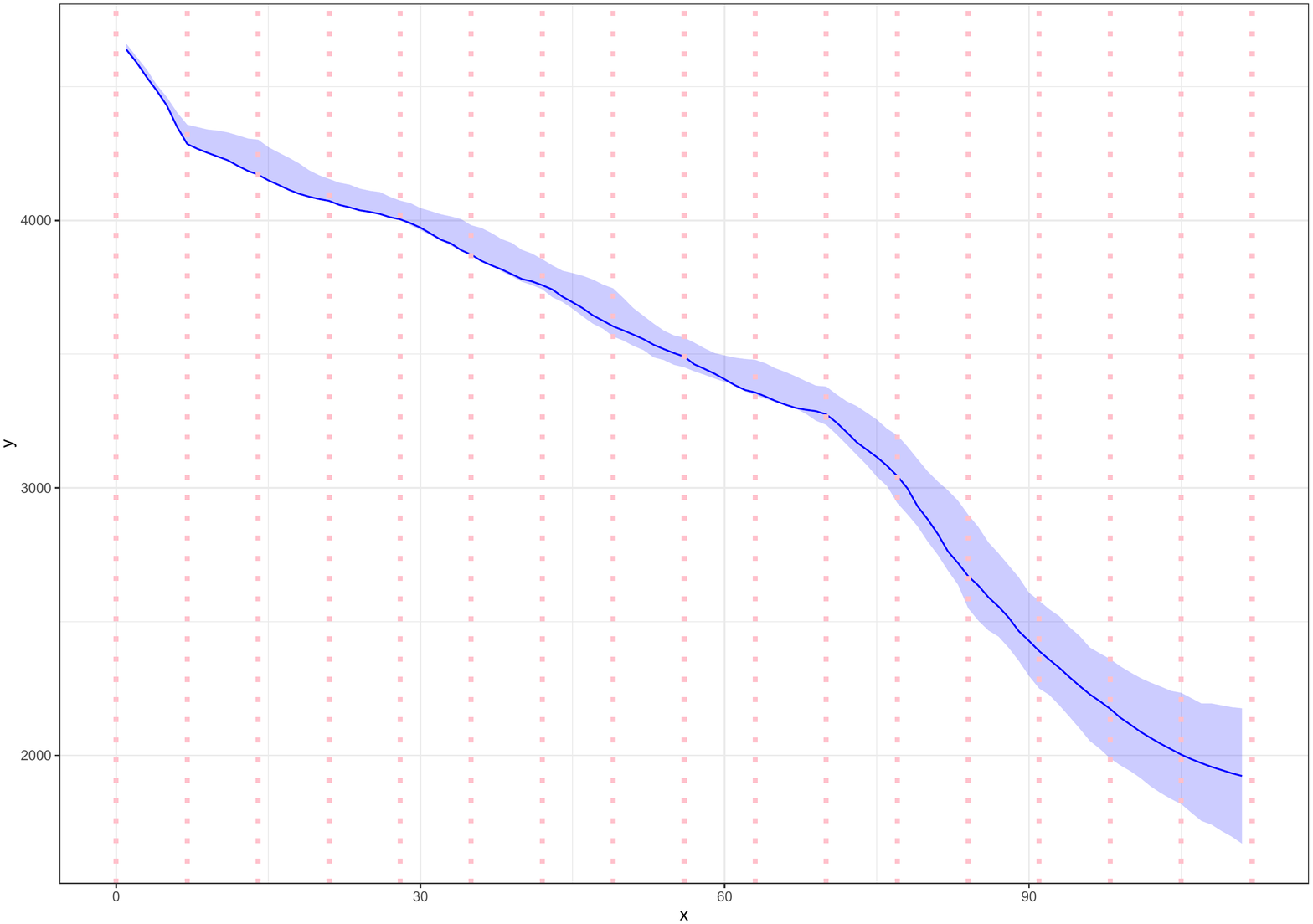}
   \caption{The estimated susceptibles aged 70+.}
  \end{subfigure}
  
  \caption{{\bf The estimated latent intensity and susceptibles (posterior median (blue line) ; 99\% CI (ribbon)) and the daily observed cases (cyan line) in Ashford.} The vertical dotted lines show the beginning of each week in the period we examine. }
  \label{EstInt_Ashford4G}
\end{figure}

\begin{figure}[!h] 
  \begin{subfigure}{7cm}
    \centering\includegraphics[width=6cm]{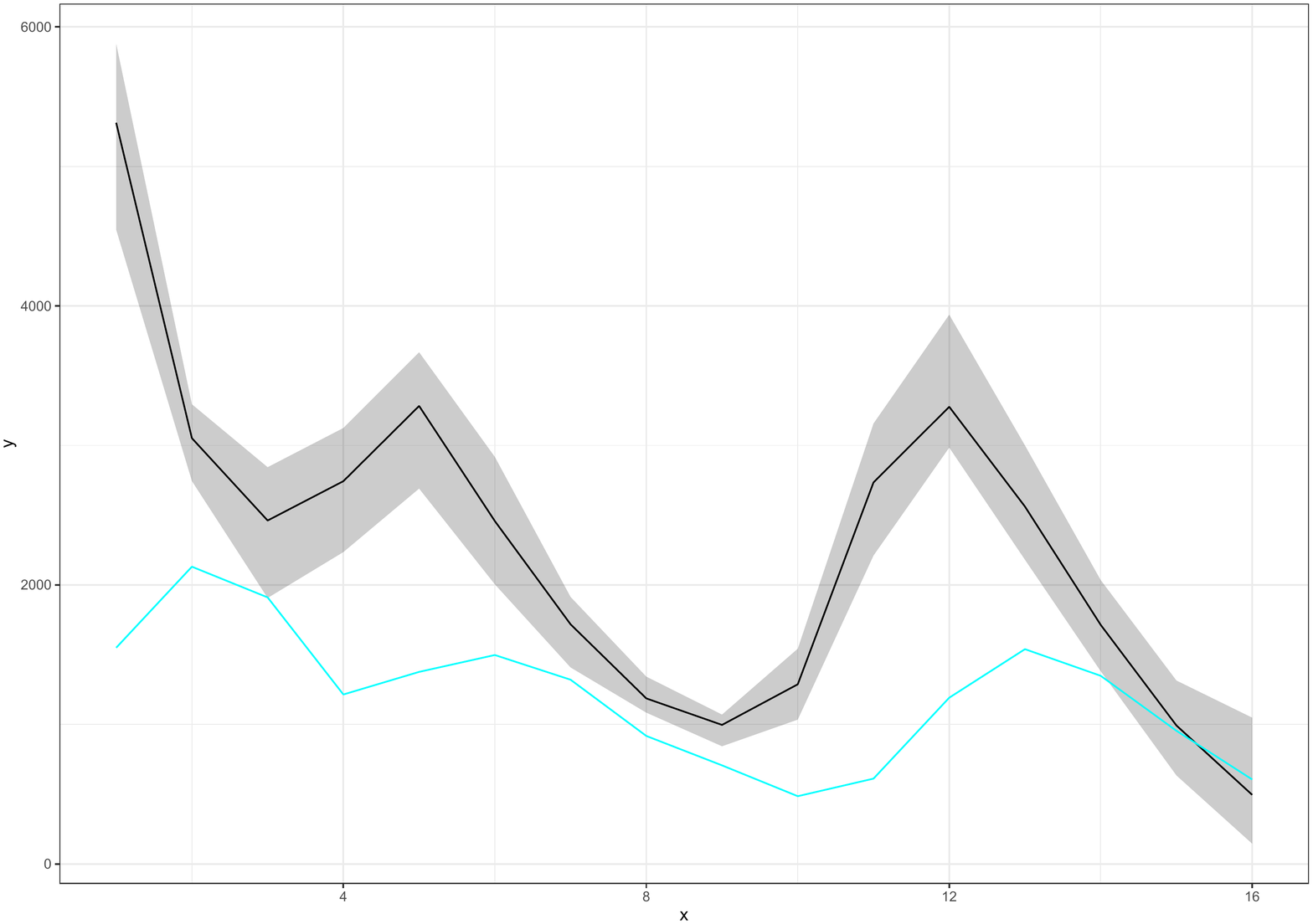}
   \caption{The estimated aggregated weekly hidden \\cases.}
  \end{subfigure}
  \begin{subfigure}{7cm}
    \centering\includegraphics[width=6cm]{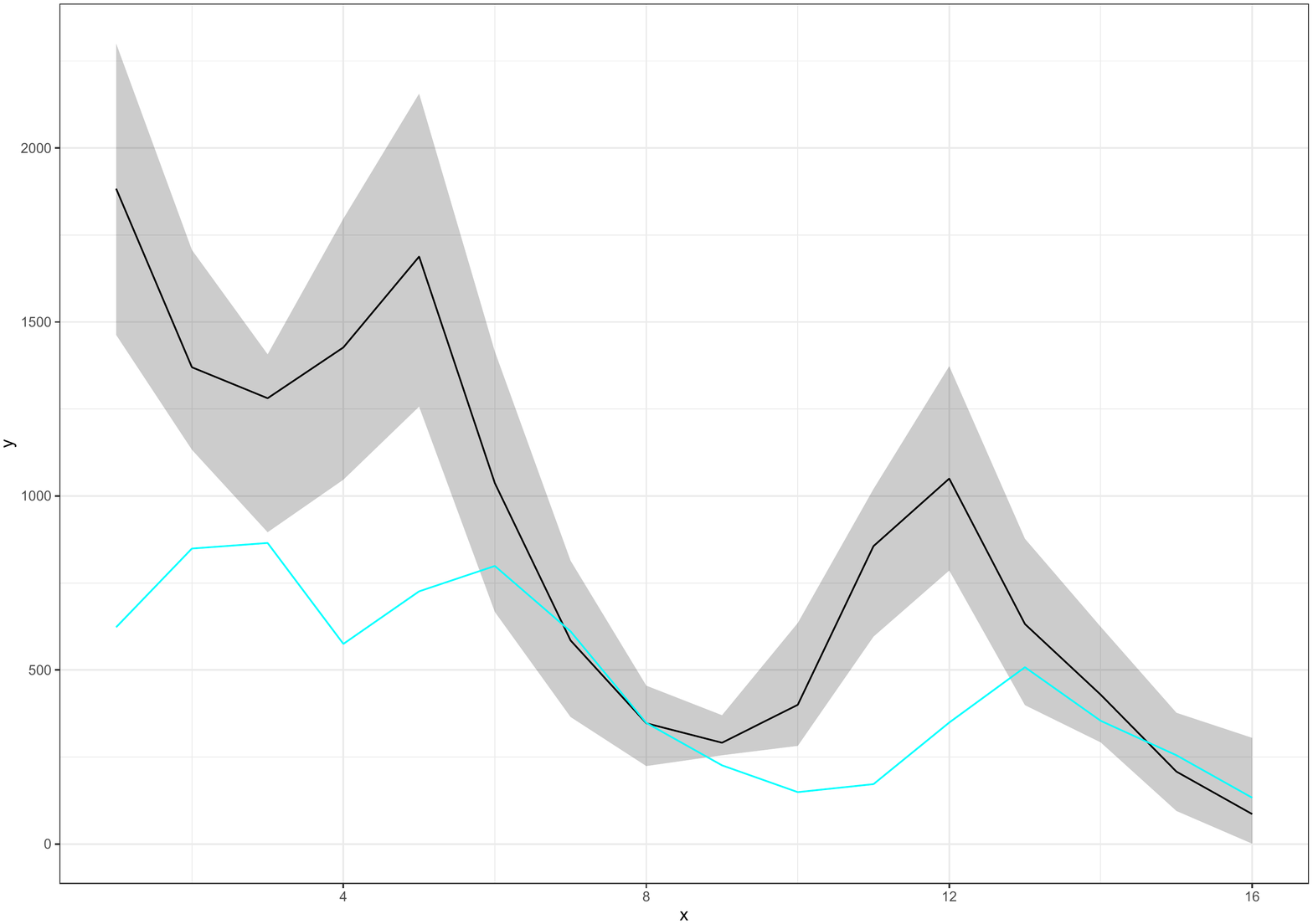}
    \caption{The estimated weekly hidden cases aged 0-29.}
  \end{subfigure}
  \begin{subfigure}{7cm}
    \centering\includegraphics[width=6cm]{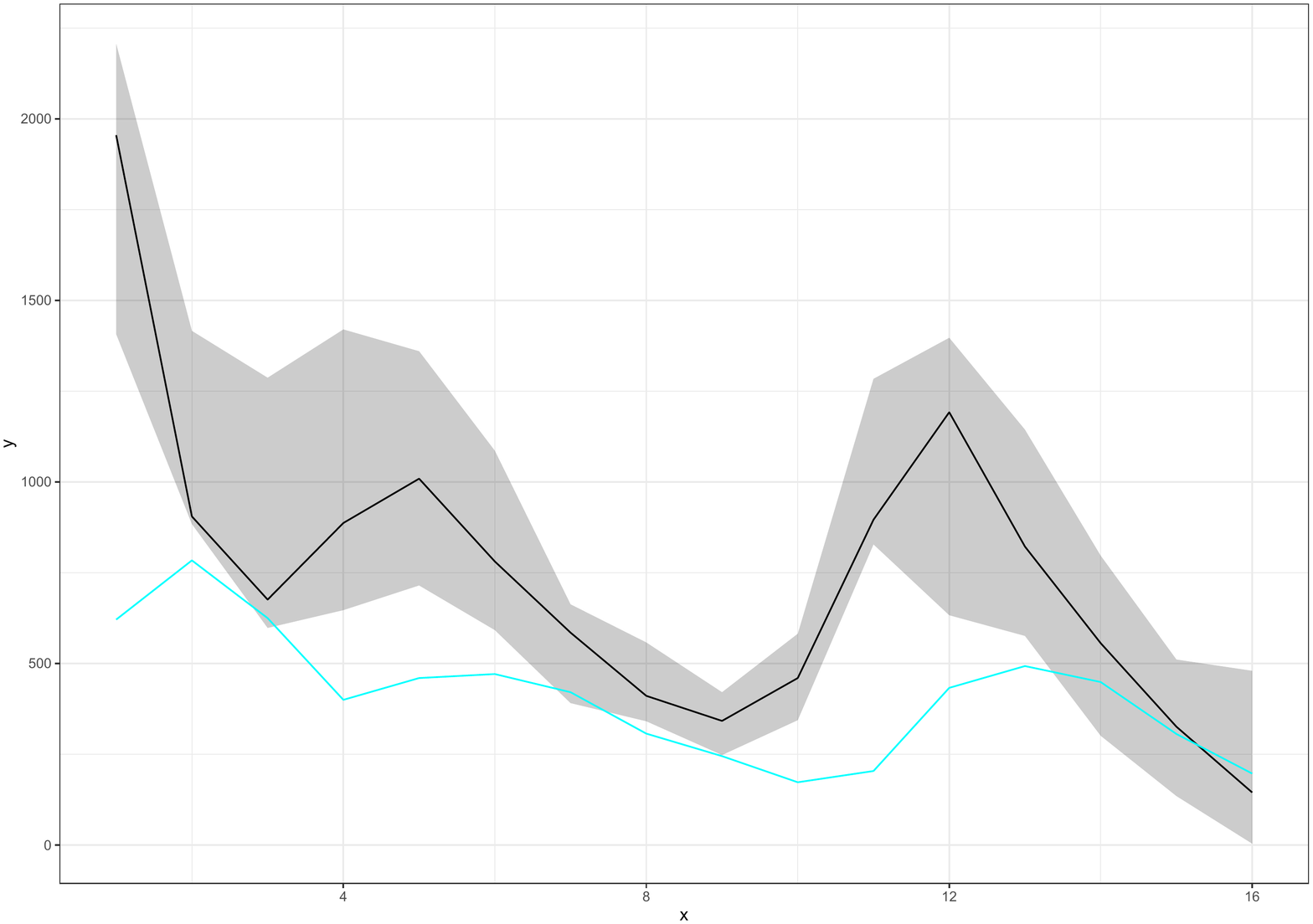}
    \caption{The estimated weekly hidden cases aged 30-49.}
  \end{subfigure}
  \begin{subfigure}{7cm}
    \centering\includegraphics[width=6cm]{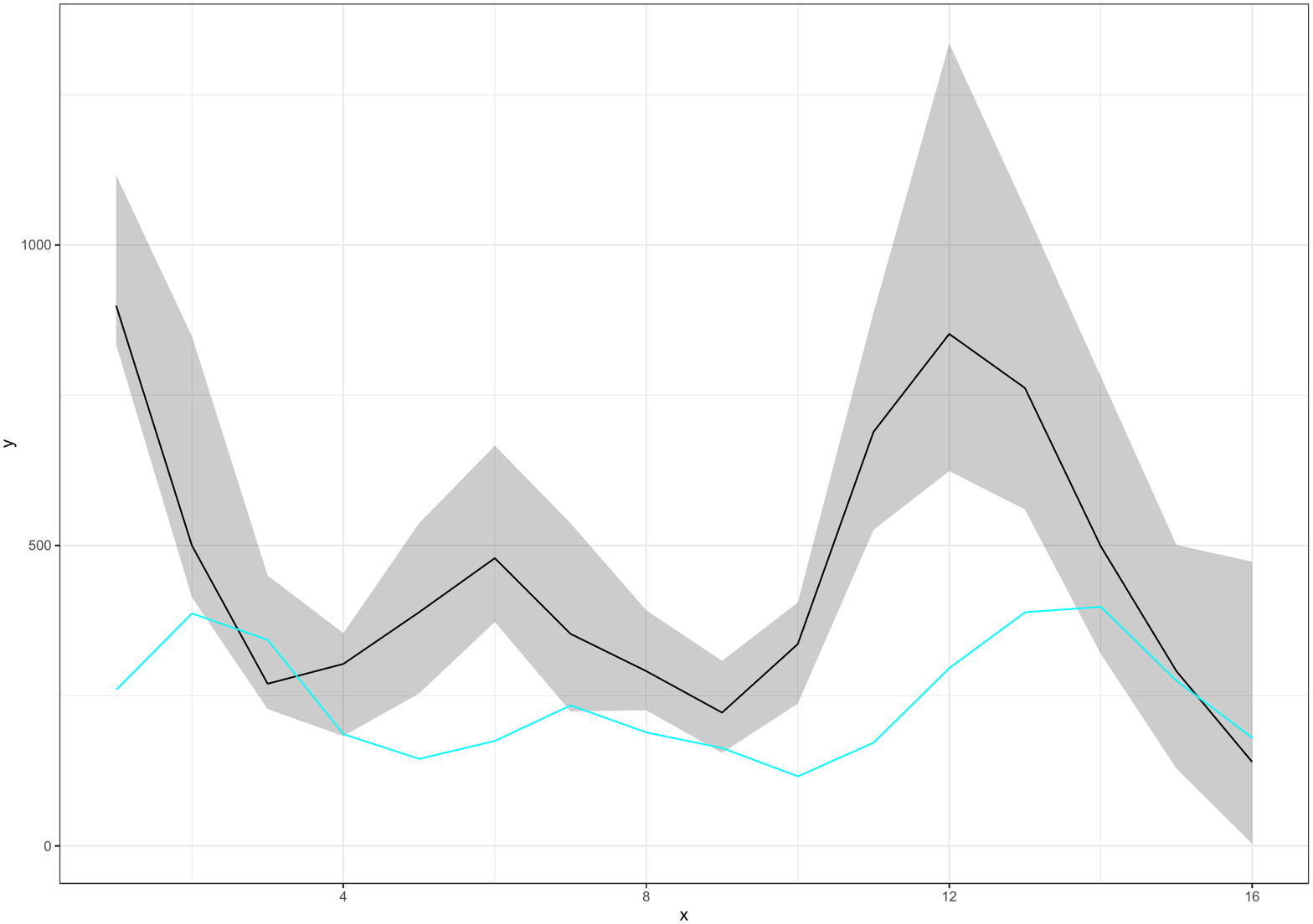}
    \caption{The estimated weekly hidden cases aged 50-69.}
  \end{subfigure}
  \begin{subfigure}{7cm}
    \centering\includegraphics[width=6cm]{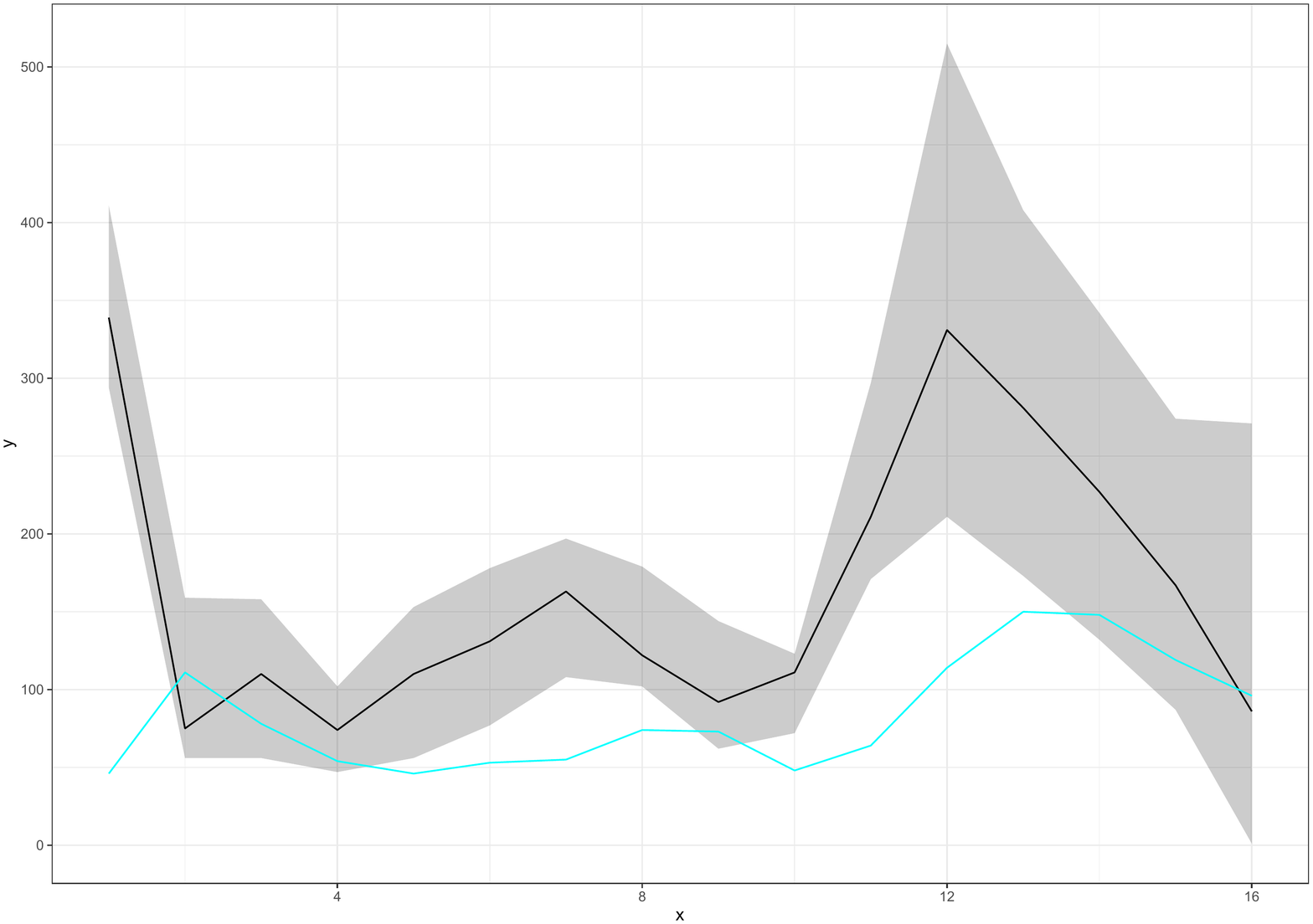}
    \caption{The estimated weekly hidden cases aged 70+.}
  \end{subfigure}
   \caption{{ \bf The estimated weekly latent cases (black line; 99\% CI (ribbon)) and the weekly observed cases (cyan line) in Ashford.}}
   \label{EHC_Ashford4G}
\end{figure}

\begin{figure}[!h] 
   \begin{subfigure}{7cm}
    \centering\includegraphics[width=6cm]{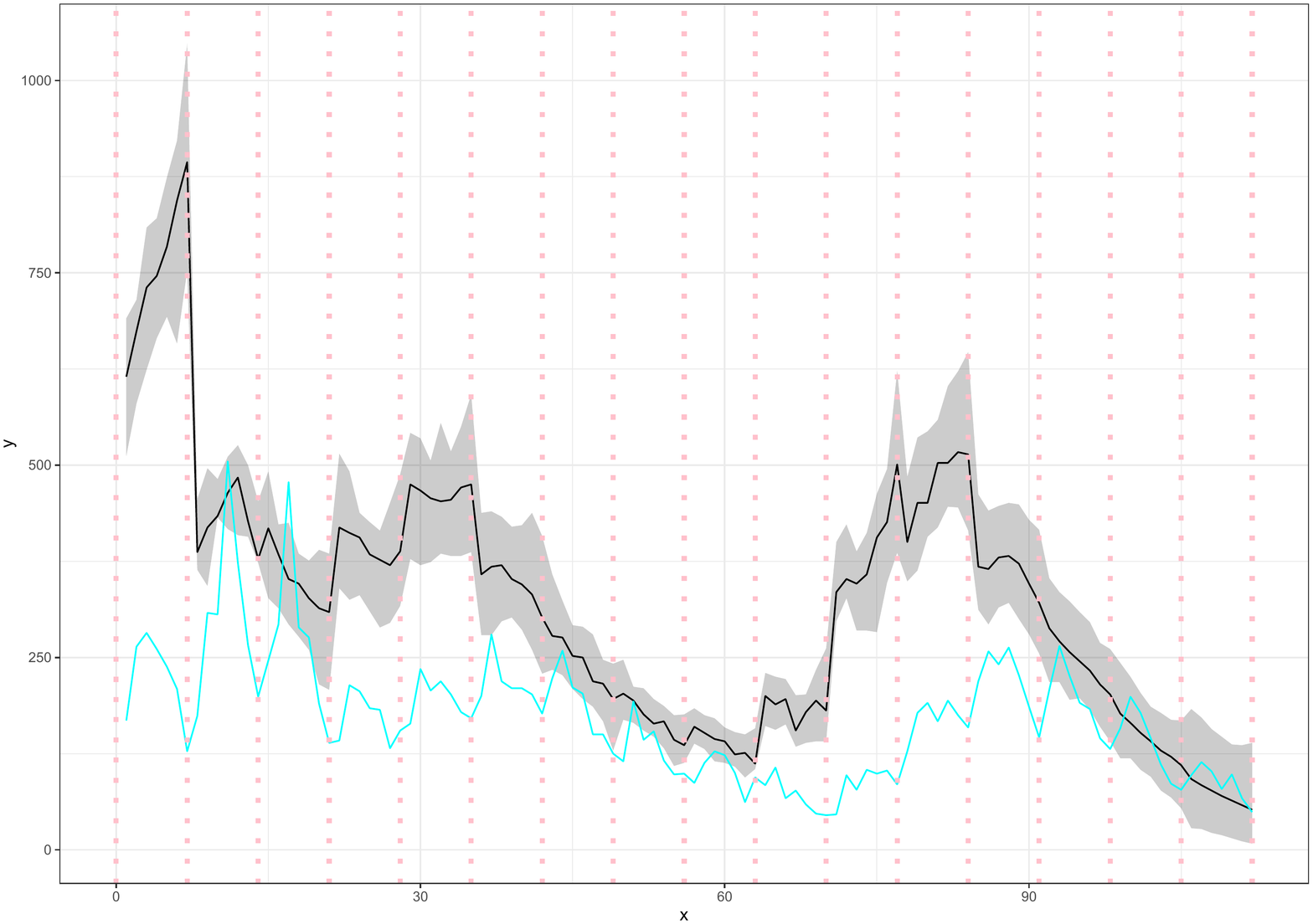}
    \caption{The estimated aggregated daily hidden cases.}
  \end{subfigure}
   \begin{subfigure}{7cm}
    \centering\includegraphics[width=6cm]{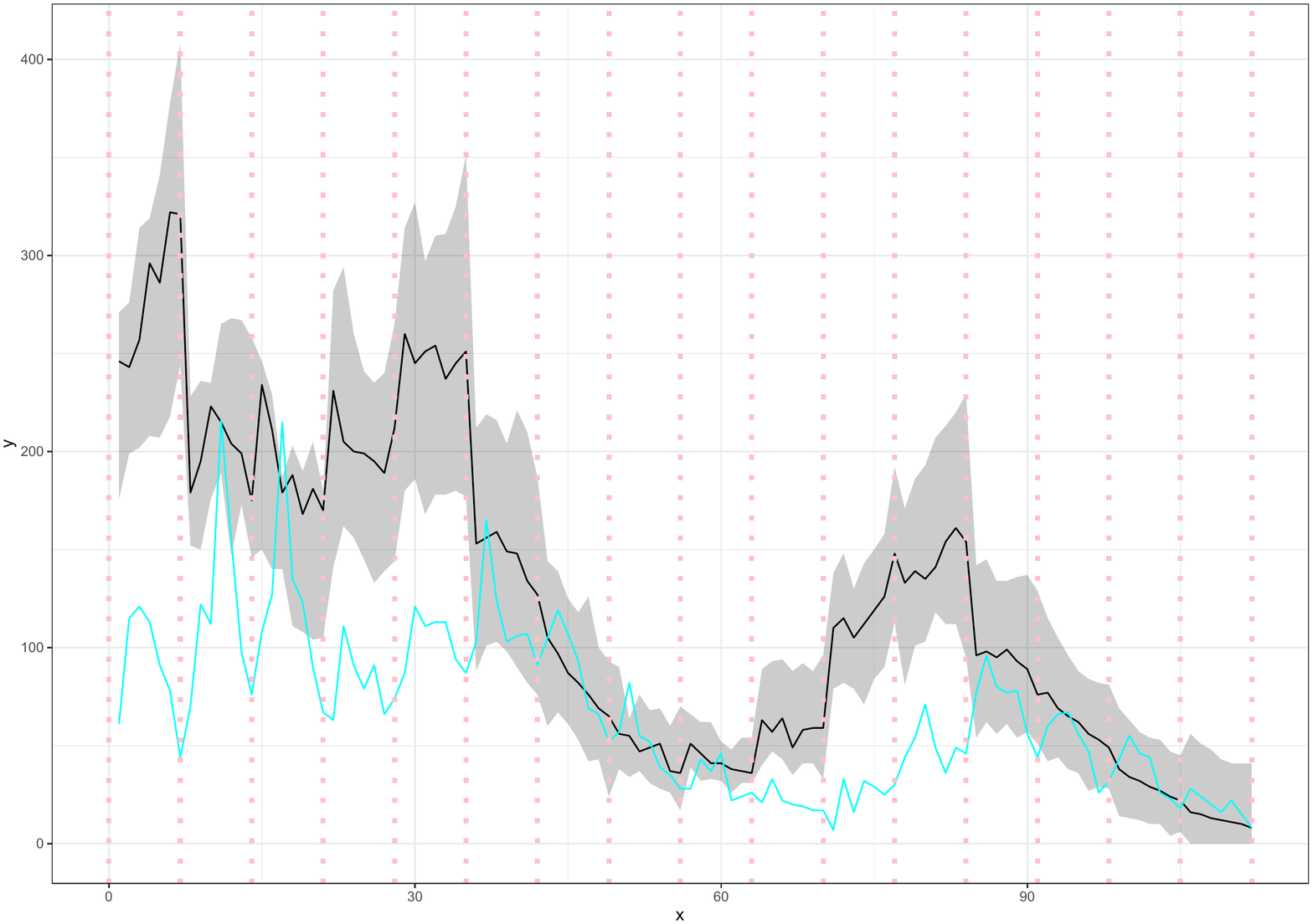}
    \caption{The estimated daily hidden cases aged 0-29.}
  \end{subfigure}
  \begin{subfigure}{7cm}
    \centering\includegraphics[width=6cm]{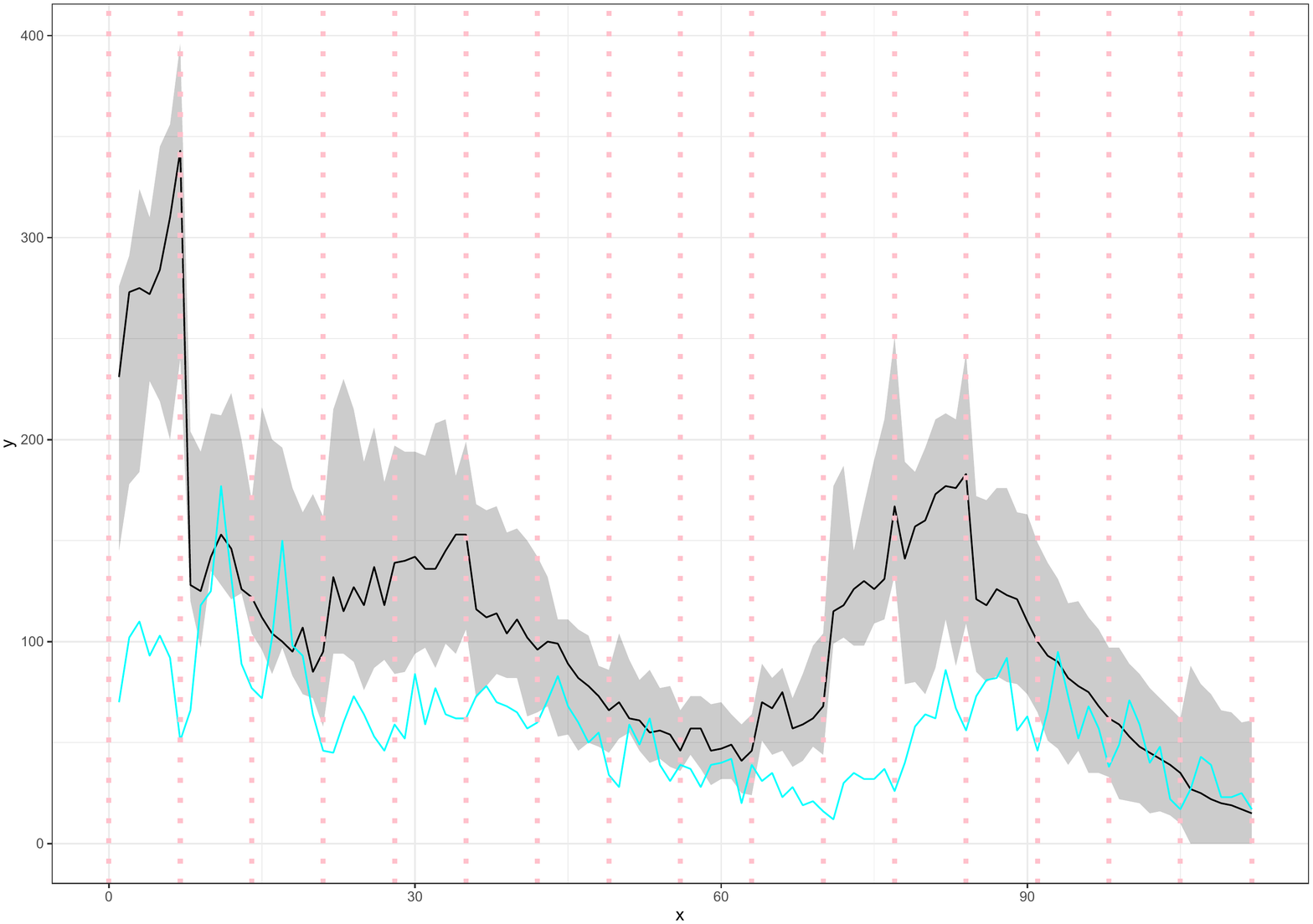}
    \caption{The estimated daily hidden cases aged 30-49.}
  \end{subfigure}
  \begin{subfigure}{7cm}
    \centering\includegraphics[width=6cm]{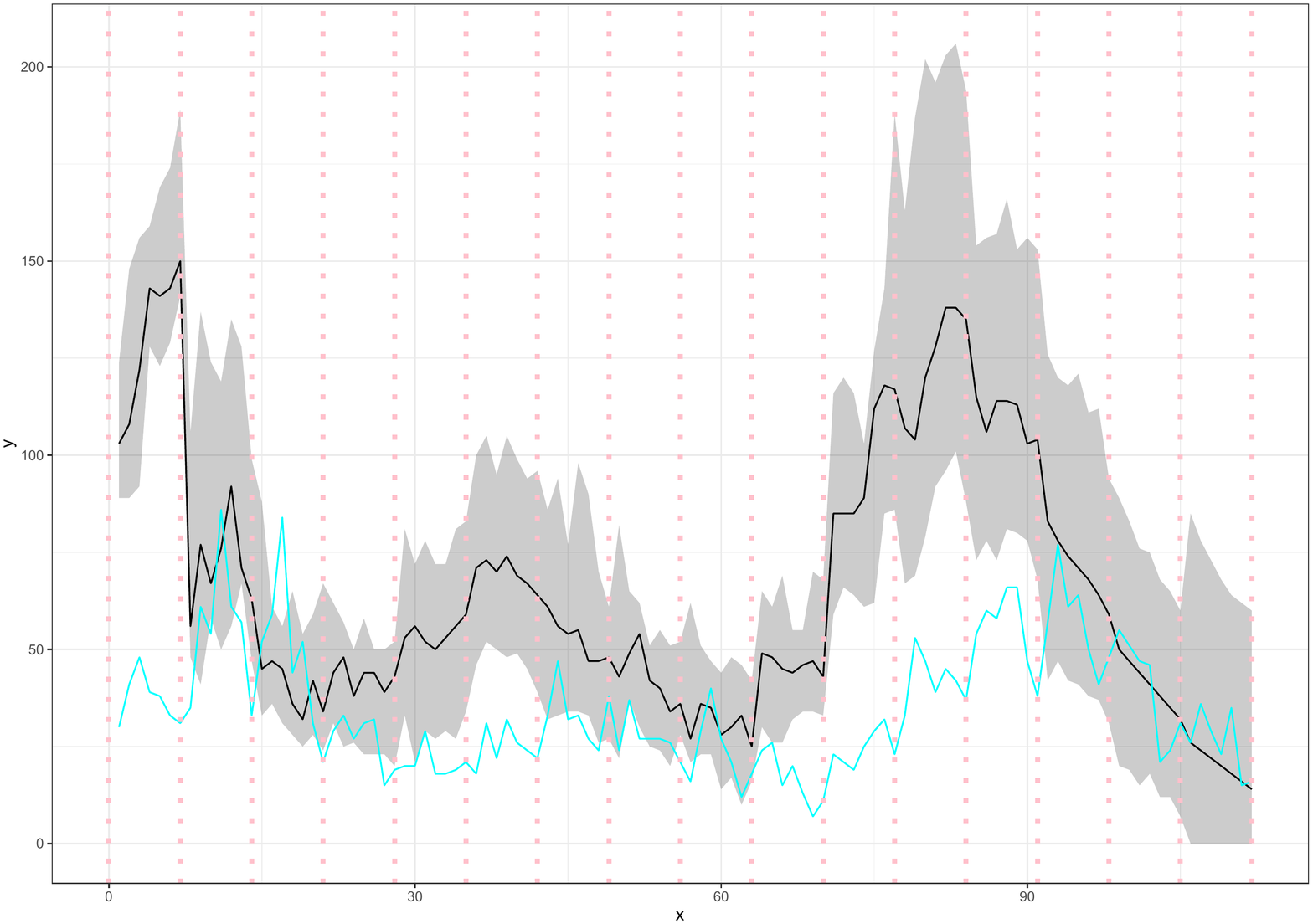}
    \caption{The estimated daily hidden cases aged 50-69.}
  \end{subfigure}
  \begin{subfigure}{7cm}
    \centering\includegraphics[width=6cm]{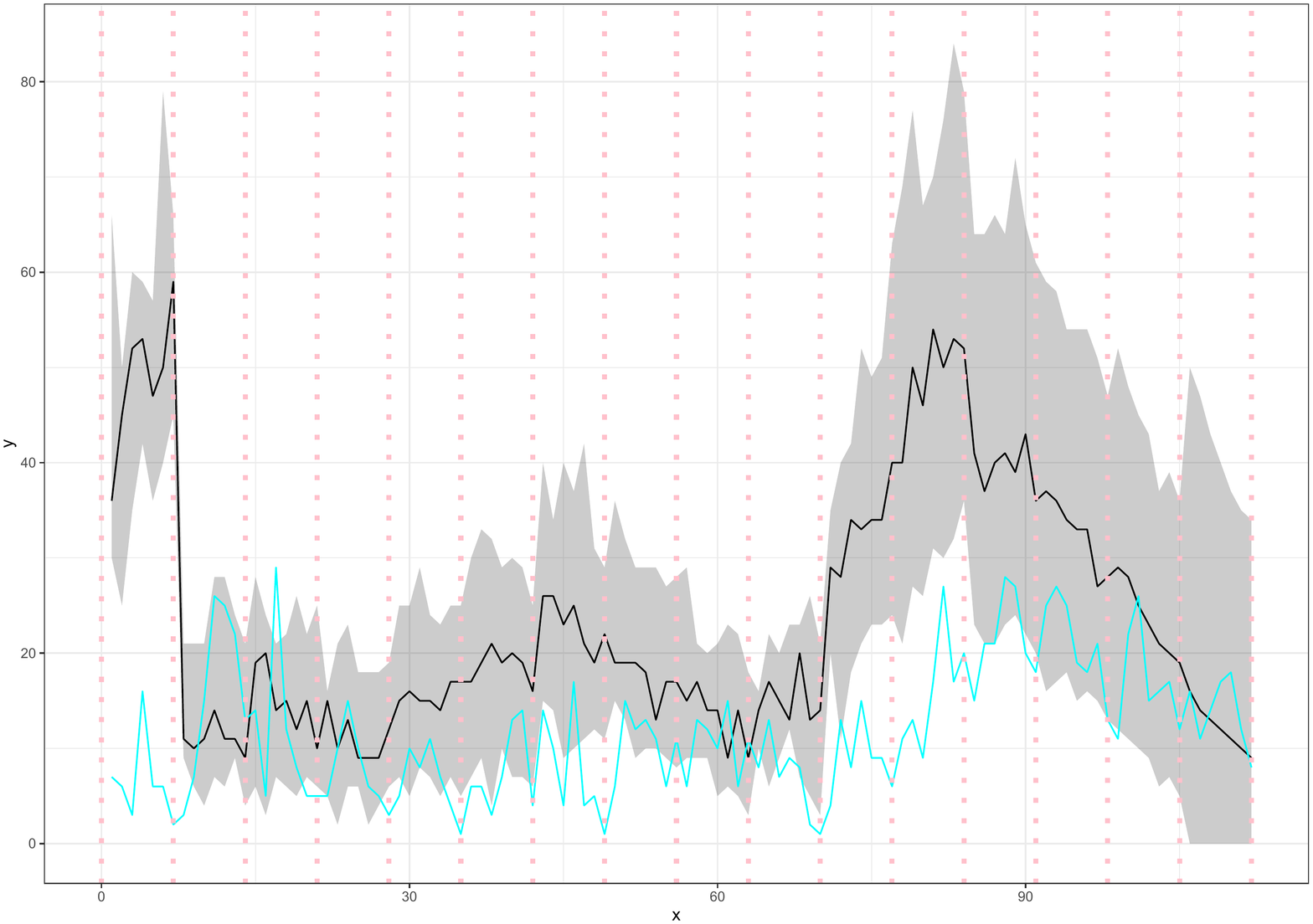}
    \caption{The estimated daily hidden cases aged 70+.}
  \end{subfigure}
   \caption{ {\bf The estimated daily latent cases (posterior median (black line); 99\% CI (ribbon)) and the daily observed cases (cyan line) in Ashford.} The vertical dotted lines show the beginning of each week in the period we examine.}
   \label{EHDC_Ashford4G}
\end{figure}

\begin{figure}[!h] 
 \begin{subfigure}{7cm}
    \centering\includegraphics[width=6cm]{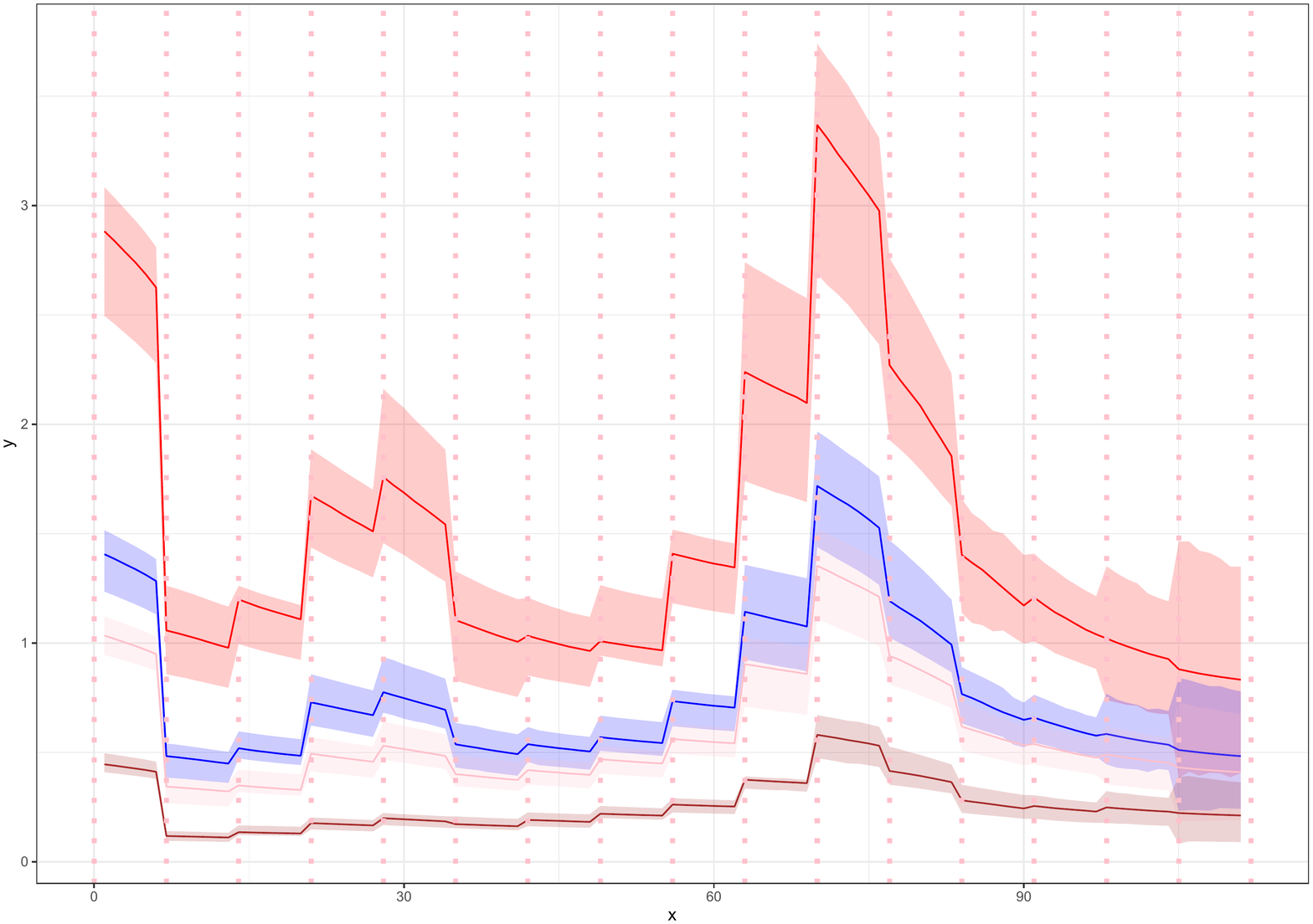}
     \caption{The per age group instantaneous reproduction numbers.}
  \end{subfigure}
  \begin{subfigure}{7cm}
    \centering\includegraphics[width=6cm]{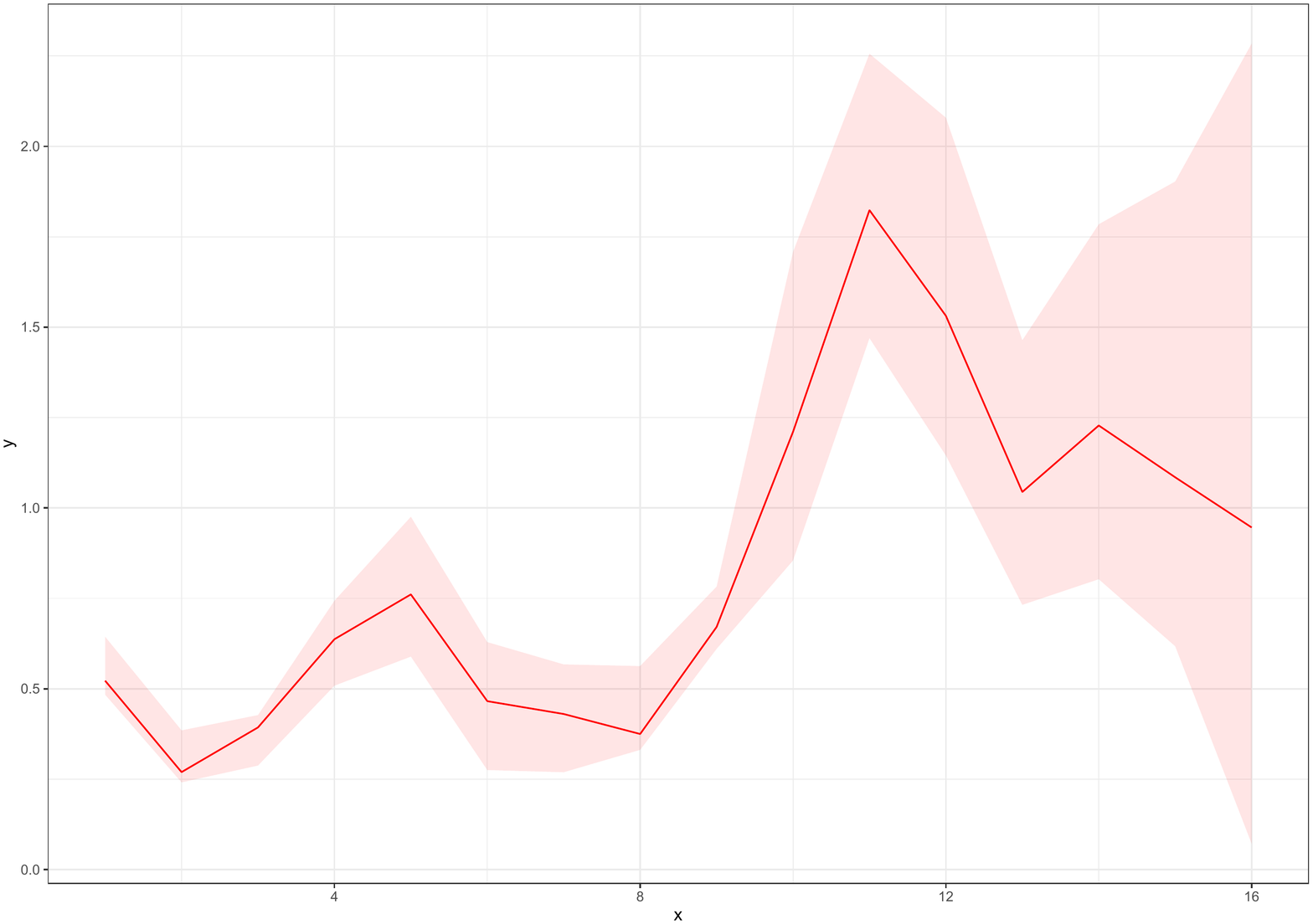}
   \caption{The estimated weights $\{\gamma_{i,0-29}\}_{i=1}^{16}$.}
  \end{subfigure}
  \begin{subfigure}{7cm}
    \centering\includegraphics[width=6cm]{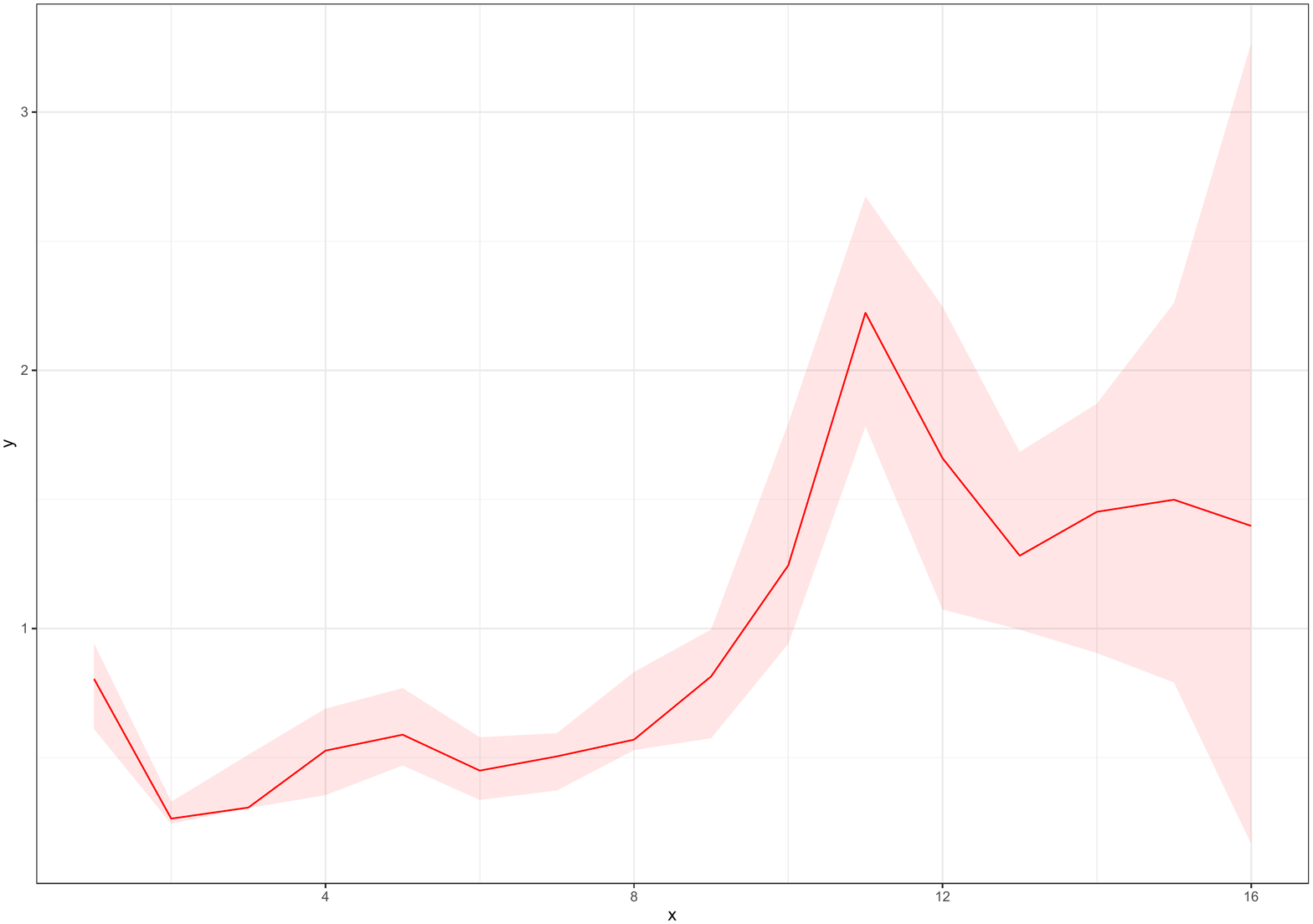}
     \caption{The estimated weights $\{\gamma_{i,30-49}\}_{i=1}^{16}$.}
  \end{subfigure}
  \begin{subfigure}{7cm}
    \centering\includegraphics[width=6cm]{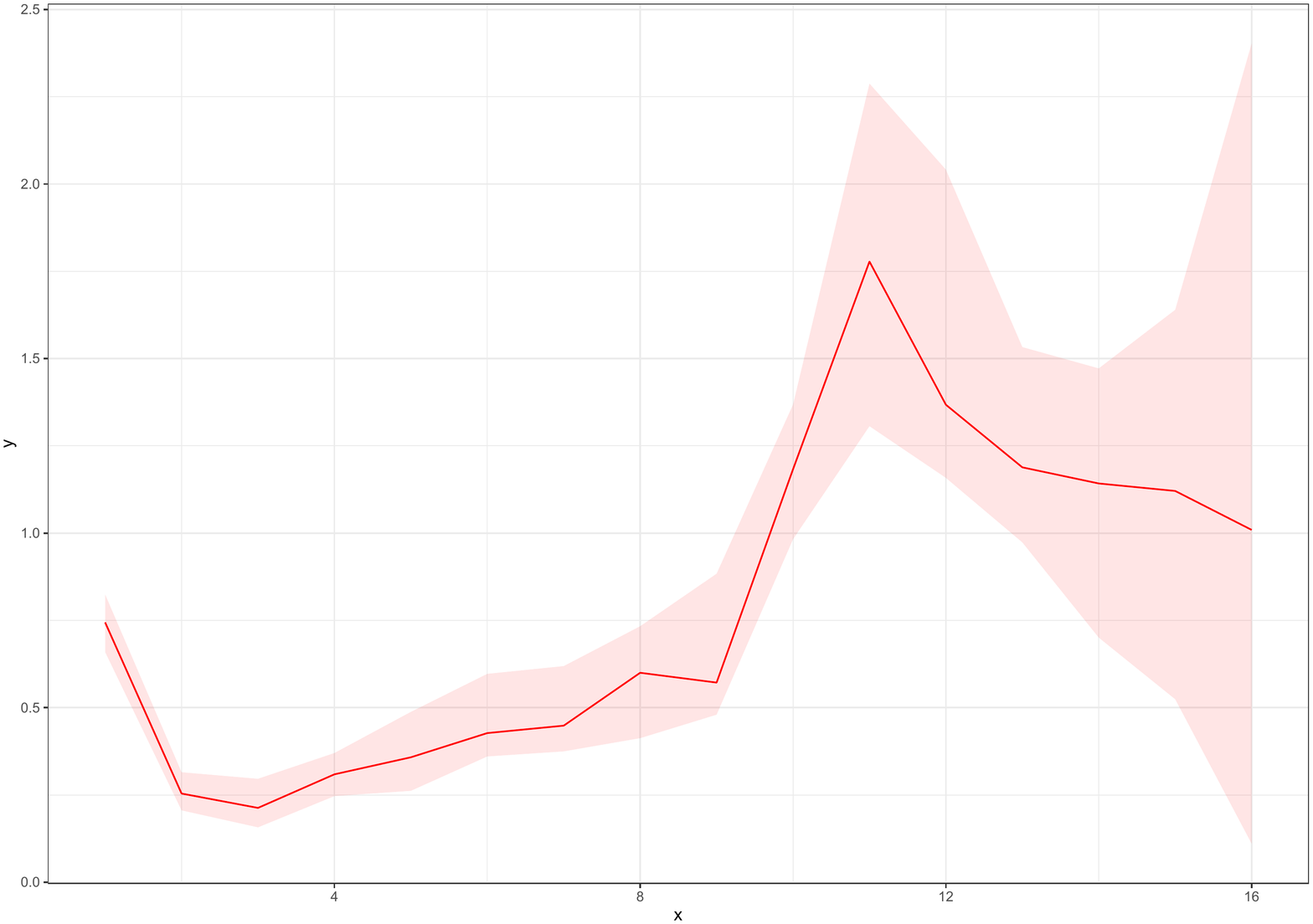}
    \caption{The estimated weights $\{\gamma_{i,50-69}\}_{i=1}^{16}$.}
  \end{subfigure}
   \begin{subfigure}{7cm}
    \centering\includegraphics[width=6cm]{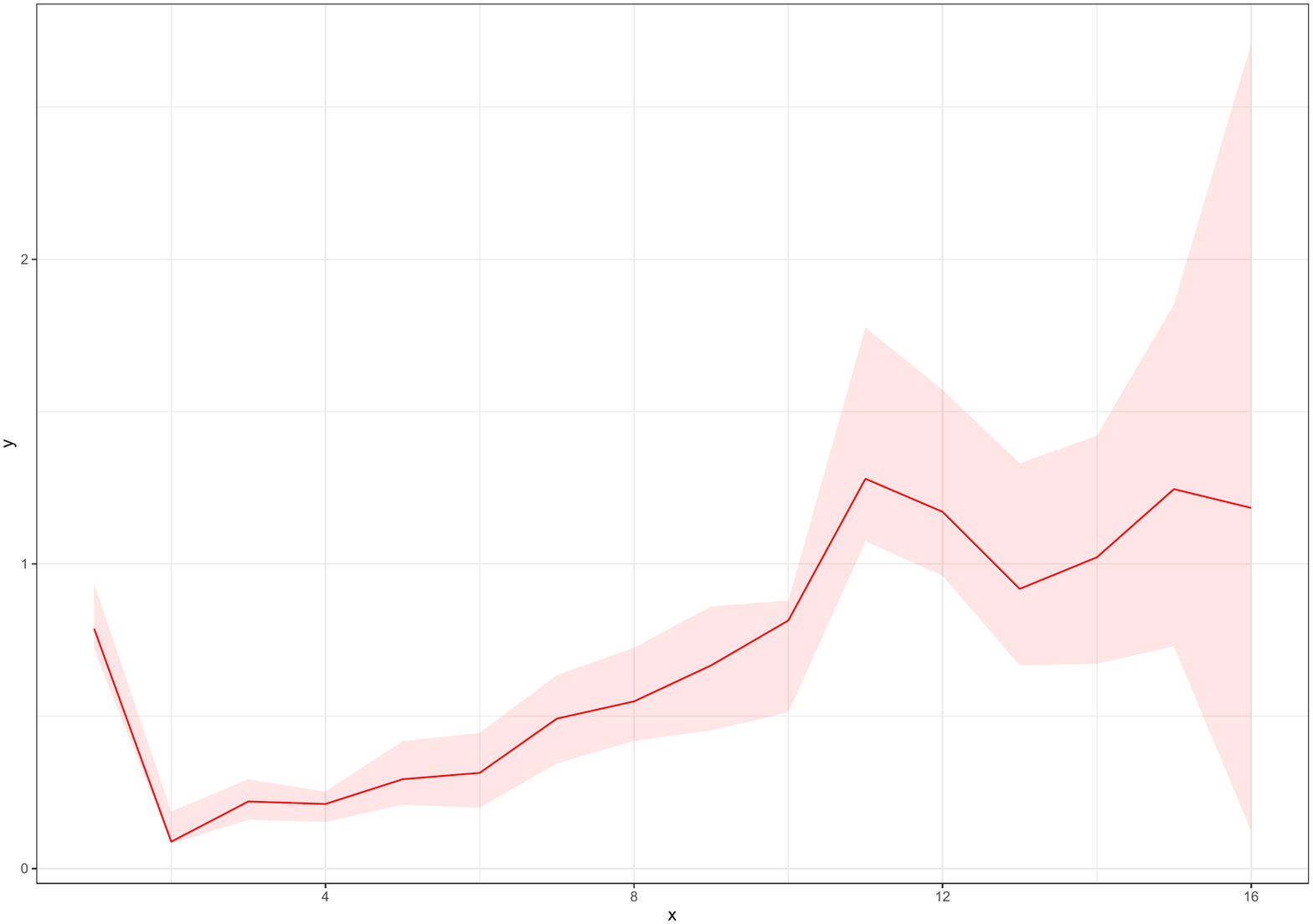}
     \caption{The estimated weights $\{\gamma_{i,70+}\}_{i=1}^{16}$.}
  \end{subfigure}

   \caption{\bf The posterior median estimate of instantaneous reproduction number per age group (0-29 (red line), 30-49 (blue line), 50-69 (pink line) and 70+ (brown line)), the posterior median estimate of weights $\{\mathbf{\{\gamma_{na}\}_{n=1}^{16}}\}_a$ (red line) and the 99\% CIs (ribbon) for Ashford.}
   \label{ER_Ashford4G}
\end{figure}

\begin{figure}[!h]
  \begin{subfigure}{7cm}
    \centering\includegraphics[width=6cm]{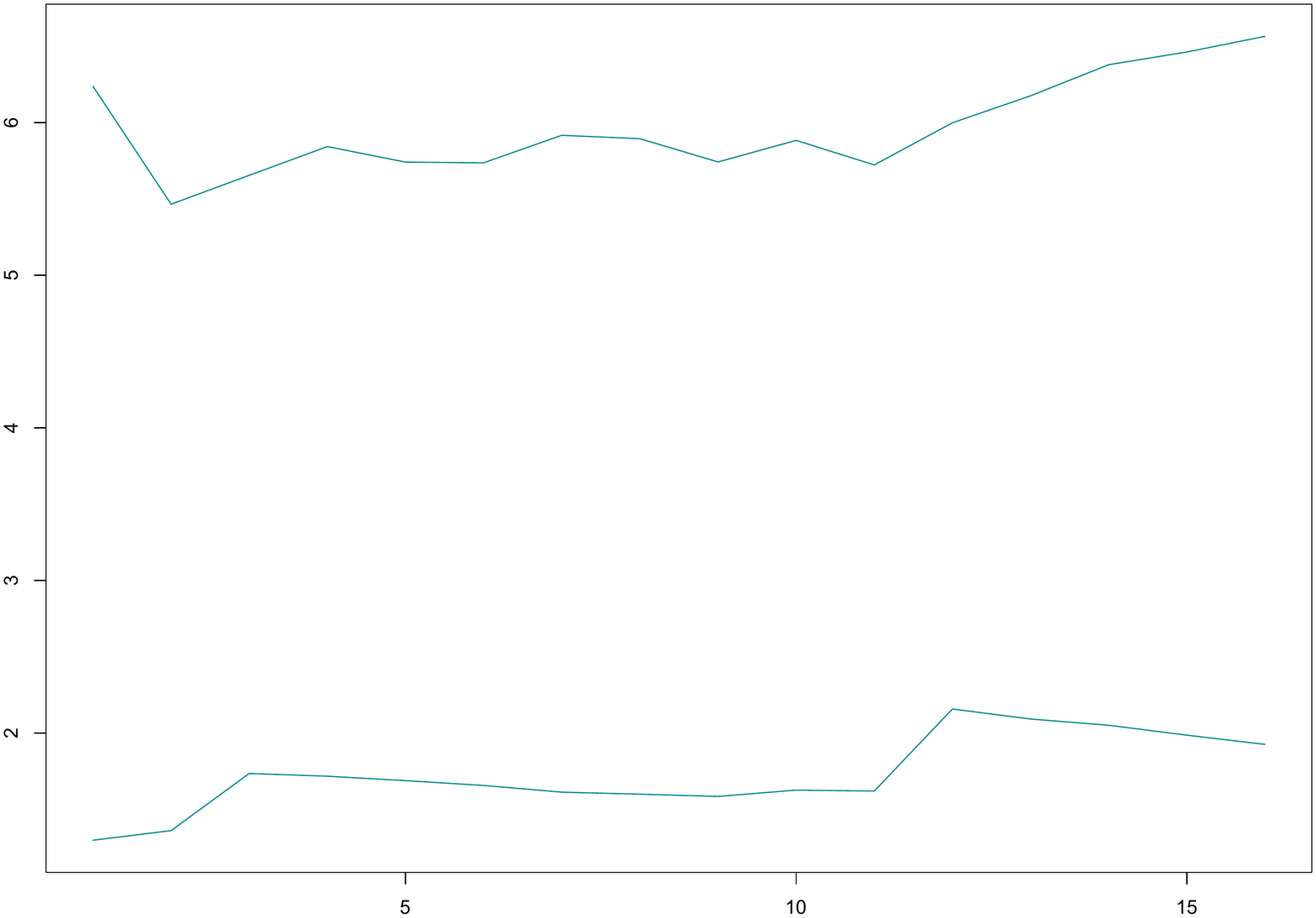}
    \caption{ $d$}
  \end{subfigure}
  \begin{subfigure}{7cm}
    \centering\includegraphics[width=6cm]{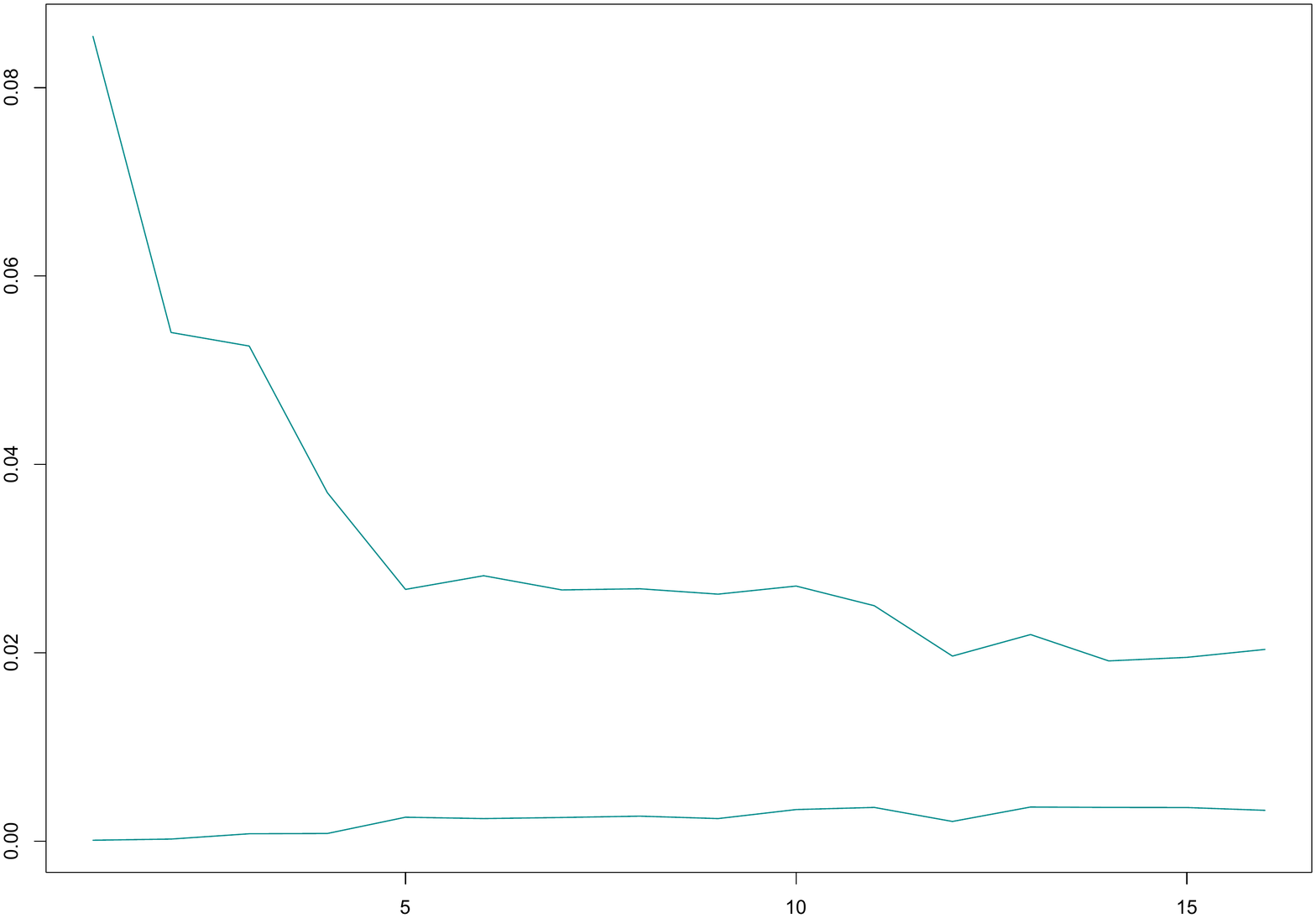}
    \caption{ $v$}
  \end{subfigure}
   \caption{\bf The $99\%$ CIs of time-constant parameters for Ashford.}
   \label{EHC2_Ashford4G}
\end{figure}

\begin{figure}[!h] 
  \begin{subfigure}{7cm}
    \centering\includegraphics[width=6cm]{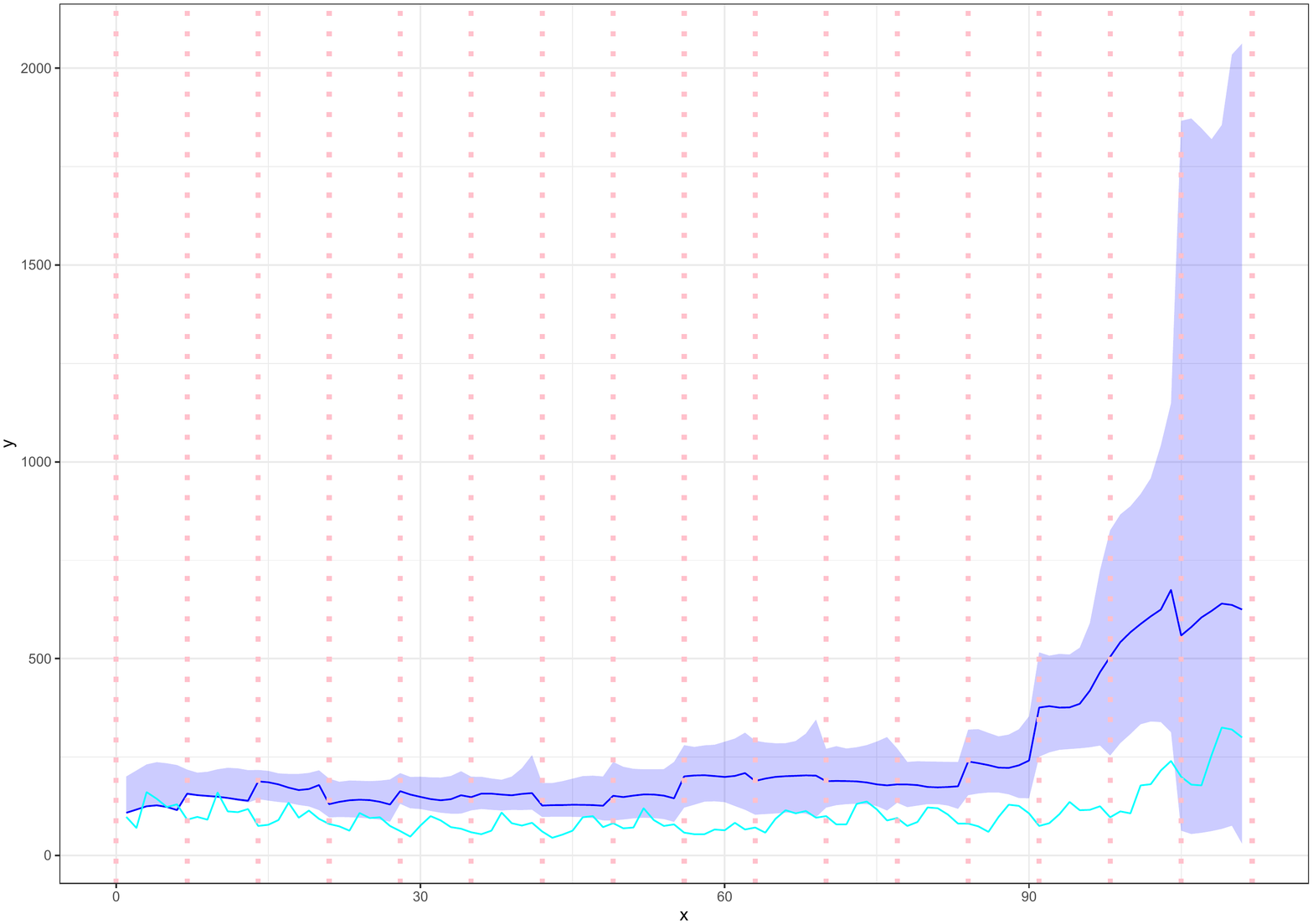}
   \caption{The estimated intensity of latent cases aged 0-29.}
  \end{subfigure}
  \begin{subfigure}{7cm}
    \centering\includegraphics[width=6cm]{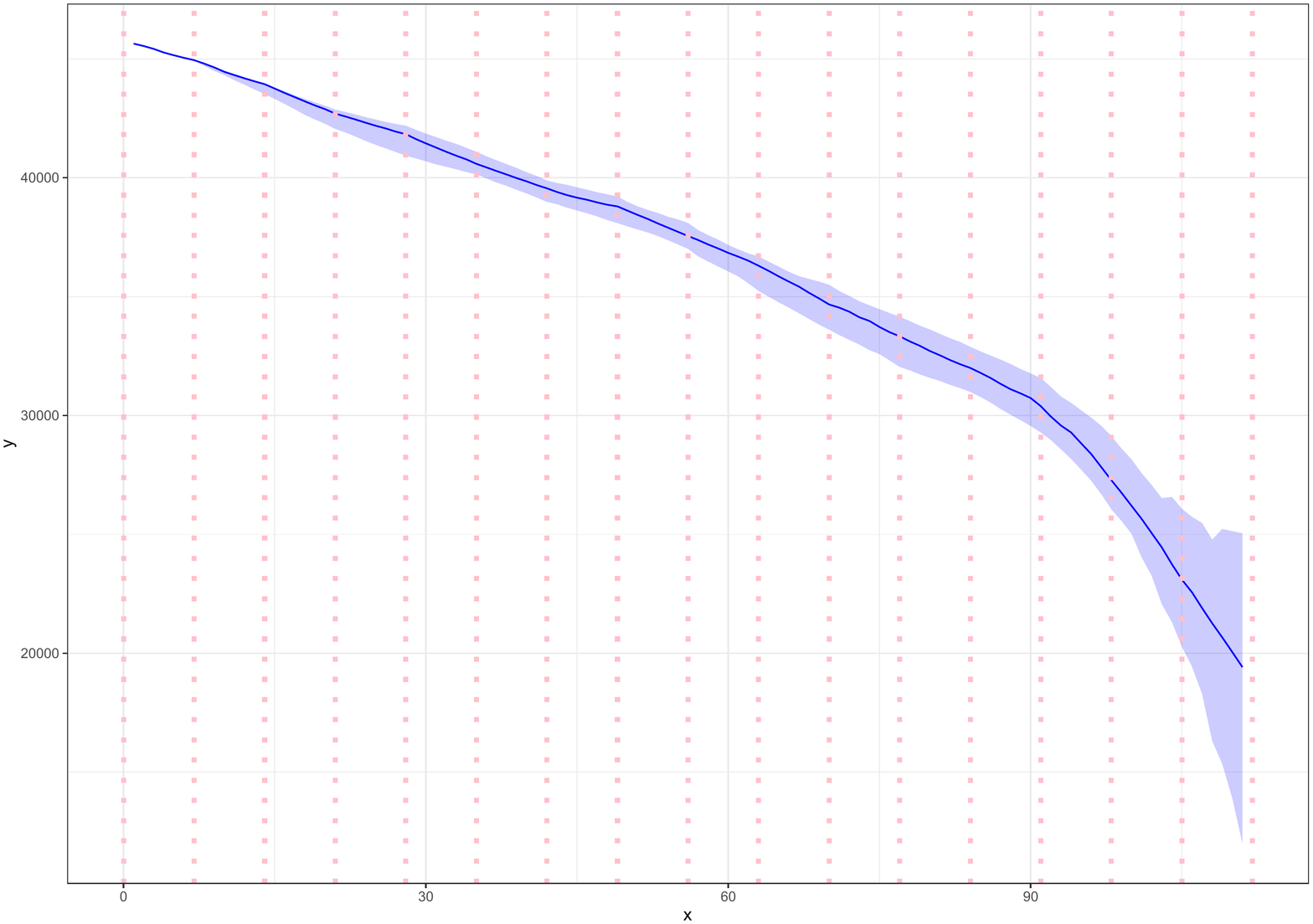}
   \caption{The estimated susceptibles aged 0-29.}
  \end{subfigure}
  \begin{subfigure}{7cm}
    \centering\includegraphics[width=6cm]{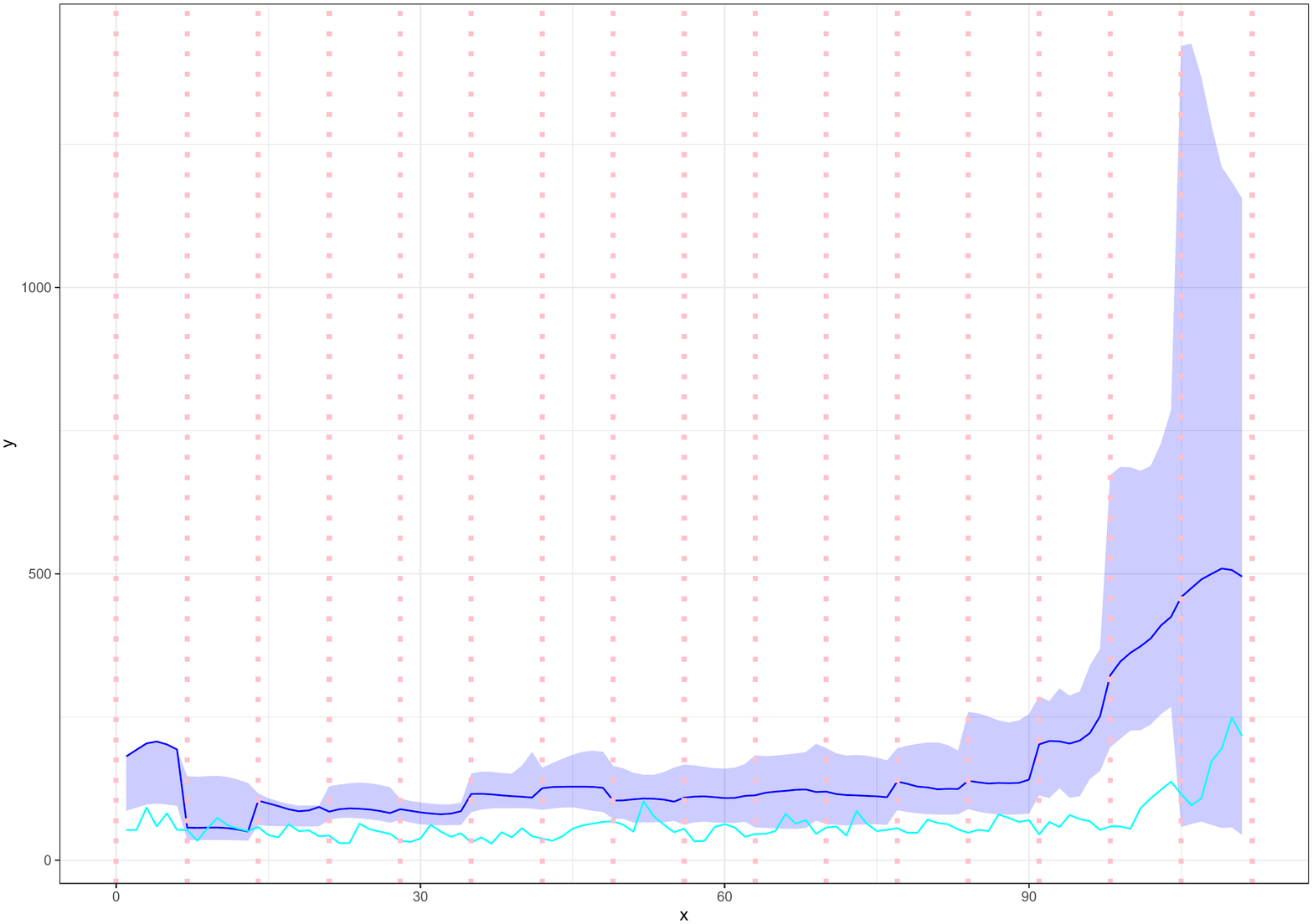}
   \caption{The estimated intensity of latent cases aged 30-49.}
  \end{subfigure}
  \begin{subfigure}{7cm}
    \centering\includegraphics[width=6cm]{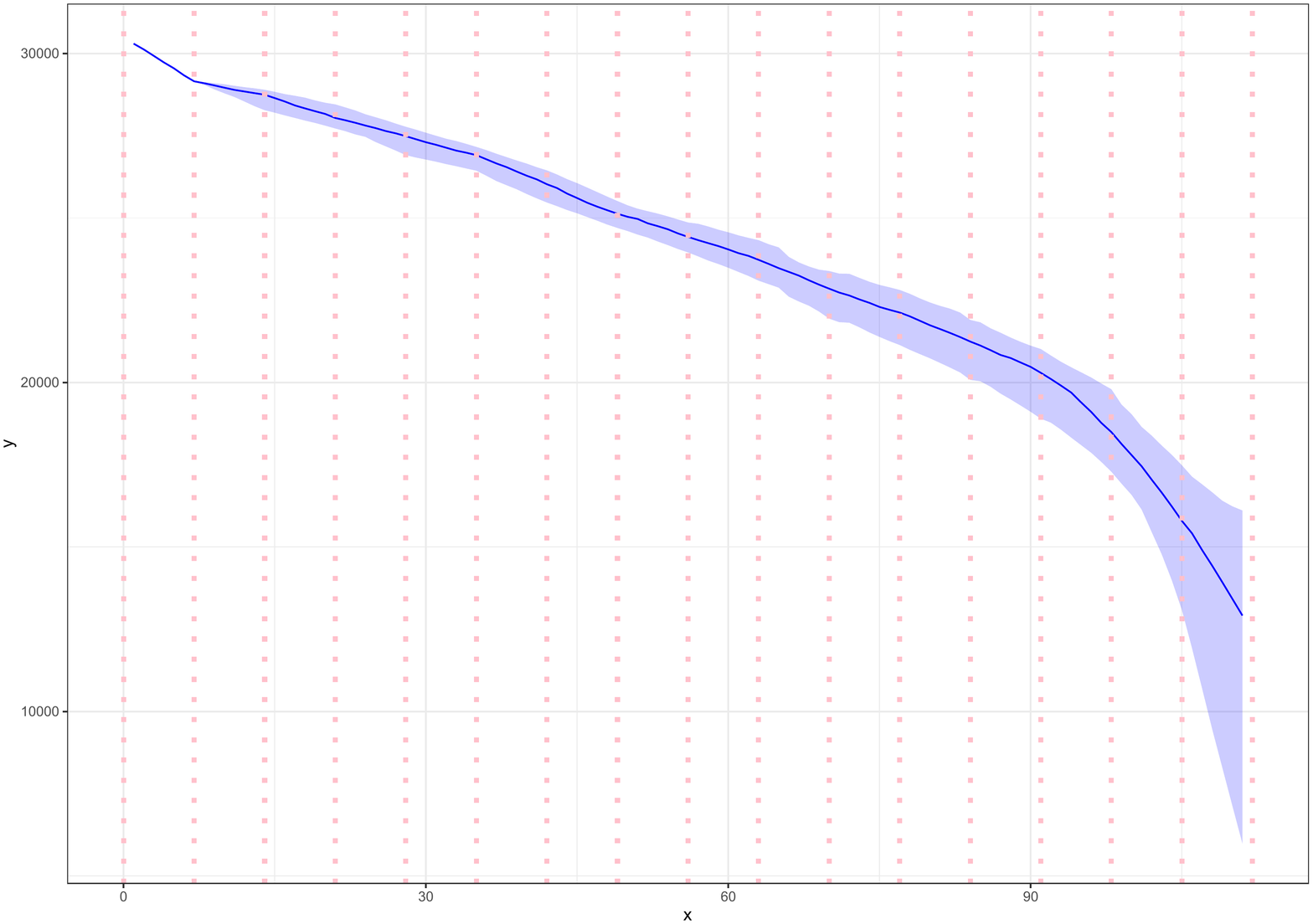}
   \caption{The estimated susceptibles aged 30-49.}
  \end{subfigure}
   \begin{subfigure}{7cm}
    \centering\includegraphics[width=6cm]{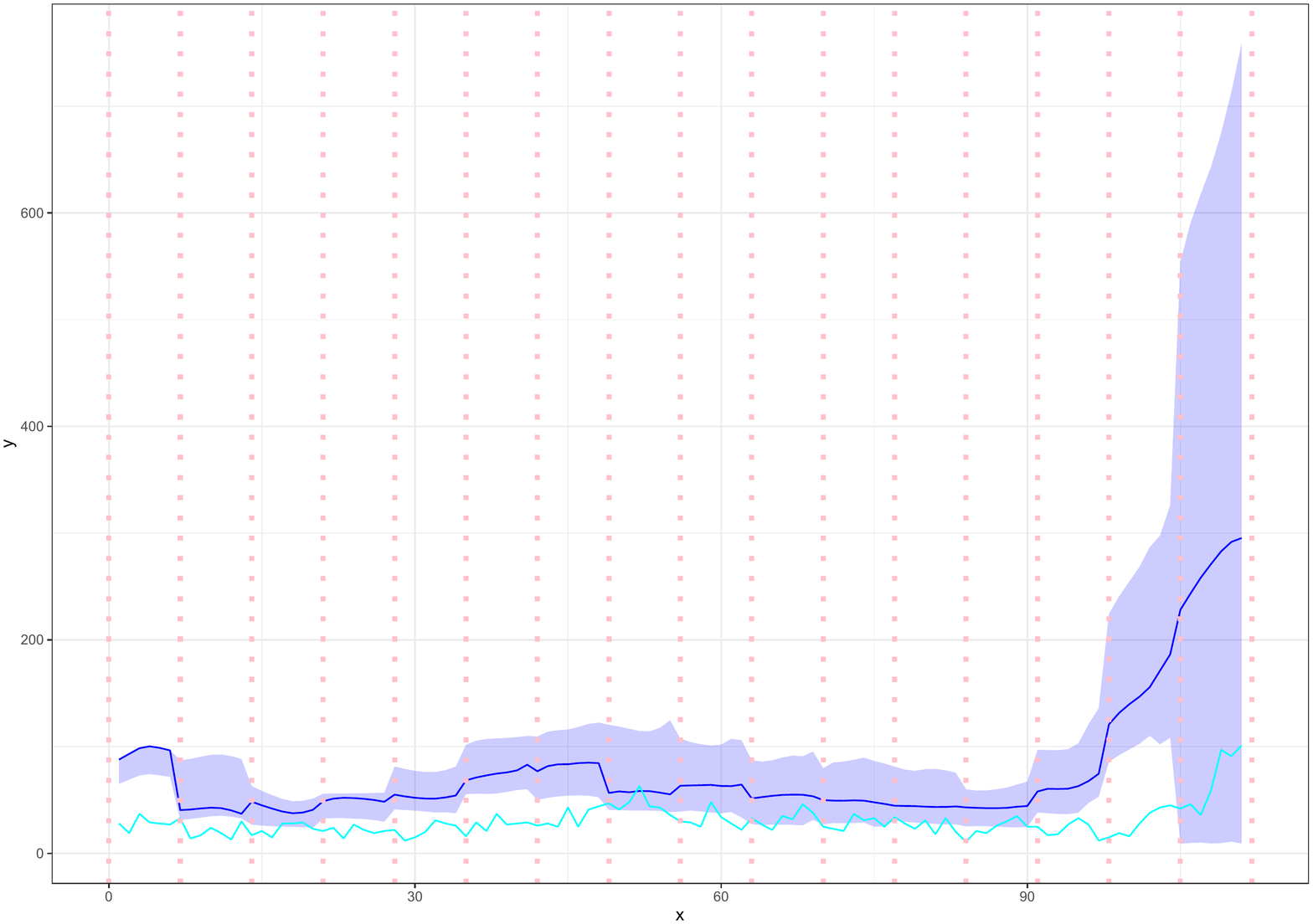}
   \caption{The estimated intensity of latent cases aged 50-69.}
  \end{subfigure}
  \begin{subfigure}{7cm}
    \centering\includegraphics[width=6cm]{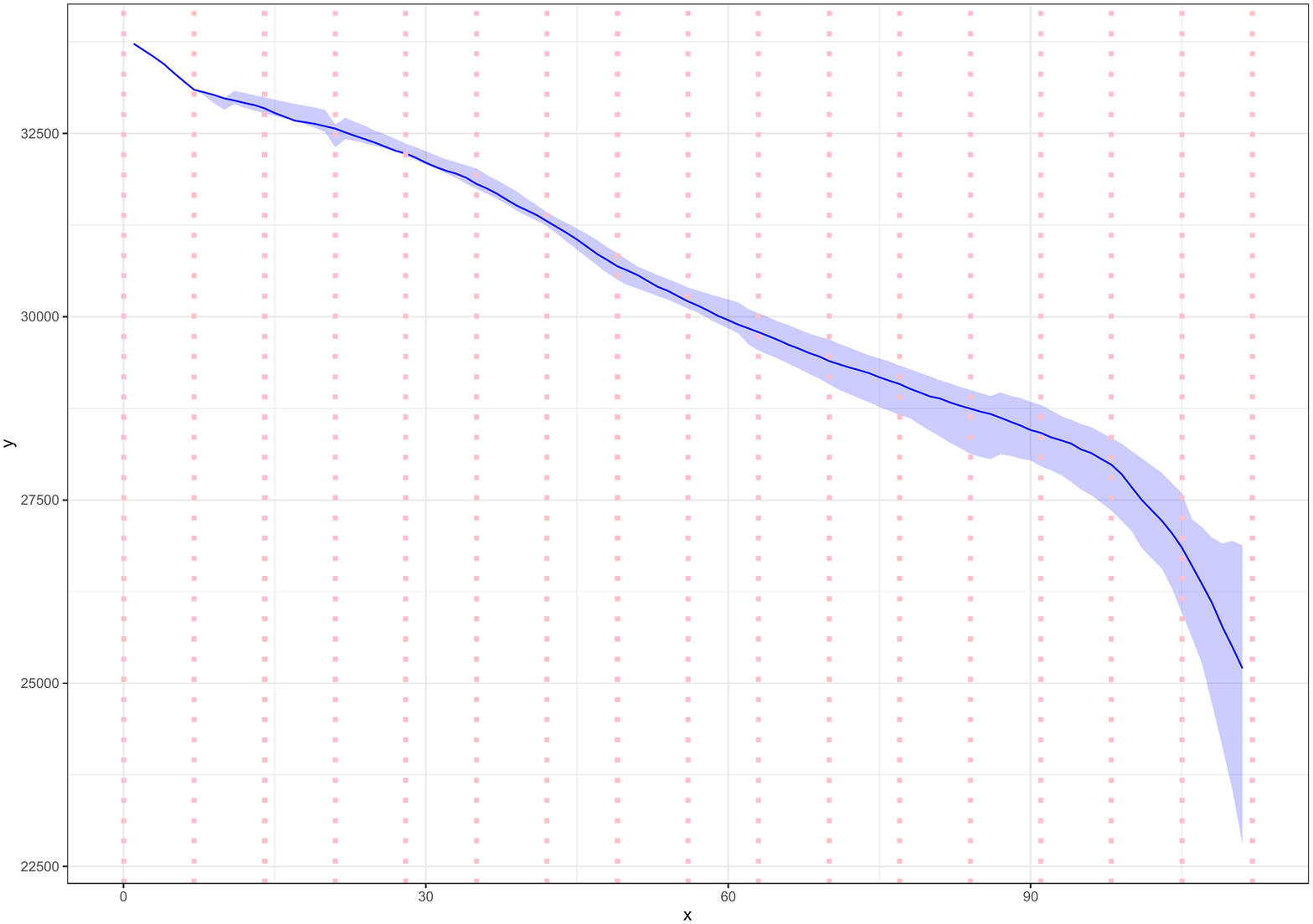}
   \caption{The estimated susceptibles aged 50-69.}
  \end{subfigure}
   \begin{subfigure}{7cm}
    \centering\includegraphics[width=6cm]{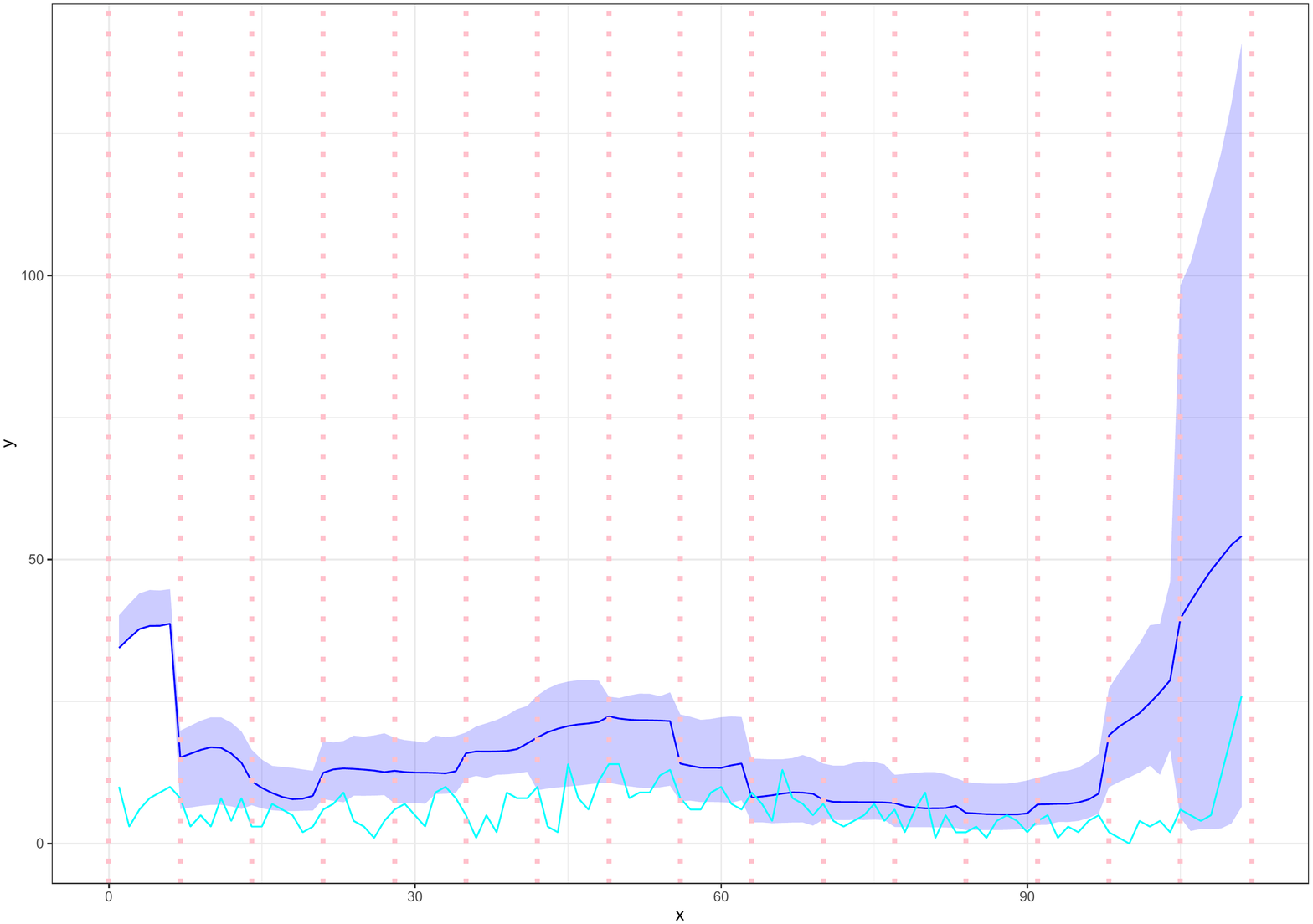}
   \caption{The estimated intensity of latent cases aged 70+.}
  \end{subfigure}
  \begin{subfigure}{7cm}
    \centering\includegraphics[width=6cm]{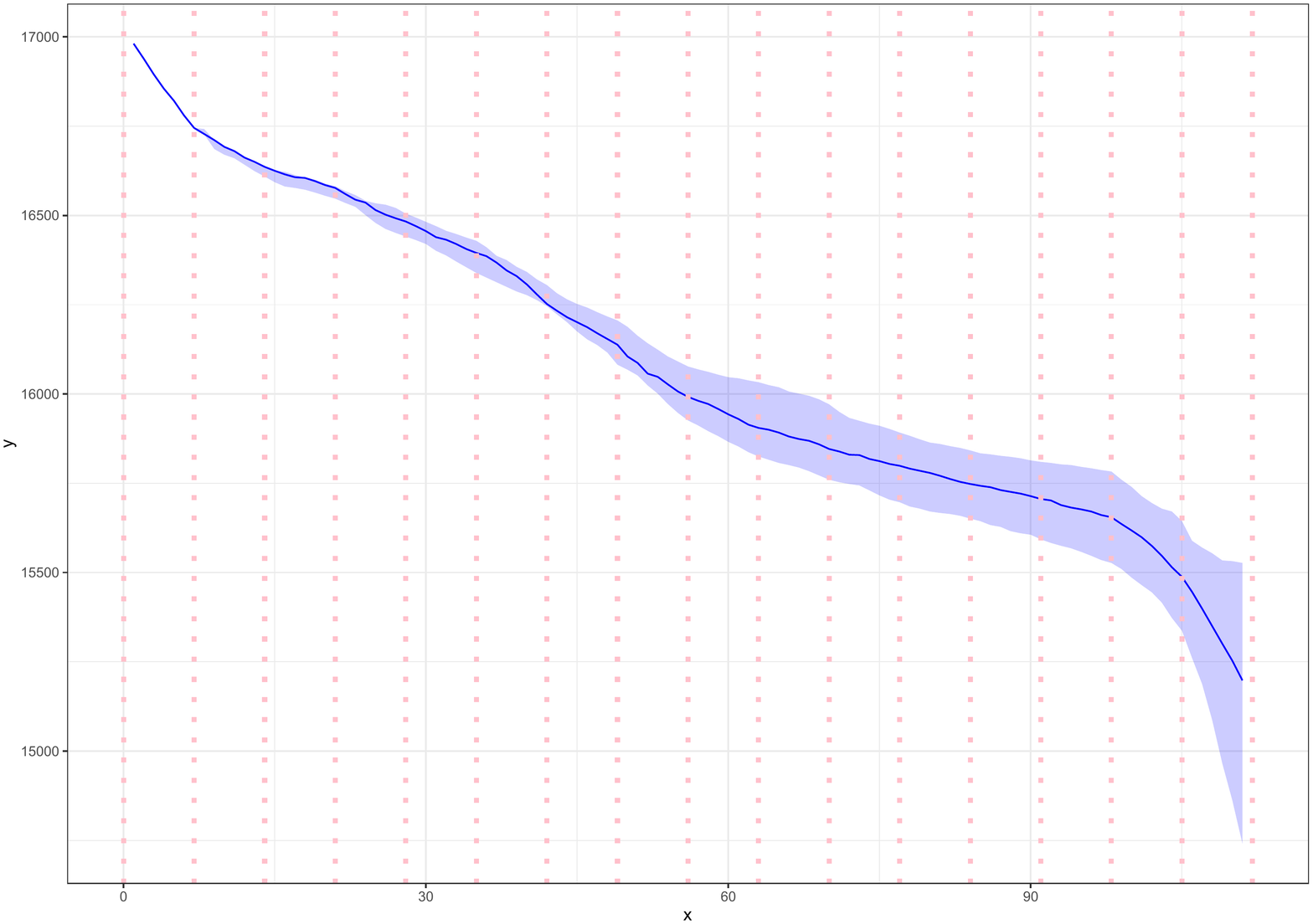}
   \caption{The estimated susceptibles aged 70+.}
  \end{subfigure}
  
  \caption{{\bf The estimated latent intensity and susceptibles (posterior median (blue line) ; 99\% CI (ribbon)) and the daily observed cases (cyan line) in Leicester.} The vertical dotted lines show the beginning of each week in the period we examine. }
  \label{EstInt_Leicester4G}
\end{figure}

\begin{figure}[!h] 
  \begin{subfigure}{7cm}
    \centering\includegraphics[width=6cm]{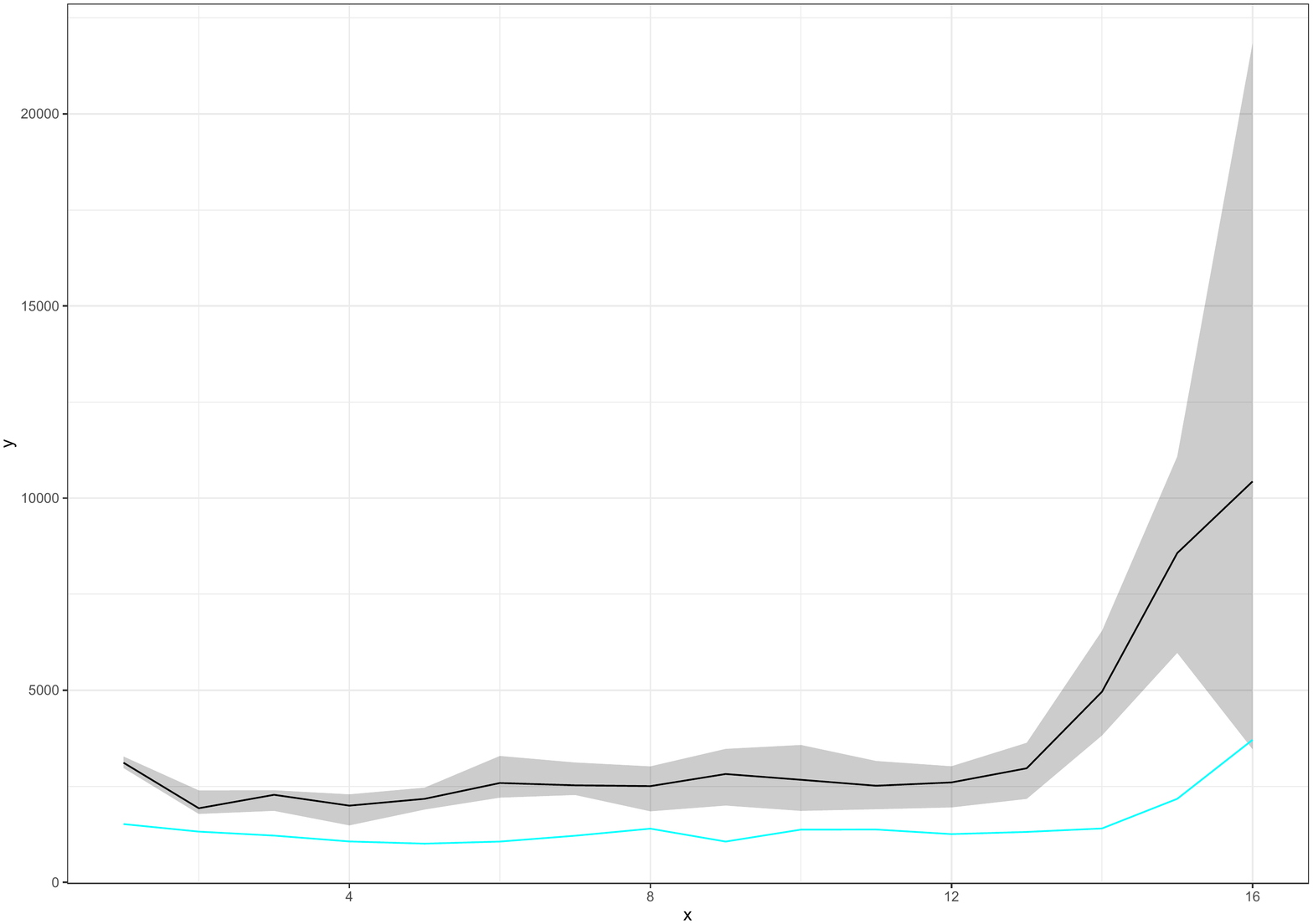}
   \caption{The estimated aggregated weekly hidden \\cases.}
  \end{subfigure}
  \begin{subfigure}{7cm}
    \centering\includegraphics[width=6cm]{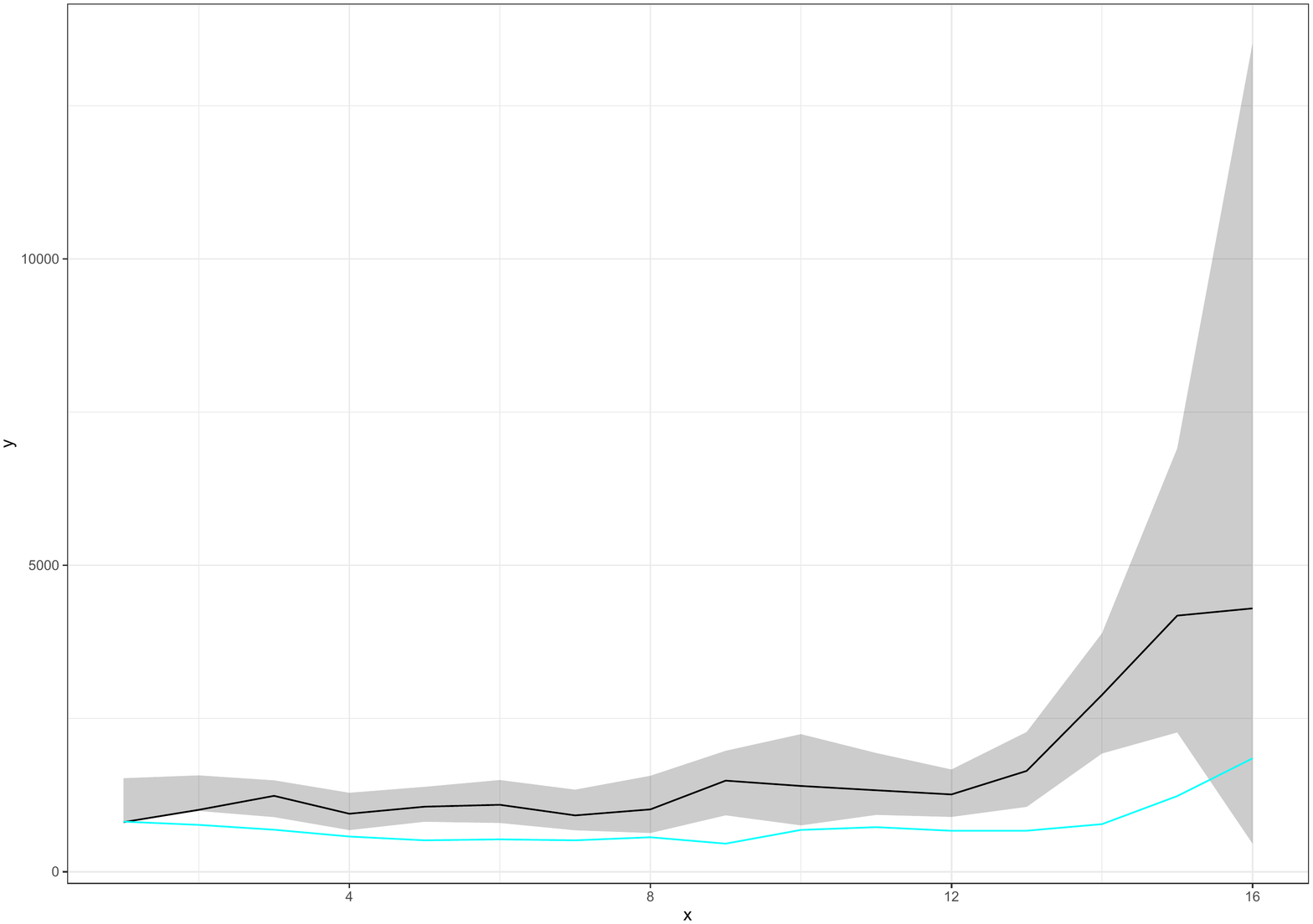}
    \caption{The estimated weekly hidden cases aged 0-29.}
  \end{subfigure}
  \begin{subfigure}{7cm}
    \centering\includegraphics[width=6cm]{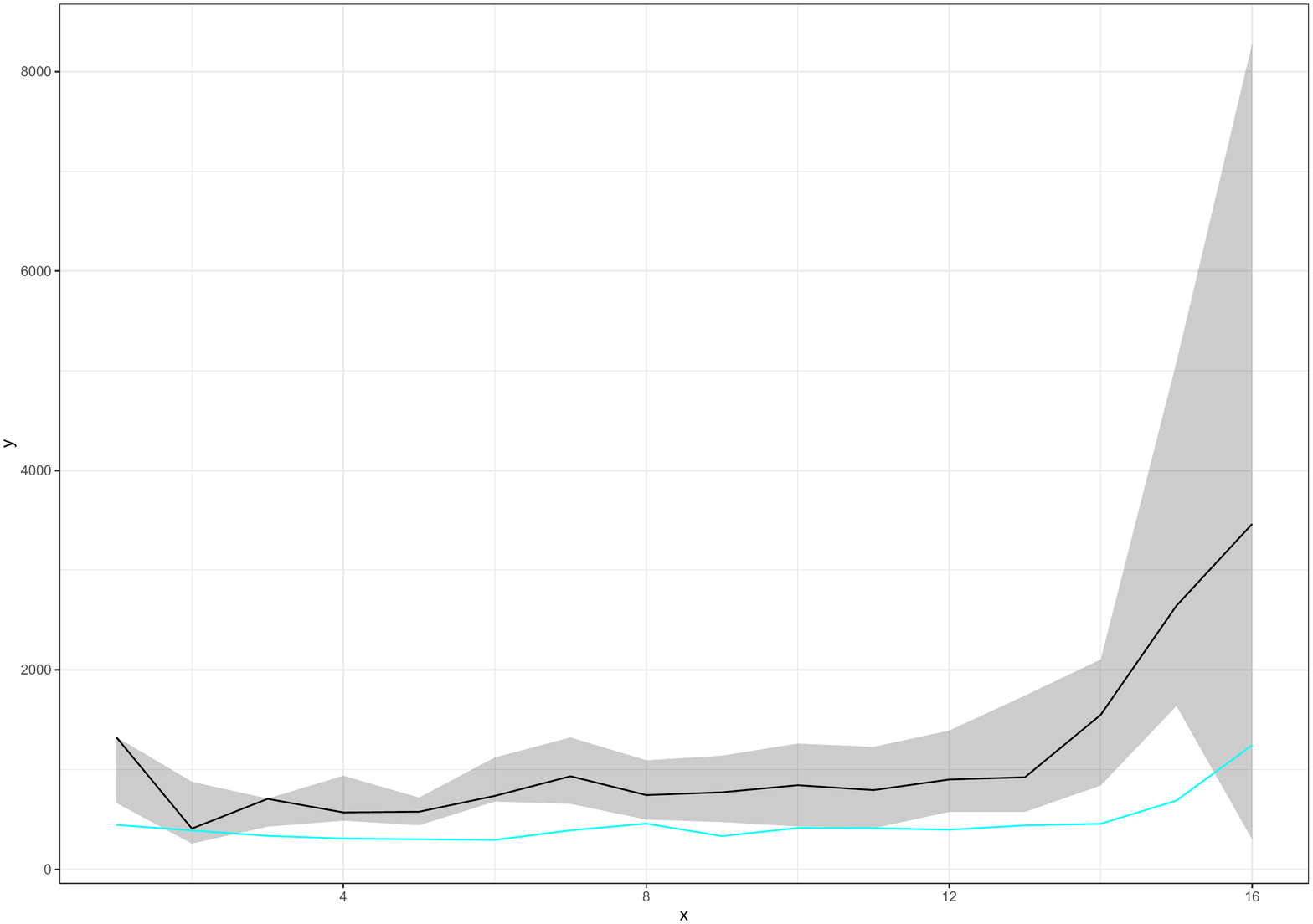}
    \caption{The estimated weekly hidden cases aged 30-49.}
  \end{subfigure}
  \begin{subfigure}{7cm}
    \centering\includegraphics[width=6cm]{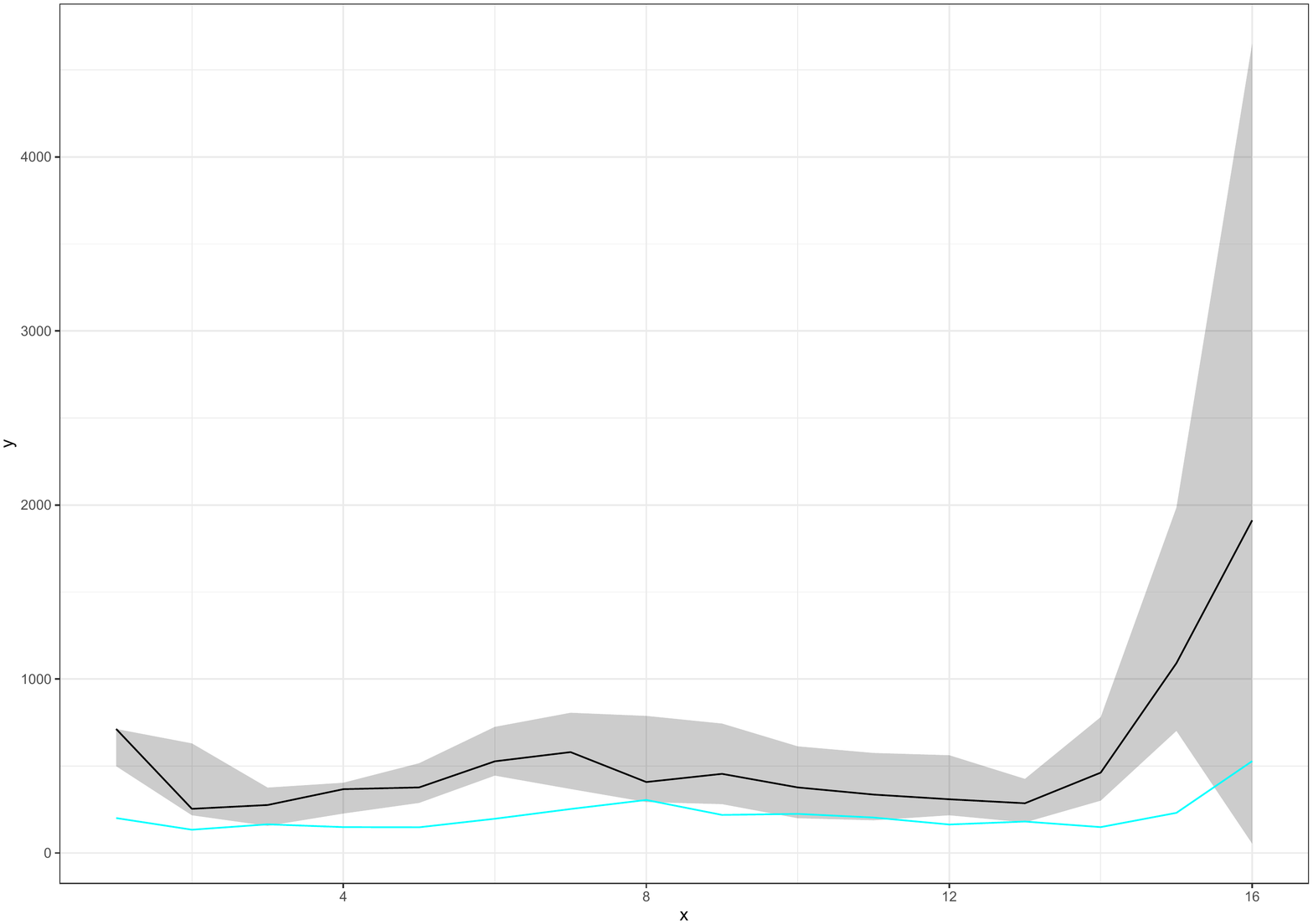}
    \caption{The estimated weekly hidden cases aged 50-69.}
  \end{subfigure}
  \begin{subfigure}{7cm}
    \centering\includegraphics[width=6cm]{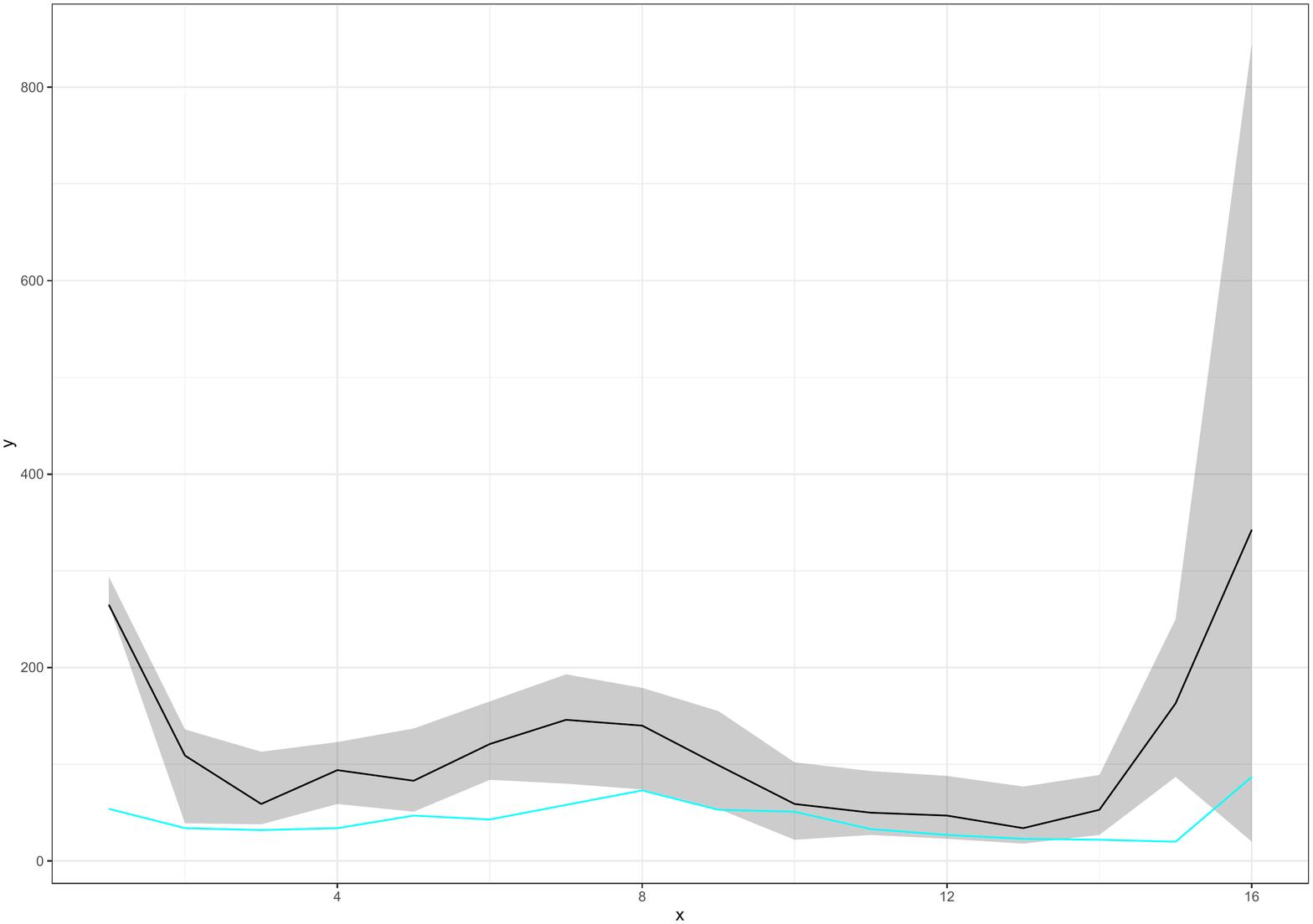}
    \caption{The estimated weekly hidden cases aged 70+.}
  \end{subfigure}
   \caption{\bf The estimated weekly latent cases (black line; 99\% CI (ribbon)) and the weekly observed cases (cyan line) in Leicester.}
   \label{EHC_Leicester4G}
\end{figure}

\begin{figure}[!h] 
   \begin{subfigure}{7cm}
    \centering\includegraphics[width=6cm]{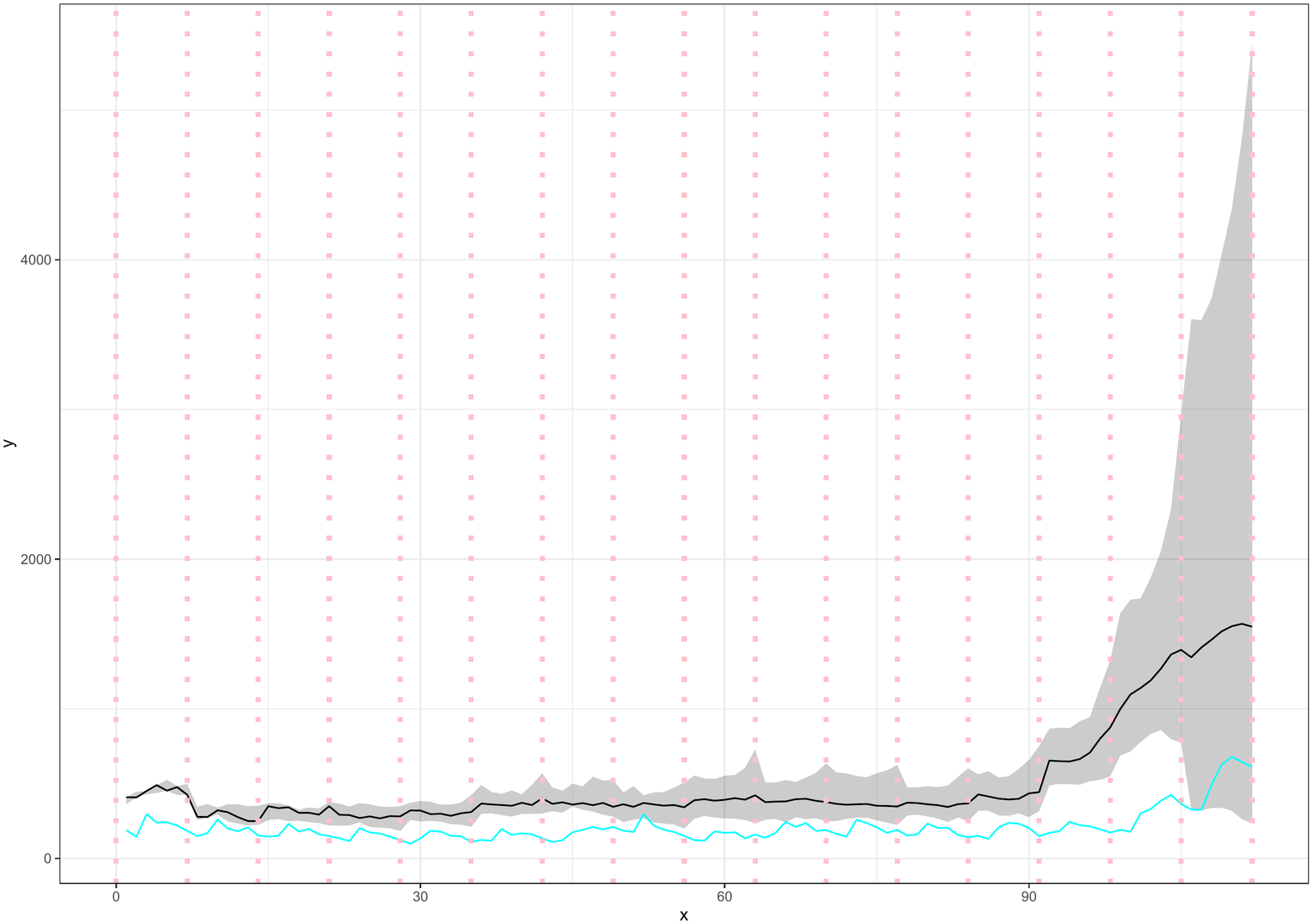}
    \caption{The estimated aggregated daily hidden cases.}
  \end{subfigure}
   \begin{subfigure}{7cm}
    \centering\includegraphics[width=6cm]{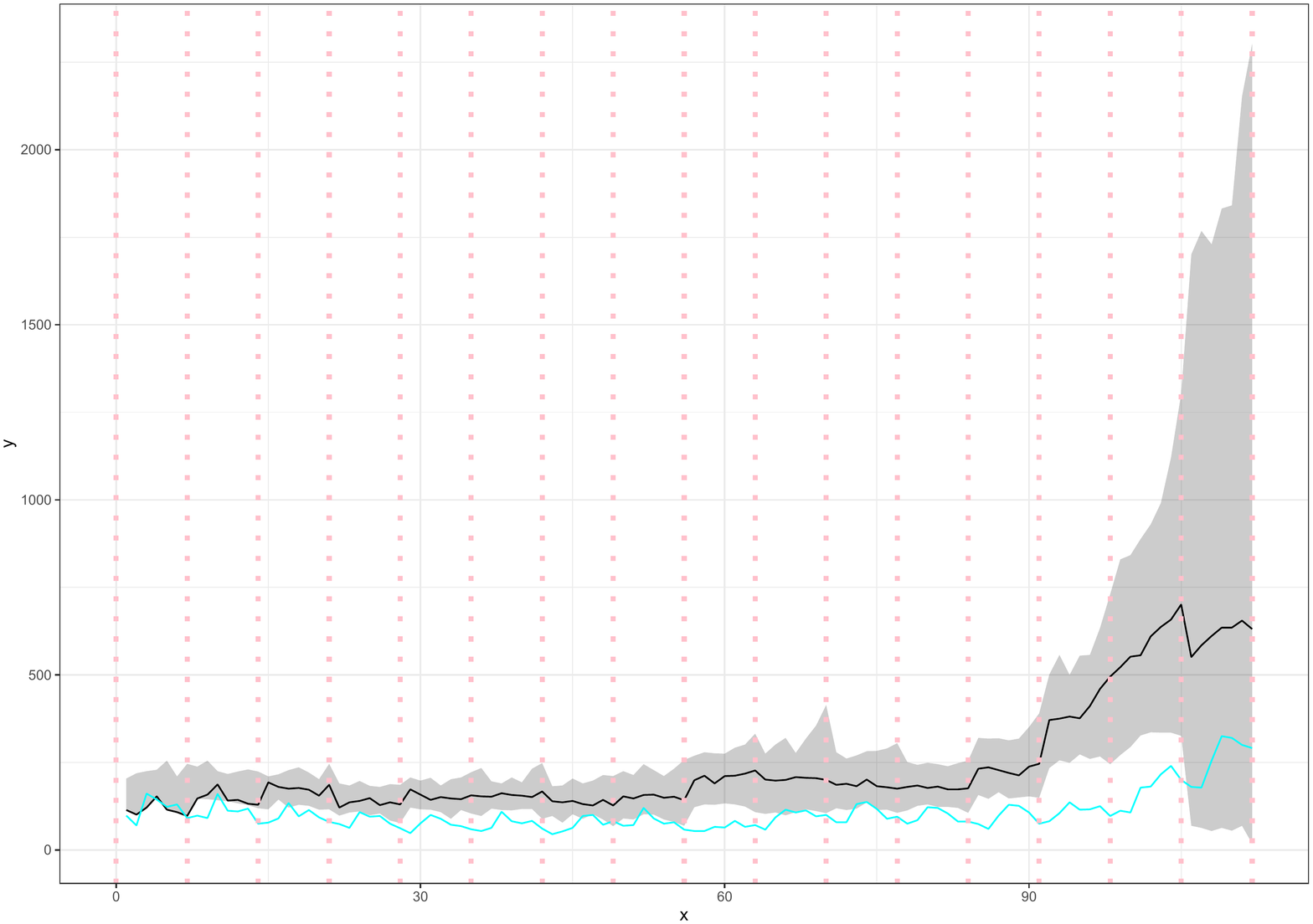}
    \caption{The estimated daily hidden cases aged 0-29.}
  \end{subfigure}
  \begin{subfigure}{7cm}
    \centering\includegraphics[width=6cm]{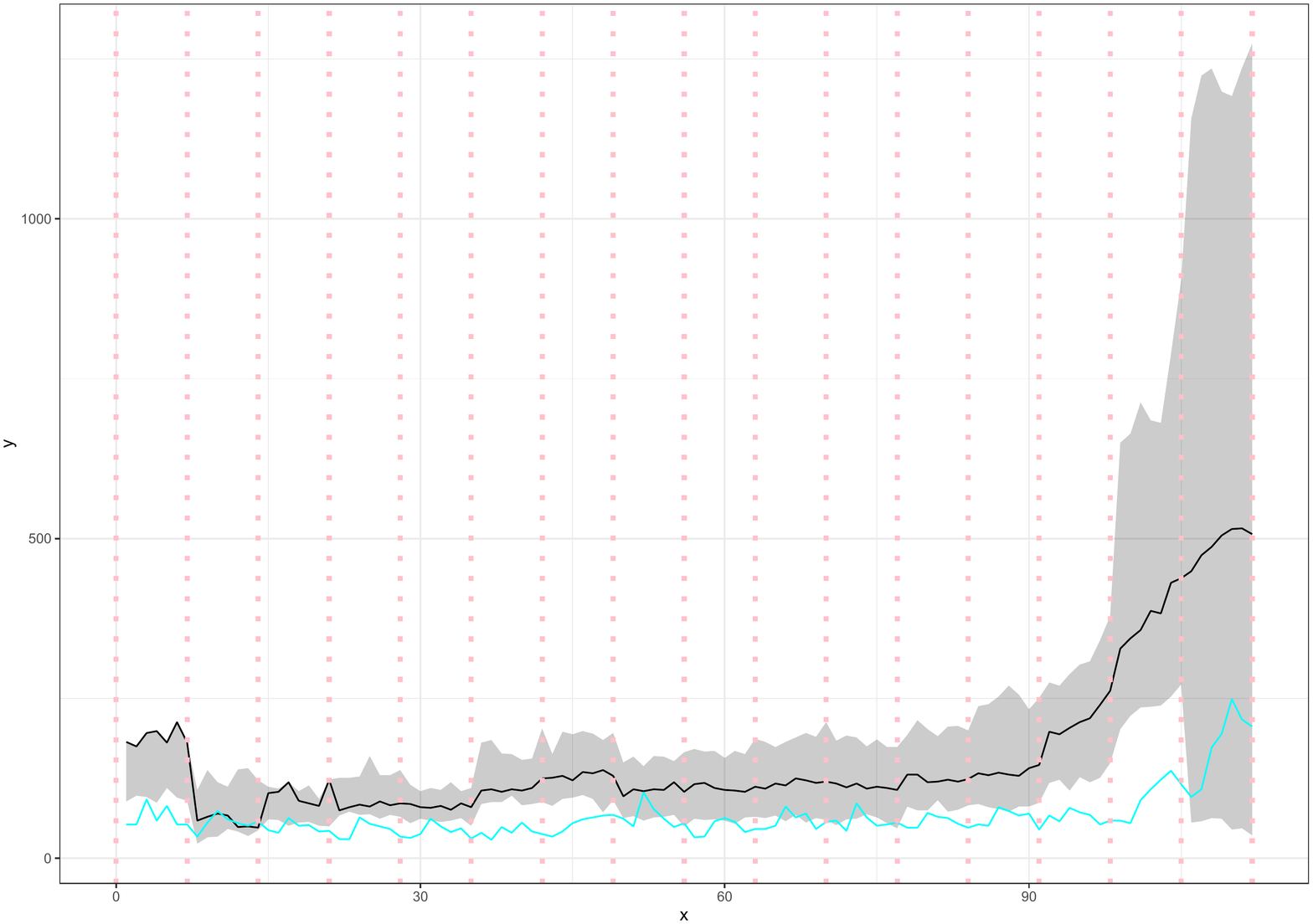}
    \caption{The estimated daily hidden cases aged 30-49.}
  \end{subfigure}
  \begin{subfigure}{7cm}
    \centering\includegraphics[width=6cm]{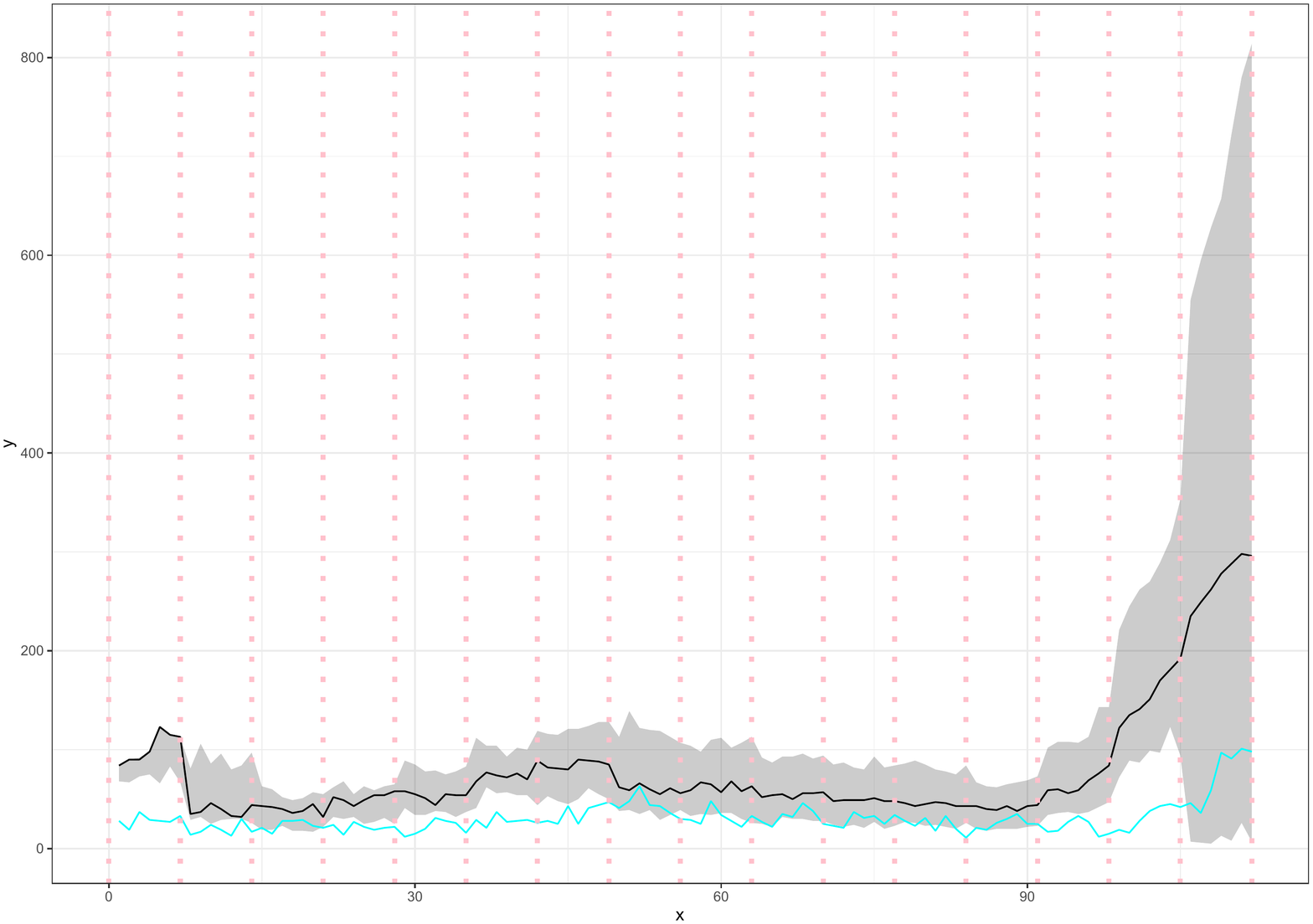}
    \caption{The estimated daily hidden cases aged 50-69.}
  \end{subfigure}
  \begin{subfigure}{7cm}
    \centering\includegraphics[width=6cm]{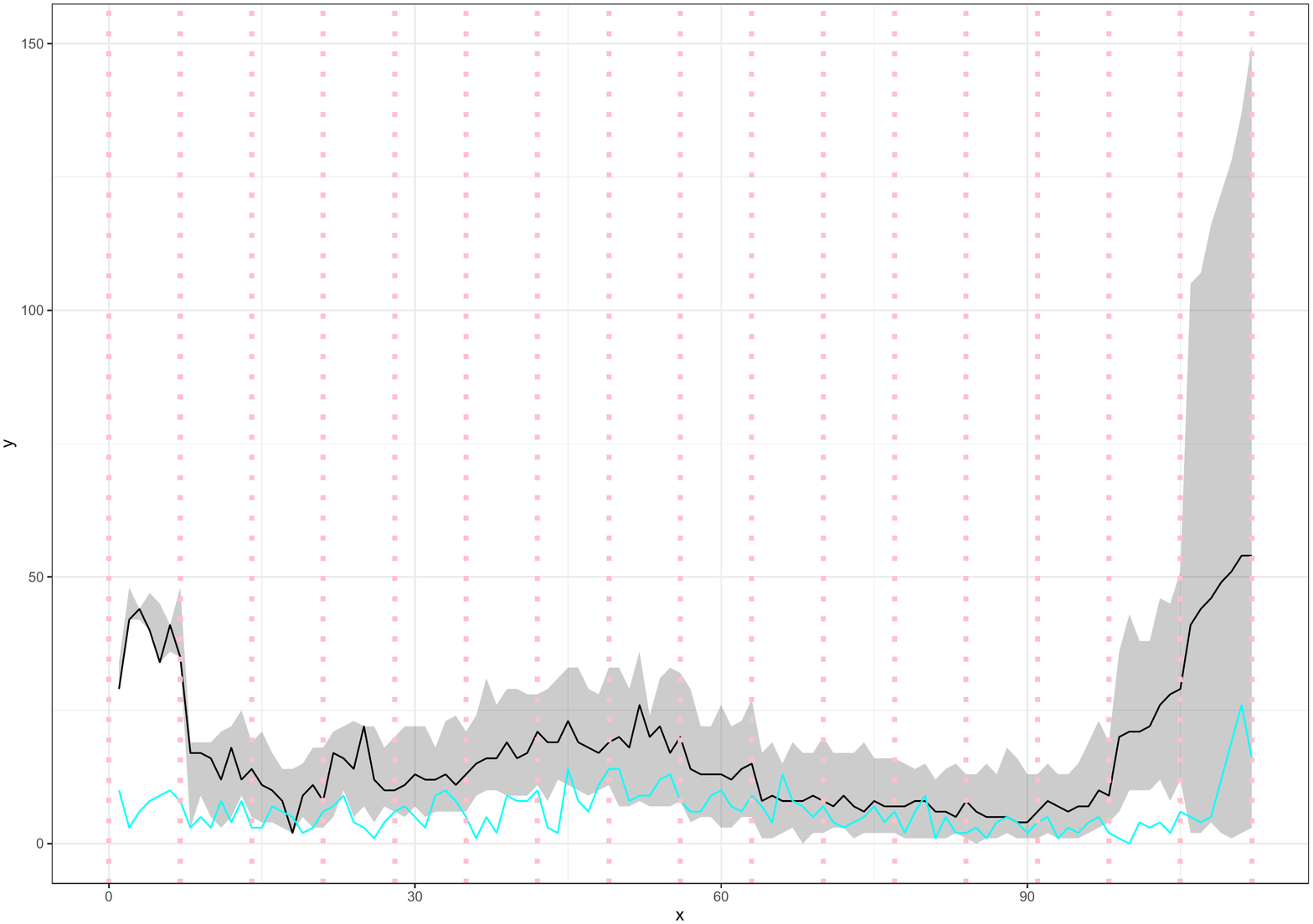}
    \caption{The estimated daily hidden cases aged 70+.}
  \end{subfigure}
   \caption{ {\bf The estimated daily latent cases (posterior median (black line); 99\% CI (ribbon)) and the daily observed cases (cyan line) in Leicester.} The vertical dotted lines show the beginning of each week in the period we examine.}
   \label{EHDC_Leicester4G}
\end{figure}

\begin{figure}[!h] 
 \begin{subfigure}{7cm}
    \centering\includegraphics[width=6cm]{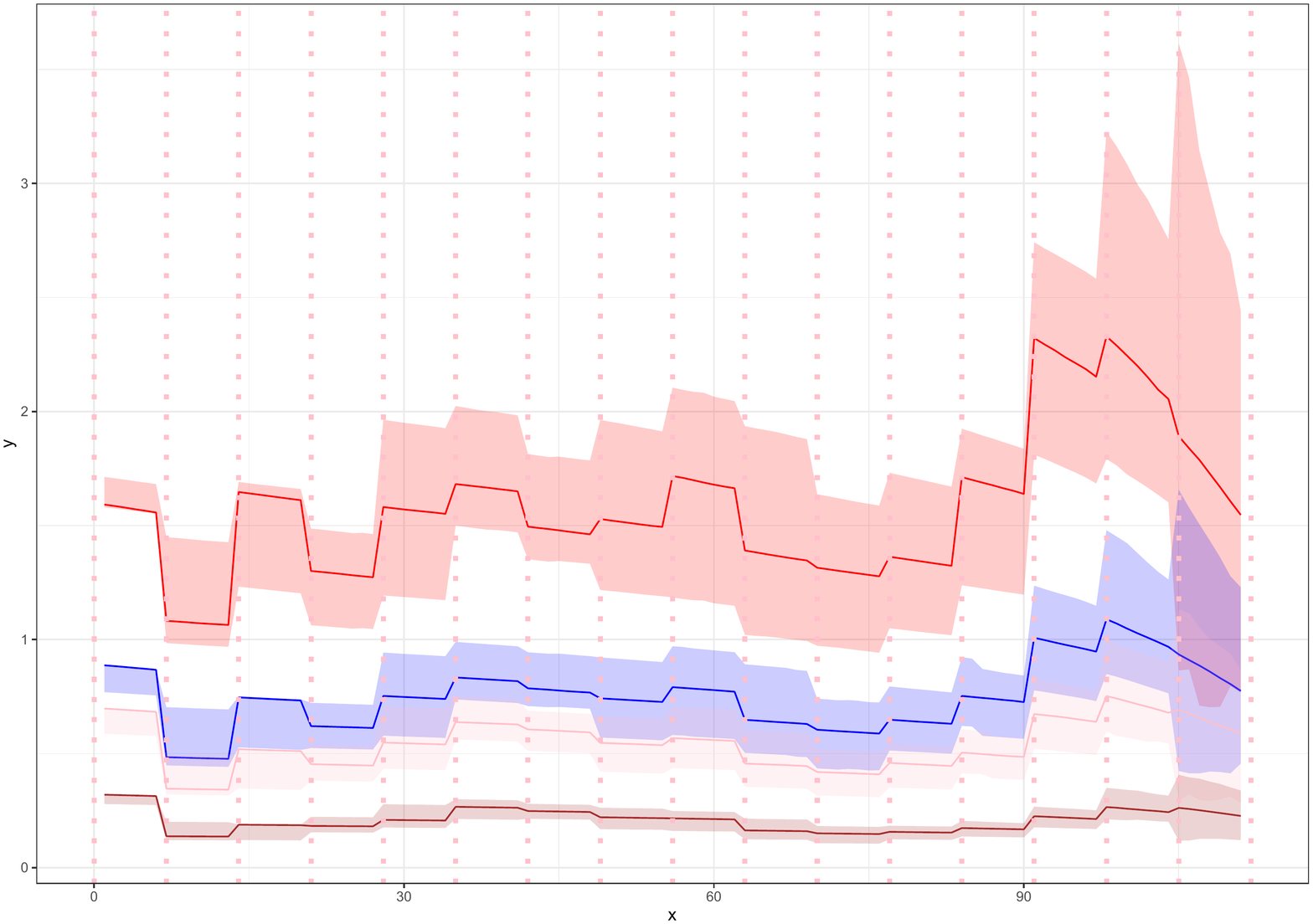}
     \caption{The per age group instantaneous reproduction numbers.}
  \end{subfigure}
  \begin{subfigure}{7cm}
    \centering\includegraphics[width=6cm]{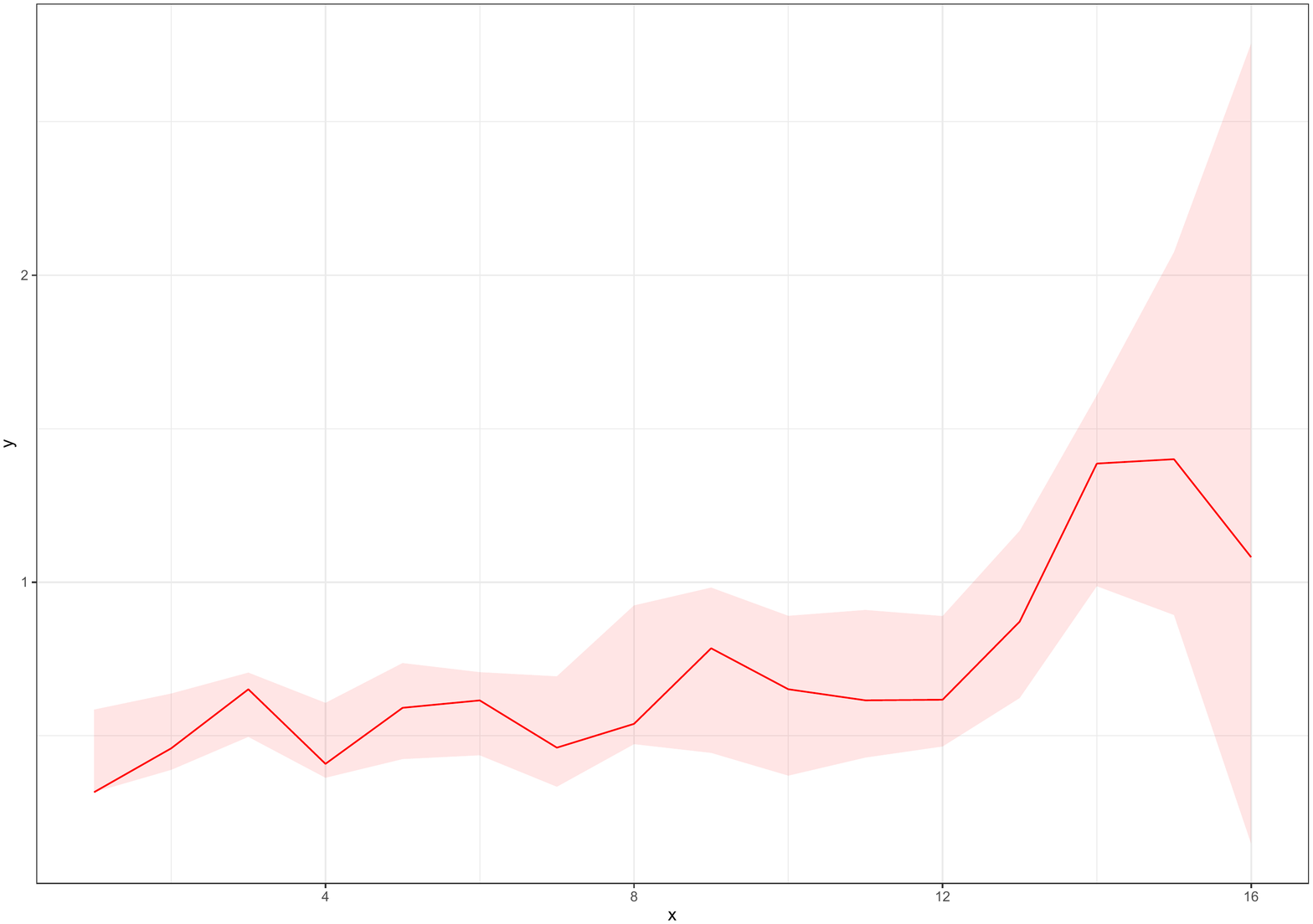}
   \caption{The estimated weights $\{\gamma_{i,0-29}\}_{i=1}^{16}$.}
  \end{subfigure}
  \begin{subfigure}{7cm}
    \centering\includegraphics[width=6cm]{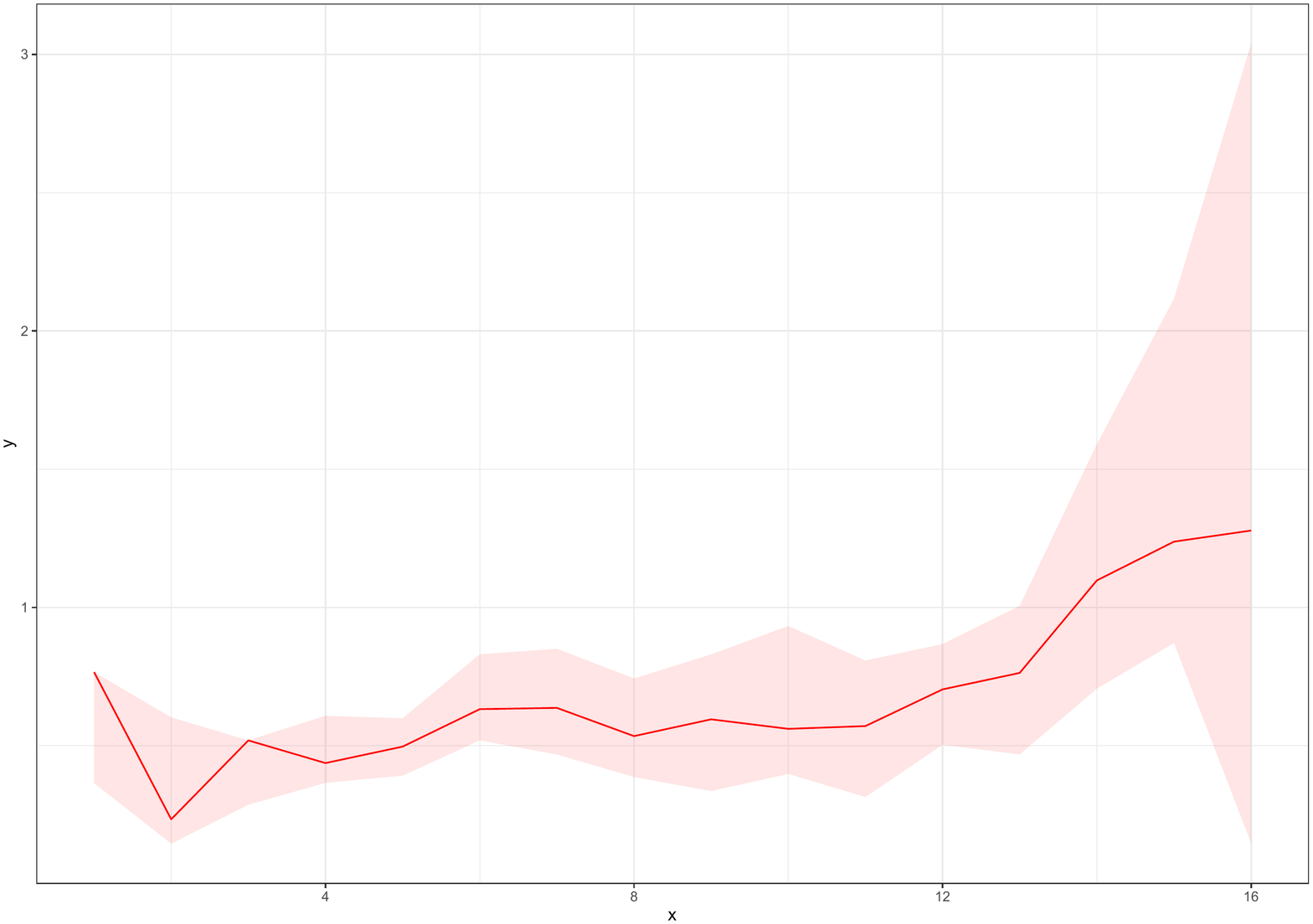}
     \caption{The estimated weights $\{\gamma_{i,30-49}\}_{i=1}^{16}$.}
  \end{subfigure}
  \begin{subfigure}{7cm}
    \centering\includegraphics[width=6cm]{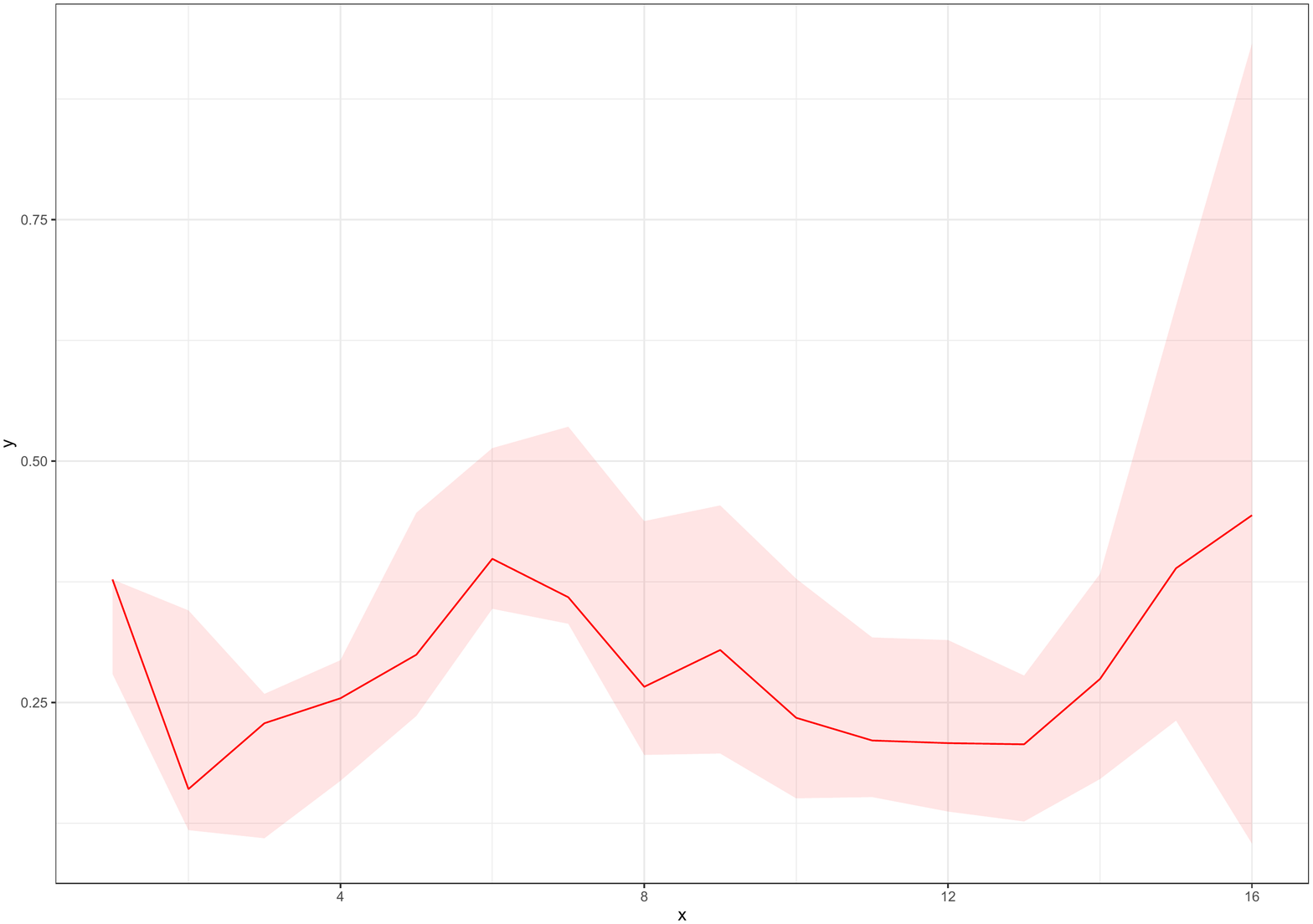}
    \caption{The estimated weights $\{\gamma_{i,50-69}\}_{i=1}^{16}$.}
  \end{subfigure}
   \begin{subfigure}{7cm}
    \centering\includegraphics[width=6cm]{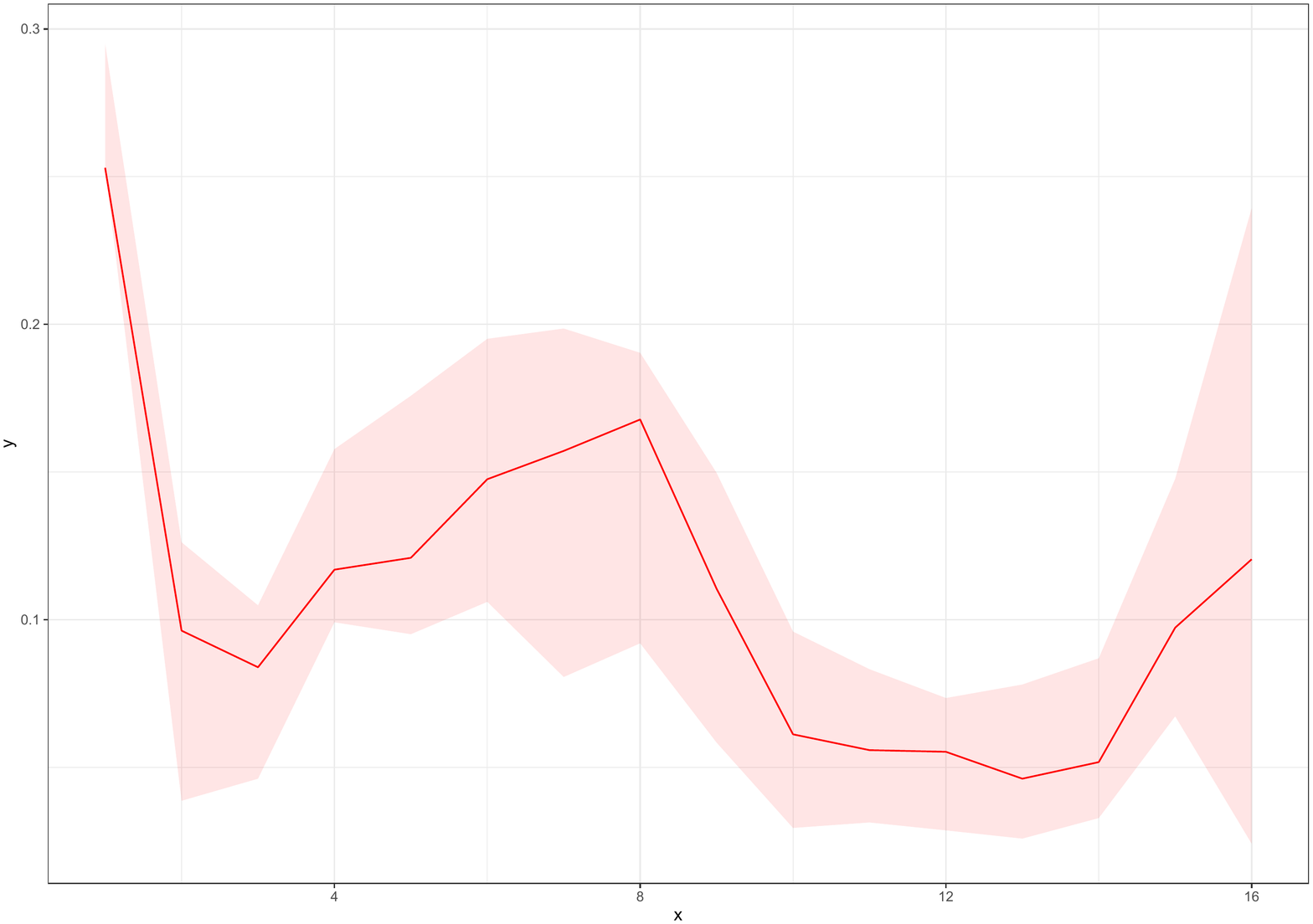}
     \caption{The estimated weights $\{\gamma_{i,70+}\}_{i=1}^{16}$.}
  \end{subfigure}

   \caption{\bf The posterior median estimate of instantaneous reproduction number per age group (0-29 (red line), 30-49 (blue line), 50-69 (pink line) and 70+ (brown line)), the posterior median estimate of weights $\{\{\gamma_{na}\}_{n=1}^{16}\}_a$ (red line) and the 99\% CIs (ribbon) for Leicester.}
   \label{ER_Leicester4G}
\end{figure}

\begin{figure}[!h] 
  \begin{subfigure}{7cm}
    \centering\includegraphics[width=6cm]{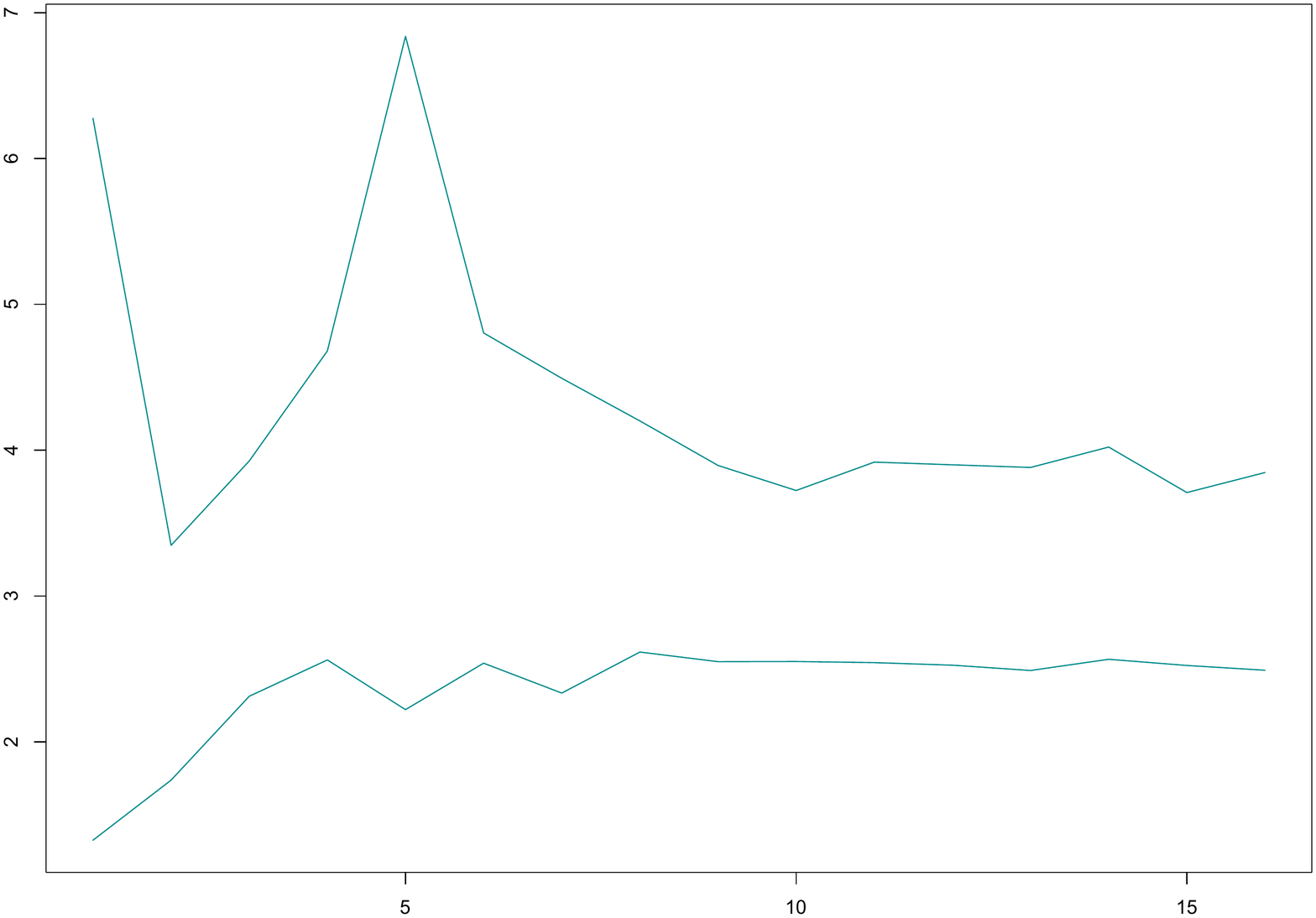}
    \caption{$d$}
  \end{subfigure}
  \begin{subfigure}{7cm}
    \centering\includegraphics[width=6cm]{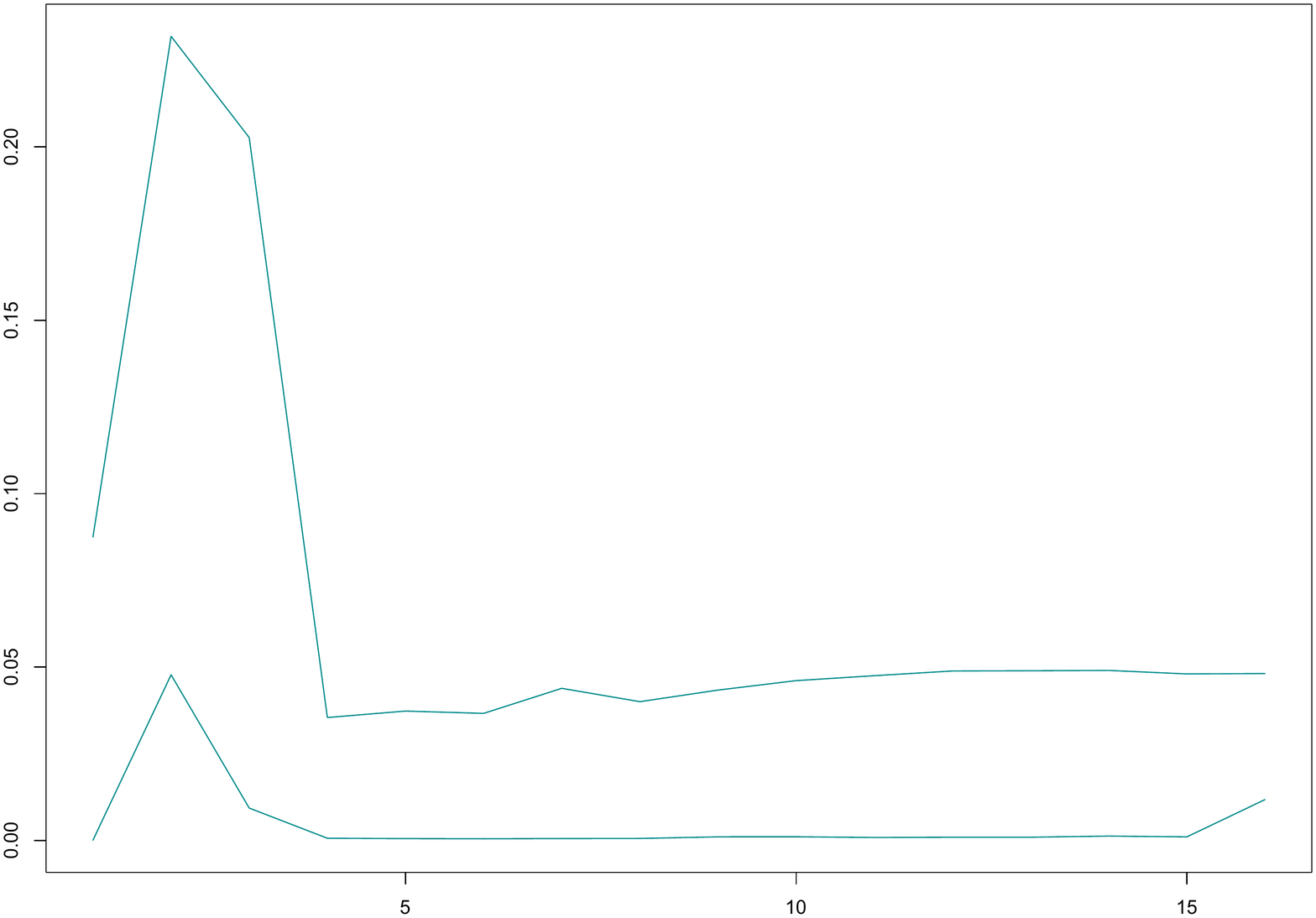}
    \caption{$v$}
  \end{subfigure}
   \caption{\bf The $99\%$ CIs of time-constant parameters for Leicester.}
   \label{EHC2_Leicester4G}
\end{figure}

Figures \ref{EstInt_Kings4G}-\ref{ER_Kings4G}, \ref{EHC2_Kings4G}- \ref{EHC2_Ashford4G}, \ref{EstInt_Leicester4G}-\ref{EHC2_Leicester4G}  show the estimated intensity, the estimated weekly and daily hidden cases, the estimated susceptibles, the estimated instantaneous reproduction number and the estimated weights $\{\gamma_{na}\}_{n=1}^{16}$ per age group $a$, and the $99\%$ CIs of time-constant parameters using 40000 particles. We observe that the estimated latent intensity and the estimated weekly latent cases of age $a$ are consistent with the weekly observed cases of age $a$. The analysis demonstrates that the instantaneous reproduction numbers $\{R_a(t)\}_a$ and $R(t)$ reflect the progress of the pandemic and capture the changes. As expected, the estimates of last days are more uncertain. 

\paragraph*{Comparing the age-stratified model (model A) with the unstructured homogeneously mixing model (model U) introduced by Lamprinakou et al.~\cite{https://doi.org/10.48550/arxiv.2208.07340}}Model A includes model U. Following Pellis et al.~\cite{pellis2020systematic}, we assume model A reflects better the reality and measures the other model's deviation from it.

We run model U considering all reported infections and separately for each age band to estimate the aggregated and per age group latent intensity,  the aggregated and per age group weekly and daily hidden cases, and the instantaneous reproduction number. We assume that $d_{min}=1$, $d_{max}=10$, $\left(\alpha,b\right)=\left(0.5,3\right)$ for Ashford, $\left(\alpha,b\right)=\left(0.1,3.5\right)$ for Kingston upon Thames and $\left(\alpha,b\right)=\left(0.1,3\right)$ for Leicester. Inspired by Pei et al.~\cite{pei2020differential}, we evaluate the goodness of fit of model U using the metric 
\begin{equation*}
    PAE=\frac{\sum_{i=1}^k |A_i-U_i|}{\sum_{i=1}^kA_i},
\end{equation*} where $A_i$ and $U_i$ are the estimations of models A and U via posterior median associated with the interval $\mathcal{T}_i$.  Table \ref{tableGOF_AM} illustrates the goodness fit of the model using the metric PAE.

\begin{figure}[!h] 
 \begin{subfigure}{7cm}
    \centering\includegraphics[width=6cm]{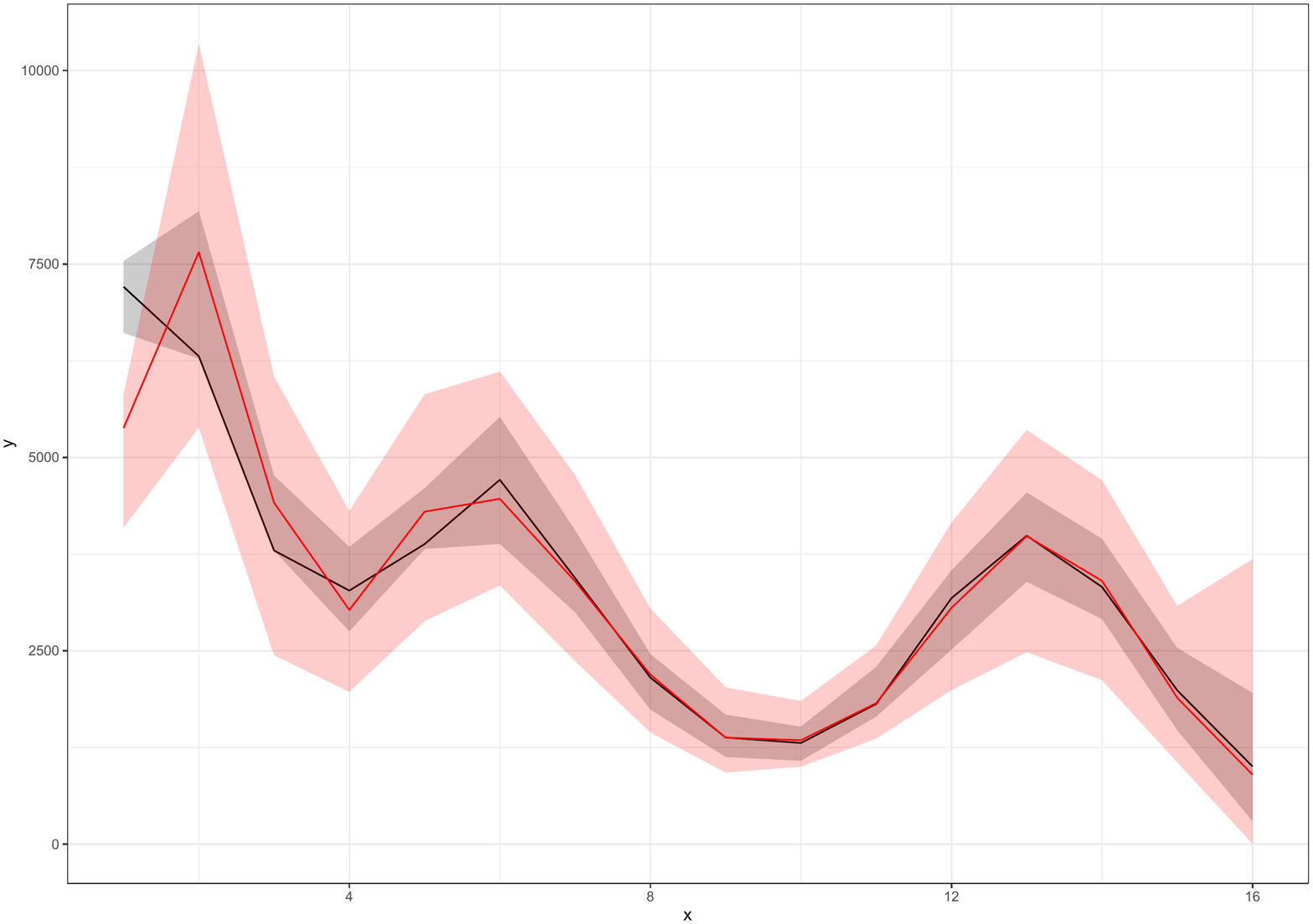}
     \caption{Aggregated weekly hidden cases.}
  \end{subfigure}
  \begin{subfigure}{7cm}
    \centering\includegraphics[width=6cm]{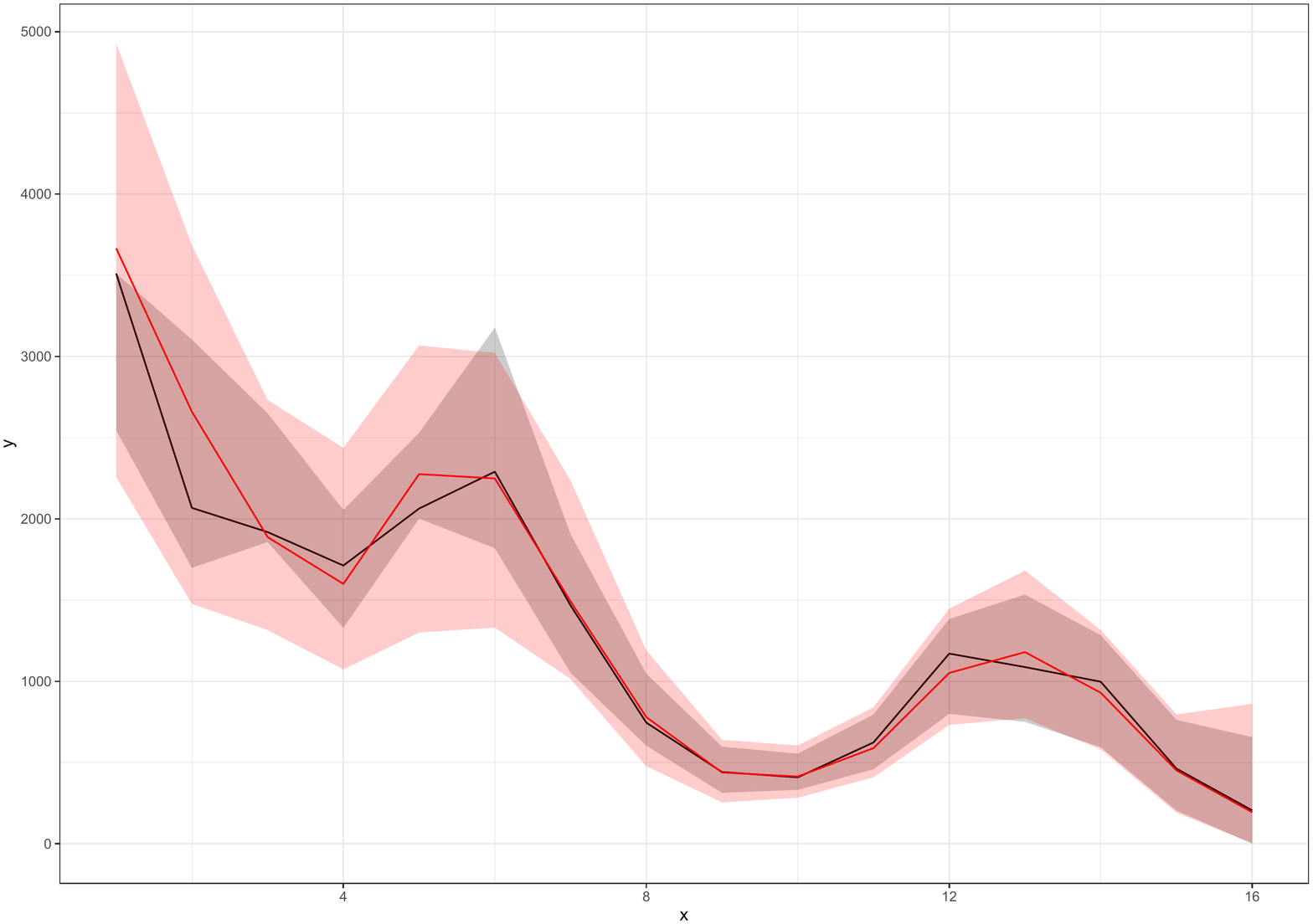}
   \caption{Aged 0-29.}
  \end{subfigure}
  \begin{subfigure}{7cm}
    \centering\includegraphics[width=6cm]{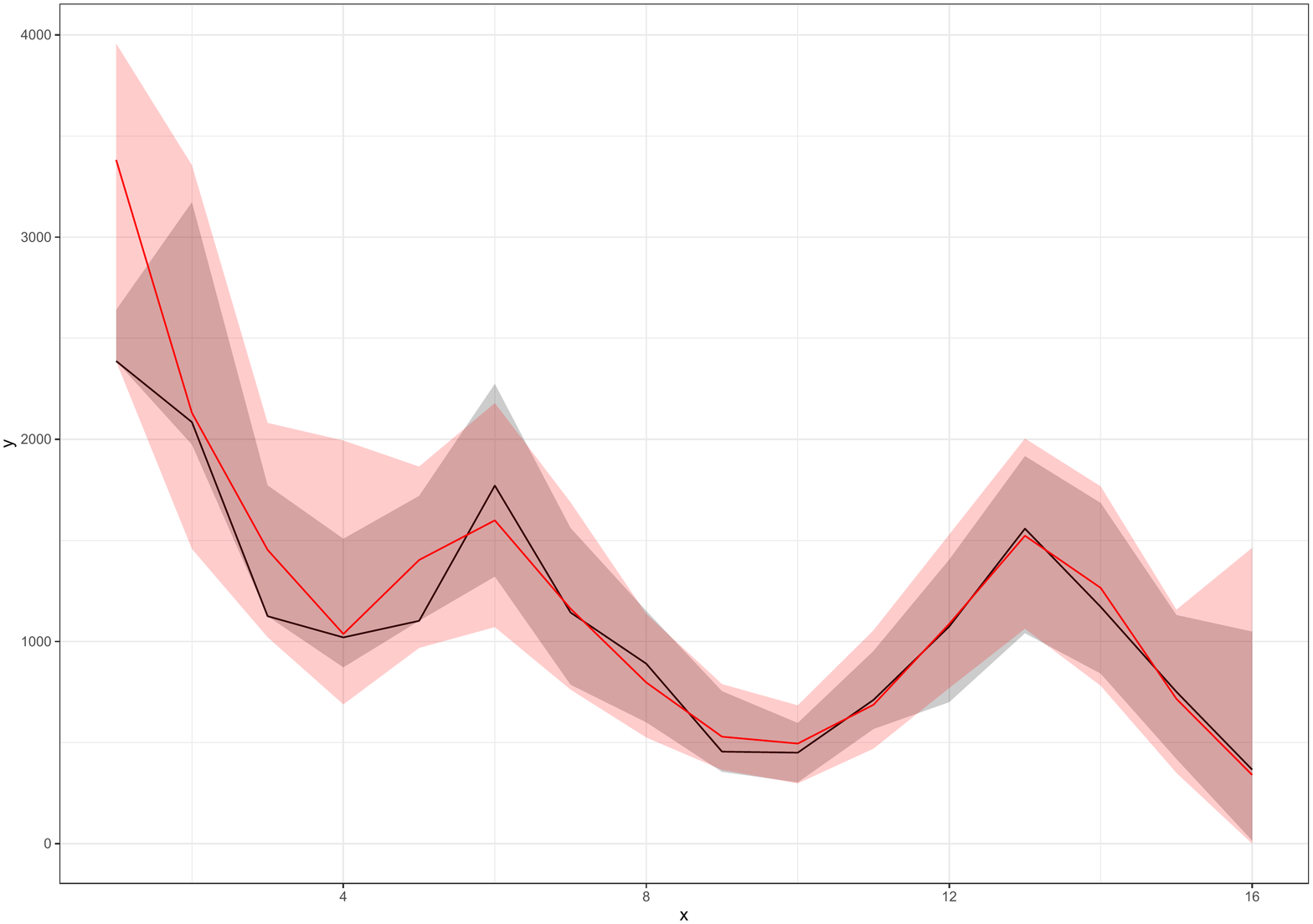}
     \caption{Aged 30-49.}
  \end{subfigure}
  \begin{subfigure}{7cm}
    \centering\includegraphics[width=6cm]{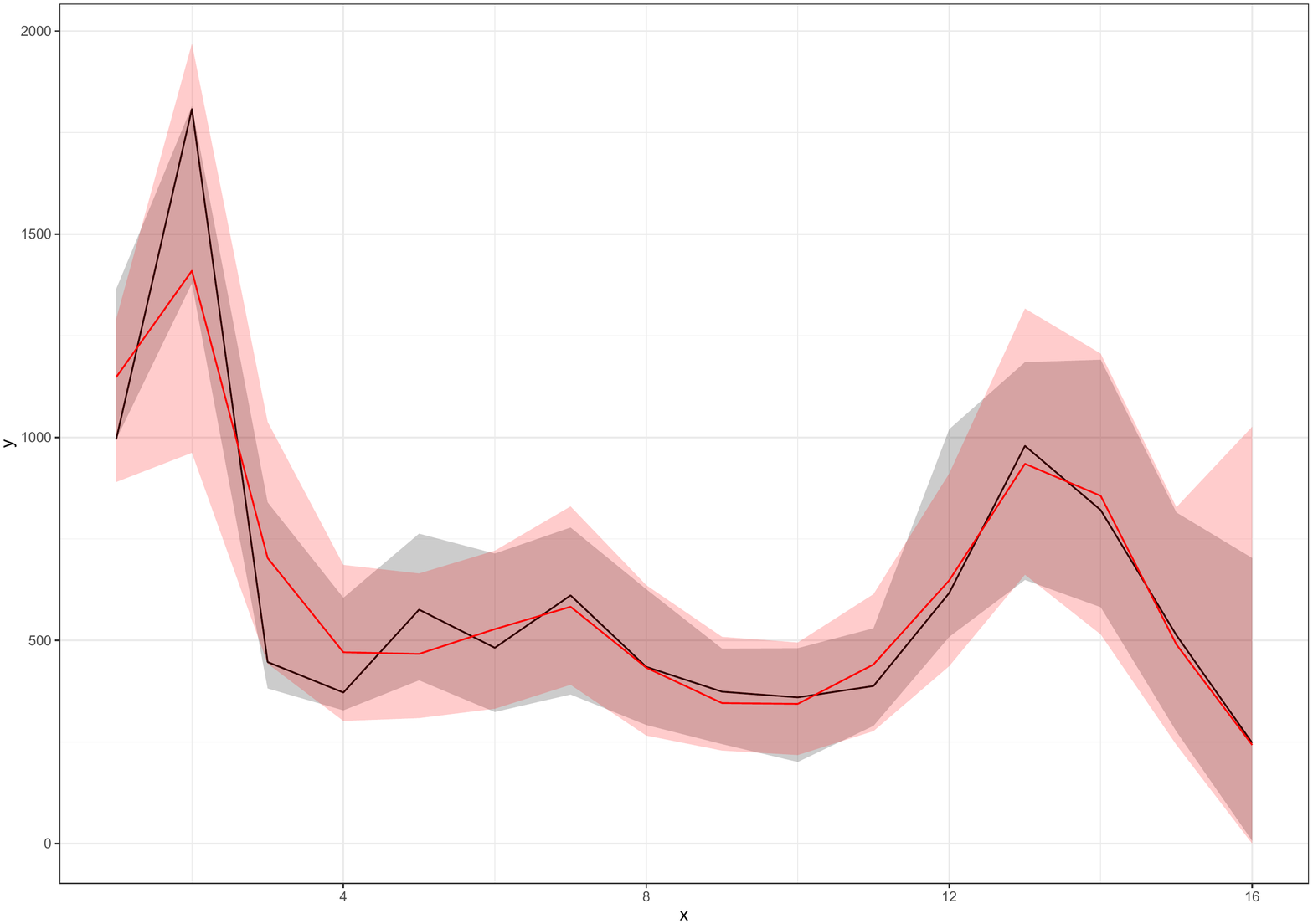}
    \caption{Aged 50-69.}
  \end{subfigure}
   \begin{subfigure}{7cm}
    \centering\includegraphics[width=6cm]{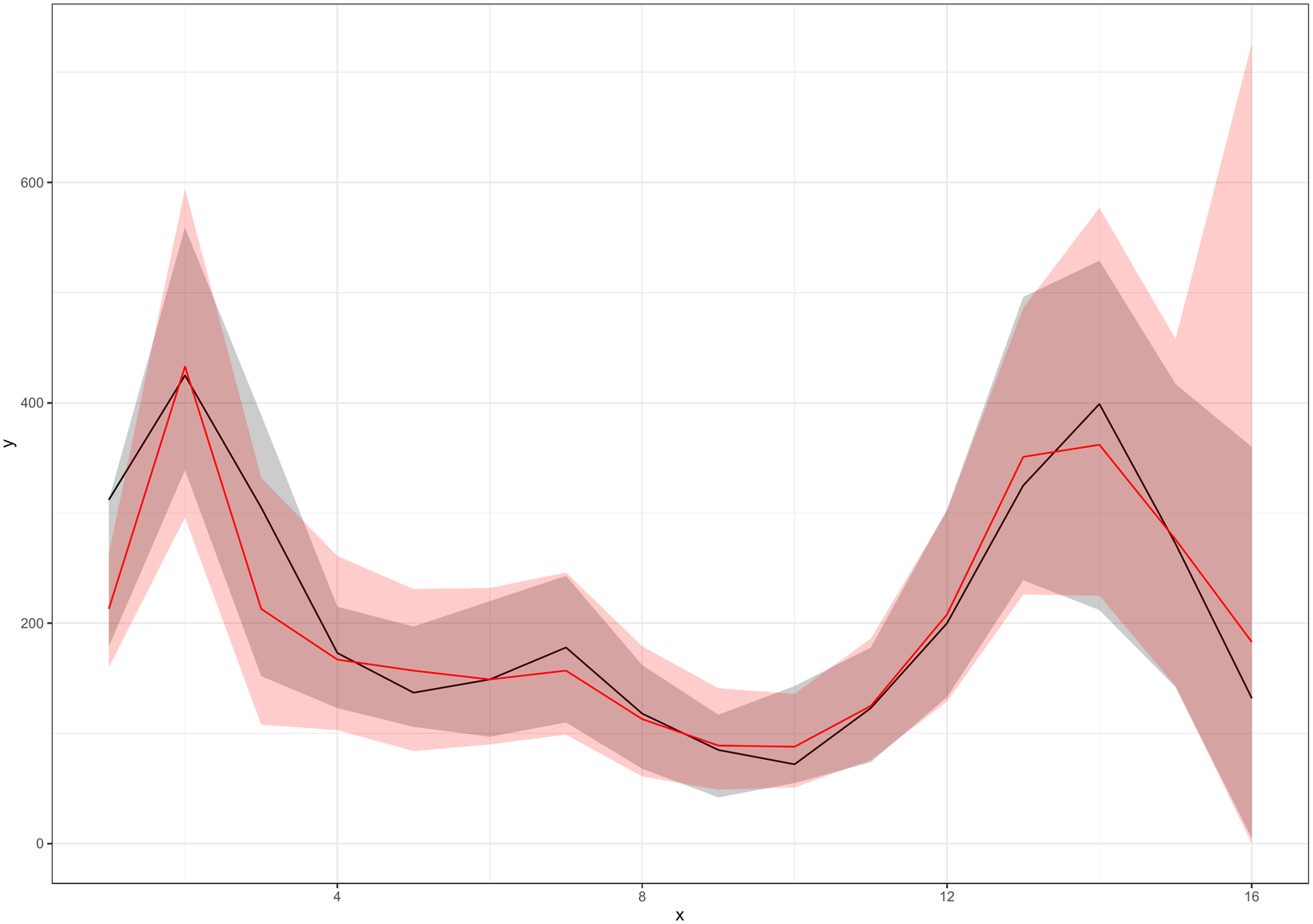}
     \caption{Aged 70+.}
  \end{subfigure}

   \caption{\bf The posterior median estimate of weekly hidden cases of model A (black line) and model U (red line), and the 99\% CIs (ribbon) in Kingston.}
   \label{CompWHC_Kings4G}
\end{figure}

\begin{figure}[!h] 
 \begin{subfigure}{7cm}
    \centering\includegraphics[width=6cm]{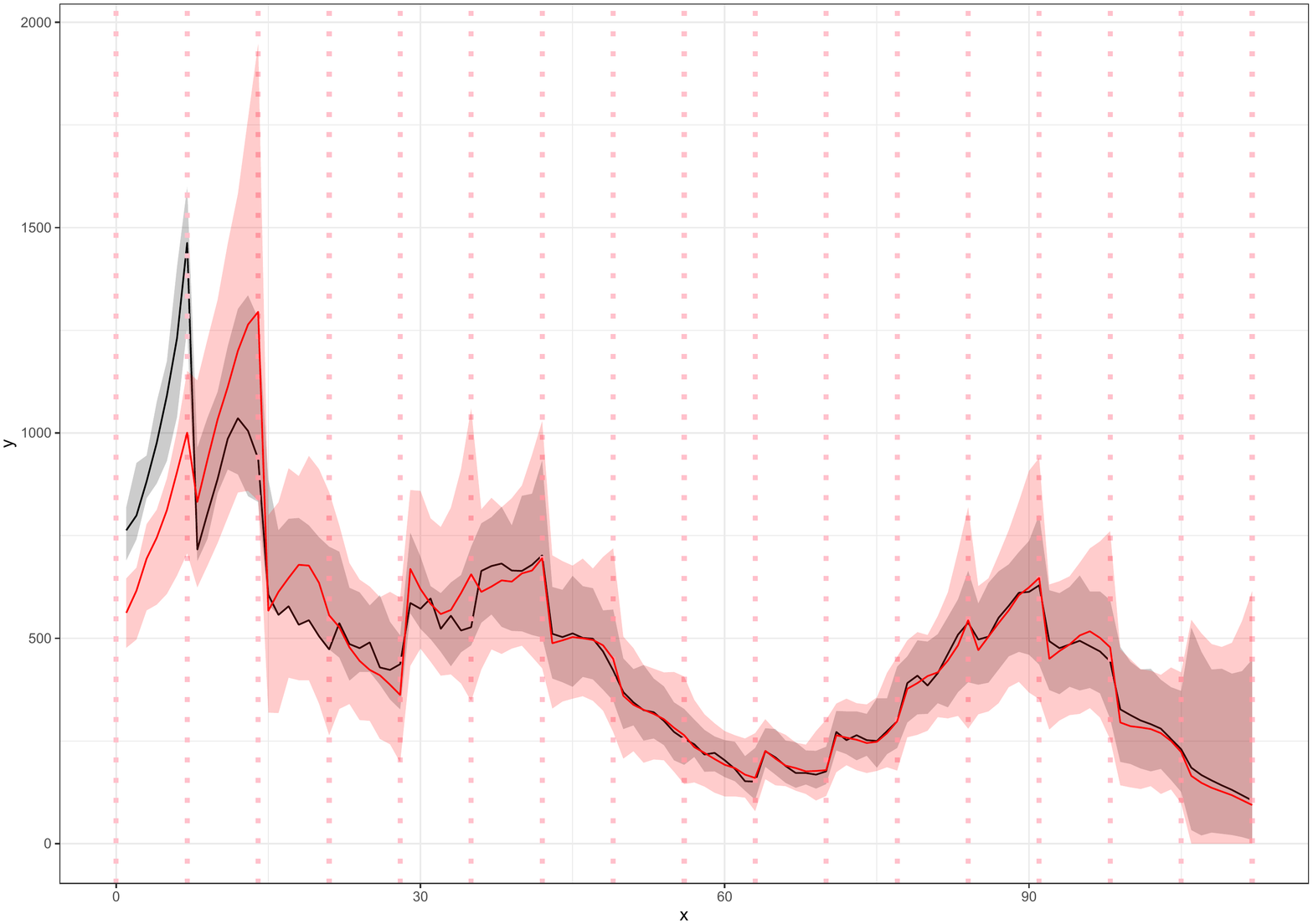}
     \caption{Aggregated daily hidden cases.}
  \end{subfigure}
  \begin{subfigure}{7cm}
    \centering\includegraphics[width=6cm]{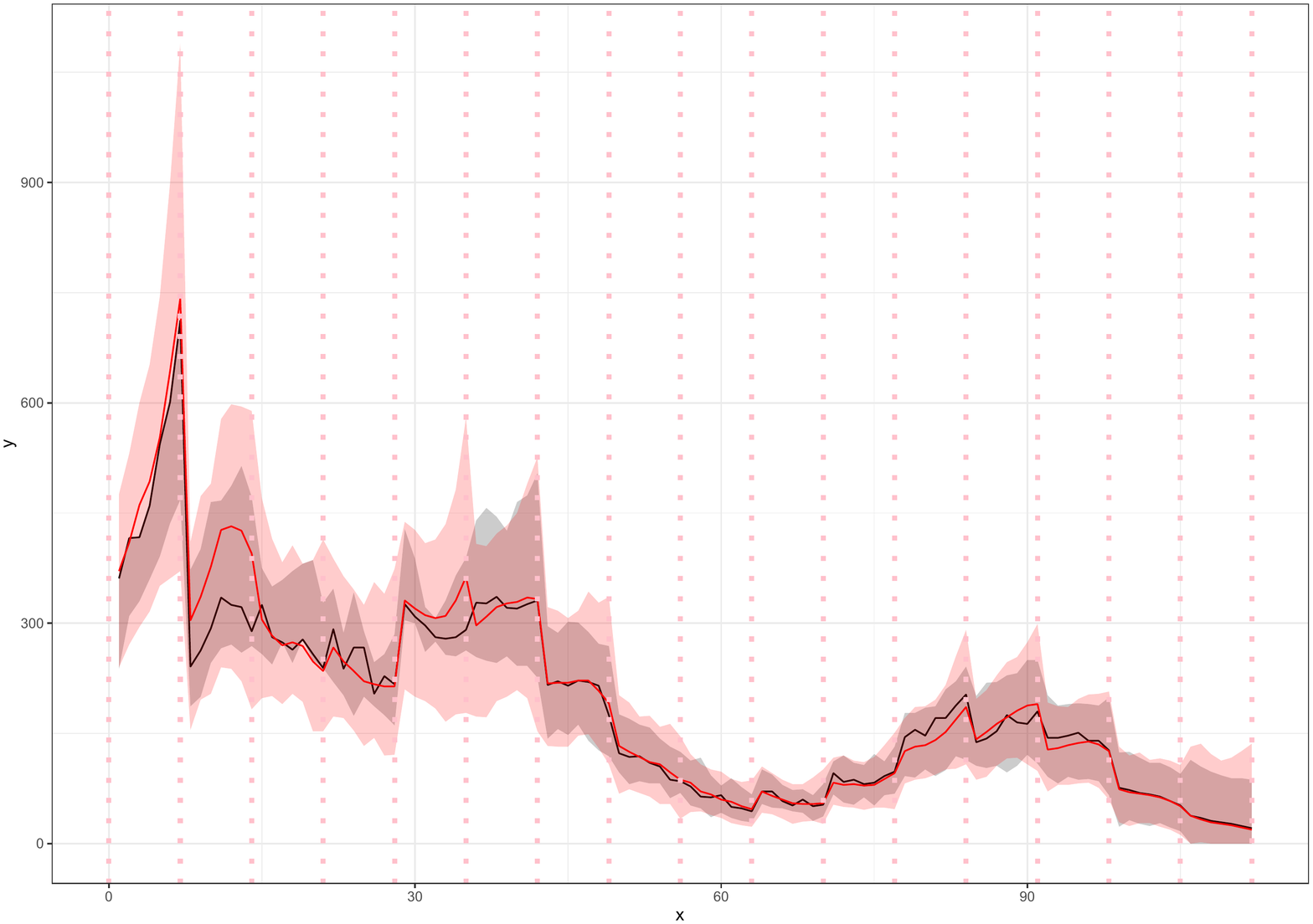}
   \caption{Aged 0-29.}
  \end{subfigure}
  \begin{subfigure}{7cm}
    \centering\includegraphics[width=6cm]{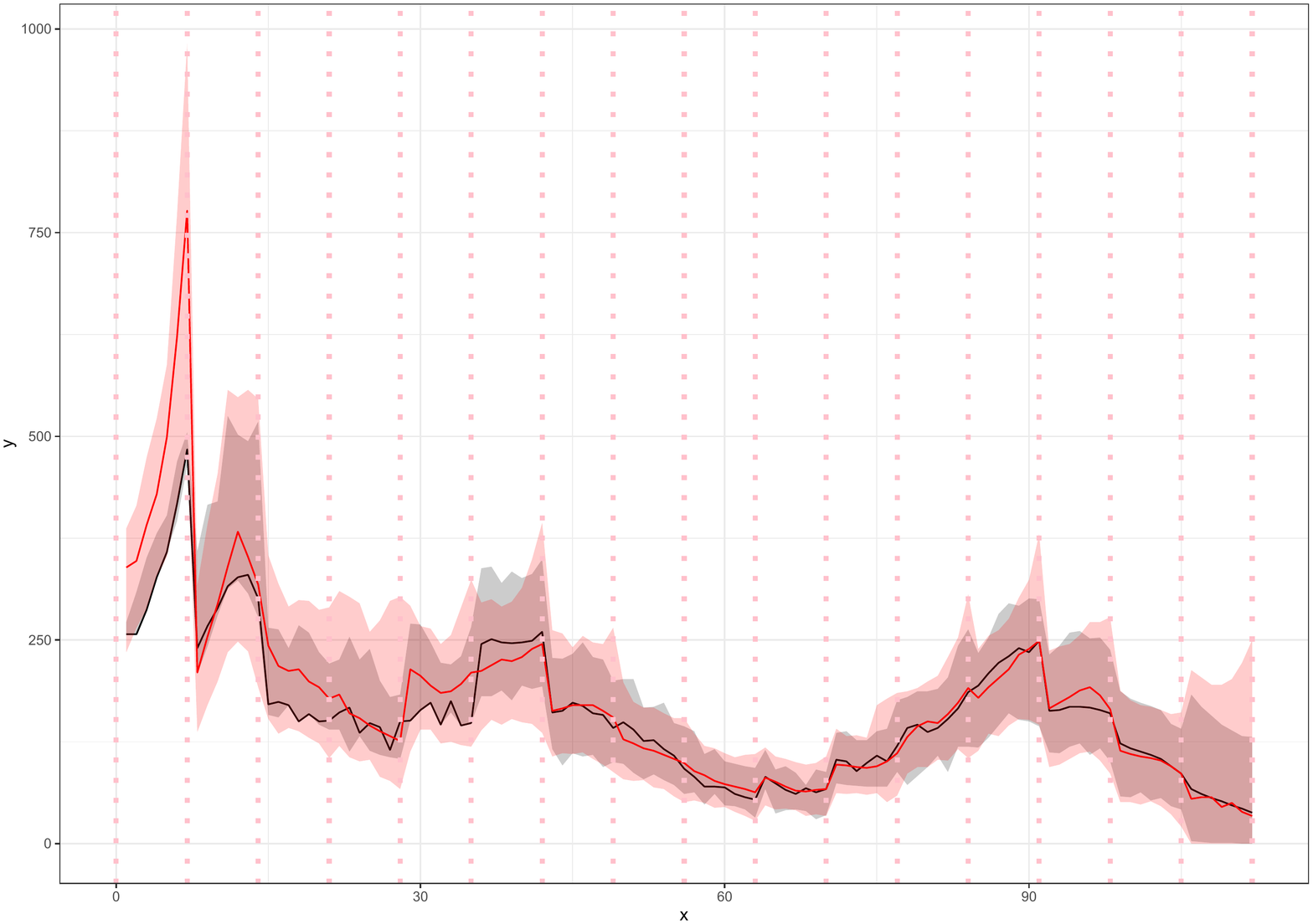}
     \caption{Aged 30-49.}
  \end{subfigure}
  \begin{subfigure}{7cm}
    \centering\includegraphics[width=6cm]{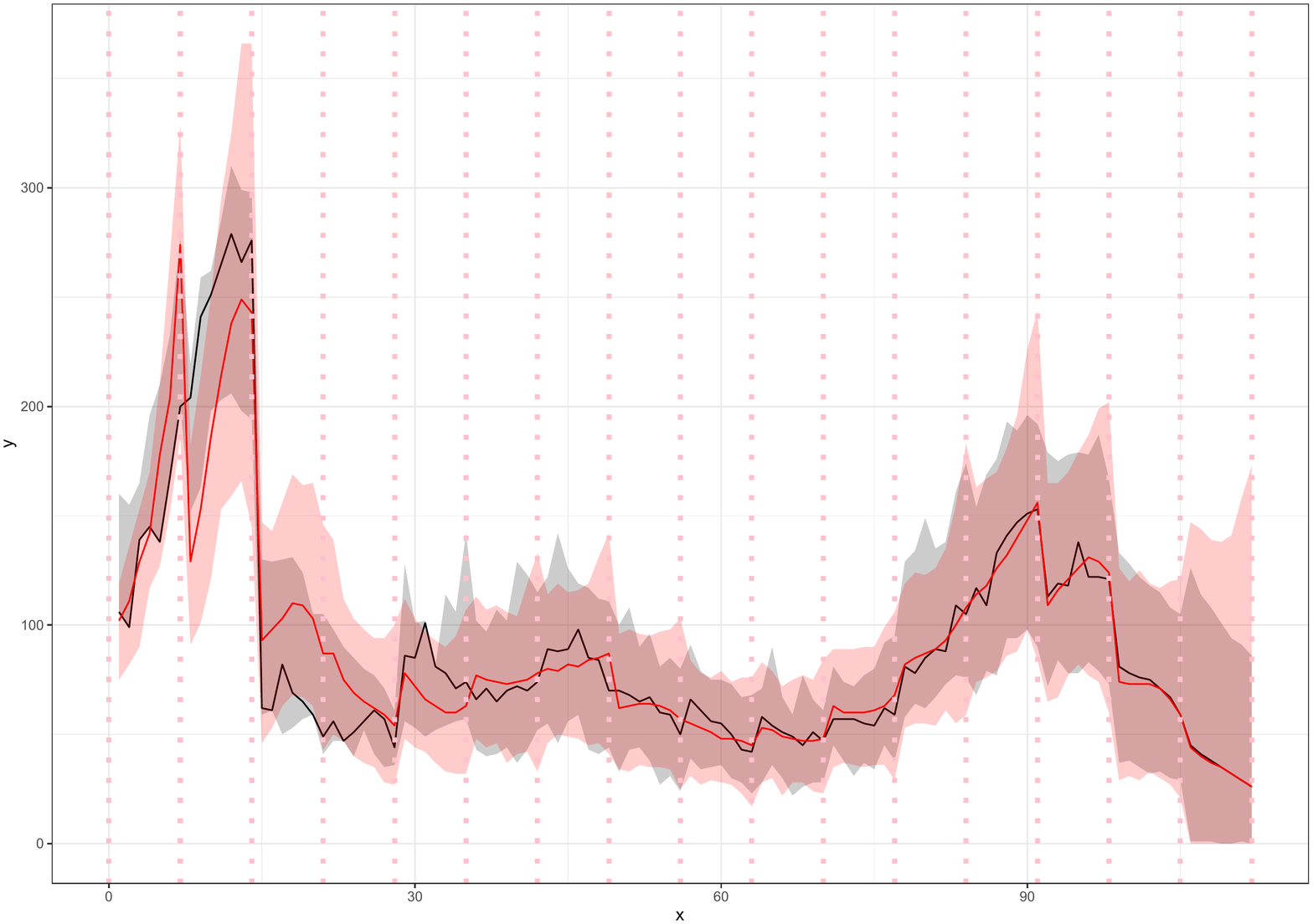}
    \caption{Aged 50-69.}
  \end{subfigure}
   \begin{subfigure}{7cm}
    \centering\includegraphics[width=6cm]{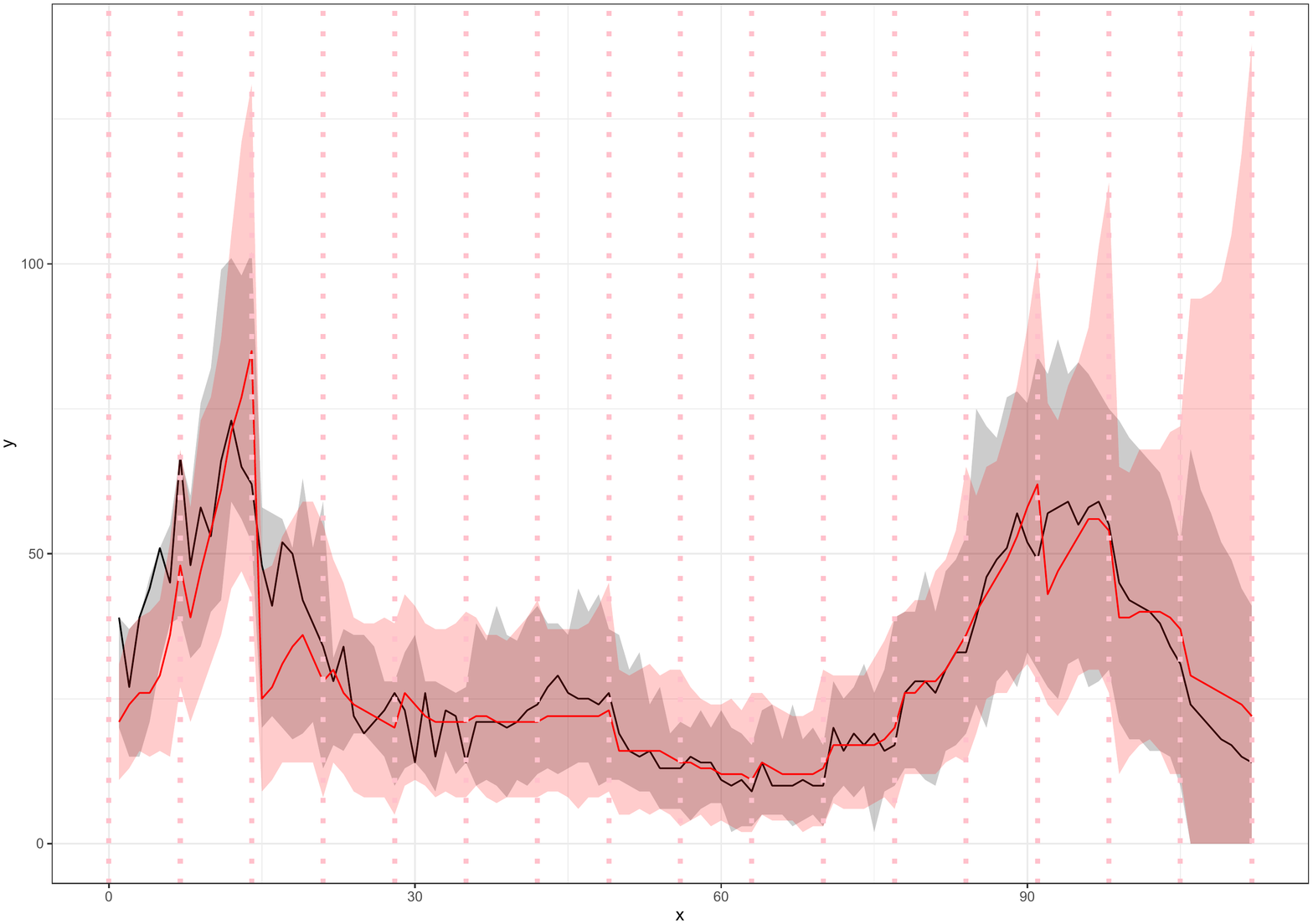}
     \caption{Aged 70+.}
  \end{subfigure}

   \caption{{\bf The posterior median estimate of daily hidden cases of model A (black line) and model U (red line), and the 99\% CIs (ribbon) in Kingston.} The vertical dotted lines show the beginning of each week in the period we examine.}
   \label{CompDHC_Kings4G}
\end{figure}

\begin{figure}[!h] 
 \begin{subfigure}{7cm}
    \centering\includegraphics[width=6cm]{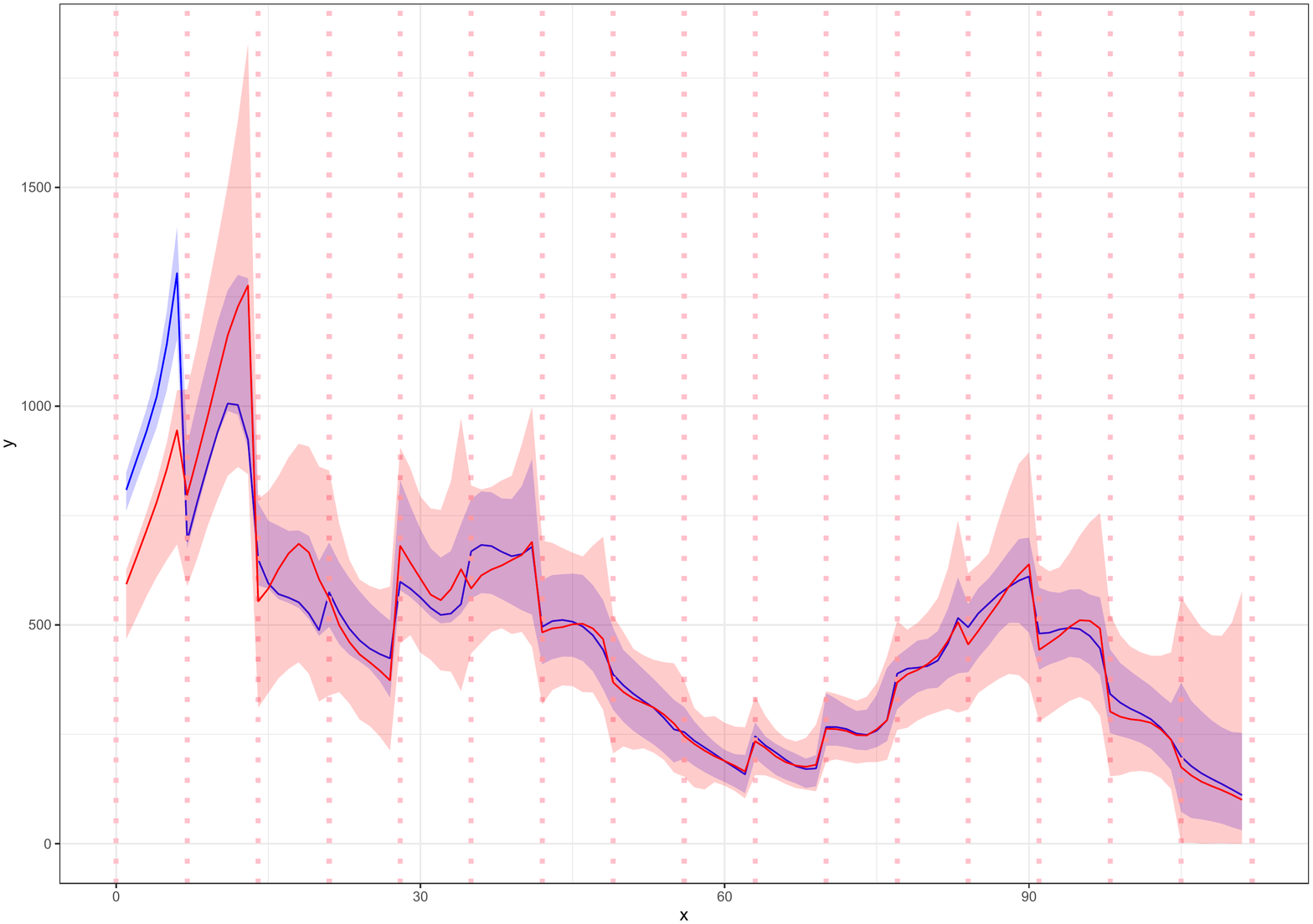}
     \caption{Aggregated latent intensity.}
  \end{subfigure}
  \begin{subfigure}{7cm}
    \centering\includegraphics[width=6cm]{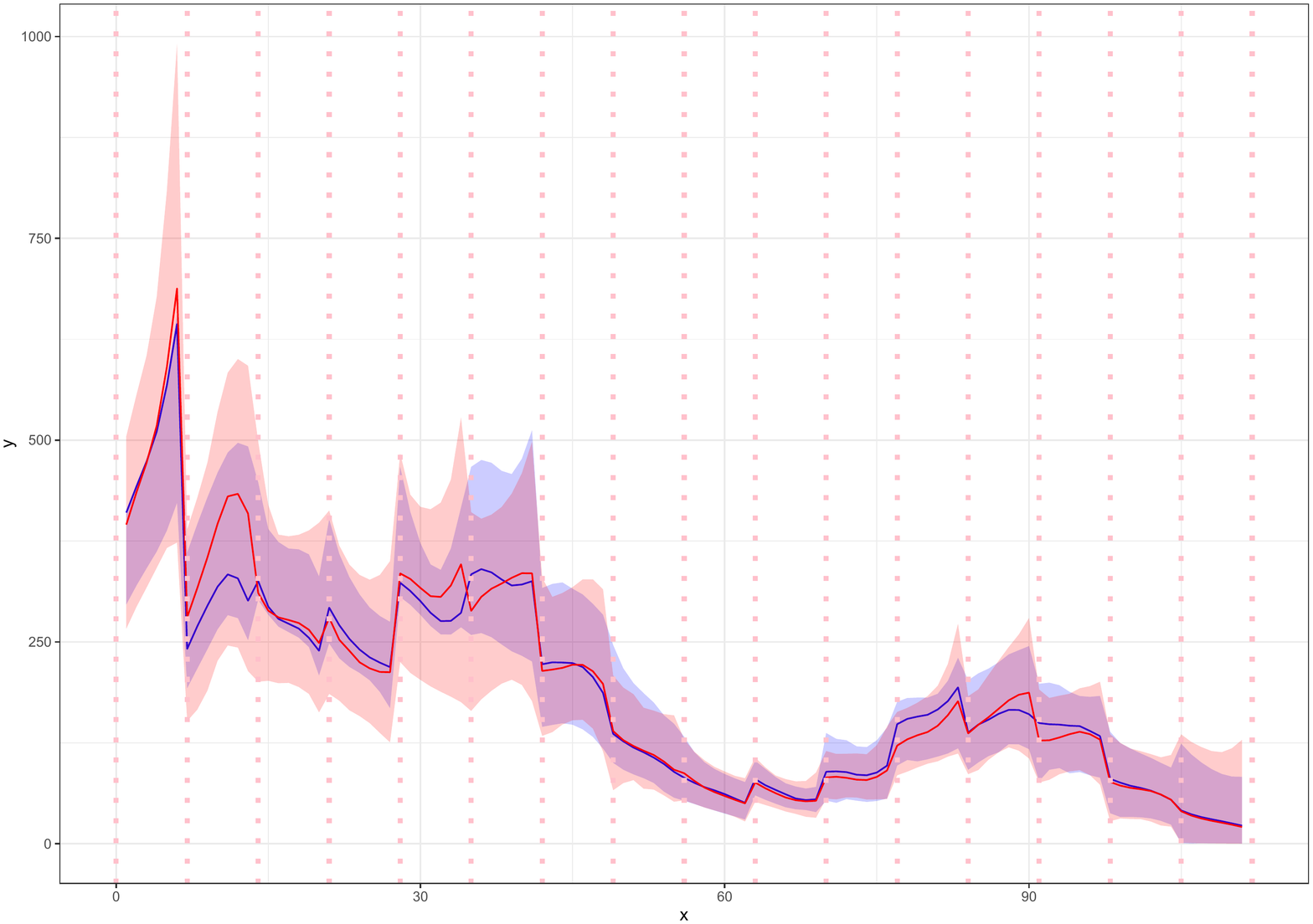}
   \caption{Aged 0-29.}
  \end{subfigure}
  \begin{subfigure}{7cm}
    \centering\includegraphics[width=6cm]{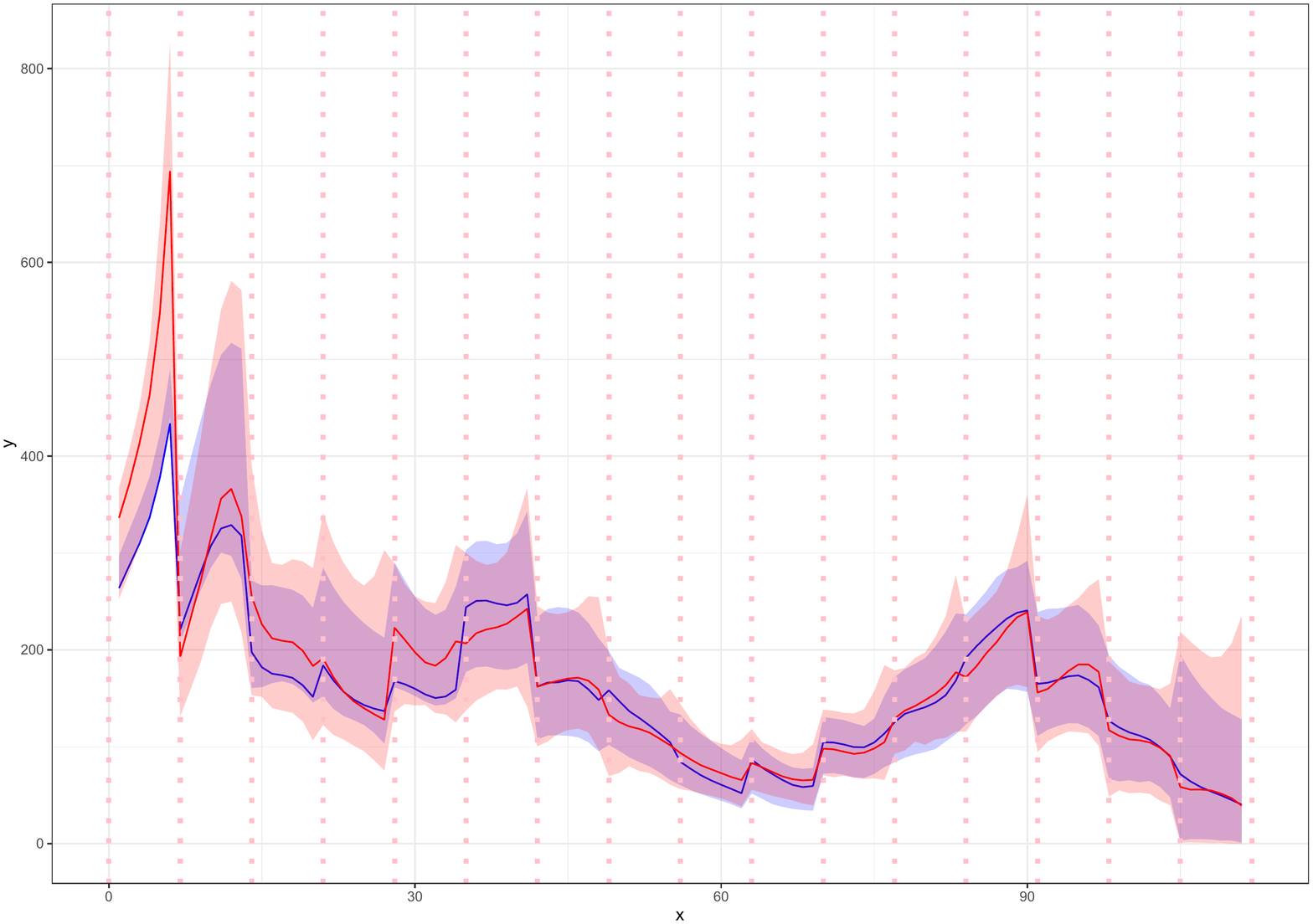}
     \caption{Aged 30-49.}
  \end{subfigure}
  \begin{subfigure}{7cm}
    \centering\includegraphics[width=6cm]{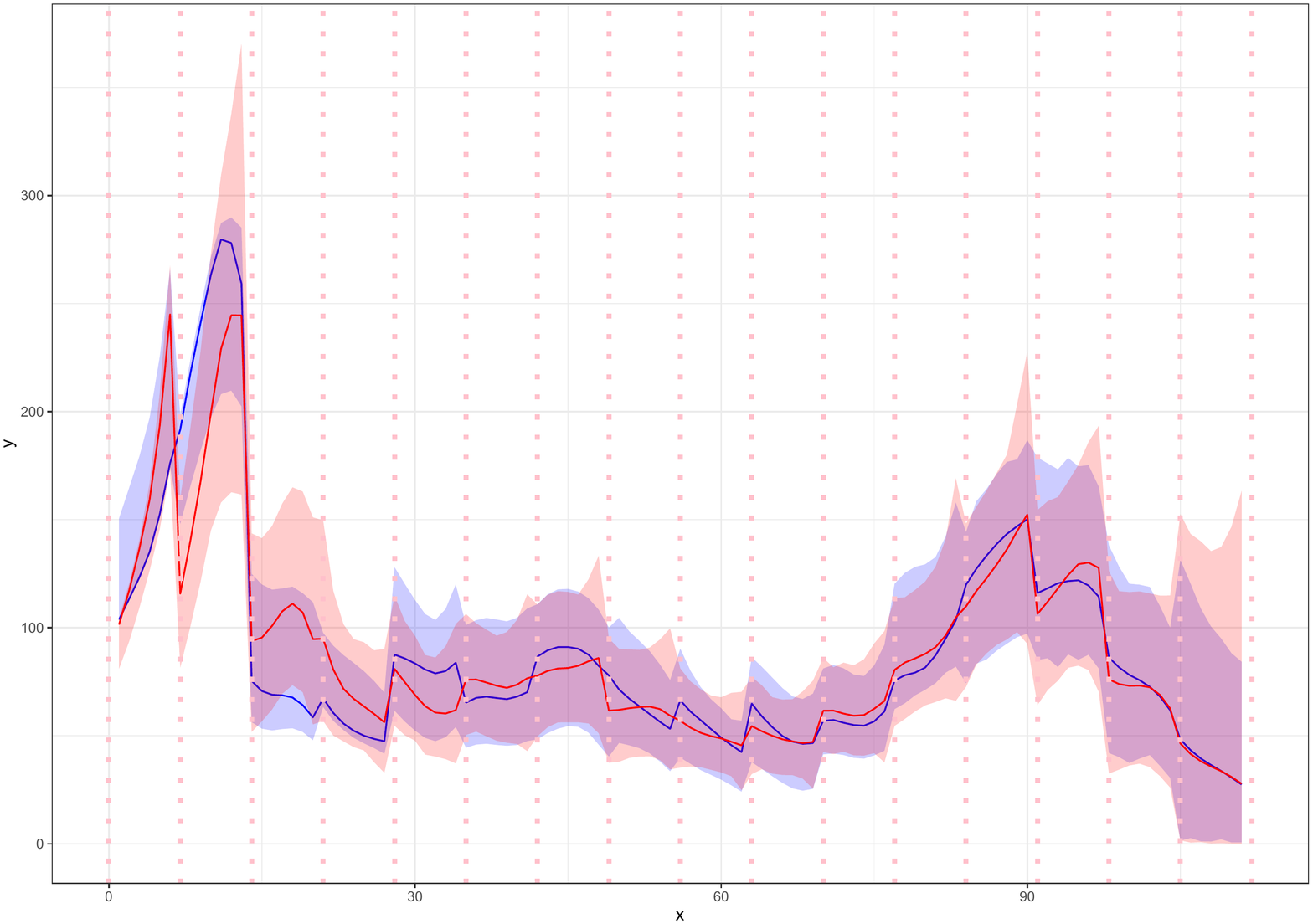}
    \caption{Aged 50-69.}
  \end{subfigure}
   \begin{subfigure}{7cm}
    \centering\includegraphics[width=6cm]{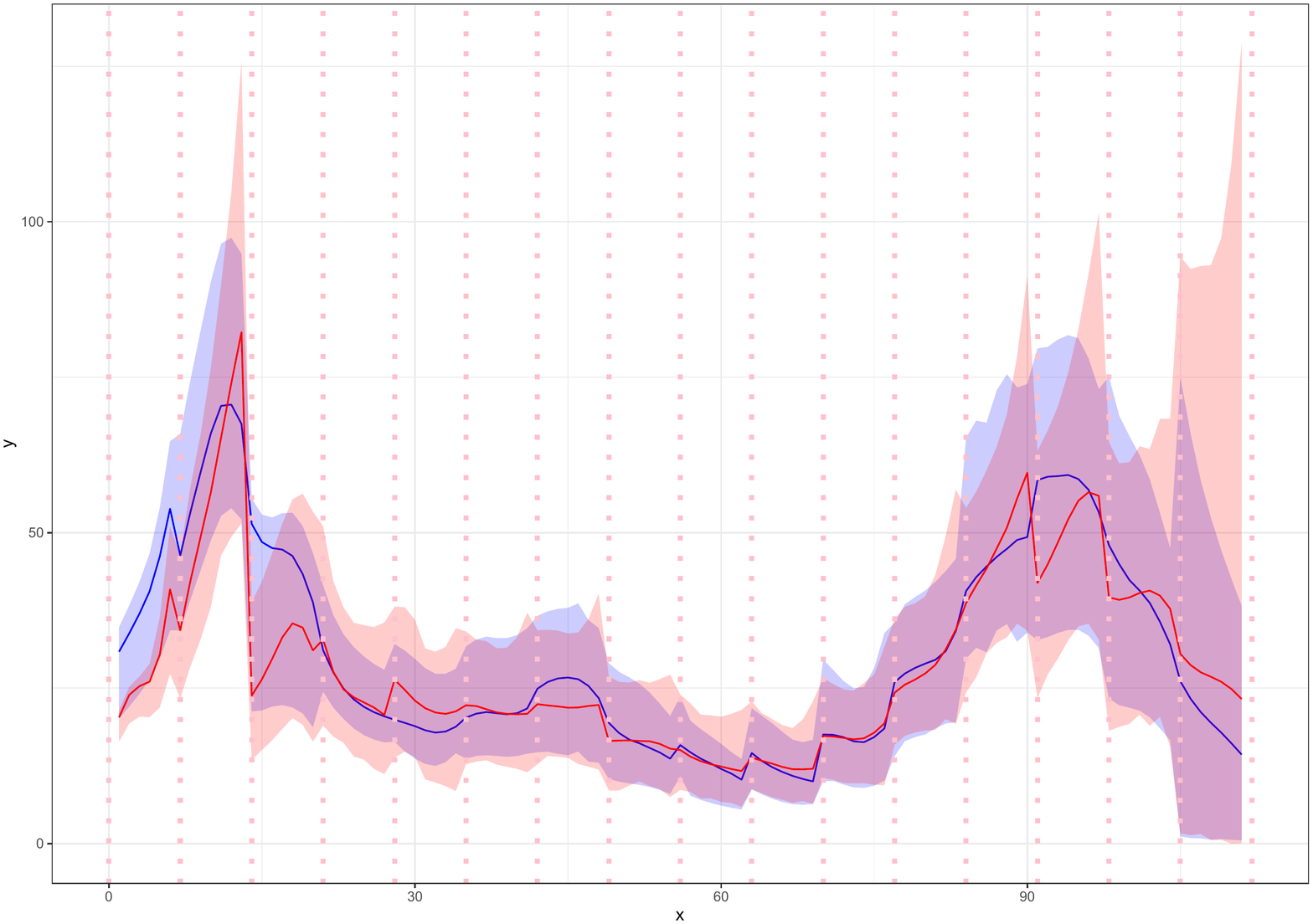}
     \caption{Aged 70+.}
  \end{subfigure}

   \caption{{\bf The posterior median estimate of latent intensity of model A (blue line) and model U (red line), and the 99\% CIs (ribbon) in Kingston.} The vertical dotted lines show the beginning of each week in the period we examine.}
   \label{CompLambda_Kings4G}
\end{figure}

\begin{figure}[!h] 
    \centering\includegraphics[width=6cm]{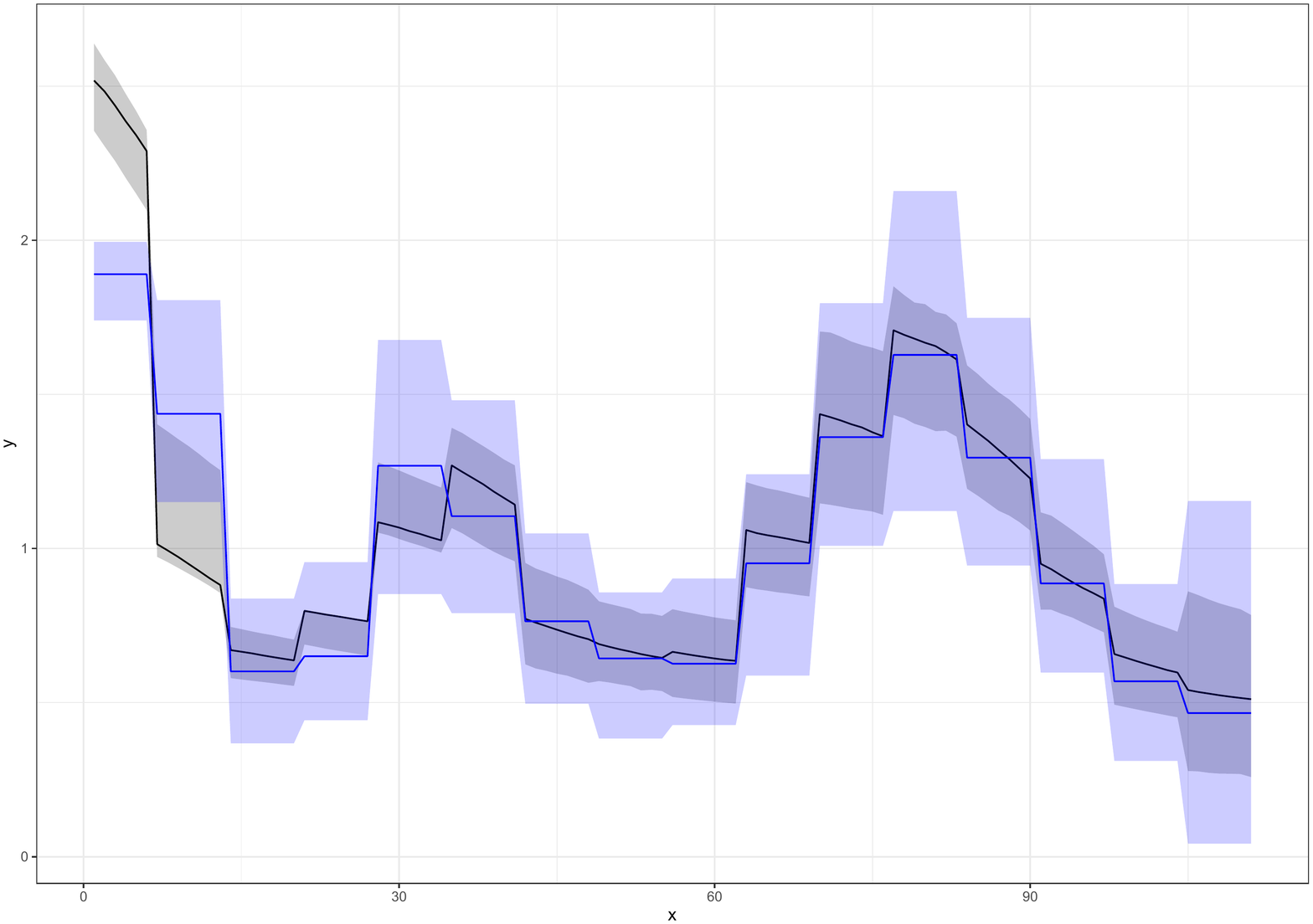}
  \caption{\bf The posterior median estimate of instantaneous reproduction number of model A (black line) and model U (blue line), and the 99\% CIs (ribbon) in Kingston.}
   \label{CompR_Kings4G}
\end{figure}

\begin{figure}[!h] 
 \begin{subfigure}{7cm}
    \centering\includegraphics[width=6cm]{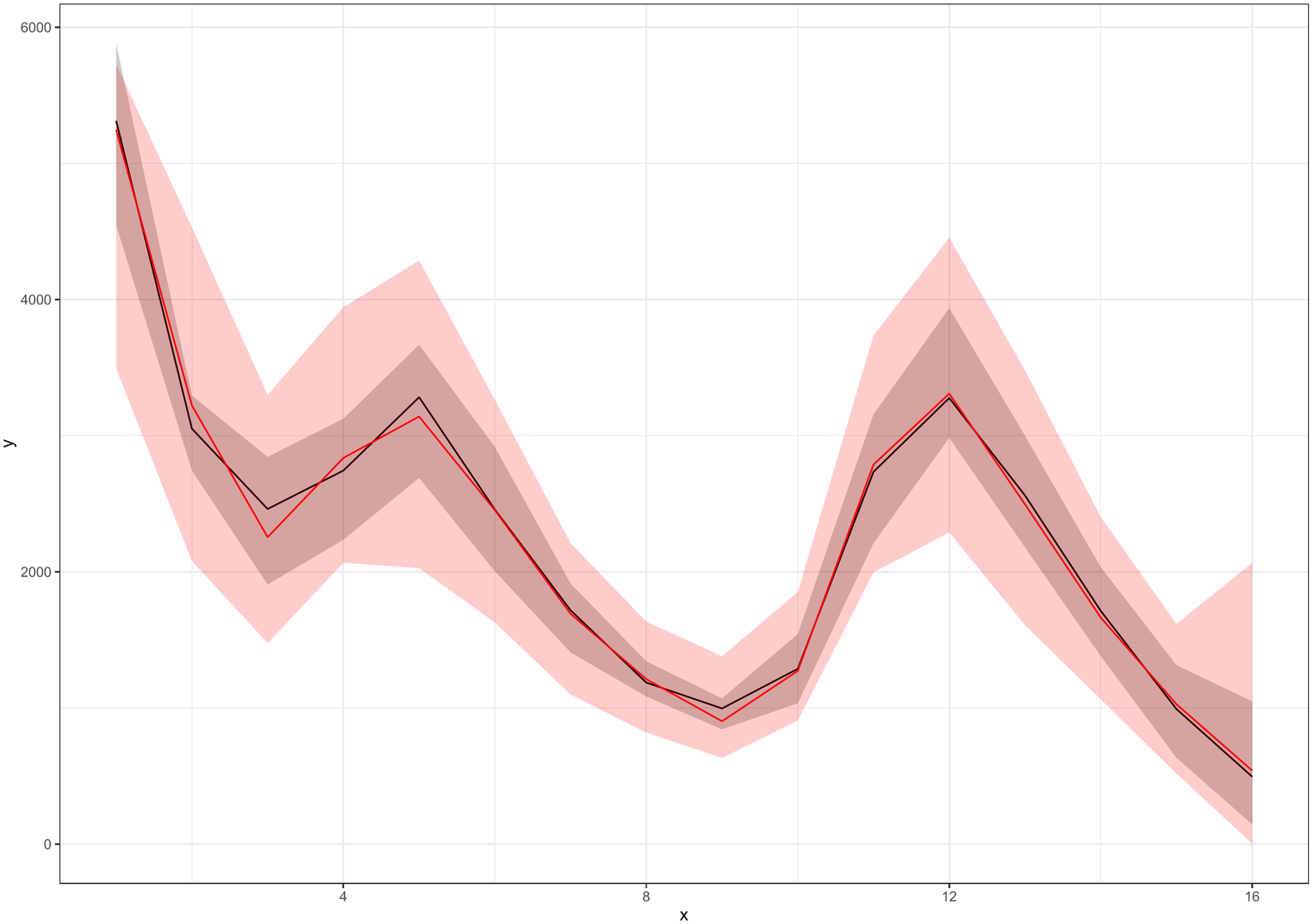}
     \caption{Aggregated weekly hidden cases.}
  \end{subfigure}
  \begin{subfigure}{7cm}
    \centering\includegraphics[width=6cm]{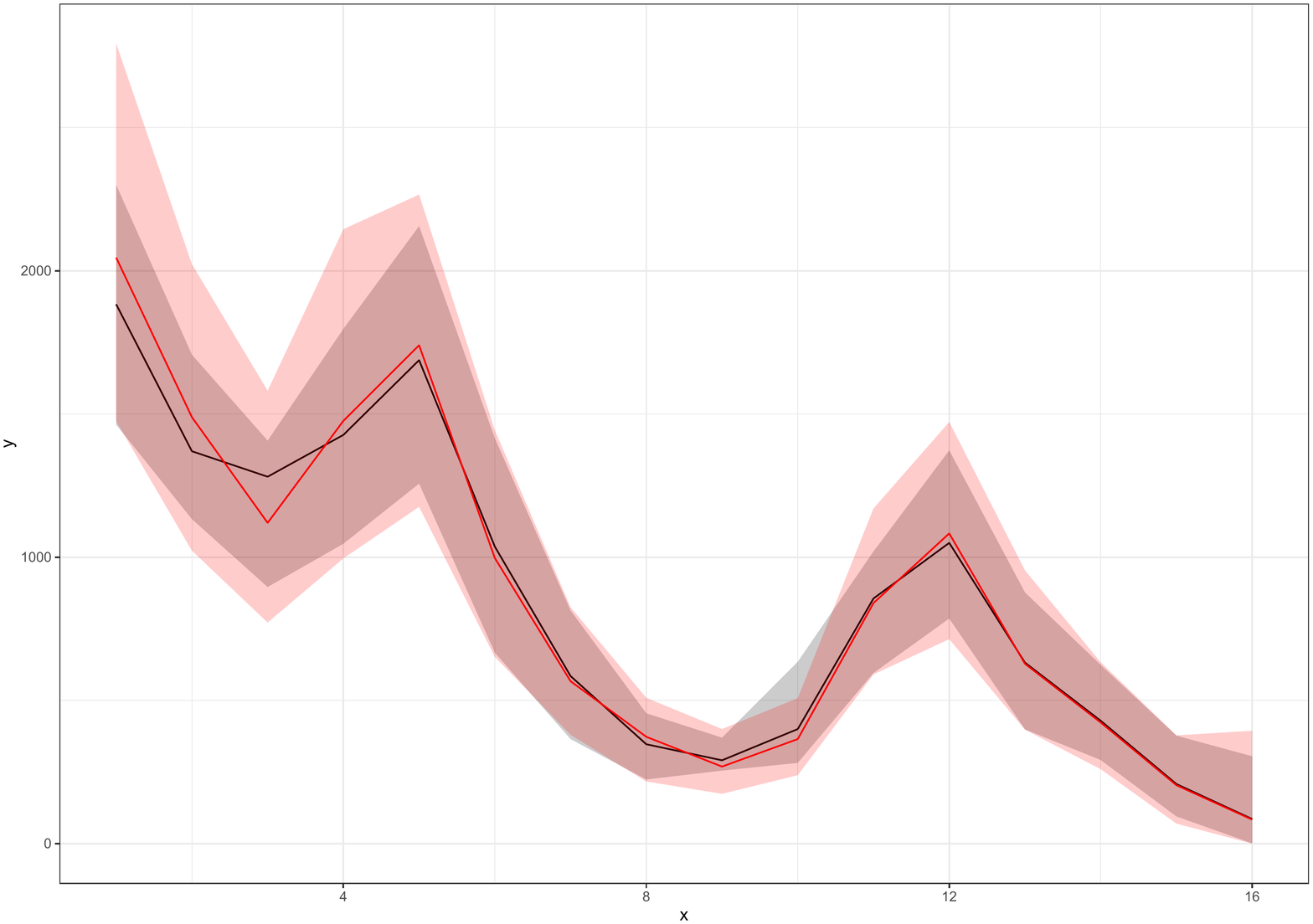}
   \caption{Aged 0-29.}
  \end{subfigure}
  \begin{subfigure}{7cm}
    \centering\includegraphics[width=6cm]{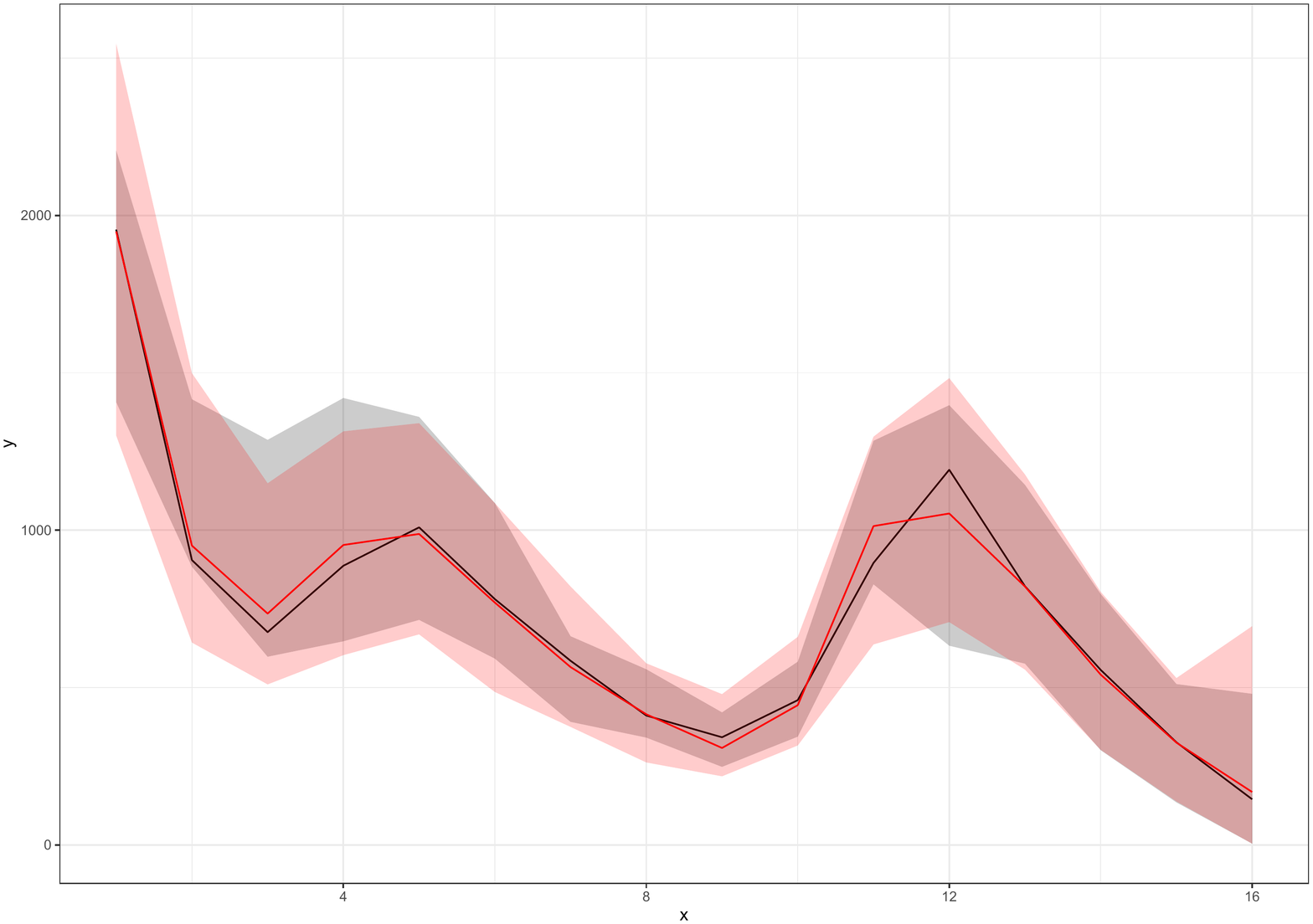}
     \caption{Aged 30-49.}
  \end{subfigure}
  \begin{subfigure}{7cm}
    \centering\includegraphics[width=6cm]{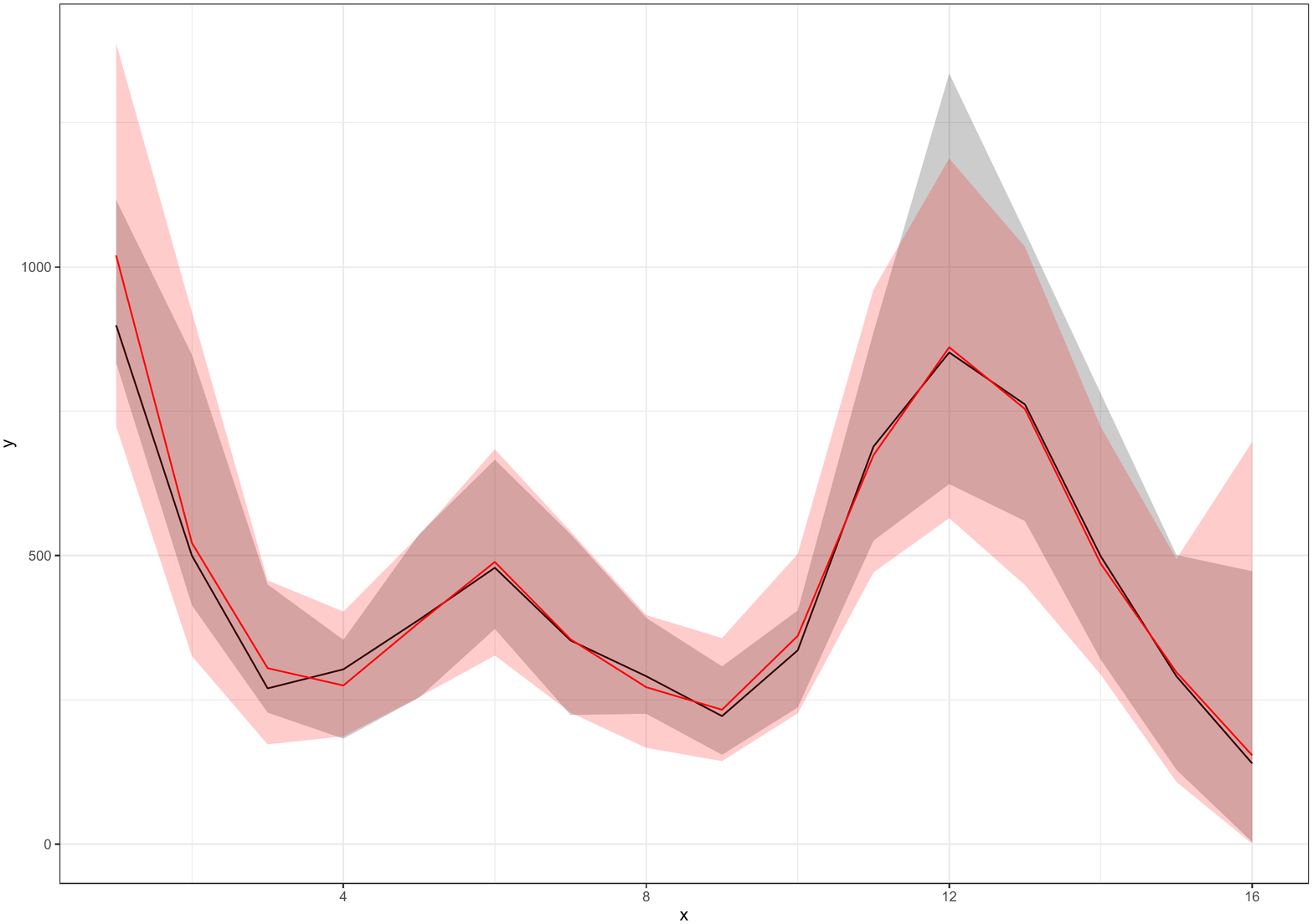}
    \caption{Aged 50-69.}
  \end{subfigure}
   \begin{subfigure}{7cm}
    \centering\includegraphics[width=6cm]{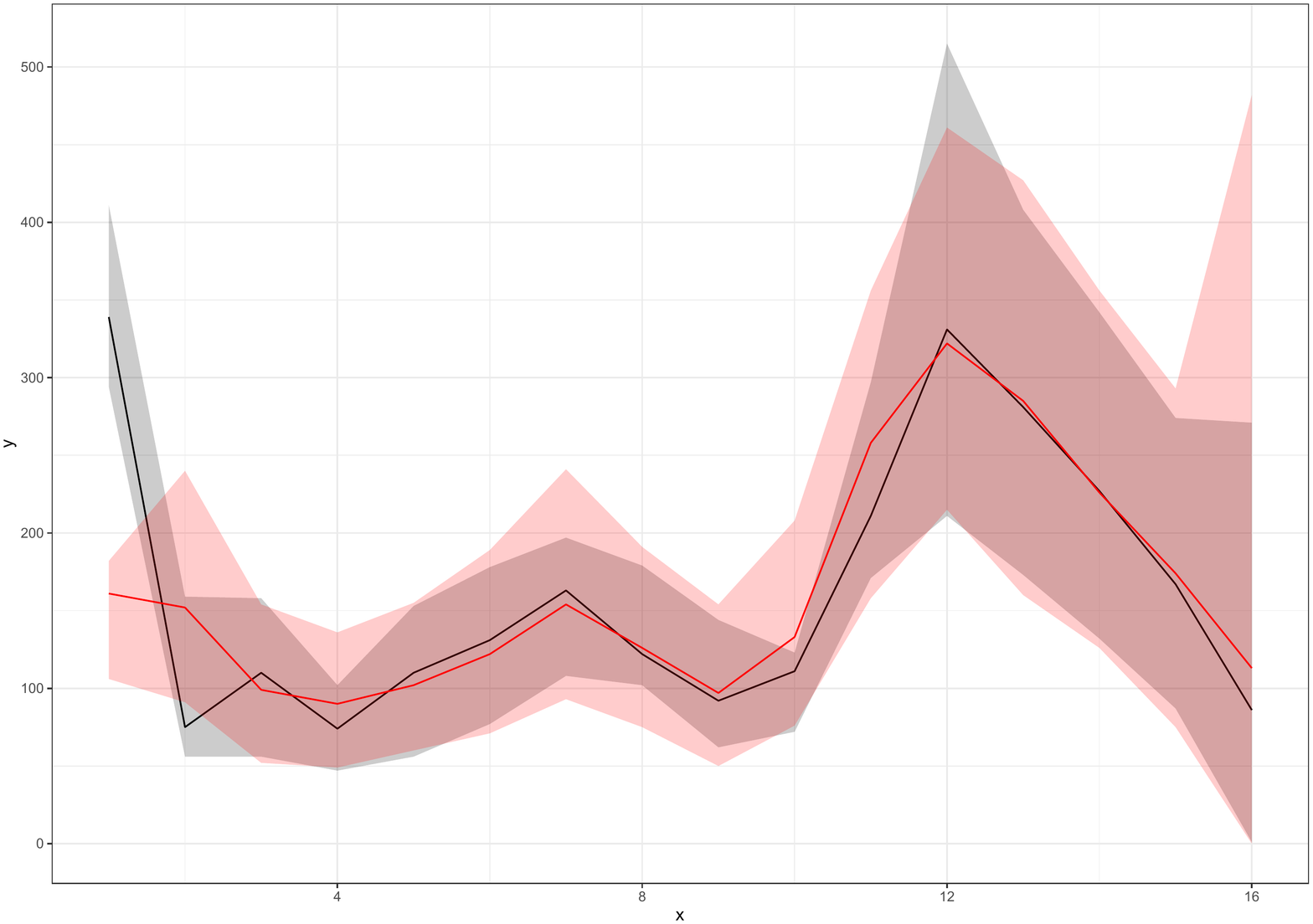}
     \caption{Aged 70+.}
  \end{subfigure}

   \caption{\bf The posterior median estimate of weekly hidden cases of model A (black line) and model U (red line), and the 99\% CIs (ribbon) in Ashford.}
   \label{CompWHC_Ashford4G}
\end{figure}

\begin{figure}[!h] 
 \begin{subfigure}{7cm}
    \centering\includegraphics[width=6cm]{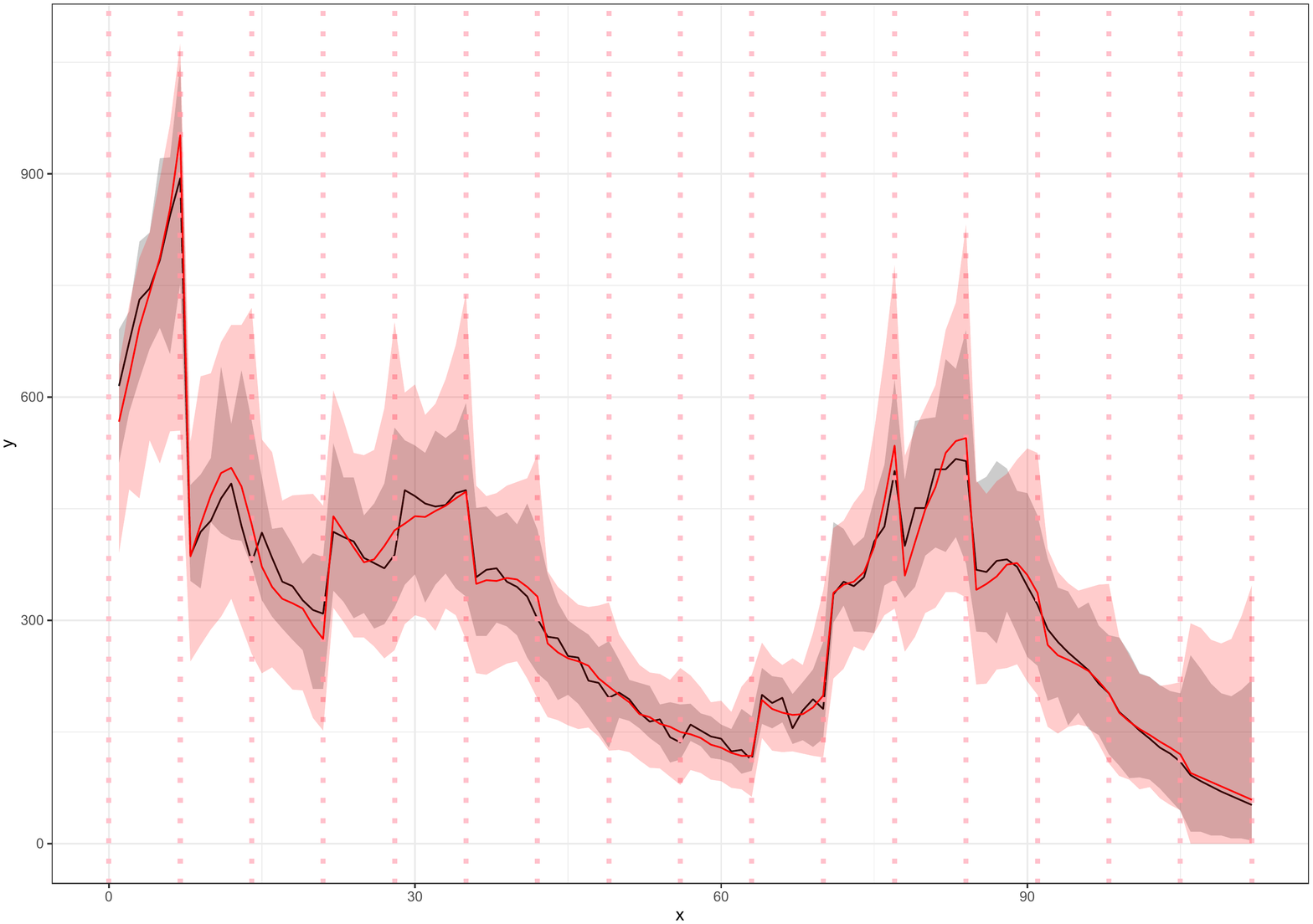}
     \caption{Aggregated daily hidden cases.}
  \end{subfigure}
  \begin{subfigure}{7cm}
    \centering\includegraphics[width=6cm]{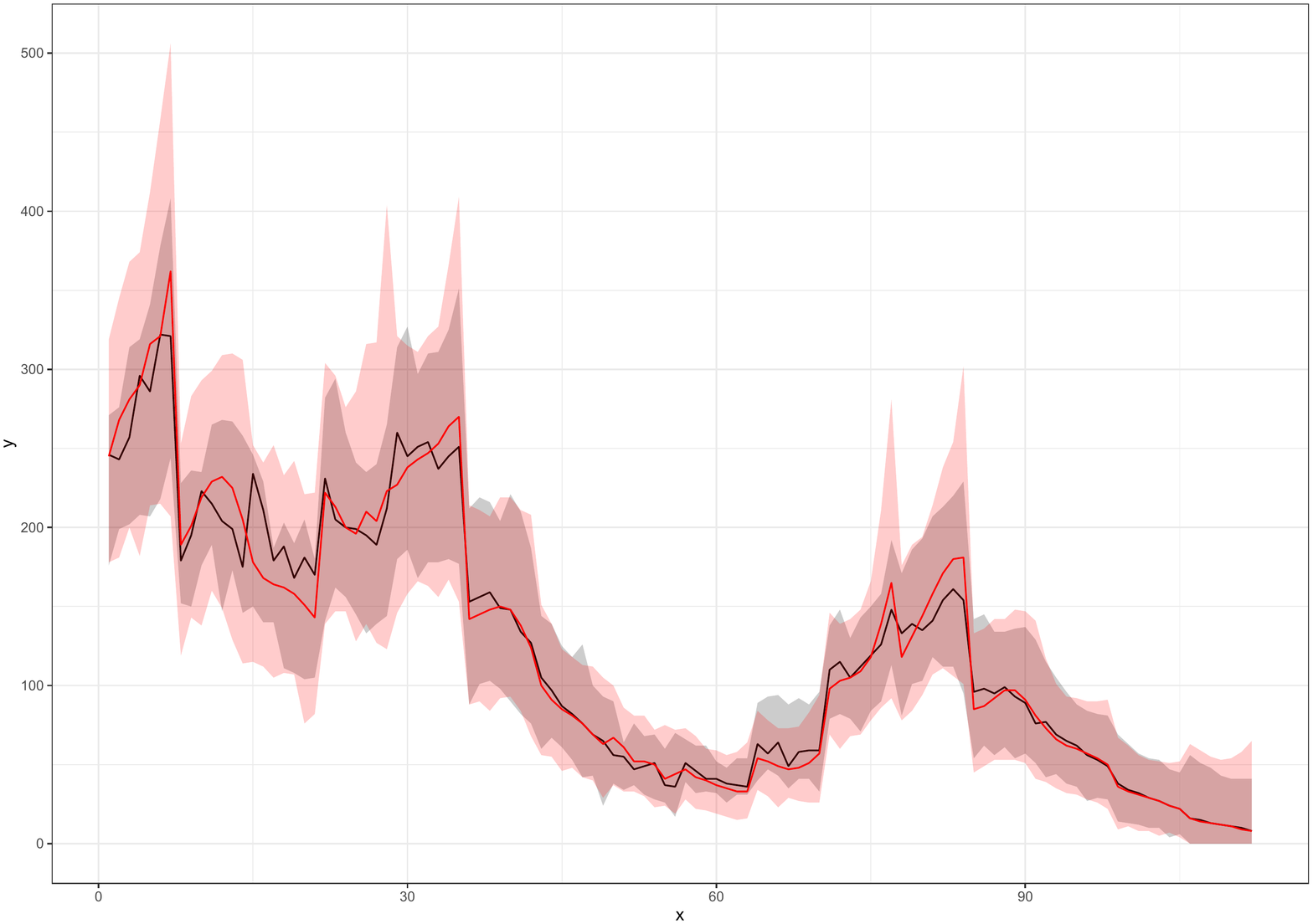}
   \caption{Aged 0-29.}
  \end{subfigure}
  \begin{subfigure}{7cm}
    \centering\includegraphics[width=6cm]{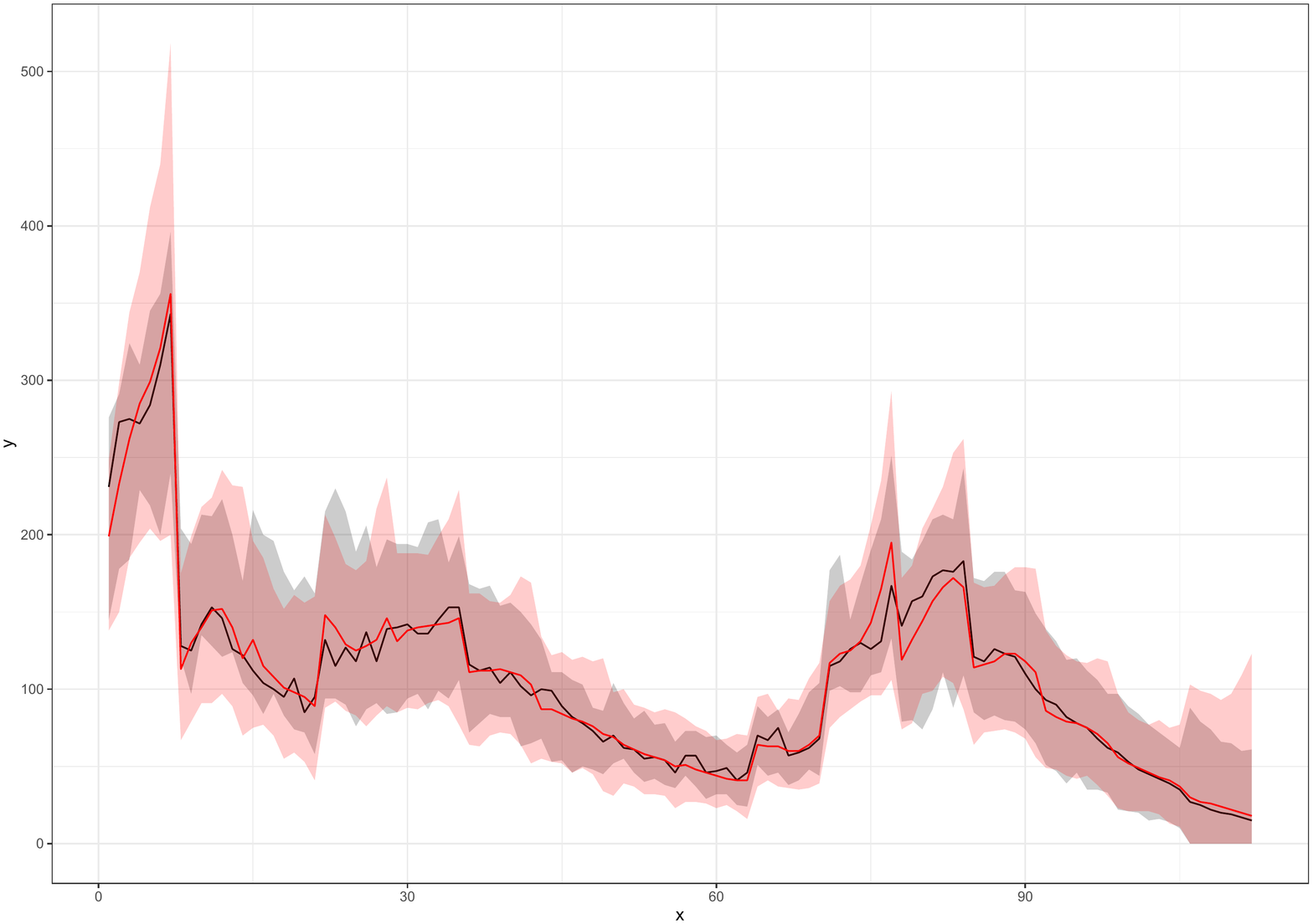}
     \caption{Aged 30-49.}
  \end{subfigure}
  \begin{subfigure}{7cm}
    \centering\includegraphics[width=6cm]{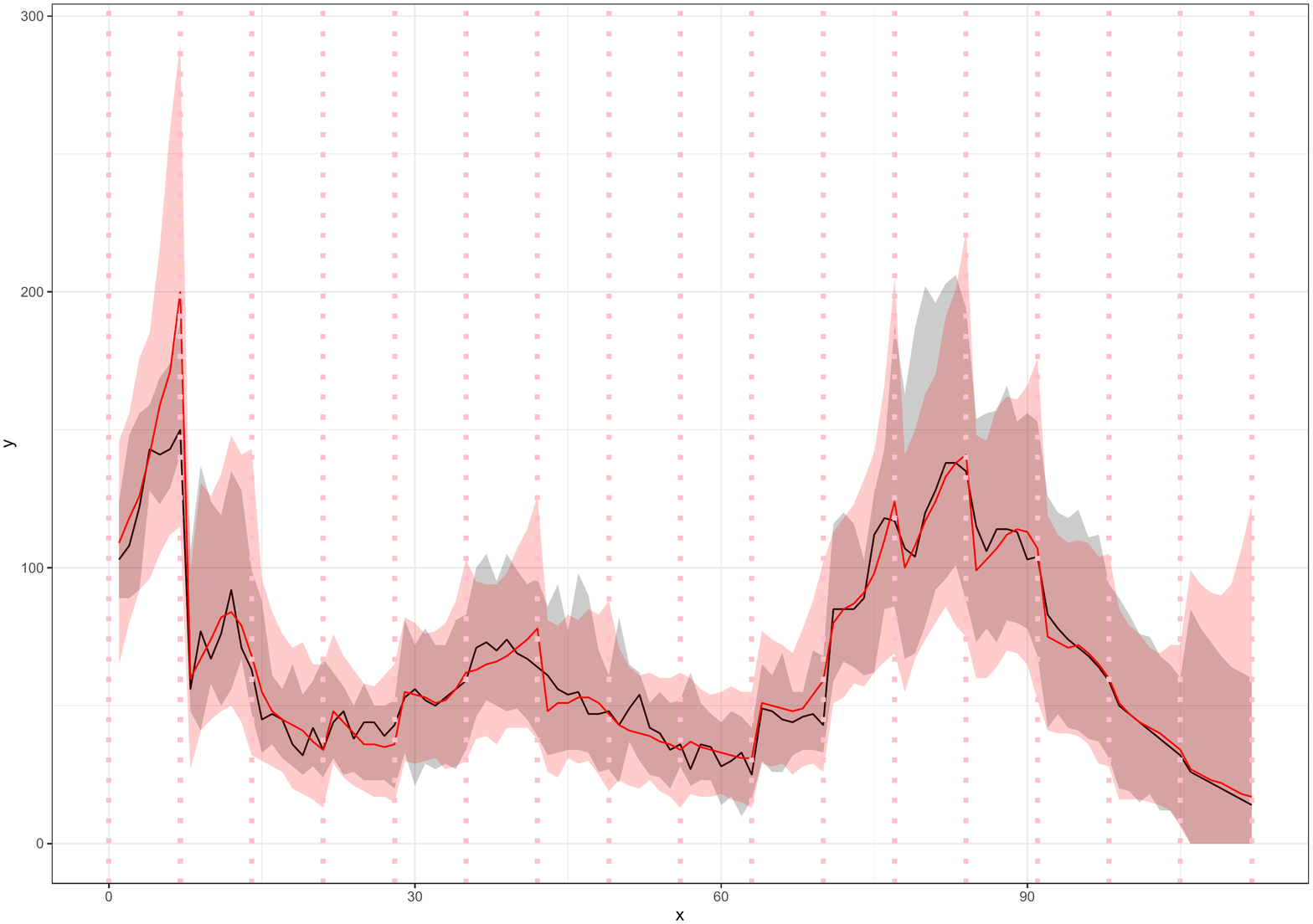}
    \caption{Aged 50-69.}
  \end{subfigure}
   \begin{subfigure}{7cm}
    \centering\includegraphics[width=6cm]{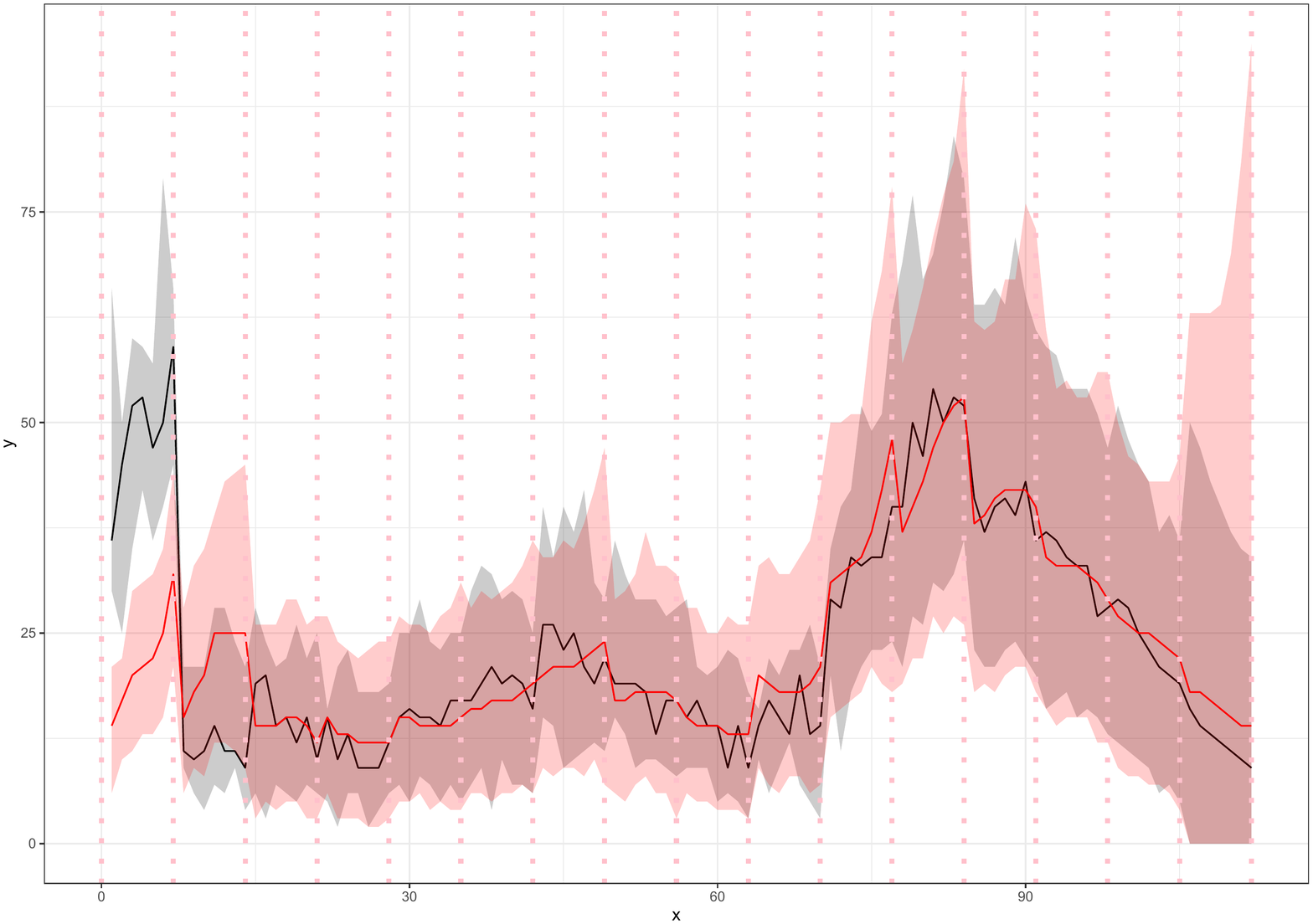}
     \caption{Aged 70+.}
  \end{subfigure}

   \caption{{\bf The posterior median estimate of daily hidden cases of model A (black line) and model U (red line), and the 99\% CIs (ribbon) in Ashford.} The vertical dotted lines show the beginning of each week in the period we examine.}
   \label{CompDHC_Ashford4G}
\end{figure}

\begin{figure}[!h] 
 \begin{subfigure}{7cm}
    \centering\includegraphics[width=6cm]{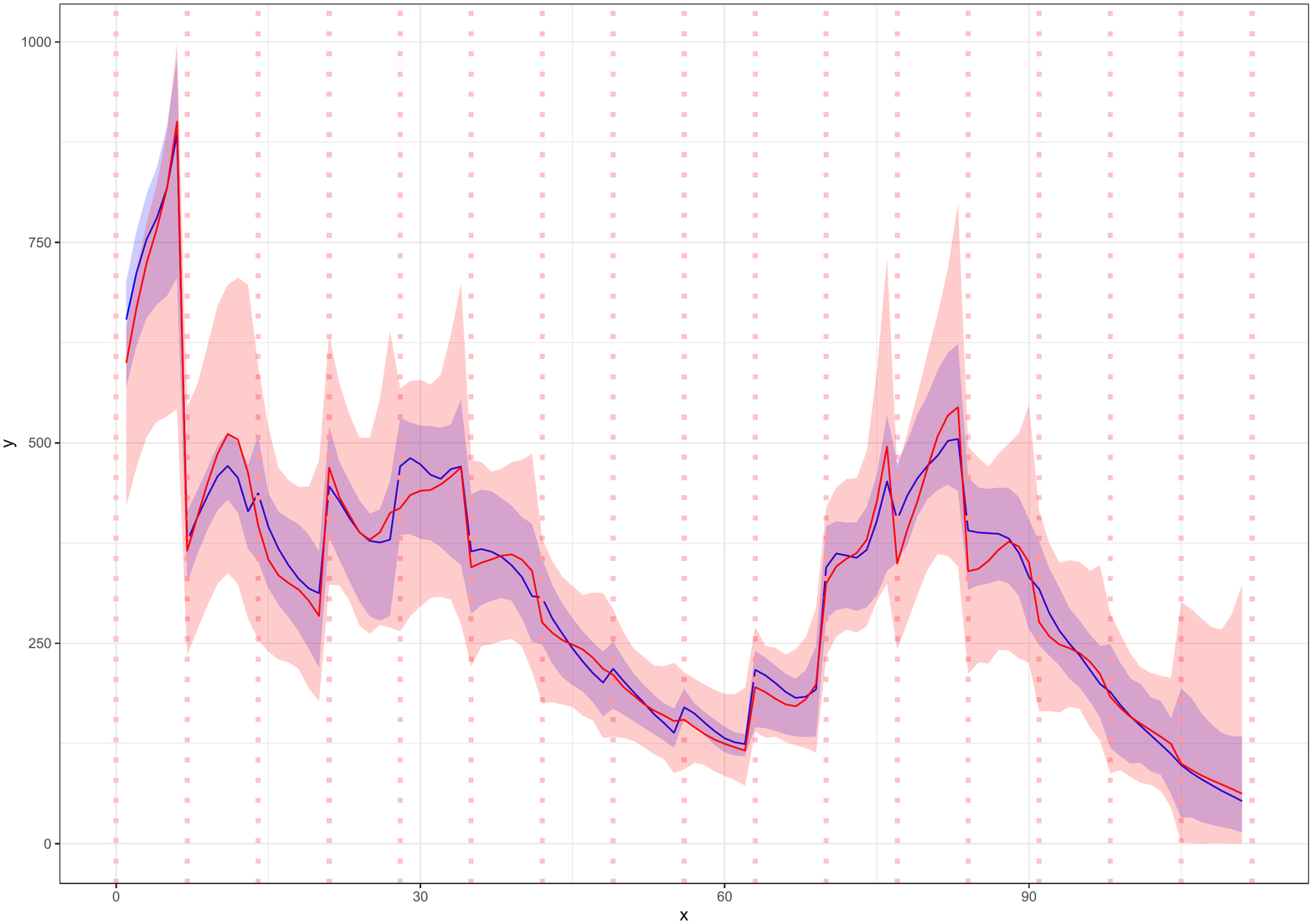}
     \caption{Aggregated latent intensity.}
  \end{subfigure}
  \begin{subfigure}{7cm}
    \centering\includegraphics[width=6cm]{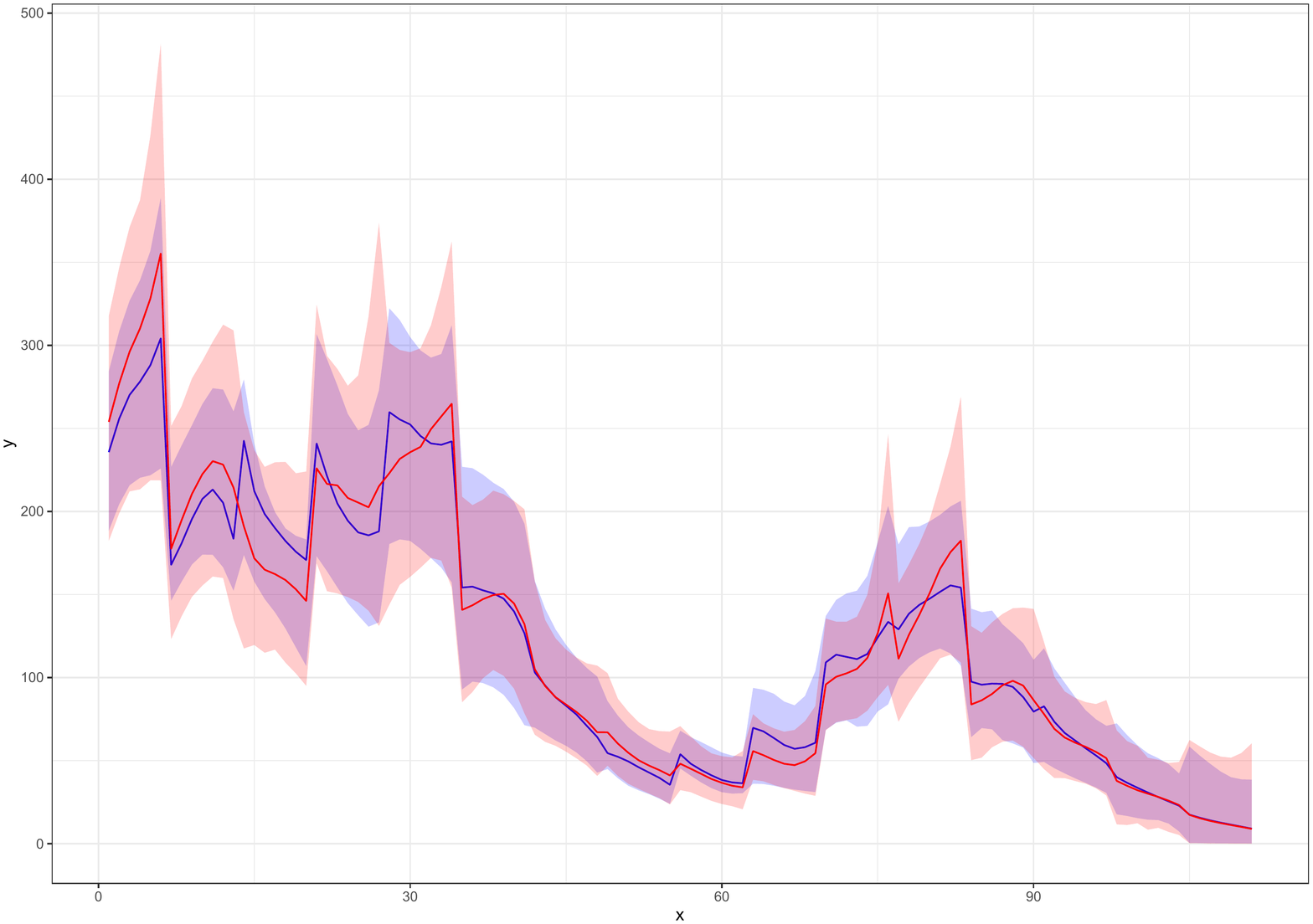}
   \caption{Aged 0-29.}
  \end{subfigure}
  \begin{subfigure}{7cm}
    \centering\includegraphics[width=6cm]{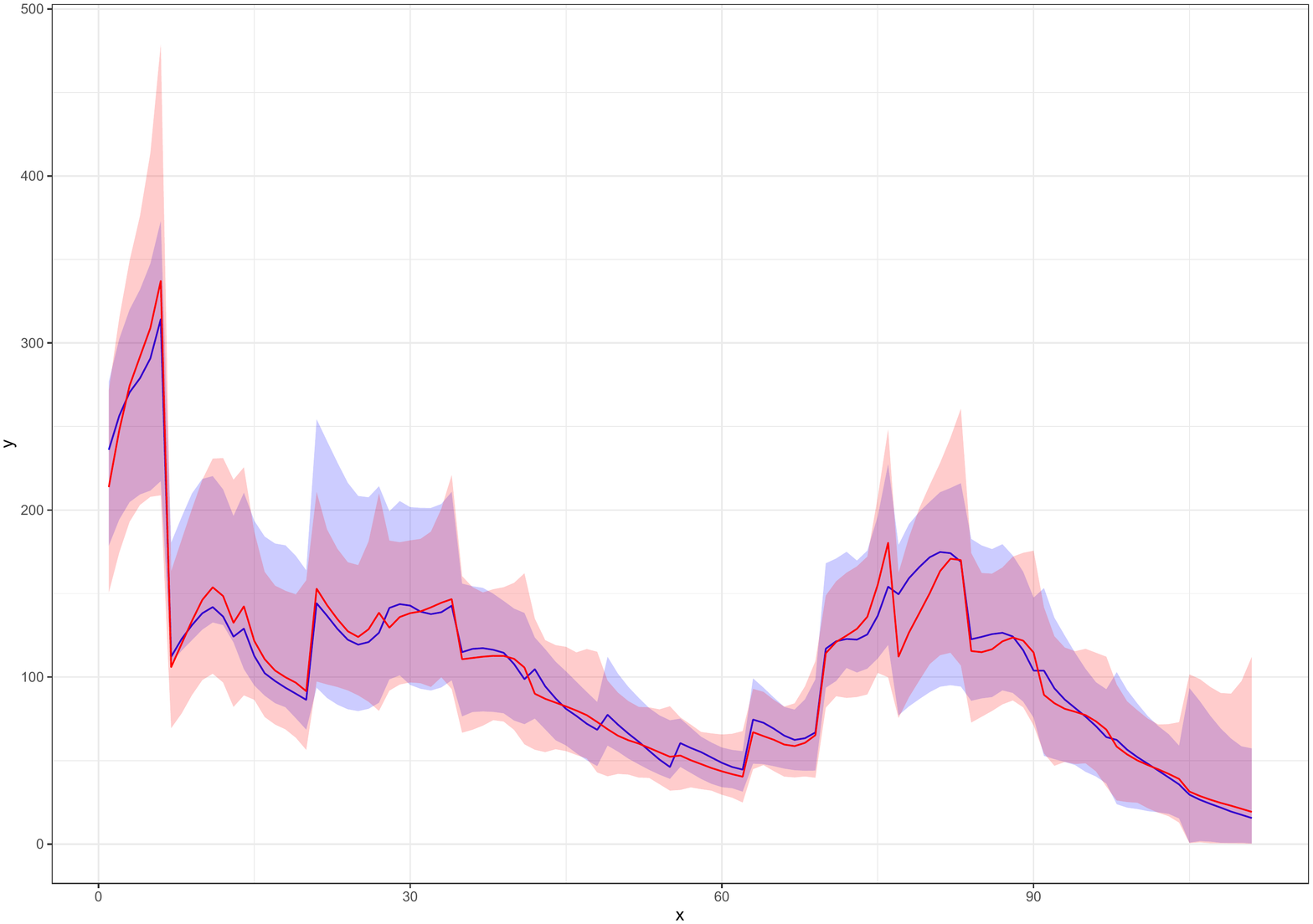}
     \caption{Aged 30-49.}
  \end{subfigure}
  \begin{subfigure}{7cm}
    \centering\includegraphics[width=6cm]{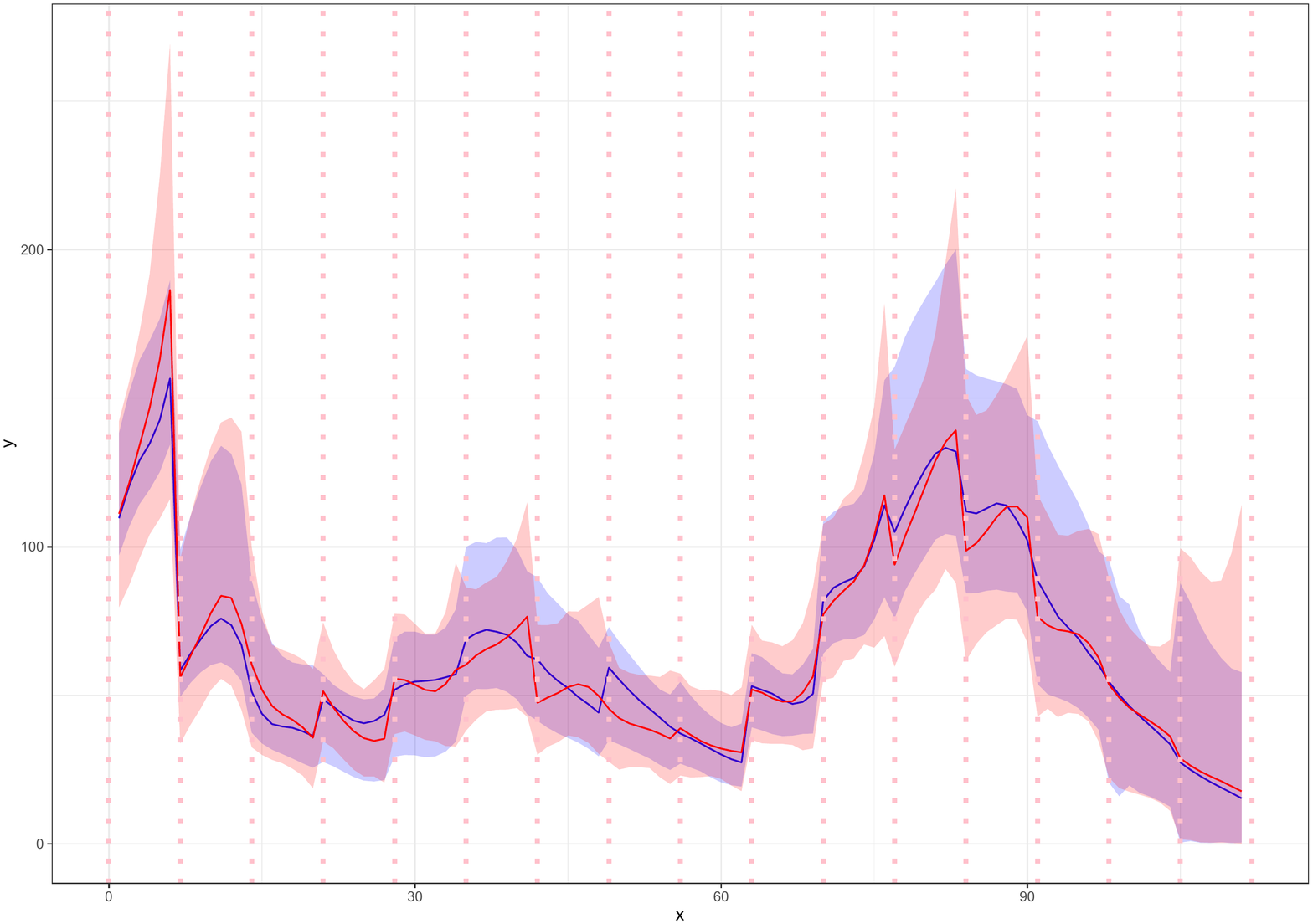}
    \caption{Aged 50-69.}
  \end{subfigure}
   \begin{subfigure}{7cm}
    \centering\includegraphics[width=6cm]{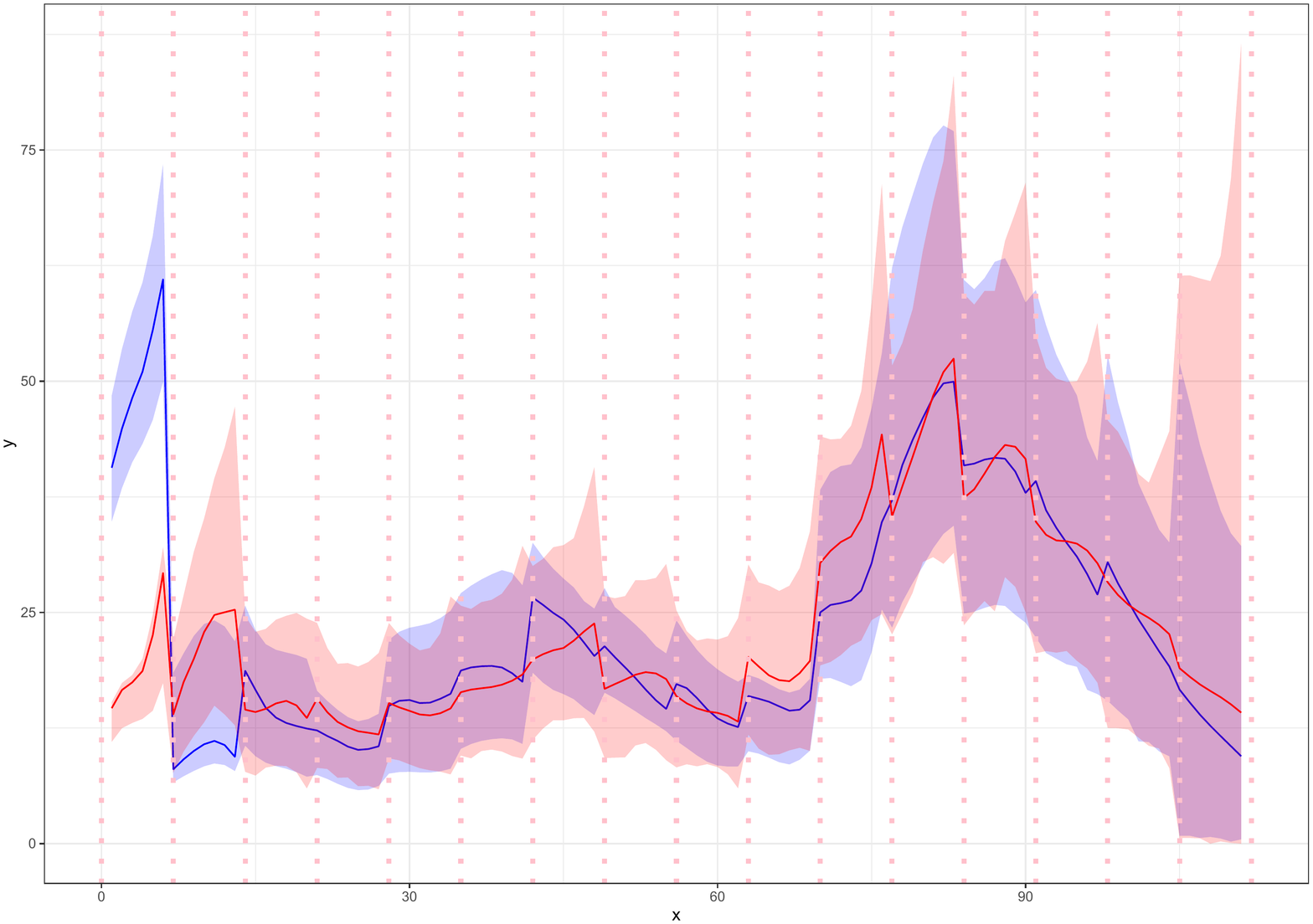}
     \caption{Aged 70+.}
  \end{subfigure}

   \caption{{\bf The posterior median estimate of latent intensity of model A (blue line) and model U (red line), and the 99\% CIs (ribbon) in Ashford.} The vertical dotted lines show the beginning of each week in the period we examine.}
   \label{CompLambda_Ashford4G}
\end{figure}

\begin{figure}[!h] 
    \centering\includegraphics[width=6cm]{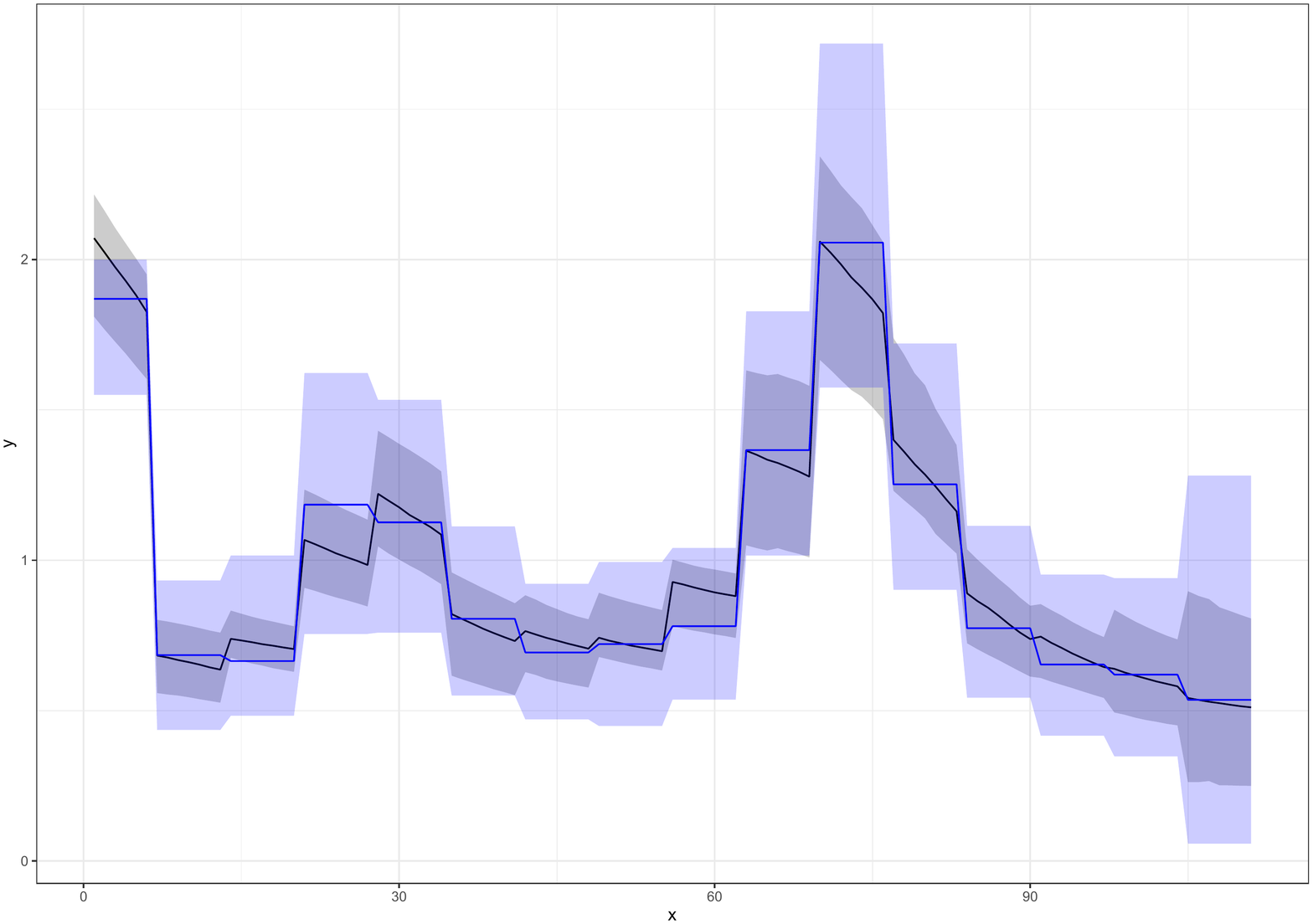}
  \caption{\bf The posterior median estimate of instantaneous reproduction number of model A (black line) and model U (blue line), and the 99\% CIs (ribbon) in Ashford.}
   \label{CompR_Ashford4G}
\end{figure}

\begin{figure}[!h] 
 \begin{subfigure}{7cm}
    \centering\includegraphics[width=6cm]{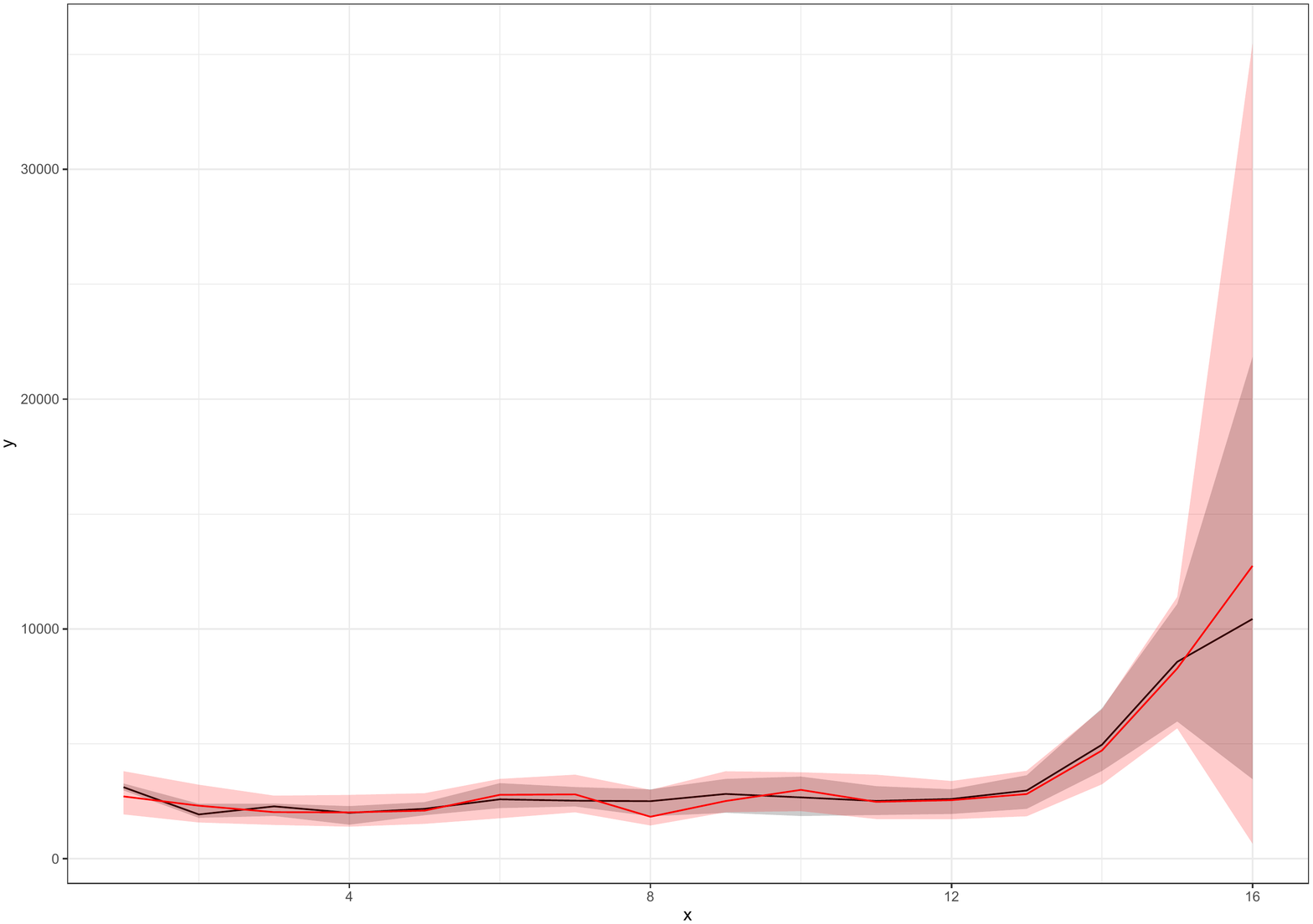}
     \caption{Aggregated weekly hidden cases.}
  \end{subfigure}
  \begin{subfigure}{7cm}
    \centering\includegraphics[width=6cm]{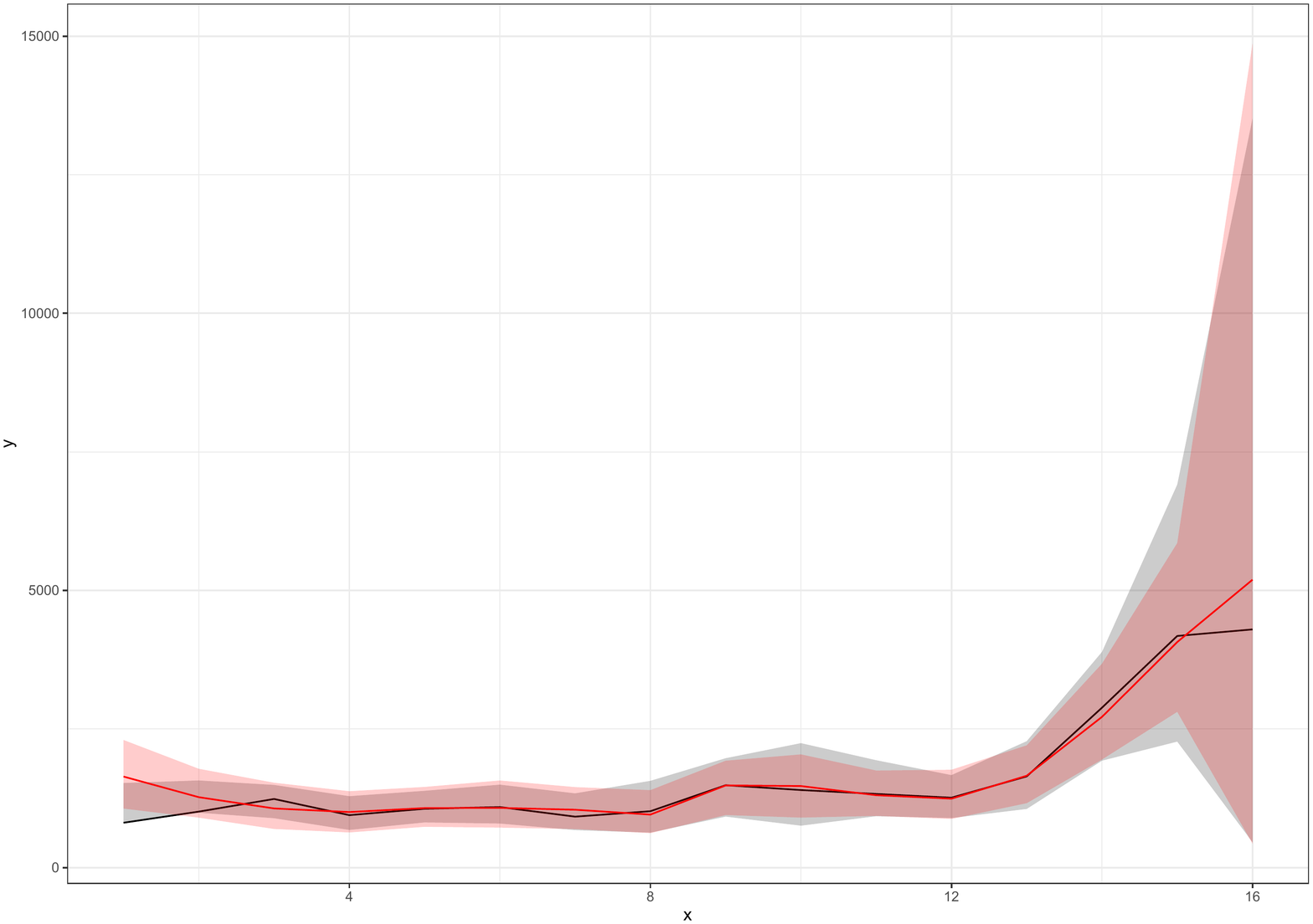}
   \caption{Aged 0-29.}
  \end{subfigure}
  \begin{subfigure}{7cm}
    \centering\includegraphics[width=6cm]{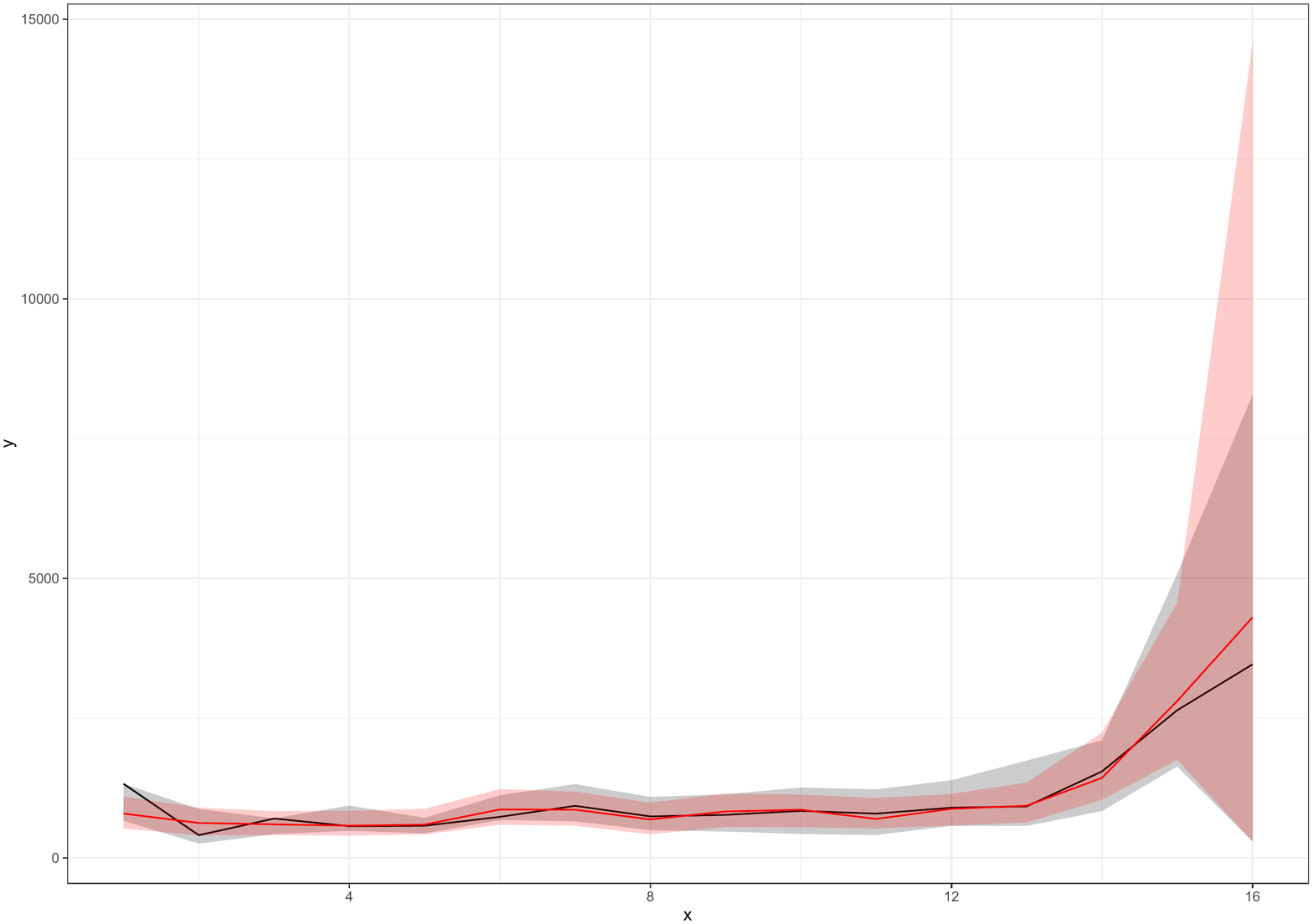}
     \caption{Aged 30-49.}
  \end{subfigure}
  \begin{subfigure}{7cm}
    \centering\includegraphics[width=6cm]{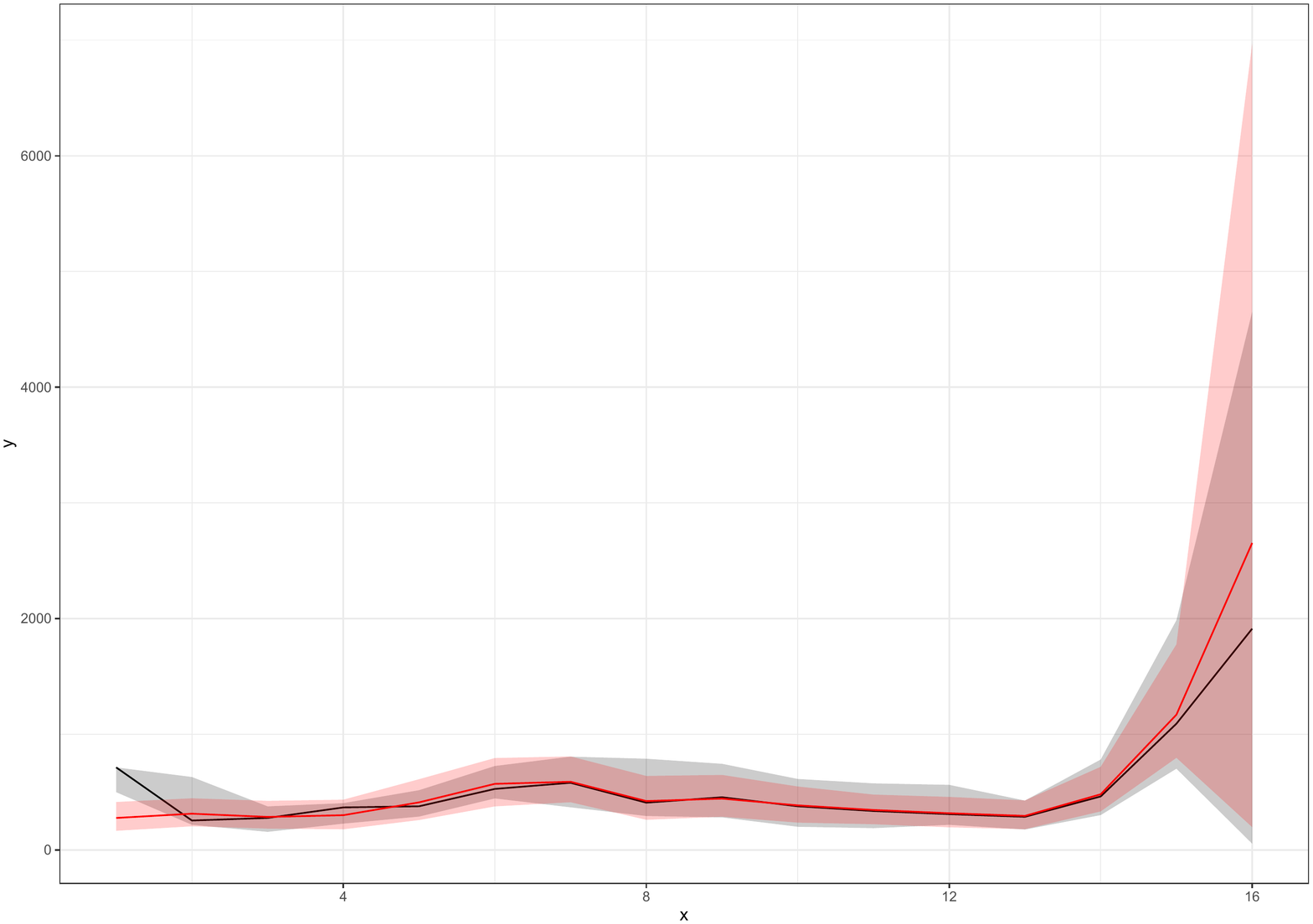}
    \caption{Aged 50-69.}
  \end{subfigure}
   \begin{subfigure}{7cm}
    \centering\includegraphics[width=6cm]{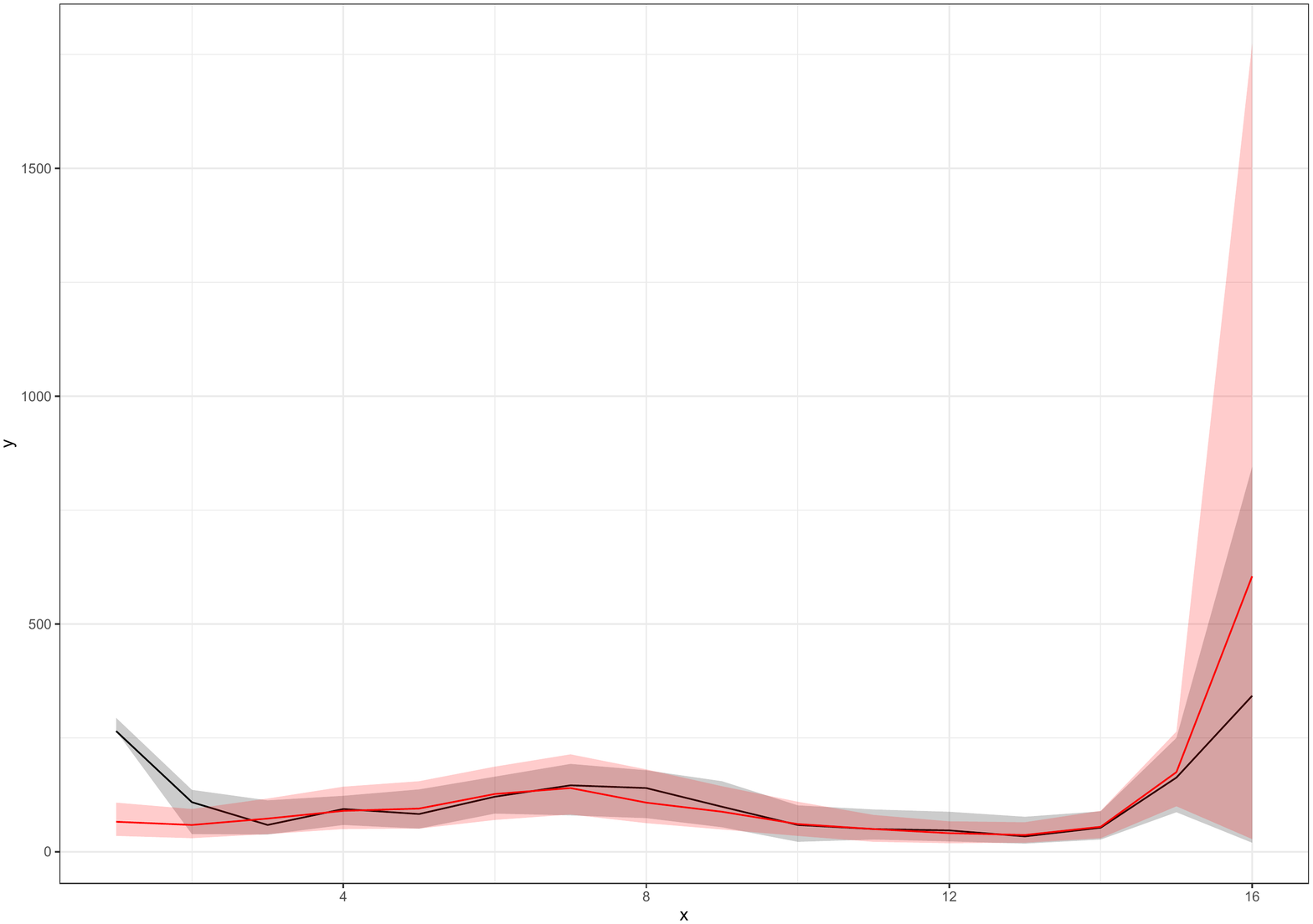}
     \caption{Aged 70+.}
  \end{subfigure}

   \caption{\bf The posterior median estimate of weekly hidden cases of model A (black line) and model U (red line), and the 99\% CIs (ribbon) in Leicester.}
   \label{CompWHC_Leicester4G}
\end{figure}

\begin{figure}[!h] 
 \begin{subfigure}{7cm}
    \centering\includegraphics[width=6cm]{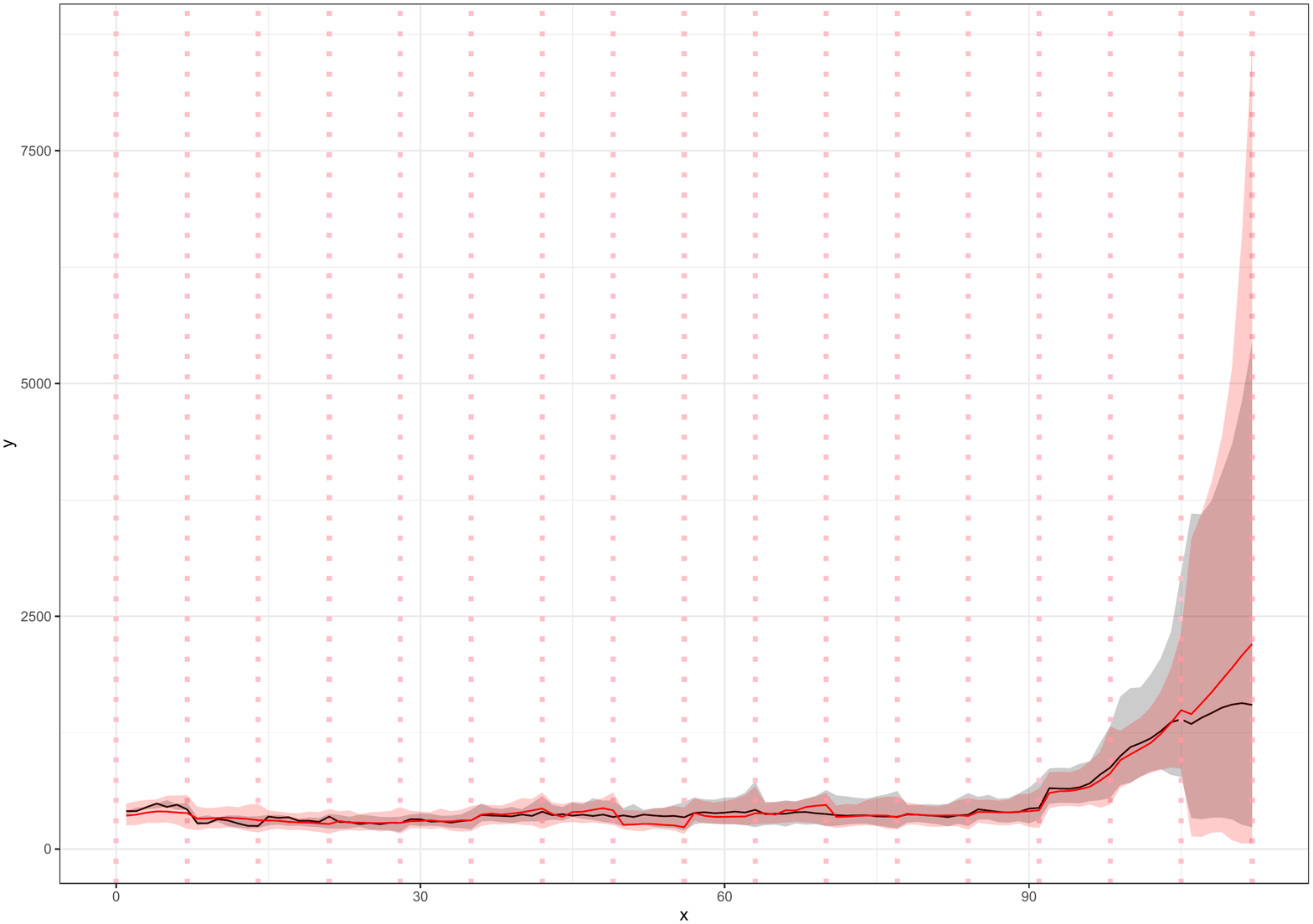}
     \caption{Aggregated daily hidden cases.}
  \end{subfigure}
  \begin{subfigure}{7cm}
    \centering\includegraphics[width=6cm]{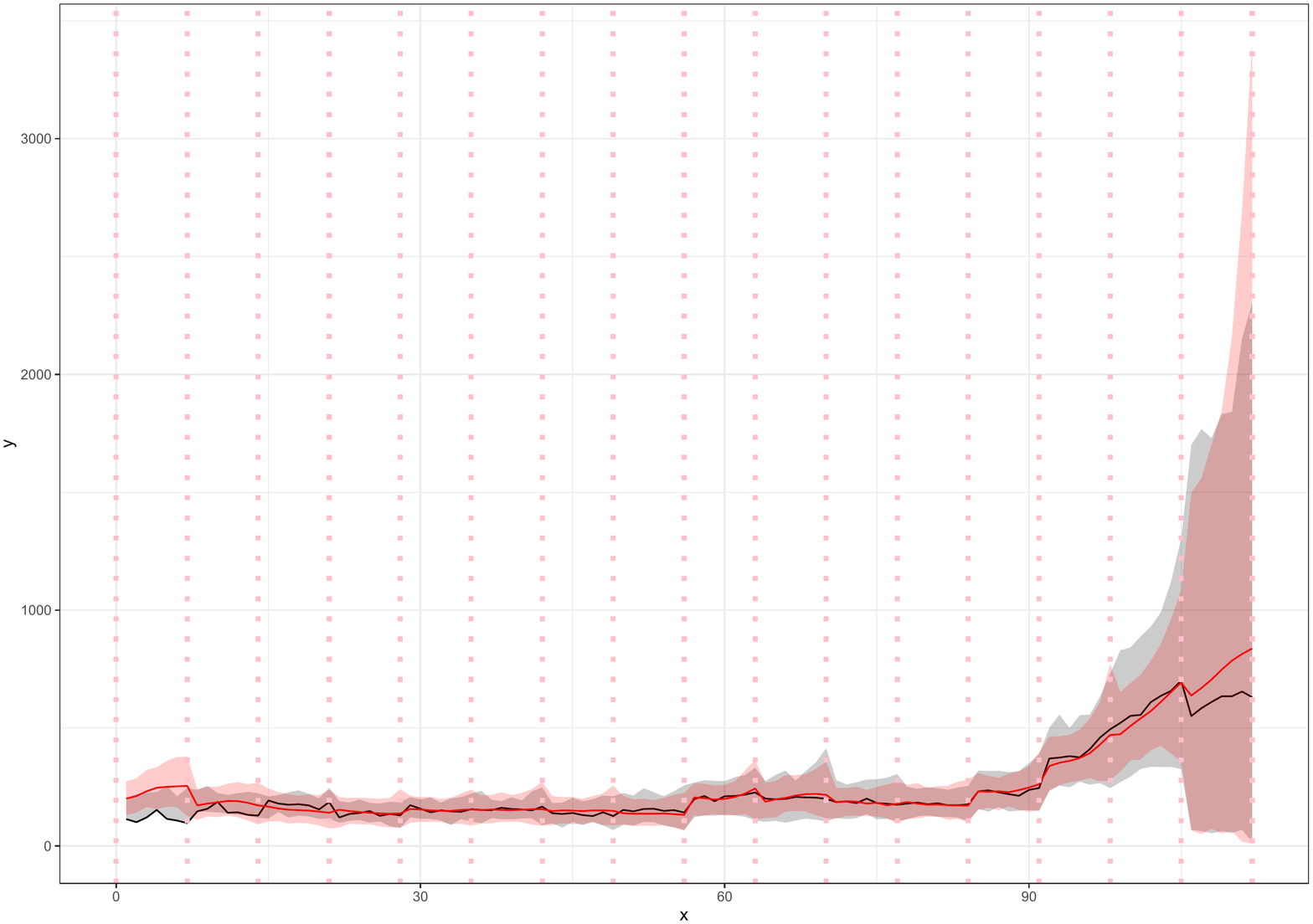}
   \caption{Aged 0-29.}
  \end{subfigure}
  \begin{subfigure}{7cm}
    \centering\includegraphics[width=6cm]{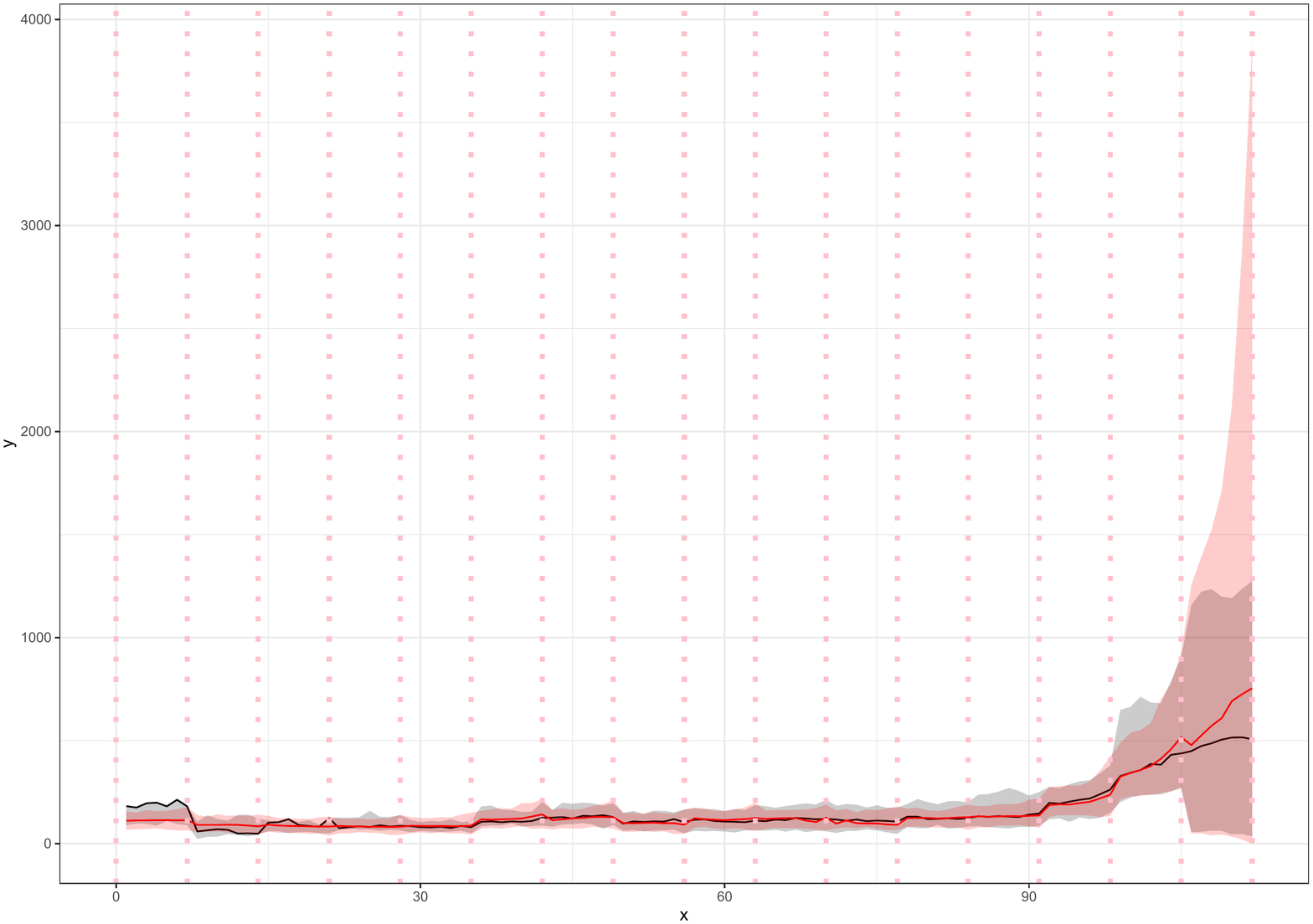}
     \caption{Aged 30-49.}
  \end{subfigure}
  \begin{subfigure}{7cm}
    \centering\includegraphics[width=6cm]{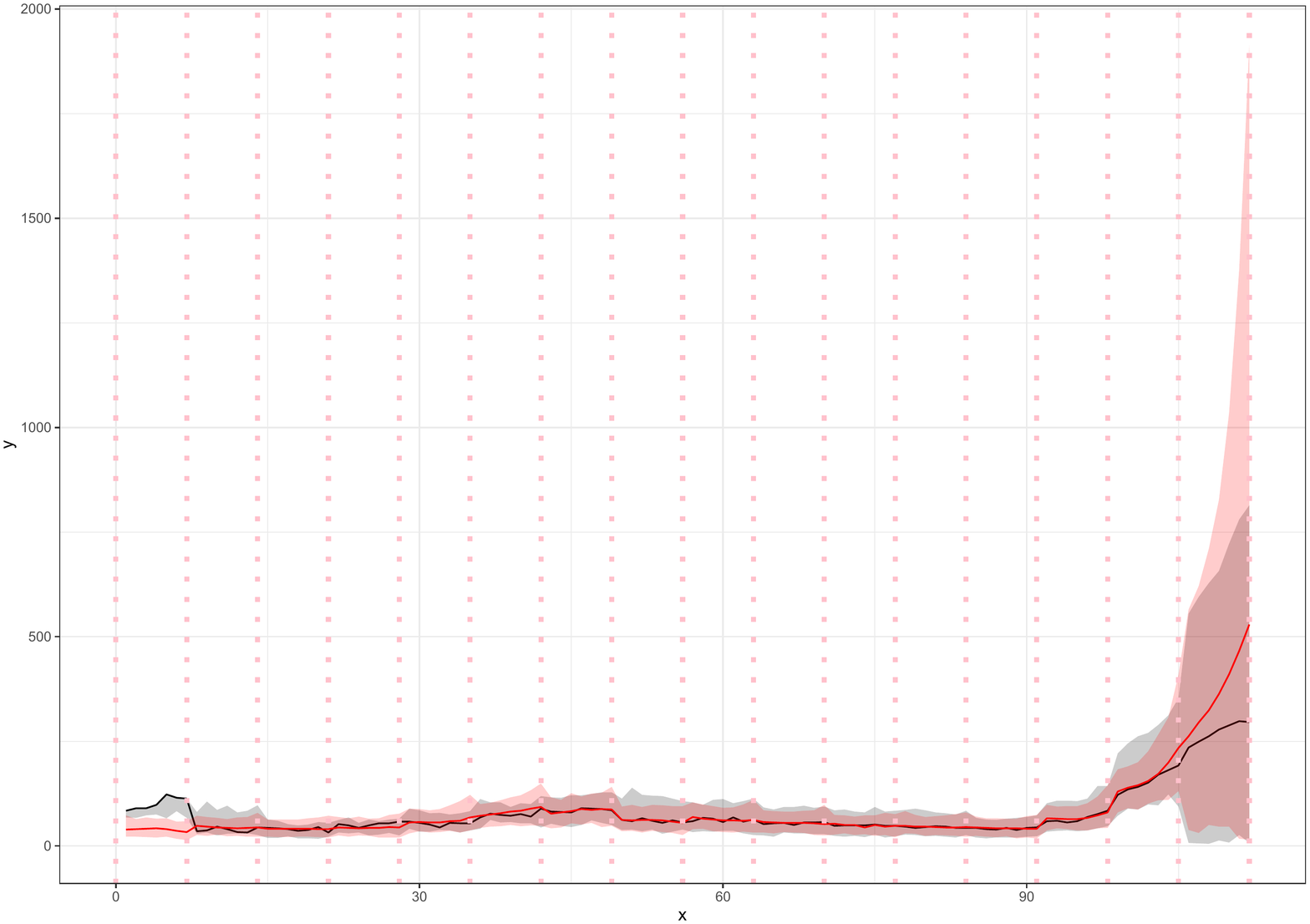}
    \caption{Aged 50-69.}
  \end{subfigure}
   \begin{subfigure}{7cm}
    \centering\includegraphics[width=6cm]{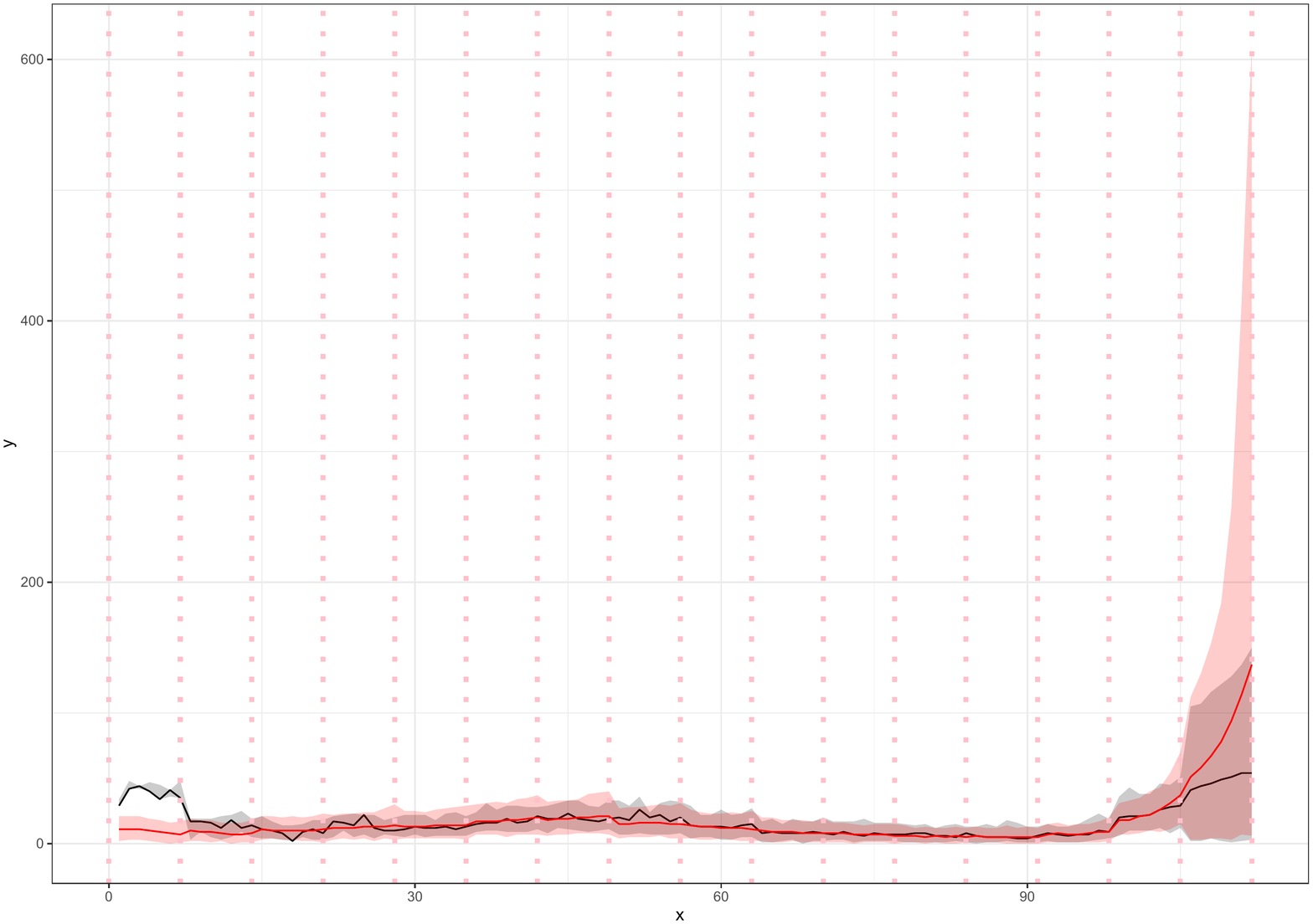}
     \caption{Aged 70+.}
  \end{subfigure}

   \caption{{\bf The posterior median estimate of daily hidden cases of model A (black line) and model U (red line), and the 99\% CIs (ribbon) in Leicester.} The vertical dotted lines show the beginning of each week in the period we examine.}
   \label{CompDHC_Leicester4G}
\end{figure}

\begin{figure}[!h] 
 \begin{subfigure}{7cm}
    \centering\includegraphics[width=6cm]{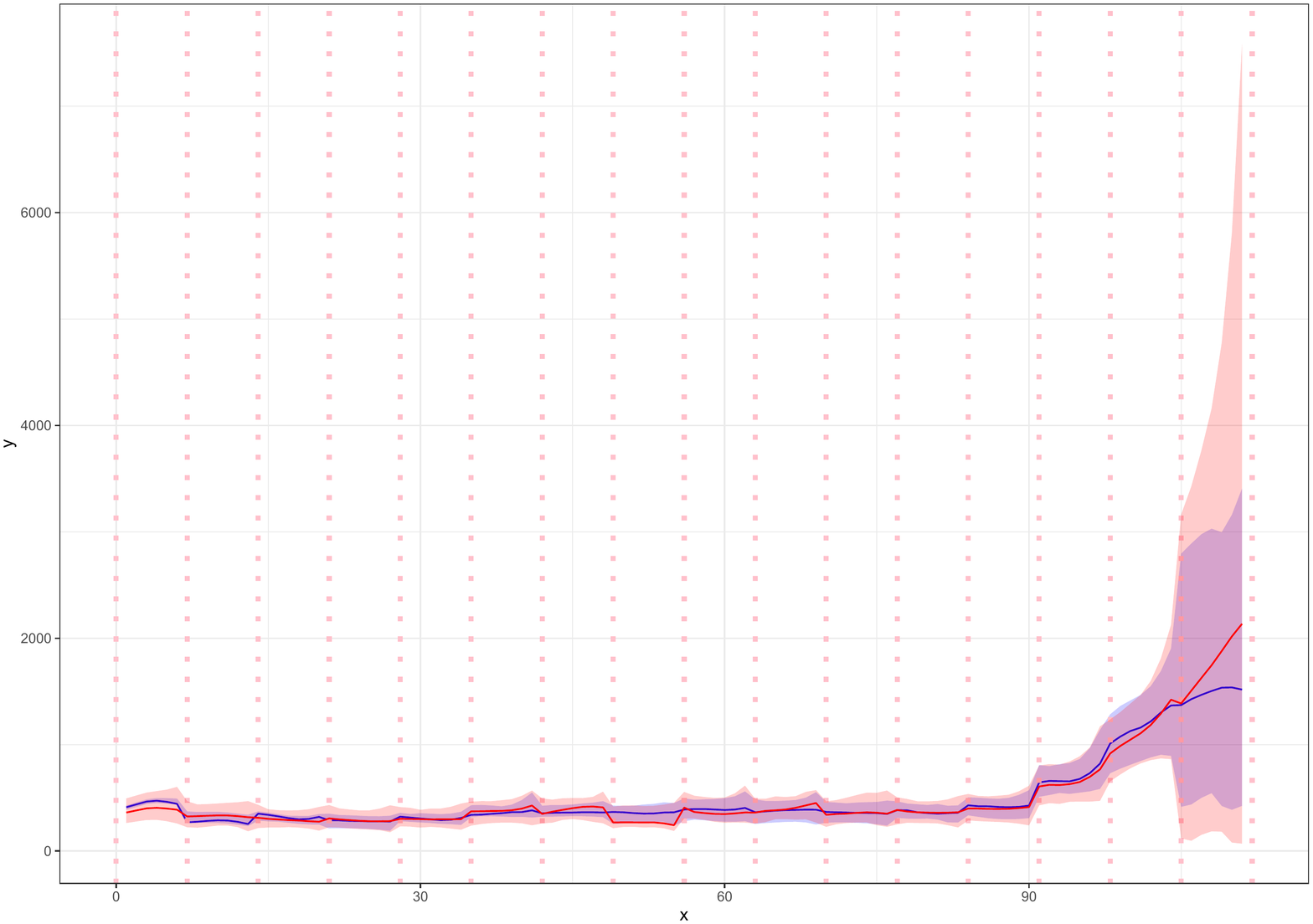}
     \caption{Aggregated latent intensity.}
  \end{subfigure}
  \begin{subfigure}{7cm}
    \centering\includegraphics[width=6cm]{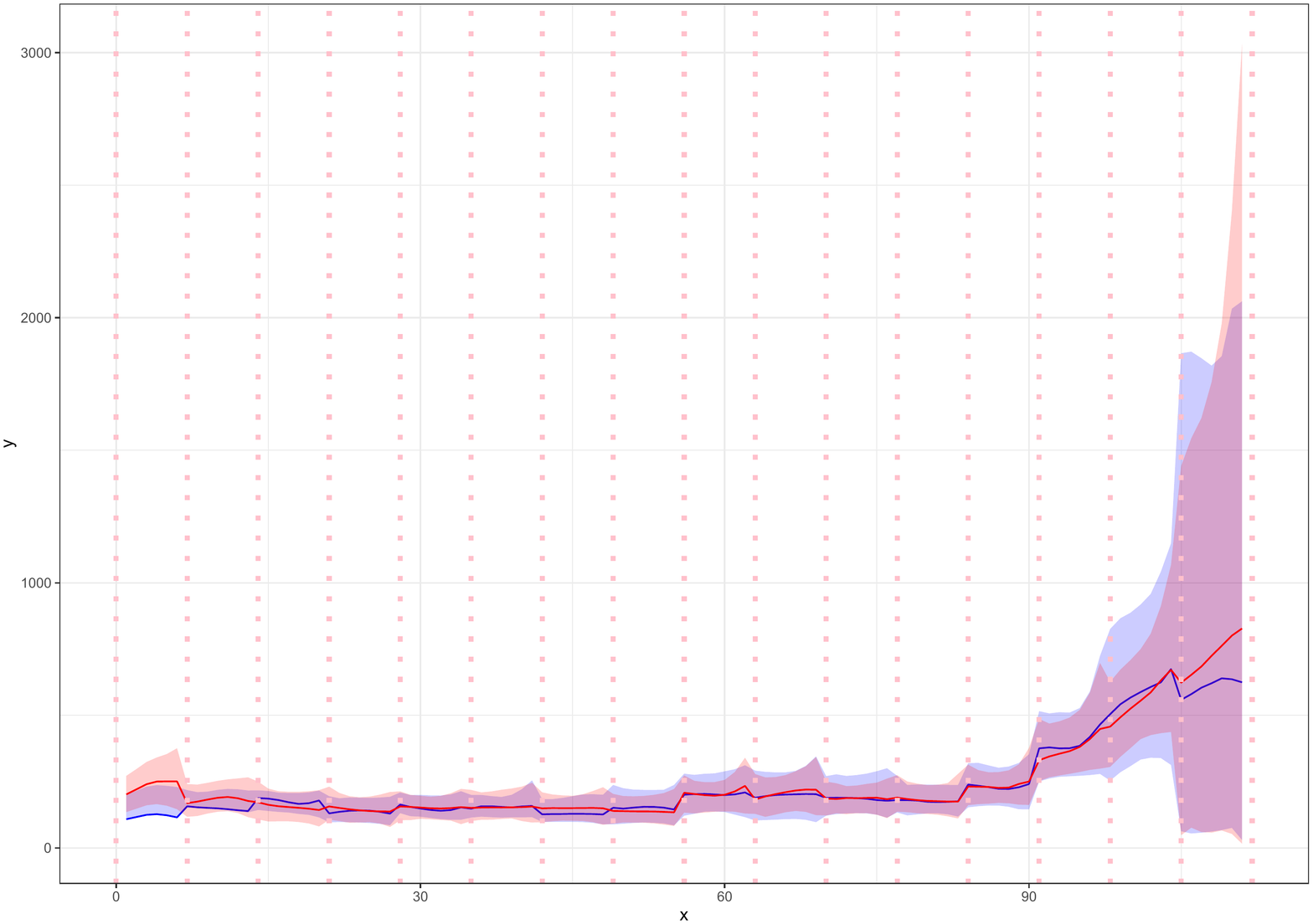}
   \caption{Aged 0-29.}
  \end{subfigure}
  \begin{subfigure}{7cm}
    \centering\includegraphics[width=6cm]{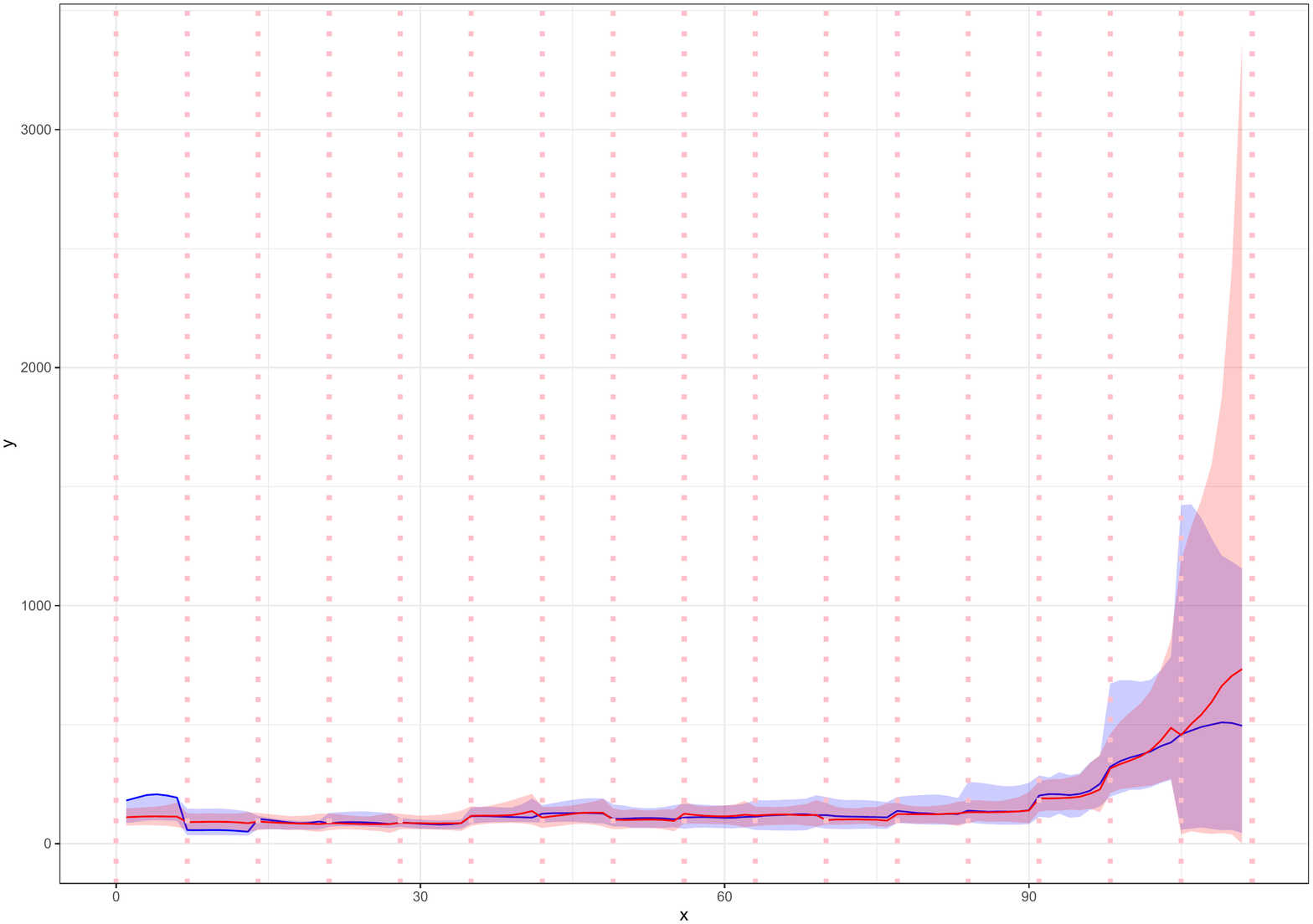}
     \caption{Aged 30-49.}
  \end{subfigure}
  \begin{subfigure}{7cm}
    \centering\includegraphics[width=6cm]{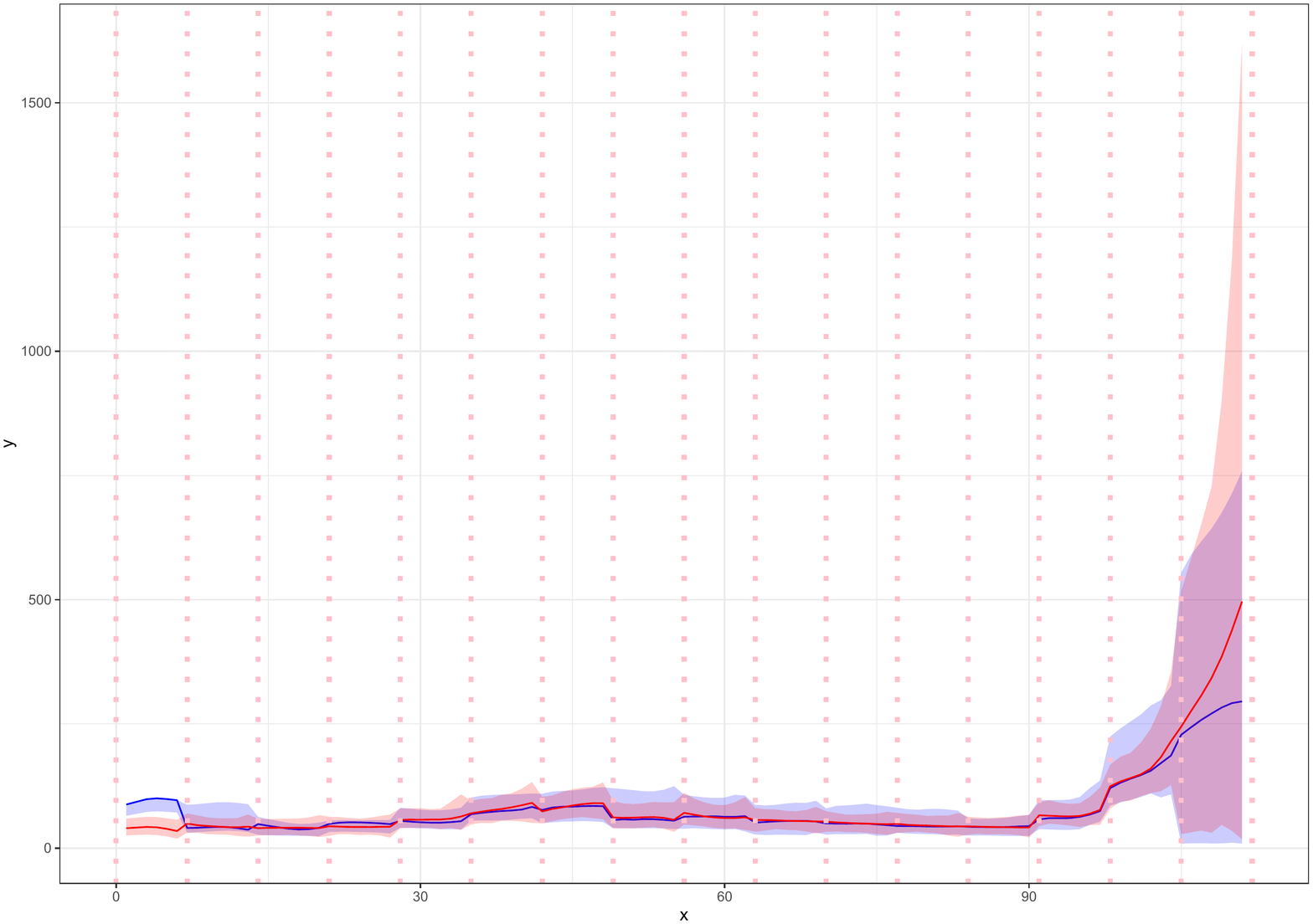}
    \caption{Aged 50-69.}
  \end{subfigure}
   \begin{subfigure}{7cm}
    \centering\includegraphics[width=6cm]{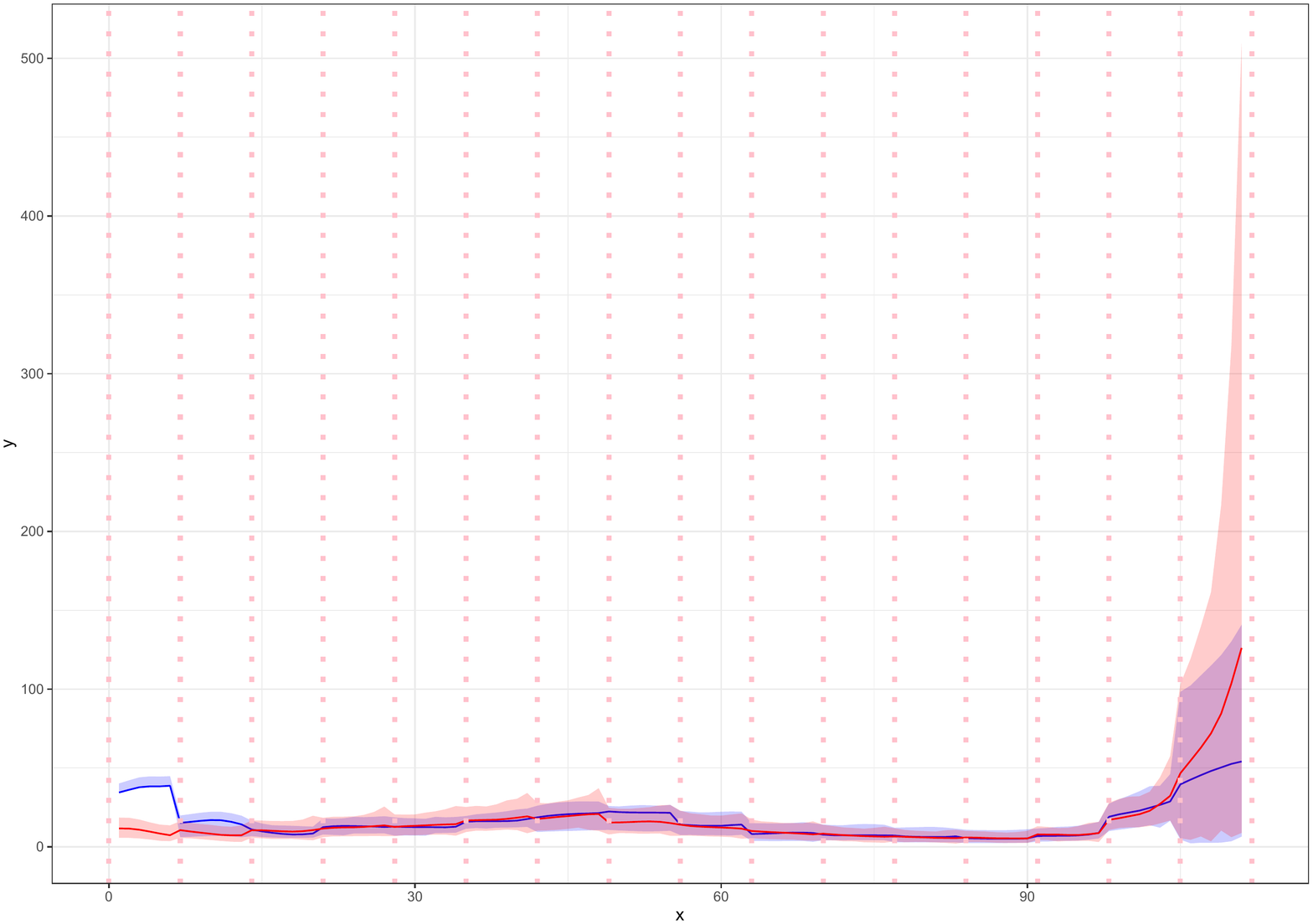}
     \caption{Aged 70+.}
  \end{subfigure}

   \caption{{\bf The posterior median estimate of latent intensity of model A (blue line) and model U (red line), and the 99\% CIs (ribbon) in Leicester.} The vertical dotted lines show the beginning of each week in the period we examine.}
   \label{CompLambda_Leicester4G}
\end{figure}

\begin{figure}[!h] 
    \centering\includegraphics[width=6cm]{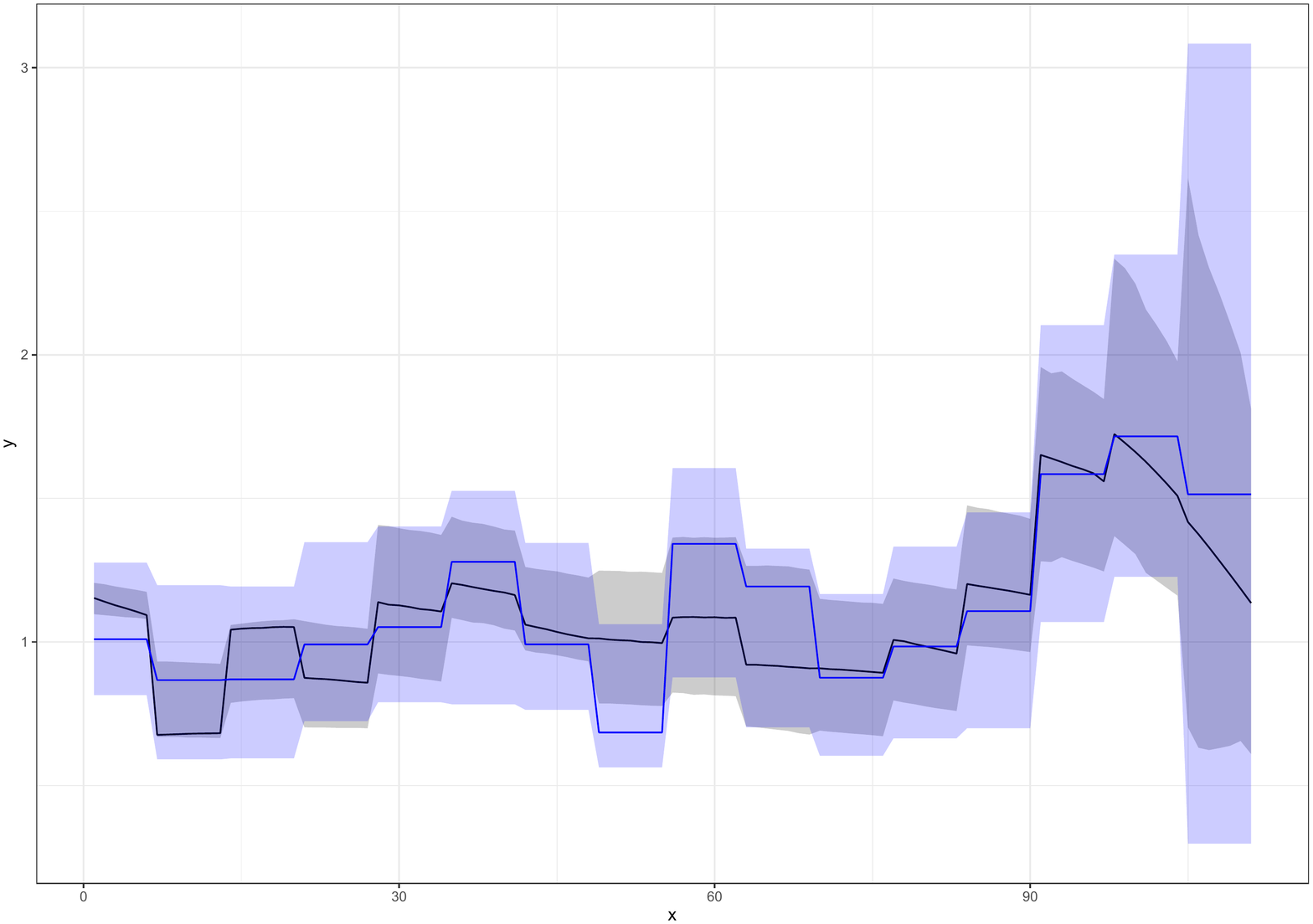}
  \caption{\bf The posterior median estimate of instantaneous reproduction number of model A (black line) and model U (blue line), and the 99\% CIs (ribbon) in Leicester.}
   \label{CompR_Leicester4G}
\end{figure}

The instantaneous reproduction number's posterior medians of both models follow the same pattern in general lines (see Figures \ref{CompR_Ashford4G}, \ref{CompR_Kings4G}, \ref{CompR_Leicester4G}). The estimated aggregated and per age group latent intensity, weekly and daily hidden cases via posterior median given by model A are similar to the ones of model U (see Figures \ref{CompWHC_Ashford4G}-\ref{CompLambda_Ashford4G}, \ref{CompWHC_Kings4G}-\ref{CompLambda_Kings4G}, \ref{CompWHC_Leicester4G}-\ref{CompLambda_Leicester4G}).  Model U derives similar estimates to those of model A, as the latent intensity and cases per age band are strongly dependent on the infections of the associated age group and independent of the other age groups' infections. The simulation analysis shows that considering the individual inhomogeneity in age and finite population, the width of CIs decreases.

The comparison between models A and U demonstrates that model U provides estimates close to the reality for the latent intensity, weekly and daily hidden cases, and a rough approximation of the instantaneous reproduction number. The possible noticeable differences in both models' estimates during the first three weeks are due to different initializations. Table \ref{tableGOF_AM} illustrates that the metric is higher than 0.1 when models' medians present differences in the first weeks. We suggest running both models considering the former $\eta$ days of the horizon we are interested in, where $\eta$ is the delay in which the transition kernel of latent cases is negligible; $\eta=21 \text {days or $3$ weeks}$ for COVID-19 data. The analysis also shows that model A decreases the uncertainty of estimates and infers the reproduction number per age group. Model U cannot be applied to infer the reproduction number per age group number and investigate the age groups' behaviours in an epidemic indicating the importance of model A. The instantaneous reproduction numbers per age group provide a real-time measurement of interventions and behavioural changes.

\begin{table}[!ht]
\centering
\caption{\bf The goodness of fit of model U for instantaneous reproduction number (R), weekly hidden cases (WHC), daily hidden cases (DHC) and latent intensity (LI) using the metric PAE.}
\begin{tabular}{ |l|l|l|l| } \hline
 \multicolumn{4}{|l|}{ \bf Goodness of fit of Model U}  \\
 \thickhline
  PAE & Kingston & Ashford & Leicester\\
 \hline
 R& 0.12 & 0.06 &0.12\\ \hline
 WHC aged 0-29 & 0.07&0.06 & 0.11   \\ \hline
 WHC aged 30-49 & 0.13 &0.05 &0.14  \\ \hline
 WHC aged 50-69 & 0.13 &0.05 &0.18    \\
 WHC aged 70+ & 0.12 &  0.17&0.33 \\ \hline
 aggregated WHC & 0.1 & 0.03 &0.11   \\ \hline
 DHC aged 0-29 & 0.08 &0.07 &0.12   \\ \hline
 DHC aged 30-49 & 0.14 & 0.07& 0.14 \\ \hline
 DHC aged 50-69 & 0.14 & 0.08 &  0.19\\ \hline
 DHC aged 70+ & 0.16 &0.2 & 0.36 \\ \hline
 aggregated DHC & 0.1 &0.05  &0.11   \\ \hline
 LI aged 0-29 & 0.08 & 0.09&  0.11 \\ \hline
LI aged 30-49 & 0.13 & 0.07&  0.13\\ \hline
 LI aged 50-69 & 0.14 & 0.07 &   0.16\\ \hline
 LI aged 70+ & 0.15 &0.2 &  0.3\\ \hline
 aggregated LI & 0.1 &0.06  & 0.1   \\ 
 \hline
\end{tabular}
\label{tableGOF_AM}
\end{table}

\paragraph*{Future prediction}
Applying the steps outlined in Algorithm \ref{AMAlgPred}, it is also possible to forecast with relative accuracy how many new infections would be reported during the next week, $\mathcal{T}_{17}$ (see Tables \ref{AMtableKingston},\ref{AMtableAshford}, \ref{AMtableLeicester}). Our forecast is subject to similar levels of uncertainty as last week's estimates. However, the algorithm can be applied to forecast the epidemic's future trajectory-whether it will be upward, downward or stable-using the posterior mean and median.

\begin{table}[!ht]
\begin{adjustwidth}{-1.5in}{0in} 
\centering
\caption{\bf The true number of reported infections in $\mathcal{T}_{16}$ and $\mathcal{T}_{17}$, and the posterior median, the posterior mean and the 95\% CIs of the estimated infections in $\mathcal{T}_{17}$ in Kingston. }
\begin{tabular}{ |l|l||l|l|l|l|}\hline
 \multicolumn{6}{|l|}{Proposed Method} \\
 \thickhline
 Reported infections & Posterior Mean & Posterior Median  &  $95\%$ CIs & True Number ($\mathcal{T}_{17}$) & True Number ($\mathcal{T}_{16}$) \\
 \hline
aggregated & 664 & 649 & (427, 921) & 667 & \\ \hline
aged 0-29 & 154 & 145 & (64, 258) & 139  & 292\\ \hline
aged 30-49 & 250 & 238 & (112, 404) & 246  & 428\\ \hline
aged 50-69 & 172 & 164& (77, 279) & 187  & 295\\ \hline
aged 70+ & 89 & 85 & (39, 142) &   95 & 150\\
 \hline
\end{tabular}
\label{AMtableKingston}
\end{adjustwidth}
\end{table}

\begin{algorithm}[!h] % enter the algorithm environment
\algsetup{linenosize=\tiny}
\tiny %\small, \footnotesize, \scriptsize, or \tiny
\caption{\bf Predicting the new aggregated and per age group observed cases in near future} 
\label{AMAlgPred}
\begin{algorithmic}[1]
\STATE{Let $\hat{v}=\frac{1}{N}\sum_{j=1}^Nv_{j,16}$ and $\hat{d}=\frac{1}{N}\sum_{j=1}^Nd_{j,16}$.}
\FOR{$j=1,..,N$}
\STATE{$\gamma_{j,17,a}\sim P(\gamma_{17,a}|\gamma_{j,16,a},\hat{d})$, $\forall$ age group $a$} \\
\STATE{$\left(S_{j,17}^N, A_{j,17}^N\right) \sim P\left(S_{17}^N,A_{17}^N|S_{j,1:16}^N,A_{j,1:16}^N,\{\gamma_{j,17,a}\}_a,\mathcal{H}_0,A_0^N\right)$} \\
\STATE{ Calculate the mean of observed cases of age $a$ in the interval denoted by $\mu_{j,17,a}$, $\forall$ age group $a$.}\\
\STATE{$Y_{j,17,a}\sim \mbox{NB}(\mu_{j,17,a},\hat{v})$, $\forall$ age group $a$.}\\
\ENDFOR
\STATE{Use the sample $\{Y_{j,17,a}\}_{j=1}^N$ to find the posterior mean, the posterior median and the 95\% CI of the estimated observed cases of age a in $\mathcal{T}_{17}$.}
\STATE{Use the sample $\{\sum_aY_{j,17,a}\}_{j=1}^N$ to find the posterior mean, the posterior median and the 95\% CI of the estimated aggregated observed cases in $\mathcal{T}_{17}$.}
\end{algorithmic}
\end{algorithm}

\subsection*{Epidemic Dynamics}
The KDPF (Algorithm \ref{APAlg}) can also infer the number of directed links between age groups $a \in \mathcal{A}$ and $a' \in \mathcal{A}$ (hereafter $a \rightarrow a'$) by capturing the process' branching structure. The underlying dynamics are revealed either by saving the parent of each latent infection (Method A) or by sampling its parent from a multinomial distribution as described above (Method B). We employed 30 randomly selected particles in Method B due to computational constraints.  

We use the ONS released demographic data for Leicester~\cite{ONS::Pop} in the simulation concepts. We work with 16 hidden states $\{X_n\}_{n=1}^{16}$ and 16 subintervals $\{\mathcal{T}_n\}_{n=1}^{16}$, where each subinterval has a week-long length. We coarse the age groups of the contact matrix for reopening schools \citep{jarvis2020quantifying} to find $\{m_{aa'}\}_{a,a'\in \mathcal{A}}$. We infer the number of directed links between age groups.

\begin{figure}[!h] 
\begin{subfigure}{7cm}
    \centering\includegraphics[width=6cm]{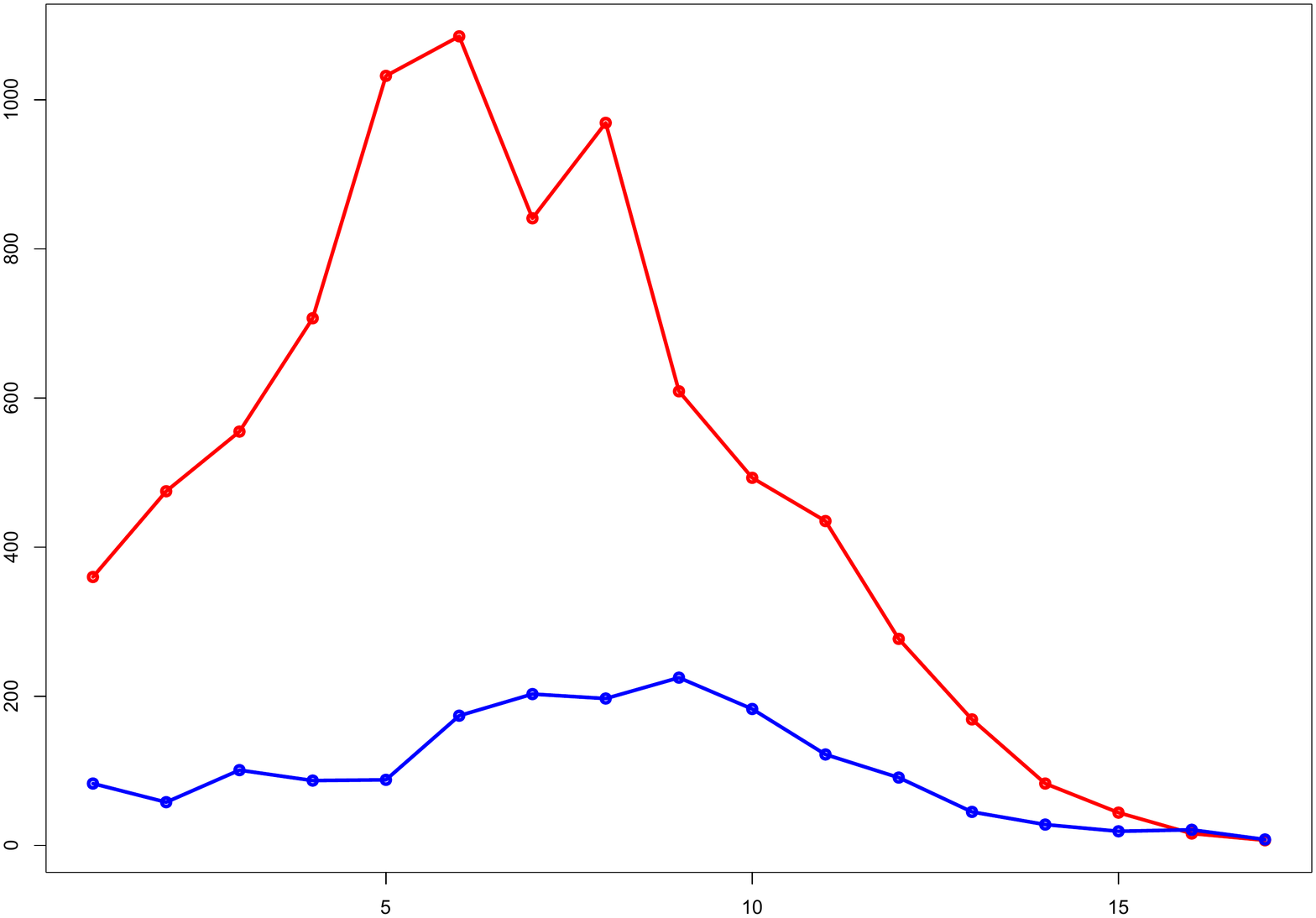}
    \caption{The weekly observed cases aged 0-59 (red line) and\\ 60+ (blue line).} 
  \end{subfigure}
  \begin{subfigure}{7cm}
    \centering\includegraphics[width=6cm]{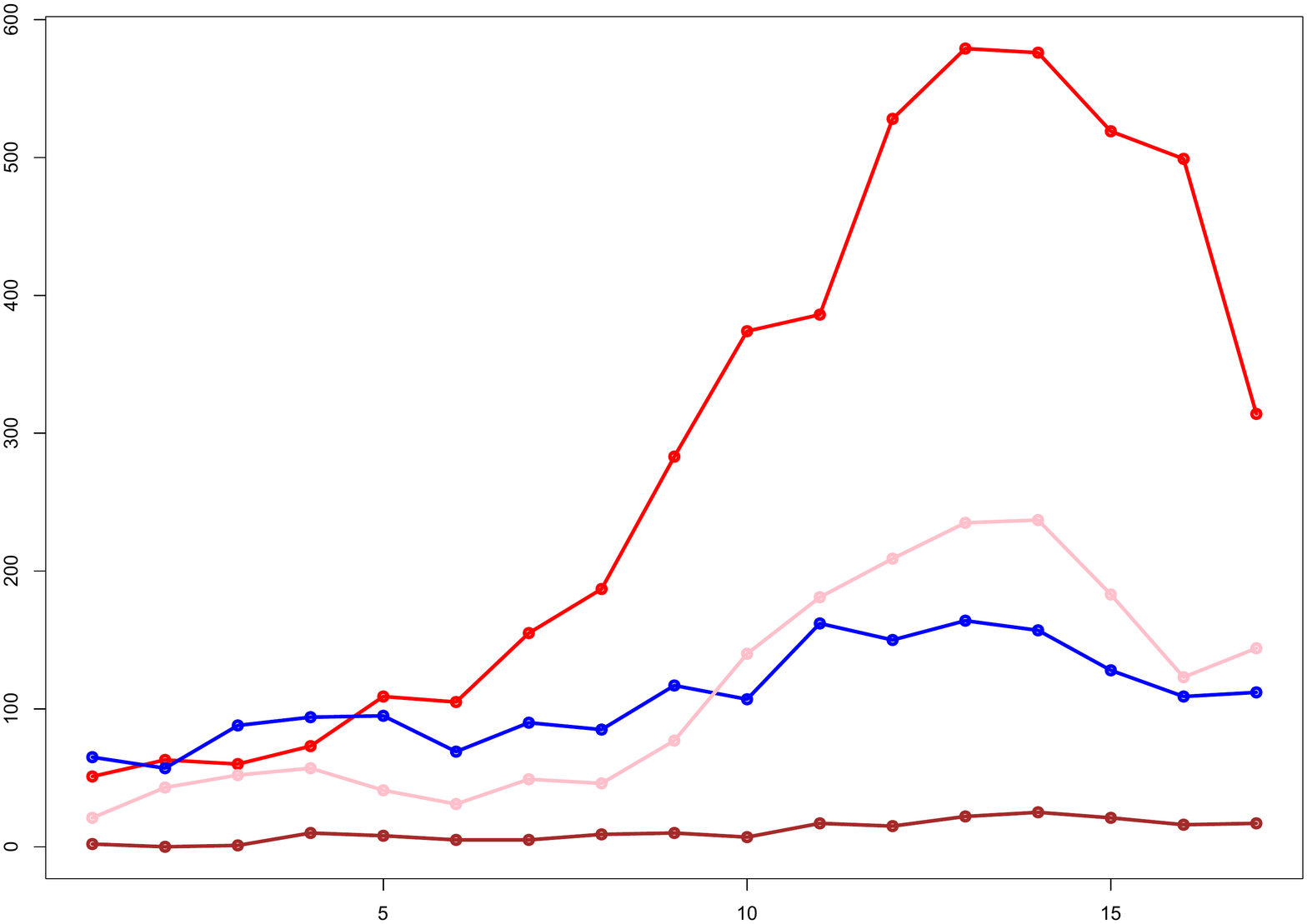}
    \caption{The weekly observed cases aged 0-29 (red line), 30-49 (blue line), 50-69 (brown line) and 70+ (pink line).} 
  \end{subfigure}
  \caption{The weekly observed cases.}
  \label{EstWOC_W}
\end{figure}

We illustrate the simulation study for 2 and 4 age groups:
\begin{itemize}
\item{\textbf{2 age groups:}  We generated weekly latent and observed cases according to the model equations (\ref{MdsAM1})-(\ref{MdsAM2}) for weeks -$2-17$ ($[21,161)$) given that the process is triggered by 1630 infectious, $56\%$ of the population is susceptible at the beginning of week -$2$ $(0-59:\ 126629,\ 60+:\  25777)$, $v=0.003$, $d=25.4$, $\beta=0.5$, $\gamma_{-2,0-59}=0.35$ and $\gamma_{-2,60+}=0.35$. The observed cases in weeks $1-17$ are 9890 $(0-59:\ 8157,\ 60+:\ 1733)$ (Figure \ref{EstWOC_W}). We ran the KDPF by assuming $\alpha=0$, $b=0.5$, $d_{min}=20$, $d_{max}=30$, $v_{min}=0.0001$, $v_{max}=0.5$ and 30000 particles.} 
\item{\textbf{4 age groups:}  We generated weekly latent and observed cases according to the model equations (\ref{MdsAM1})-(\ref{MdsAM2}) for weeks -$2-17$ ($[21,161)$) given that the process is triggered by 564 infectious, $56\%$ of the population is susceptible at the beginning of week -$2$ $(0-29:\ 72426,\ 30-49:\ 39018,\ 50-69:\ 29034,\ 70+:\  12994)$, $v=0.005$, $d=23.7$, $\beta=0.5$, $\gamma_{-2,0-29}=0.4$, $\gamma_{-2,30-49}=0.47$, $\gamma_{-2,50-69}=0.36$ and $\gamma_{-2,70+}=0.18$. The observed cases in weeks $1-17$ are 8769 $(0-29:\ 4861,\ 30-49:\ 1849,\ 50-69:\ 1869, 70+:\ 190)$ (Figure \ref{EstWOC_W}). We ran the KDPF by assuming $\alpha=0$, $b=1$, $d_{min}=20$, $d_{max}=30$, $v_{min}=0.0001$, $v_{max}=0.5$ and 40000 particles.    } 
\end{itemize} Both scenarios are consistent with Edmunds et al.~\cite{edmunds1997mixes}, i.e., older participants have more contacts with older people, and Wallinga et al.~\cite{wallinga2006using}, i.e., younger adults are more likely to catch an infection and contribute more to its transmission due to their high number of contacts.  Tables \ref{WIWA_4G}-\ref{WIWB_2G} demonstrate that the proposed algorithm approaches well the ground truth. 

\begin{table}[!ht]
\centering
\caption{ \bf The true number, the posterior median and the 99\% CI of the number of directed links between age groups ($a_1: 0-29$, $a_2:30-49$, $a_3:50-69$, $a_4:70+$).}
\begin{tabular}{|l|l|l|l| }
 \hline
 \multicolumn{4}{|l|}{\bf Estimating the number of directed links $a_i\rightarrow a_j $ using Method A}  \\
 \thickhline
  $a_i\rightarrow a_j$ & Posterior Median & $99\%$ CI & True Number\\
 \hline
  $a_1\rightarrow a_1$ & 8527 & (7635,\ 9580) &  8774 \\ \hline
 $a_1\rightarrow a_2$ & 2543 & (2271,\ 2831) &  2654  \\ \hline
  $a_1\rightarrow a_3$ &2289 & (2039,\ 2543) & 2515 \\ \hline
 $a_1\rightarrow a_4$ & 194 &  (146,\ 243) &  217 \\ \hline
 $a_2\rightarrow a_1$ &1071 &(946,\ 1207) &  1093 \\ \hline
 $a_2\rightarrow a_2$  & 686 & (573,\ 805) &  710  \\ \hline
  $a_2\rightarrow a_3$ &639 & (544,\ 735) & 662 \\ \hline
 $a_2\rightarrow a_4$ & 90 & (64,\ 119) &  106 \\ \hline
  $a_3\rightarrow a_1$& 563 & (481,\ 647) & 616 \\ \hline
 $a_3\rightarrow a_2$ &357 & (296,\ 424) & 392  \\ \hline
  $a_3\rightarrow a_3$ & 729 & (606,\ 854) & 789\\ \hline
 $a_3\rightarrow a_4$ & 95 &(67,\ 123) &  109 \\ \hline
   $a_4\rightarrow a_1$& 13 & (5,\ 23) & 16 \\ \hline
 $a_4\rightarrow a_2$ &  13& (5,\ 23) & 12  \\ \hline
  $a_4\rightarrow a_3$  & 24 &(11,\ 37) & 27 \\ \hline
 $a_4\rightarrow a_4$ & 11 & (4,\ 22) & 14 \\
 \hline
\end{tabular}
\label{WIWA_4G}
\end{table}

\begin{table}[!ht]
\centering
\caption{\bf The true number, the posterior median and the 99\% CI of the number of directed links between age groups ($a_1: 0-29$, $a_2:30-49$, $a_3:50-69$, $a_4:70+$).}
\begin{tabular}{ |l|l|l|l| }
 \hline
 \multicolumn{4}{|l|}{\bf Estimating the number of directed links $a_i\rightarrow a_j $ using Method B}  \\
 \thickhline
  $a_i\rightarrow a_j$ & Posterior Median & $99\%$ CI & True Number\\
 \hline
  $a_1\rightarrow a_1$ & 8438 & (7934,\ 9100) &  8774 \\ \hline
 $a_1\rightarrow a_2$ & 2539 & (2339,\ 2681) &  2654  \\ \hline
  $a_1\rightarrow a_3$ &2285 & (2122,\ 2445) & 2515 \\ \hline
 $a_1\rightarrow a_4$ & 189 &  (156,\ 252) &  217 \\ \hline
 $a_2\rightarrow a_1$ &1077 &(950,\ 1232) &  1093 \\ \hline
 $a_2\rightarrow a_2$  & 623 & (604,\ 787) &  710  \\ \hline
  $a_2\rightarrow a_3$ &632 & (552,\ 721) & 662 \\ \hline
 $a_2\rightarrow a_4$ & 91 & (69,\ 116) &  106 \\ \hline
  $a_3\rightarrow a_1$& 572 & (514,\ 628) & 616 \\ \hline
 $a_3\rightarrow a_2$ & 361 & (323,\ 397) & 392  \\ \hline
  $a_3\rightarrow a_3$ & 736 & (593,\ 875) & 789\\ \hline
 $a_3\rightarrow a_4$ & 94 &(77,\ 111) &  109 \\ \hline
   $a_4\rightarrow a_1$& 14 & (3,\ 20) & 16 \\ \hline
 $a_4\rightarrow a_2$ &  12& (5,\ 24) & 12  \\ \hline
  $a_4\rightarrow a_3$  & 25 &(14,\ 43) & 27 \\ \hline
 $a_4\rightarrow a_4$ & 12 & (4,\ 19) & 14 \\
 \hline
\end{tabular}

\label{WIWB_4G}
\end{table}

\begin{table}[!ht]
\centering
\caption{\bf The true number, the posterior median and the 99\% CI of the number of directed links between age groups ($a_1: 0-59$, $a_2:60+$).}
\begin{tabular}{ |l|l|l|l| }
 \hline
 \multicolumn{4}{|l|}{\bf Estimating the number of directed links $a_i\rightarrow a_j $ using Method A}  \\
 \thickhline
  $a_i\rightarrow a_j$ &  Posterior Median & $99\%$ CI & True Number\\
 \hline
  $a_1\rightarrow a_1$& 15219 & (14083,\ 16453)) &  15214 \\ \hline
 $a_1\rightarrow a_2$ & 2936 & (2638,\ 3205) &  3025  \\ \hline
  $a_2\rightarrow a_1$ & 297 & (243,\ 350) & 300 \\ \hline
 $a_2\rightarrow a_2$ & 389 & (317,\ 461) &  381 \\
 \hline
\end{tabular}
\label{WIWA_2G}
\end{table}

\begin{table}[!ht]
\centering
\caption{\bf The true number, the posterior median and the 99\% CI of the number of directed links between age groups ($a_1: 0-59$, $a_2:60+$).}
\begin{tabular}{ |l|l|l|l| }
 \hline
 \multicolumn{4}{|l|}{\bf Estimating the number of directed links $a_i\rightarrow a_j $ using Method B}  \\
 \thickhline
  $a_i\rightarrow a_j$ &  Posterior Median & $99\%$ CI & True Number\\
 \hline
  $a_1\rightarrow a_1$& 15214 & (14478,\ 16294)) &  15214 \\ \hline
 $a_1\rightarrow a_2$ & 2946 & (2697,\ 3076) &  3025  \\ \hline
  $a_2\rightarrow a_1$ & 298 & (250,\ 339) & 300 \\ \hline
 $a_2\rightarrow a_2$ & 391 & (330,\ 472) &  381 \\
 \hline
\end{tabular}
\label{WIWB_2G}
\end{table}

\section*{Discussion}
Isham and Medley~\cite{isham1996models}; Wallinga et al.~\cite{wallinga1999perspective}; Farrington at al.~\cite{farrington2001estimation} contend that it is necessary to account for individual heterogeneities while modelling the transmission of an infectious disease. The unstructured homogeneously mixing epidemic model (Model U) introduced by Lamprinakou et al.~\cite{https://doi.org/10.48550/arxiv.2208.07340} is a step toward developing epidemic models considering individual heterogeneities and revealing underlying dynamics. In this paper, working in this direction, we suggest a novel age-stratified epidemic model considering a finite population (Model A) and using a marked latent Hawes process for the infections. We propose a KDPF to infer the marked counting process and forecast the epidemic's future trajectory over short time horizons. We demonstrate the performance of the proposed algorithm on COVID-19.

The simulation analysis of synthetic data shows that the KDPF approaches well the ground truth. We demonstrate that the estimated latent cases and the latent intensity are consistent with the aggregated observed cases for each age group in various local authorities in the UK. The analysis also reveals that each age group's instantaneous reproduction number reflects the pandemic's progression and the group's evolving behavioural characteristics. The uncertainty of estimates increases in the last states of the algorithm, as the reported infections carry information about the progress of the epidemic by transferring the delay between the reported and the actual infection time. We also show how the algorithm can be employed to project the epidemic's course in the near future. However, our forecast is subject to similar levels of uncertainty as the last state's estimates. 

According to Cori et al.~\cite{cori2013new}, the size of the time window will impact the estimations of the instantaneous reproduction number. Small sizes lead to faster detection of transmission changes and higher statistical noise, whereas large sizes lead to more smoothing and reductions in statistical noise. In accordance with Cori et al.~\cite{cori2013new}, who suggest an appropriate way of choosing the time window size, we have selected a weekly time window to analyse the real data. 

Model A includes model U, and according to \cite{pellis2020systematic}, model A reflects better the spreading of the epidemic. The estimated aggregated and per age group latent intensity, weekly and daily hidden cases via posterior median given by model A are similar to the ones of model U. The instantaneous reproduction number's posterior medians of both models follow the same pattern in general lines. The analysis shows that model A derives narrower CIs, indicating that considering the individual inhomogeneity in age and finite population decreases the uncertainty of estimates. The instantaneous reproduction numbers for each age group offer a real-time gauge for interventions and behavioural changes. The inapplicability of model U to determine the reproduction number per age group and give insight into each group's behaviour also demonstrates the necessity of model A. We note that accounting for each person's age group does not increase the algorithm's complexity. 

Particle filters can have an exponential cost in the dimensionality of the hidden state to be stable~\cite{beskos2017stable}. Increasing the number of age groups raises the dimensionality of the hidden state and, by extension, the computational resources to implement the proposed algorithm. An alternative approach to remedy this issue would be to consider the process $\gamma(t, a)$ independent of the age group $a$. However, that approach decreases the flexibility of the model on the real data.

When modelling the epidemic using a marked Hawkes process, it is easy to consider more individual inhomogeneities, such as community structure (e.g. household) and location, immunization status against the disease, and medical conditions. This type of modelling reveals the dynamics of the epidemic, including who infected whom, which is challenging because of invisible transmission pathways \citep{yang2013mixture,kim2019modeling}. Future work also considers the inference of ascertainment rate, using various transition kernels for modelling the latent and the reported infection cases, as well as more sophisticated ways for initializing the set of infectious triggering the epidemic process,$\mathcal{H}_{0a}$ and the number of susceptibles at the beginning of the process, $S_{T_0a}$, for each age group $a \in \mathcal{A}$.

\section*{Supporting information}
\paragraph*{S1: 4 age groups} 
\label{Appendix4G}

ONS shows that $47.19\%$ of the population is aged under 30 years (0-29), $25.42\%$ aged 30 to 49 years (30-49), $18.92\%$ aged 50 to 69 years (50-69), and $8.47\%$ aged 70 years and over (70+). We coarse the age groups of the contact matrix for reopening schools \citep{jarvis2020quantifying} and get the matrix: \begin{equation*}
m=\begin{bmatrix}
5.82 & 1.93 & 1.05 & 0.24\\
2.75 & 1.60 & 1.00 & 0.36 \\
1.54 & 1.09 & 1.27 & 0.43 \\
0.60 & 0.71 & 0.76 & 0.87
\end{bmatrix}. \end{equation*}
The process is triggered by 4963 infectious. The times of their infections, $\mathcal{H}_0$ , are uniformly allocated in 21 days ($[0,21)$). We generate weekly latent and observed cases according to the model equations (\ref{MdsAM1})-(\ref{MdsAM2}) for weeks -$2-17$ ($[21,161)$) given $\mathcal{H}_0$, $v=0.004$, $d=25.57$, $\beta=0.5$, $\gamma_{-2,0-29}=0.49$, $\gamma_{-2,30-49}=0.47$, $\gamma_{-2,50-69}=0.10$ and $\gamma_{-2,70+}=0.38$. We consider that about 36.46\% of the population is susceptible at the beginning of week -2 ; 129073 susceptibles $(0-29: 60893,\ 30-49:\  32837,\ 50-69:\ 24414,\ 70+: 10929)$. 

We are interested in inferring the latent cases in weeks $1-16$ with $\mathcal{H}_0$ being the set of times of latent infections in weeks -$2-0$. We assume $\alpha=0$, $b=0.5$, $d_{min}=20$, $d_{max}=30$, $v_{min}=0.0001$ and $v_{max}=0.5$. Using the generated observed cases in weeks -$1-1$ as described above, we estimate the latent infections with their associated age groups in weeks -$2-0$ as follows: The latent cases of age group $a_v$ on the week $i$ is equal to the number of events in age $a_v$ occurred on the week $(i + 1)$ times $1/\beta$, and are spread uniformly in $[(i+2)\times7 + 21, (i+3)\times7 +21)$ for $-2\leq i \leq 0$.

\begin{figure}[!h] 
  \begin{subfigure}{7cm}
    \centering\includegraphics[width=6cm]{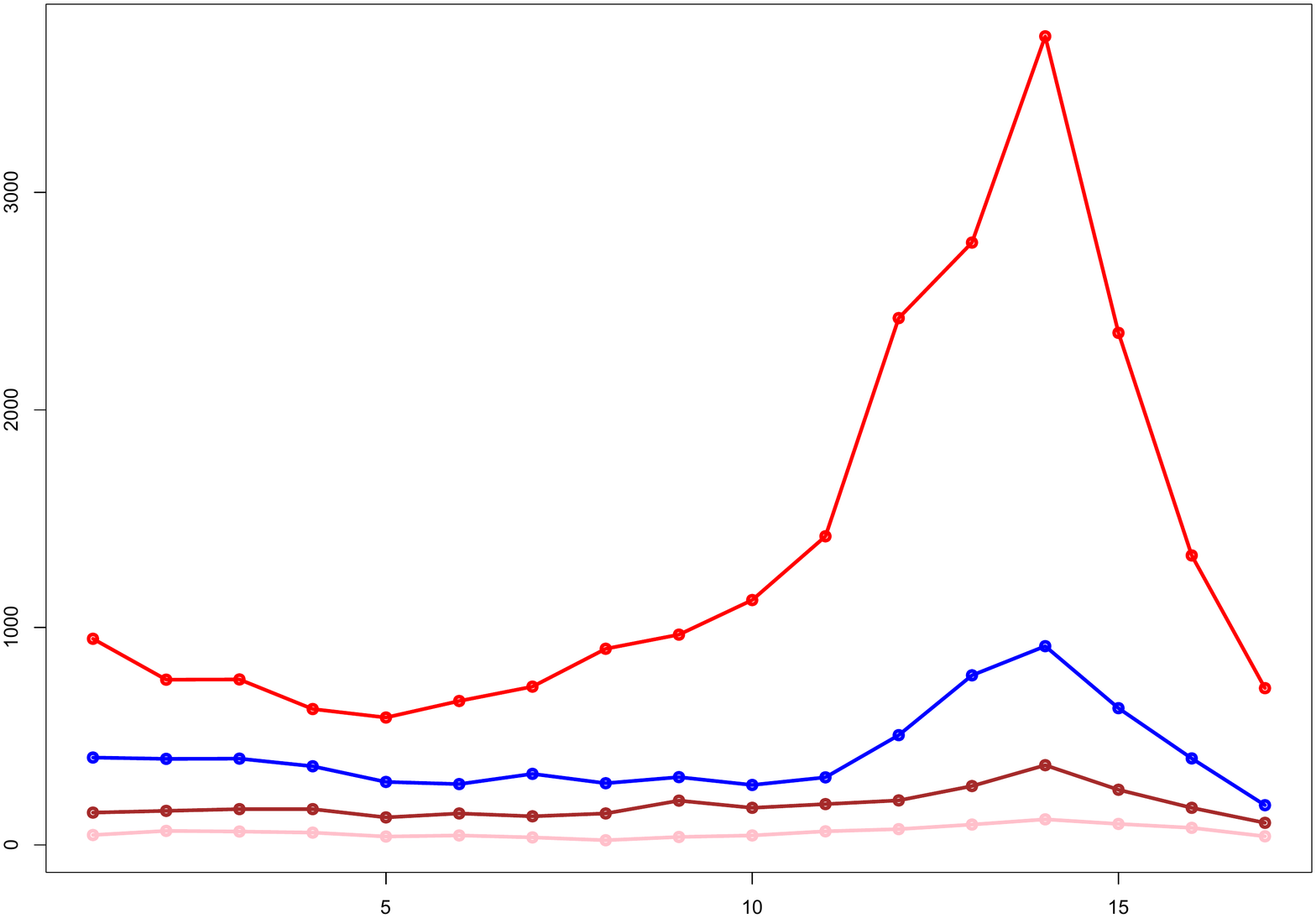}
    \caption{The weekly observed cases aged 0-29 (red line), 30-49 (blue line), 50-69 (brown line) and 70+ (pink line).} 
  \end{subfigure}
  \begin{subfigure}{7cm}
    \centering\includegraphics[width=6cm]{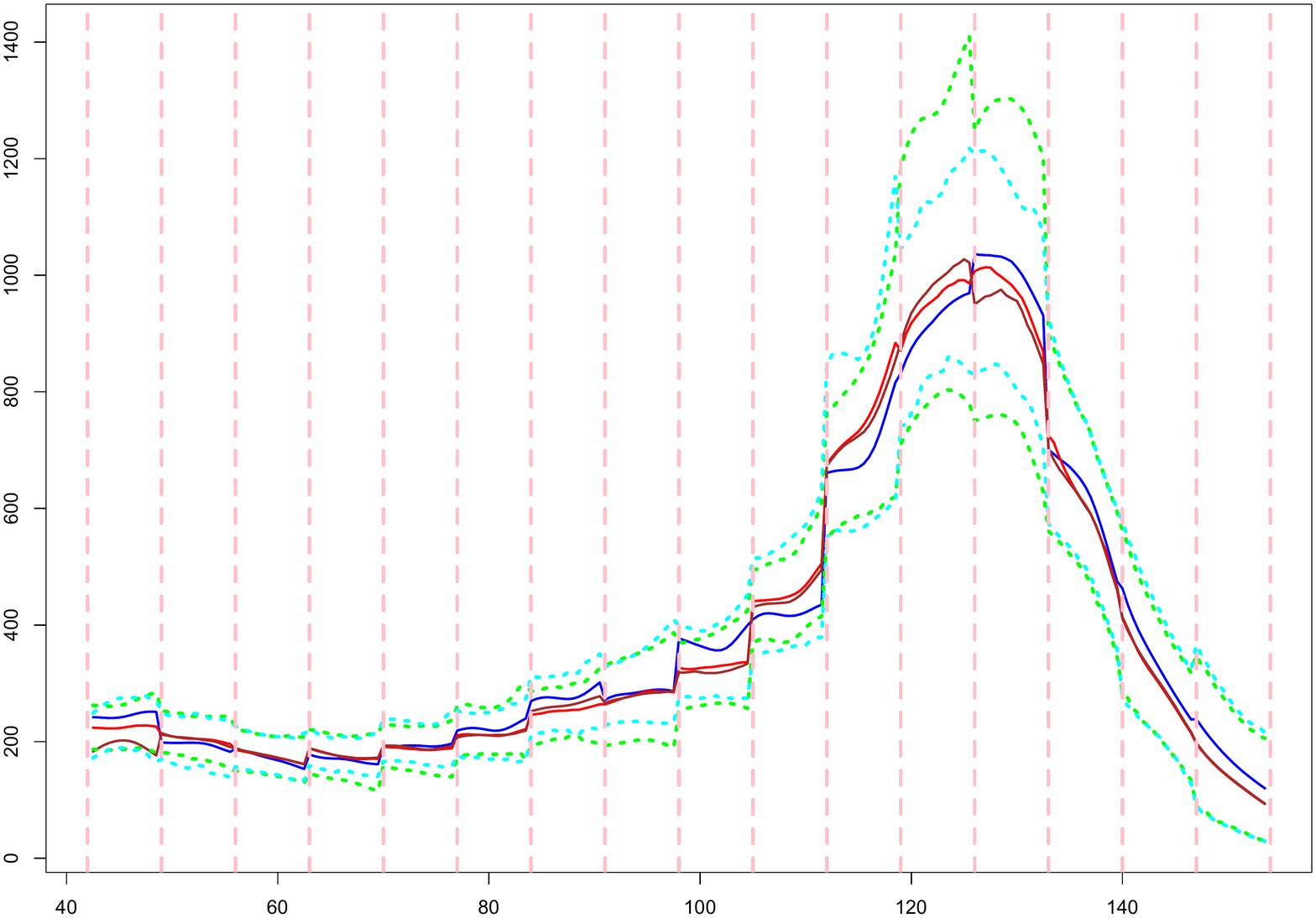}
   \caption{The estimated intensity of latent cases aged 0-29.}
  \end{subfigure}
  \begin{subfigure}{7cm}
    \centering\includegraphics[width=6cm]{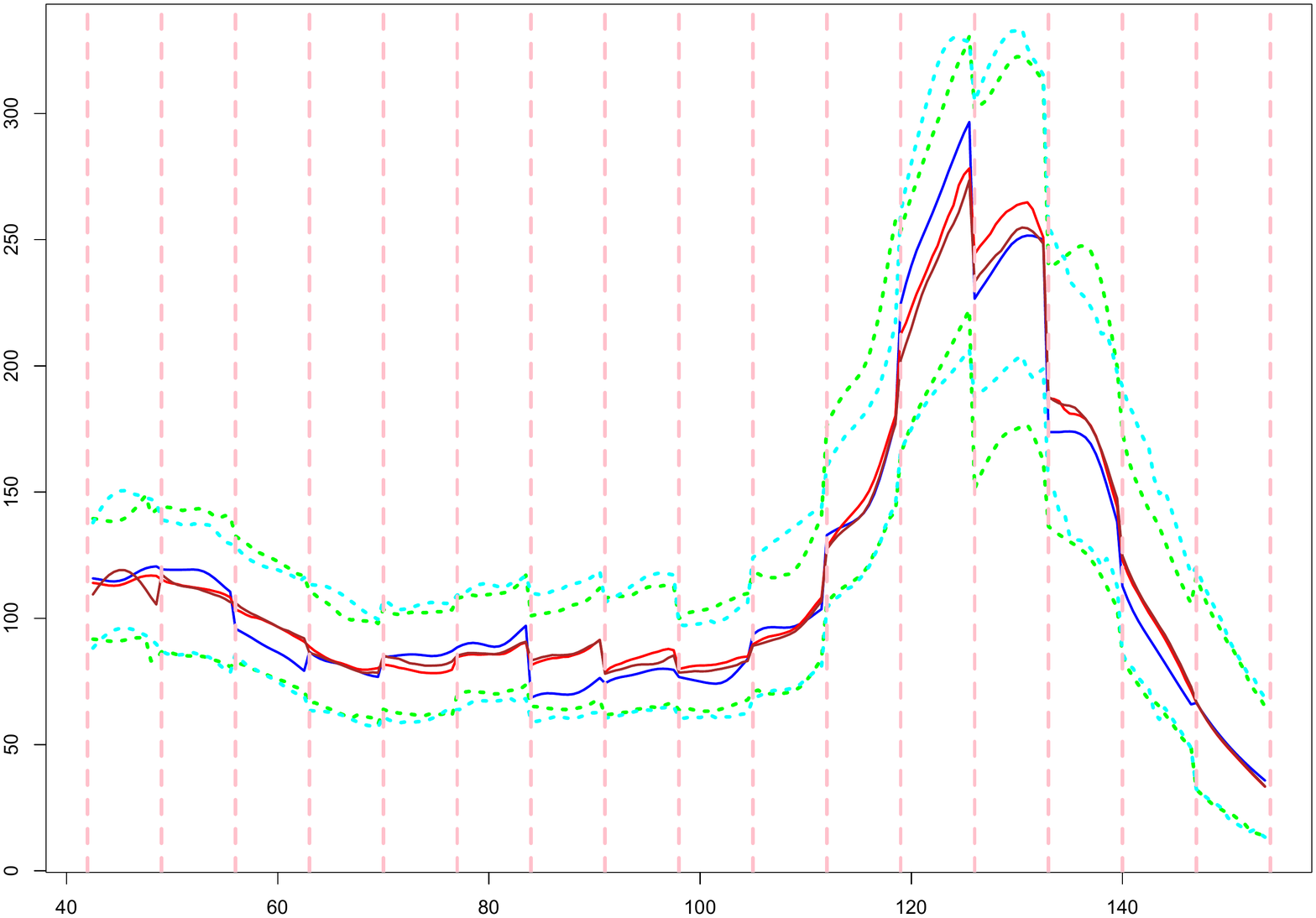}
   \caption{The estimated intensity of latent cases aged 30-49.}
  \end{subfigure}
  \begin{subfigure}{7cm}
    \centering\includegraphics[width=6cm]{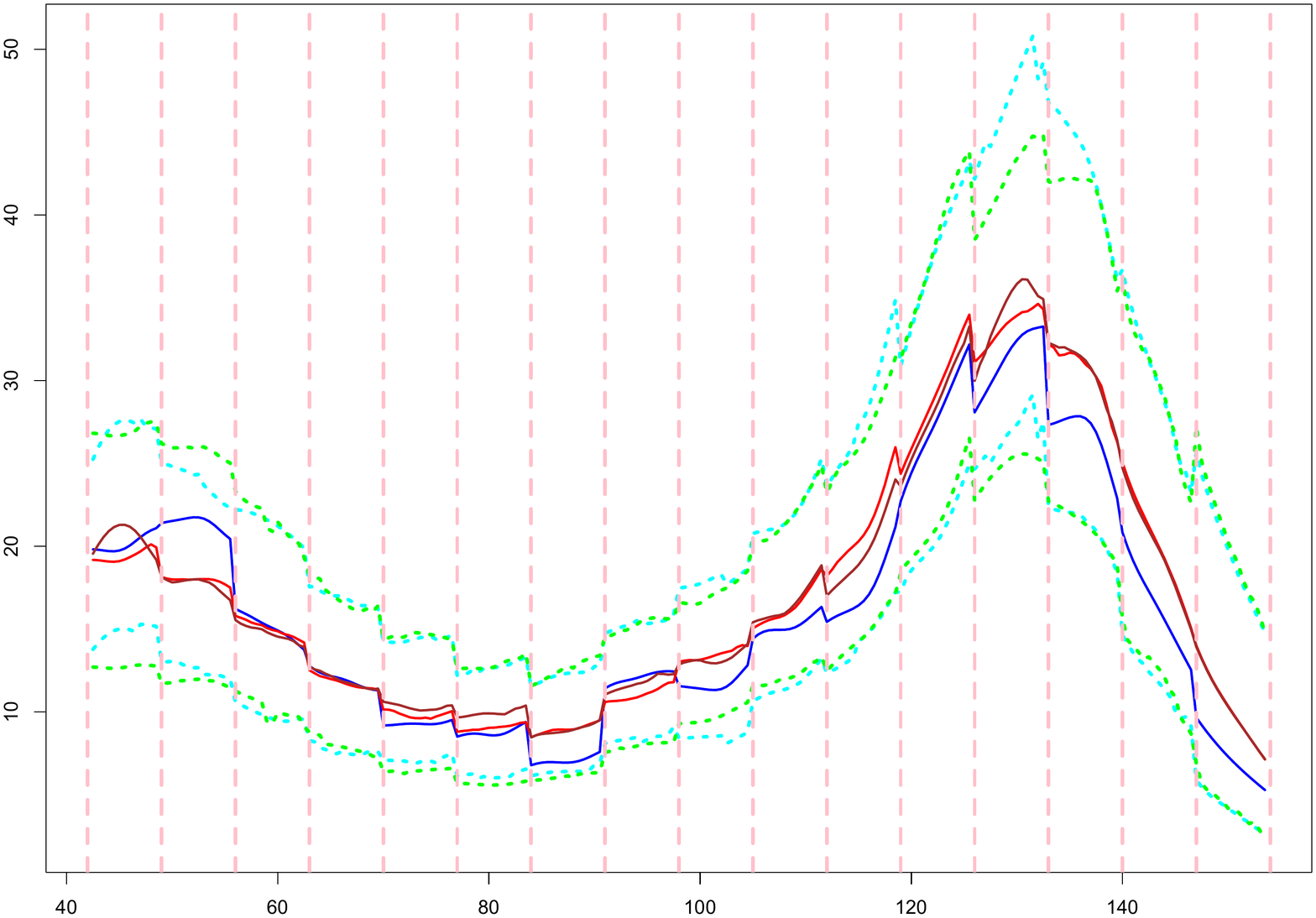}
   \caption{The estimated intensity of latent cases aged 50-69.}
  \end{subfigure}
  \begin{subfigure}{7cm}
    \centering\includegraphics[width=6cm]{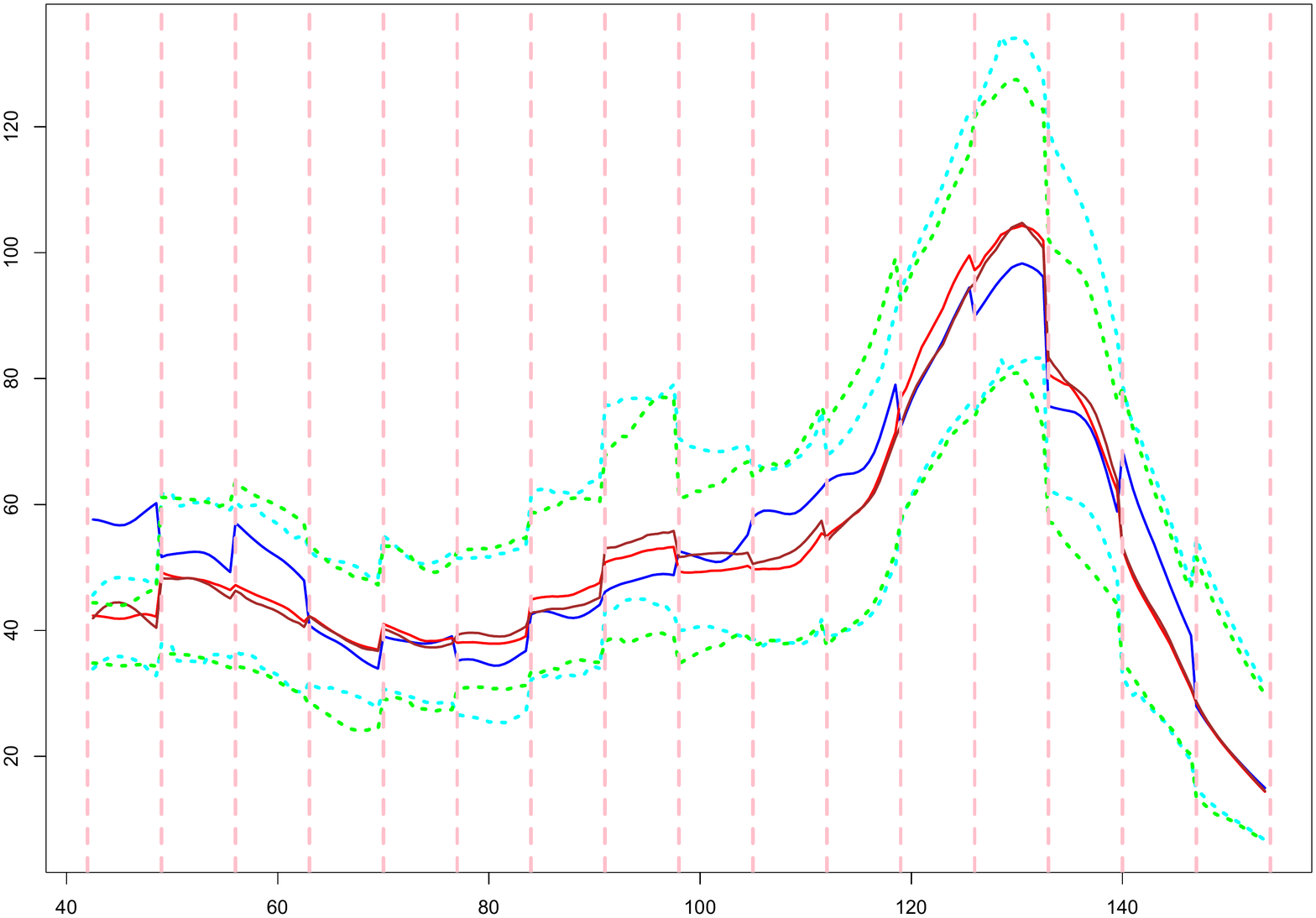}
   \caption{The estimated intensity of latent cases aged 70+.}
  \end{subfigure}
  \begin{subfigure}{7cm}
    \centering\includegraphics[width=6cm]{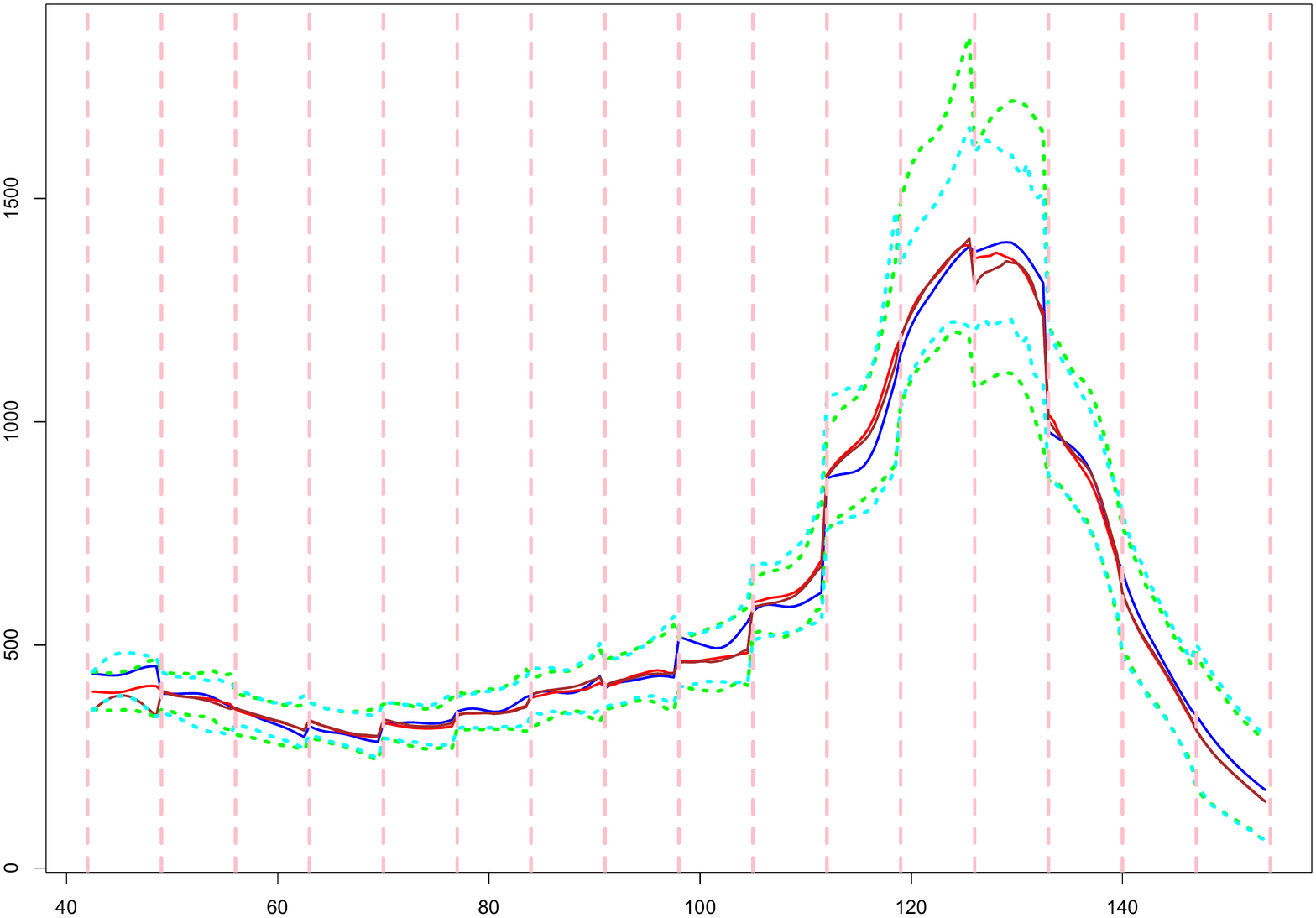}
   \caption{The aggregated estimated intensity of latent cases.}
  \end{subfigure}
  \caption{\bf{The weekly observed cases, the true latent intensities (blue line), and the estimated latent intensities considering 4 age groups (with estimated seeds (posterior median (brown line) ; 99\% CI (cyan line)), and true seeds (posterior median (red line) ; 99\% CI (green line))).}}
  \label{EstInt_AG4}
\end{figure}

The ground truth is characterized by $\mathcal{H}_0$ consisting of 7987 infections $(0-29:\ 4539,\ 30-49:\ 2309,\ 50-69:\ 281,\ 70+:\  768 )$, and 121176 susceptibles $(0-29:\ 56354,\ 30-49:\ 30528,\ 50-69:\ 24133,\ 70+:\ 10161)$ at the beginning of week 1. The estimated seeds and susceptibles are 7754 $(0-29:\ 4404,\ 40-49:\ 2254,\ 50-69:\ 392,\ 70+:\ 704)$, and 121319 $(0-29:\ 56489,\ 30-49:\ 30583,\ 50-69:\  24022,\ 70+:\ 10225)$, respectively. The observed cases in weeks $1-17$ are 33977 $(0-29:\ 22978,\ 30-49:\ 7046,\ 50-69:\ 1015,\ 70+:\ 3118)$ (Figure \ref{EstInt_AG4}).  

\begin{figure}[!h] 
  \begin{subfigure}{7cm}
    \centering\includegraphics[width=6cm]{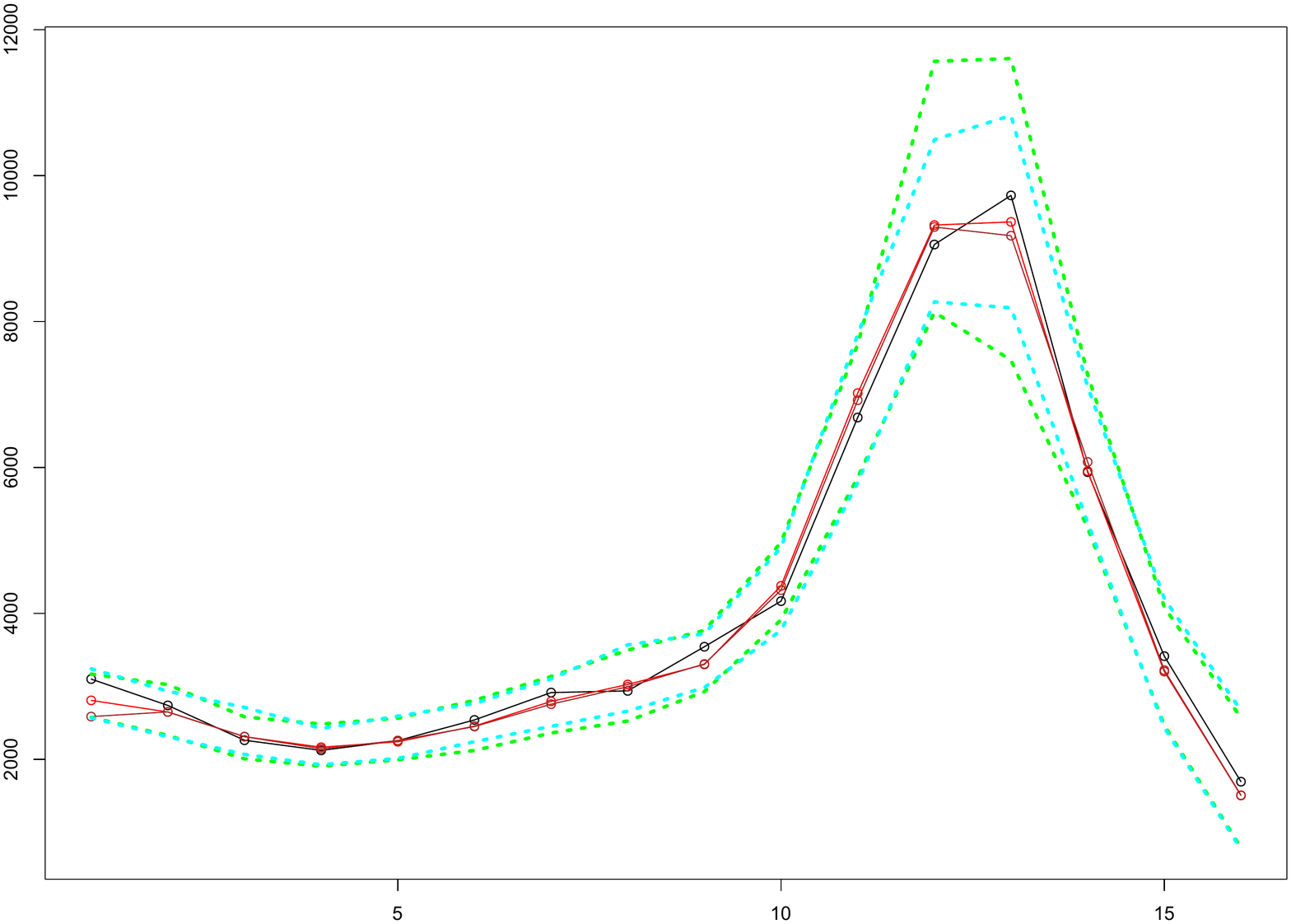}
   \caption{The aggregated estimated weekly latent cases.}
  \end{subfigure}
  \begin{subfigure}{7cm}
    \centering\includegraphics[width=6cm]{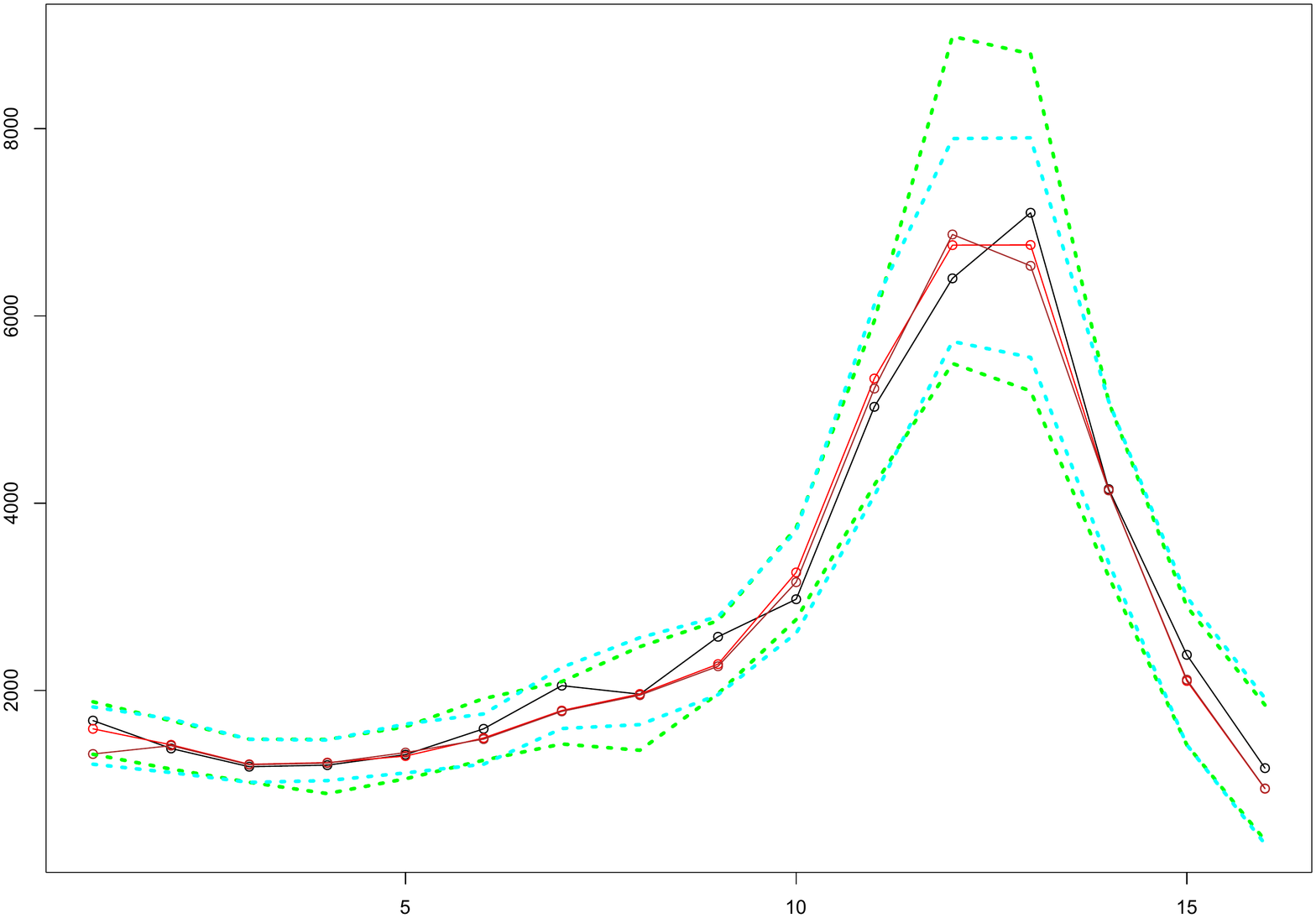}
    \caption{The estimated weekly latent cases aged 0-29.}
  \end{subfigure}
  \begin{subfigure}{7cm}
    \centering\includegraphics[width=6cm]{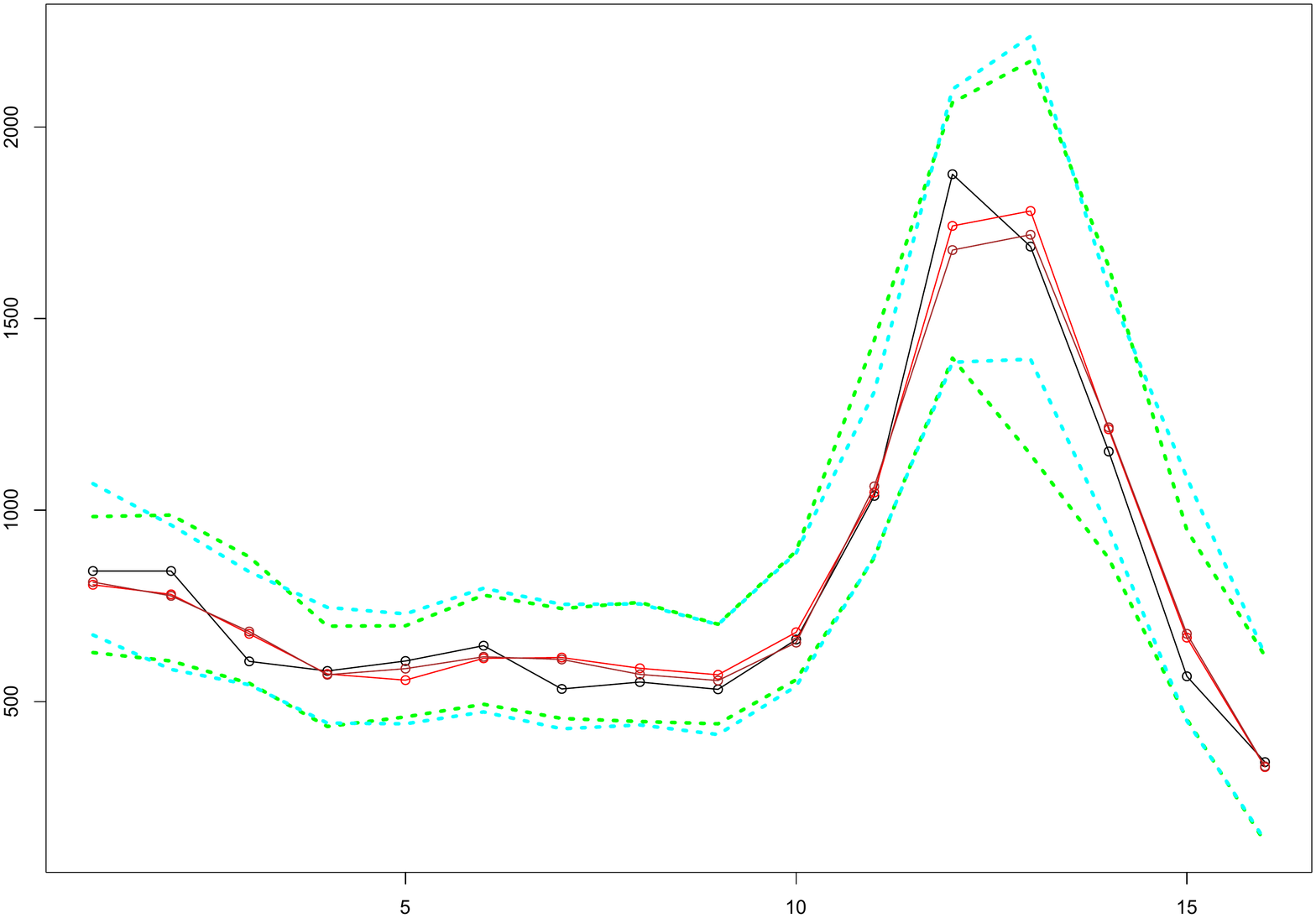}
    \caption{The estimated weekly latent cases aged 30-49.}
  \end{subfigure}
   \begin{subfigure}{7cm}
    \centering\includegraphics[width=6cm]{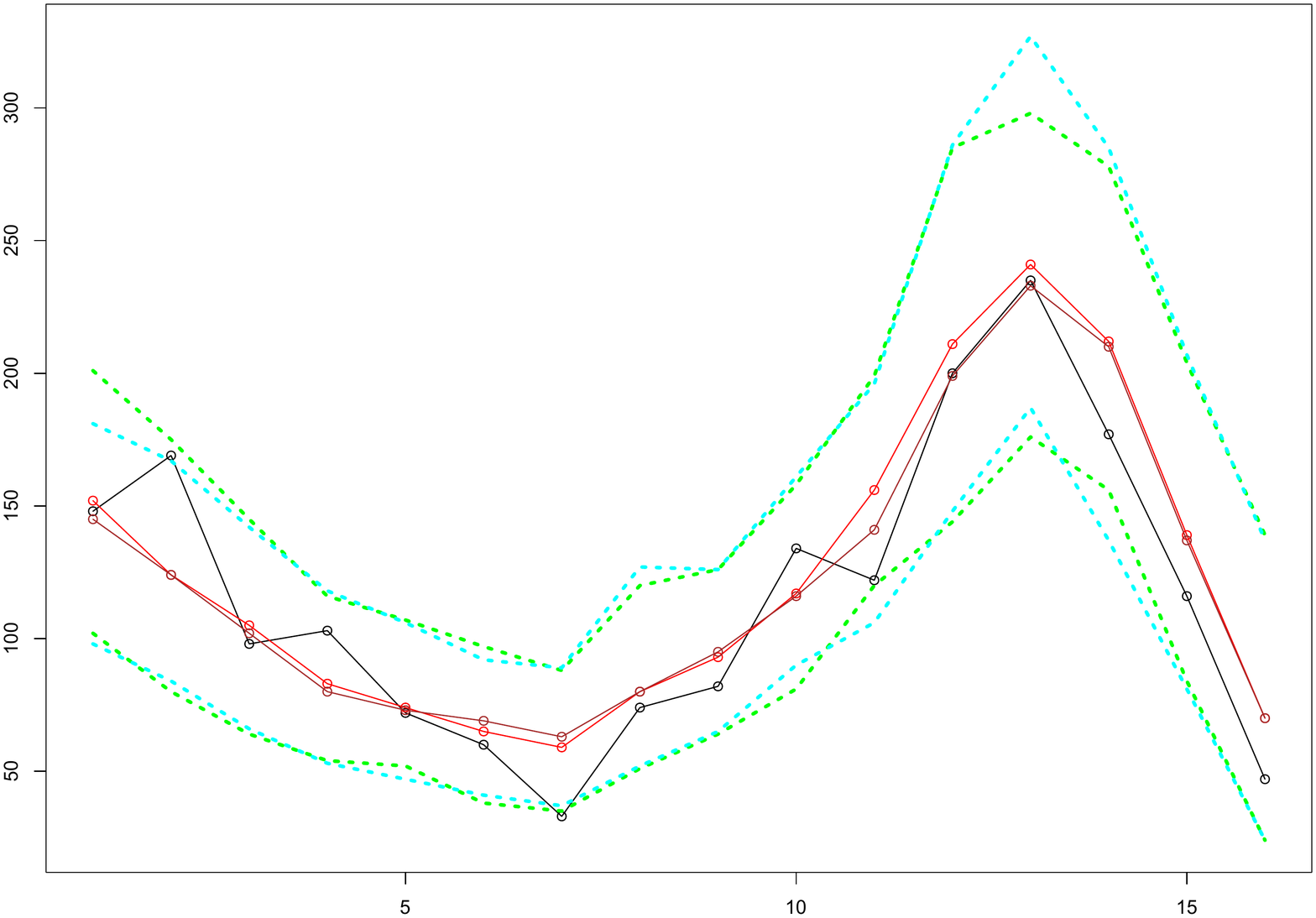}
    \caption{The estimated weekly latent cases aged 50-69.}
  \end{subfigure}
  \begin{subfigure}{7cm}
    \centering\includegraphics[width=6cm]{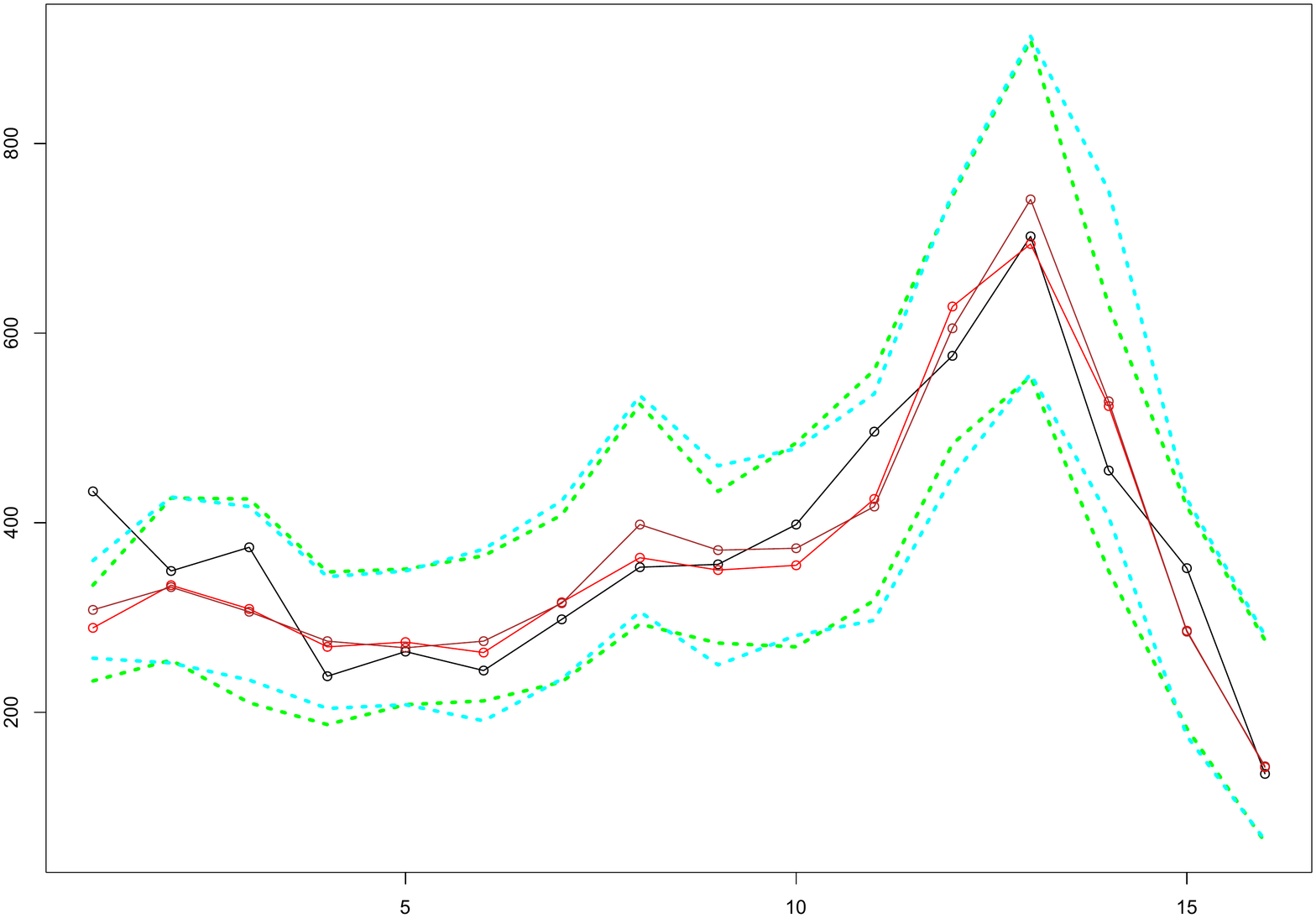}
    \caption{The estimated weekly latent cases aged 70+.}
  \end{subfigure}
   \caption{\bf{The estimated weekly latent cases (with estimated seeds (posterior median (brown line); 99\% CI (cyan line)) and  true seeds (posterior median (red line); 99\% CI (green line)))  and the true weekly hidden cases (black line) considering 4 age groups.}}
   \label{EHC_AG4}
\end{figure}

\begin{figure}[!h] 
  \begin{subfigure}{7cm}
    \centering\includegraphics[width=6cm]{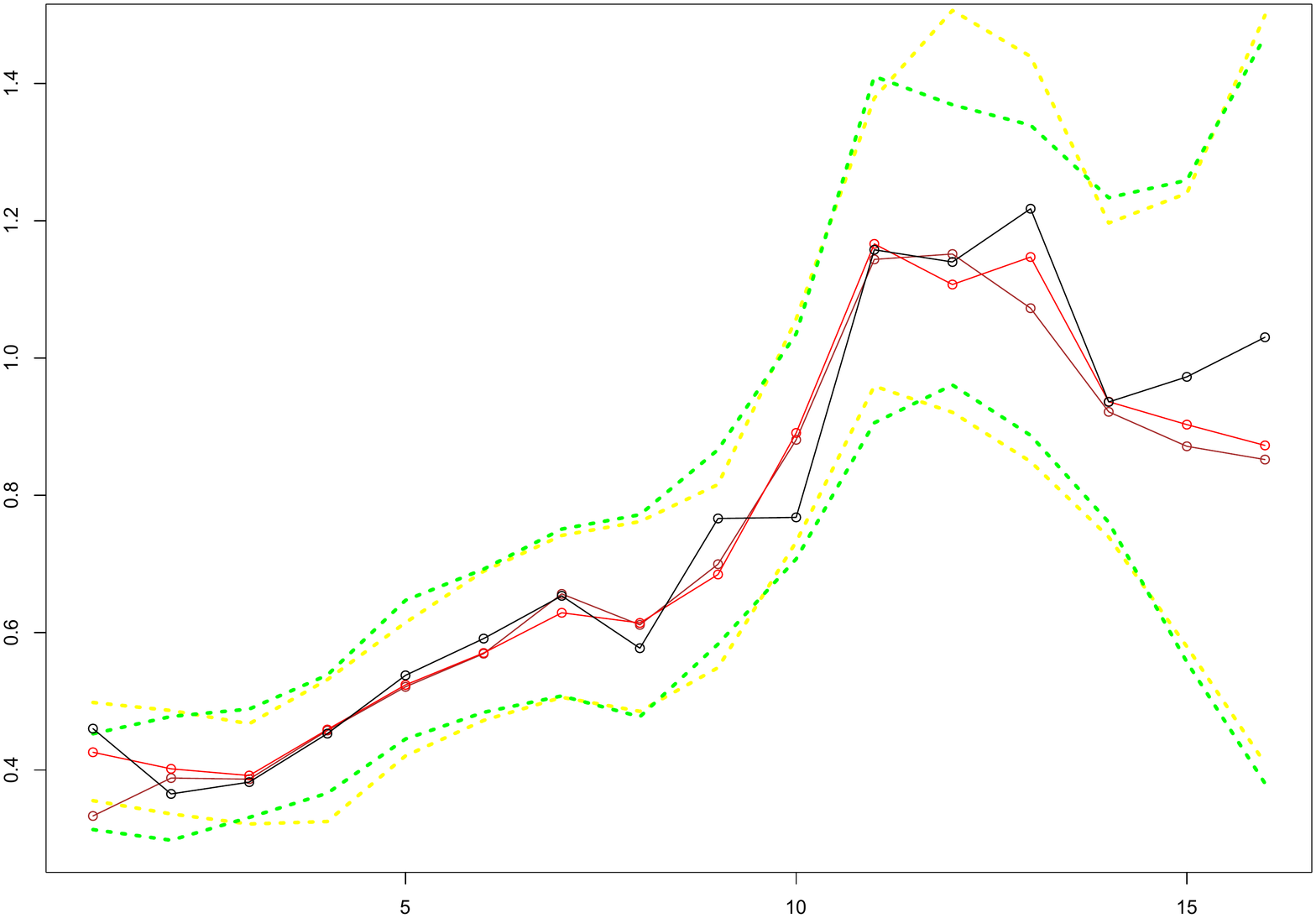}
   \caption{The estimated weights $\{\gamma_{i,0-29}\}_{i=1}^{16}$.}
  \end{subfigure}
  \begin{subfigure}{7cm}
    \centering\includegraphics[width=6cm]{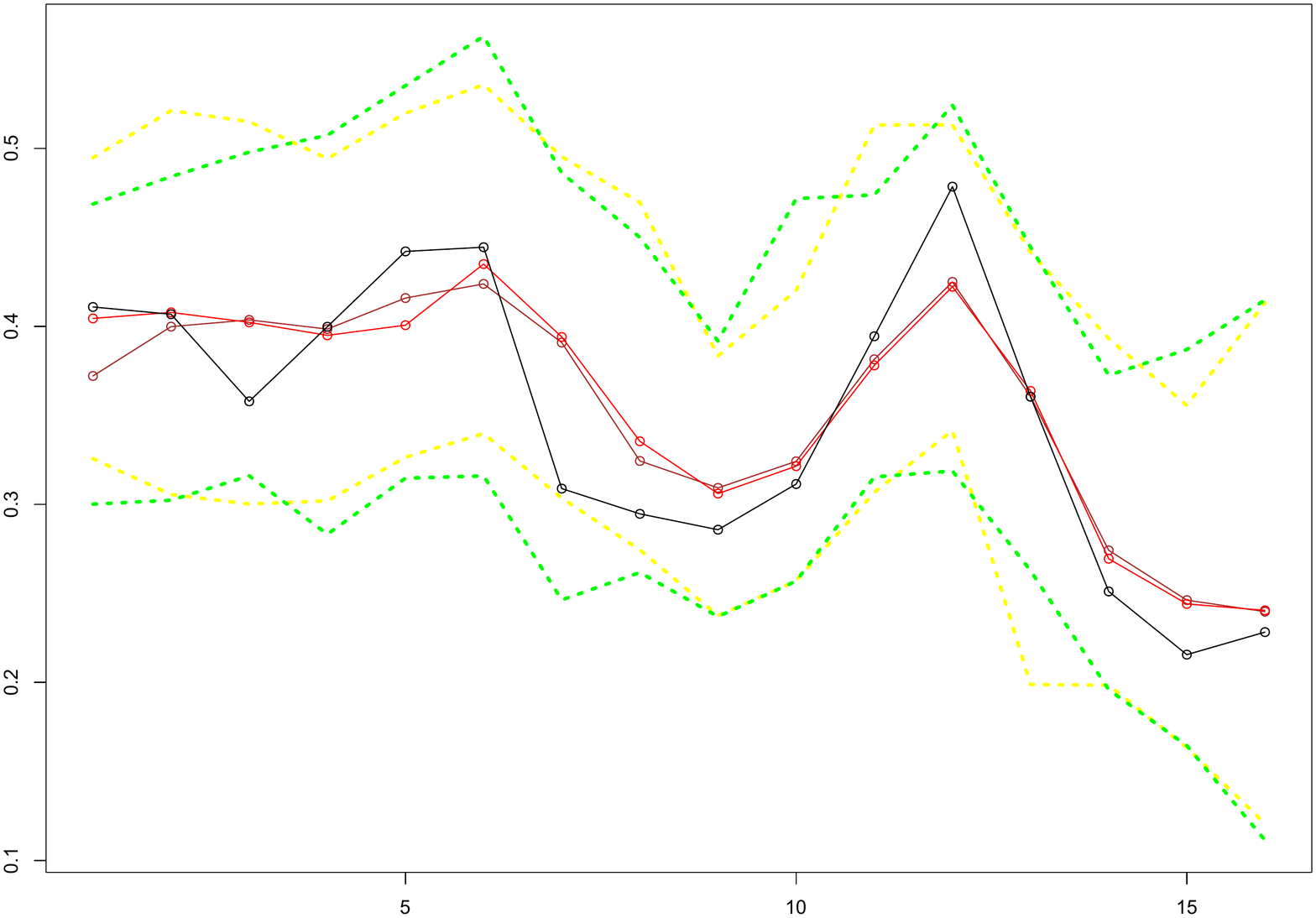}
     \caption{The estimated weights $\{\gamma_{i,30-49}\}_{i=1}^{16}$.}
  \end{subfigure}
  \begin{subfigure}{7cm}
    \centering\includegraphics[width=6cm]{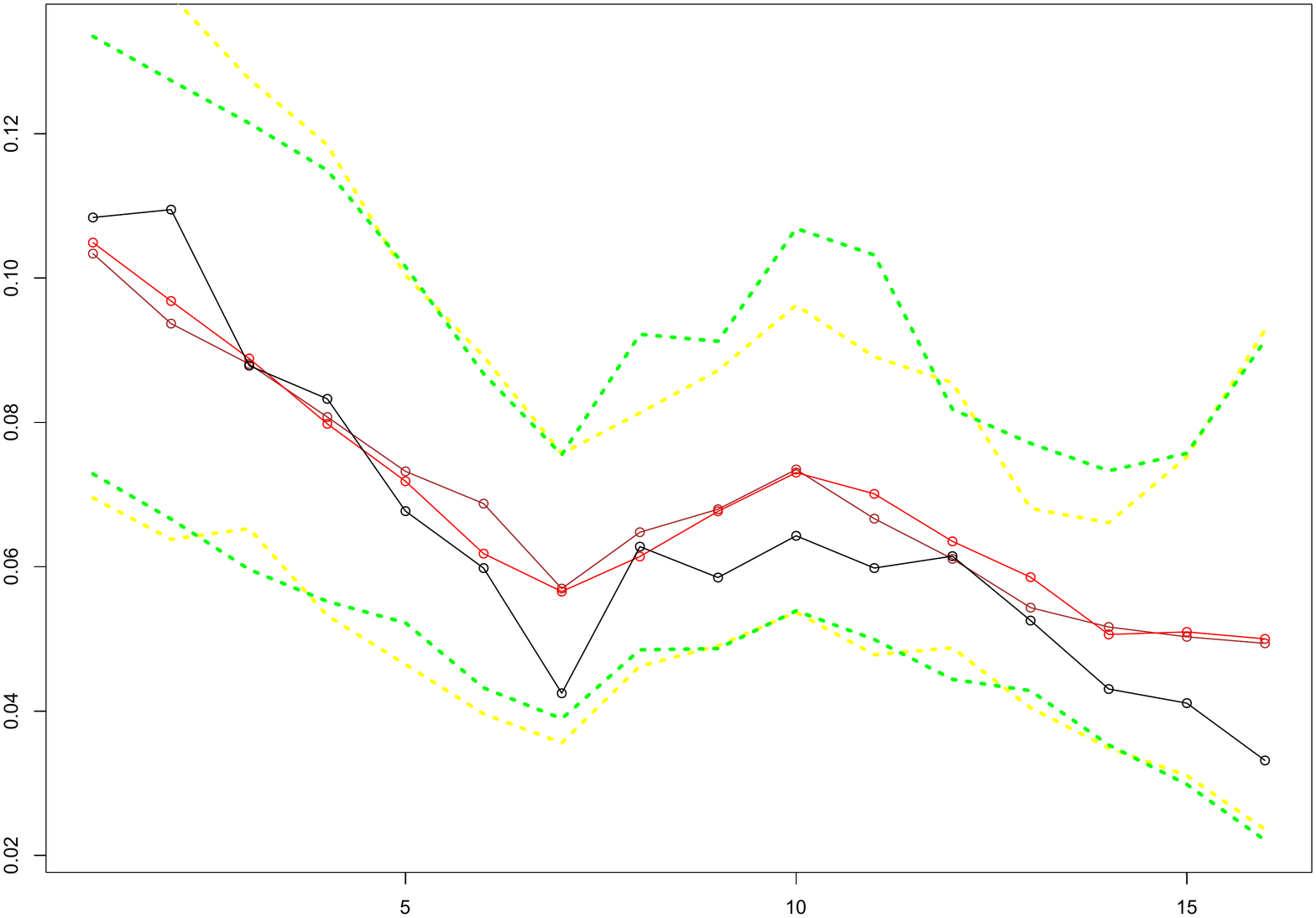}
    \caption{The estimated weights $\{\gamma_{i,50-69}\}_{i=1}^{16}$.}
  \end{subfigure}
   \begin{subfigure}{7cm}
    \centering\includegraphics[width=6cm]{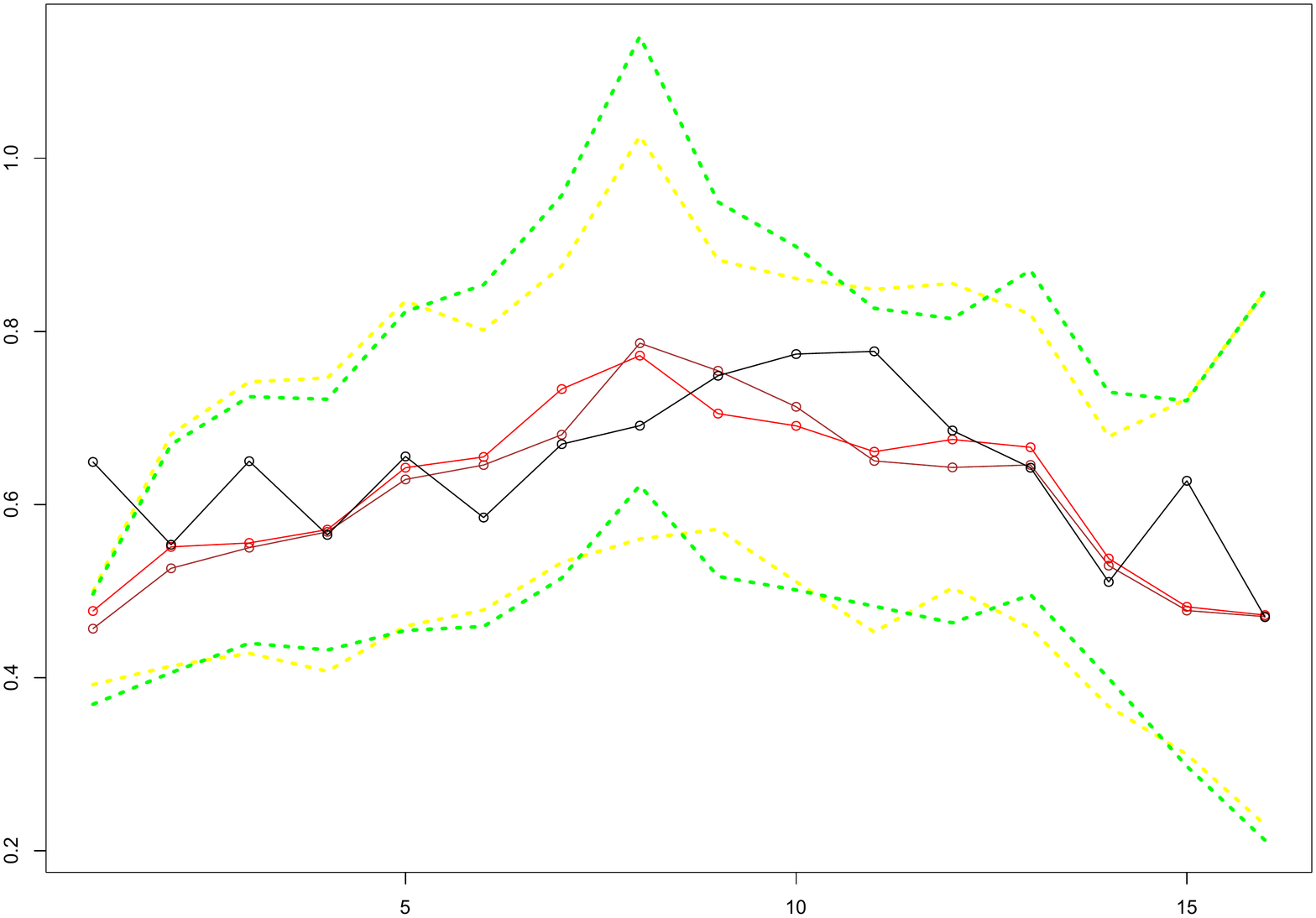}
     \caption{The estimated weights $\{\gamma_{i,70+}\}_{i=1}^{16}$.}
  \end{subfigure}

   \caption{\bf{The estimated weights $\{\gamma_{na}\}_a$ (with estimated seeds (posterior median (brown line); 99\% CI (green line)) and true seeds (red line; 99\% CI (yellow line))) and the true values (black line) considering 4 groups.}}
   \label{ER_AG4}
\end{figure}

 The figures \ref{EstInt_AG4}-\ref{ER_AG4} show the estimated intensities, the estimated weekly hidden cases and the estimated weights $\{\{\gamma_{na}\}_{n=1}^{16}\}_a$  for 30000 particles. We observe that the $99\%$ CIs do not cover the ground truth of the latent intensity in the age group 70+ during week 1 ($[42, 49)$), the weight associated with week 1 in the age group $70+$ ($\gamma_{1,70+}$) and the latent cases aged $70+$ in week 1. This might be observed because we have considered $\gamma_{1,70+}\sim \mbox{Uniform}(0,0.5)$ while the ground truth is higher than 0.5. Table \ref{tableM1} confirms the convergence of posterior estimates of weights and weekly hidden cases per age group concerning the number of particles. We note that the 99\% CIs of the time-constant parameters include the actual values of the parameters.
 
\begin{table}[!ht]
\centering
\caption{\bf{MCSEs of posterior means of weights $\{\gamma_a\}_a$ and  weekly hidden cases considering 4 groups.}}
\begin{tabular}{|l|l|l|l|l| }
 \hline
 \multicolumn{5}{|l|}{\bf{Convergence of the posterior estimates}}  \\
 \thickhline
  $MCSE$ & $N=10000$ & $N=20000$ & $N=30000$ & $N=40000$\\
 \hline
 $\gamma_1$& 0.000741 & 0.000565 & 0.000438 &  0.000428 \\ \hline
 $\gamma_2$ & 0.00043 & 0.0003 & 0.000243 &  0.000208  \\ \hline
 $\gamma_3$ & 0.000101 & 0.000077 & 0.00006 & 0.000053 \\ \hline
 $\gamma_4$ & 0.000811 & 0.00054 & 0.000444 &  0.000381 \\ \hline
 $Y_1$ & 2.4775 & 1.83933 & 1.373194 &  1.415707  \\ \hline
 $Y_2$ & 2.423559 & 2.119945 & 2.007936& 1.897117  \\ \hline
 $Y_3$ & 0.181335 & 0.127363 &  0.102718&0.094132 \\ \hline
 $Y_4$ & 0.462724 & 0.302392 & 0.260452 &0.213064 \\
 \hline
\end{tabular}

\label{tableM1}
\end{table}

\paragraph*{S2: Local Authorities} \label{LA} We illustrate the estimated intensity, the estimated weekly and daily hidden cases, the estimated susceptibles, the estimated instantaneous reproduction number and the estimated weights $\{\gamma_{na}\}_{n=1}^{16}$ per age group $a$ using 40000 particles in Ashford and Kingston upon Thames. We also figure out the $99\%$ CIs of time-constant parameters.

\begin{table}[H]
\begin{adjustwidth}{-1.5in}{0in} 
\centering
\caption{ \bf {The true number of reported infections in $\mathcal{T}_{16}$ and $\mathcal{T}_{17}$, and the posterior median, the posterior mean and the $95\%$ CIs of the estimated infections in $\mathcal{T}_{17}$ in Ashford.}} 
\begin{tabular}{ |l|l|l|l|l|l|} \hline
 \multicolumn{6}{|l|}{\bf Proposed Method} \\
 \thickhline
 Reported infections & Posterior Mean & Posterior Median  &  $95\%$ CIs & True Number ($\mathcal{T}_{17}$) &  True Number ($\mathcal{T}_{16}$) \\
 \hline
aggregated & 327 & 318 & (195, 469) & 408 & \\ \hline
aged 0-29 & 67 & 62 & (22, 118) & 85 & 133 \\ \hline
aged 30-49 & 105 & 98 & (40, 178) & 107 & 197 \\ \hline
aged 50-69 & 98 & 91& (34, 172) & 134  & 180 \\ \hline
aged 70+ &57 & 54 & (20, 99) & 82  & 96\\ 
 \hline
\end{tabular}
\label{AMtableAshford}
\end{adjustwidth}
\end{table}

\begin{table}[!h]
\begin{adjustwidth}{-1.7in}{0in} 
\centering
\caption{\bf The true number of reported infections in $\mathcal{T}_{16}$ and $\mathcal{T}_{17}$, and the posterior median, the posterior mean and the 95\% CIs of the estimated infections in $\mathcal{T}_{17}$ in Leicester. }
\begin{tabular}{ |l|l|l|l|l|l|} \hline
 \multicolumn{6}{|l|}{\bf Proposed Method} \\
 \thickhline
 Reported infections & Posterior Mean & Posterior Median  &  $95\%$ CIs & True Number ($\mathcal{T}_{17}$) & True Number ($\mathcal{T}_{16}$)\\ 
 \hline
aggregated & 4969 & 4793 & (2577, 7763) & 5794 & \\ \hline
aged 0-29 & 2275 & 2090 & (681, 4325) & 2367  & 1844\\ \hline
aged 30-49 & 1659 & 1557 & (554, 3013) & 2054  & 1244\\ \hline
aged 50-69 & 881 & 819& (246, 1653) & 1112  & 528\\ \hline
aged 70+ & 154 & 142 & (38, 290) &   261 & 87\\
 \hline
\end{tabular}
\label{AMtableLeicester}
\end{adjustwidth}
\end{table}

\section*{Acknowledgments}
We thank Prof Young for their constructive comments on this paper.

\nolinenumbers

% Either type in your references using
% \begin{thebibliography}{}
% Text
% \end{thebibliography}
%
% or
%
% Compile your BiBTeX database using our plos2015.bst
% style file and paste the contents of your .bbl file
% here. See http://journals.plos.org/plosone/s/latex for 

\end{document}